\documentclass[longauth]{aa}  

\usepackage{graphicx}
\usepackage{enumitem}
\usepackage[varg]{txfonts}
\usepackage{natbib}
\usepackage{xcolor}
\begin{document} 
\renewcommand{\thefootnote}{\alph{footnote}}

\title{Water in star-forming regions: Physics and chemistry from clouds
  to disks as probed by {\it Herschel} spectroscopy}

   \subtitle{}

   \author{E.F. van Dishoeck\inst{\ref{inst1},\ref{inst2}}
     \and
     L.E. Kristensen\inst{\ref{inst3}}
     \and
     J.C. Mottram\inst{\ref{inst4}}
     \and
     A.O. Benz\inst{\ref{inst5}}
\and E.A. Bergin\inst{\ref{inst6}}
\and P. Caselli\inst{\ref{inst2}}
\and F. Herpin\inst{\ref{inst7}}
\and M.R. Hogerheijde\inst{\ref{inst1},\ref{inst21}}
\and D. Johnstone\inst{\ref{inst8},\ref{inst9}}
\and R. Liseau\inst{\ref{inst10}}
\and B. Nisini\inst{\ref{inst11}}
\and M. Tafalla\inst{\ref{inst12}}
\and F.F.S. van der Tak\inst{\ref{inst13},\ref{inst14}}
\and F. Wyrowski\inst{\ref{inst15}}
\and A. Baudry\inst{\ref{inst7}}
\and M. Benedettini\inst{\ref{inst16}}
\and P. Bjerkeli\inst{\ref{inst10}}
\and G.A. Blake\inst{\ref{inst17}}
\and J. Braine\inst{\ref{inst7}}
\and S. Bruderer\inst{\ref{inst2},\ref{inst5}}
\and S. Cabrit\inst{\ref{inst18}}
\and J. Cernicharo\inst{\ref{inst19}}
\and Y. Choi\inst{\ref{inst13},\ref{inst20}}
\and A. Coutens\inst{\ref{inst7}}
\and Th. de Graauw\inst{\ref{inst1},\ref{inst13}}
\and C. Dominik\inst{\ref{inst21}}
\and D. Fedele\inst{\ref{inst22}}
\and M. Fich\inst{\ref{inst23}}
\and A. Fuente\inst{\ref{inst12}}
\and K. Furuya\inst{\ref{inst24}}
\and J.R. Goicoechea\inst{\ref{inst19}}
\and D. Harsono\inst{\ref{inst1}}
\and F.P. Helmich\inst{\ref{inst13},\ref{inst14}}
\and G.J. Herczeg\inst{\ref{inst25}}
\and T. Jacq\inst{\ref{inst7}}
\and A. Karska\inst{\ref{inst26}}
\and M. Kaufman\inst{\ref{inst27}}
\and E. Keto\inst{\ref{inst28}}
\and T. Lamberts\inst{\ref{inst29}}
\and B. Larsson\inst{\ref{inst30}}
\and S. Leurini\inst{\ref{inst15},\ref{inst31}}
\and D.C. Lis\inst{\ref{inst32}}
\and G. Melnick\inst{\ref{inst28}}
\and D. Neufeld\inst{\ref{inst33}}
\and L. Pagani\inst{\ref{inst18}}
\and M. Persson\inst{\ref{inst10}}
\and R. Shipman\inst{\ref{inst13}}
\and V. Taquet\inst{\ref{inst22}}
\and T.A. van Kempen\inst{\ref{inst34}}
\and C. Walsh\inst{\ref{inst35}}
\and S.F. Wampfler\inst{\ref{inst36}}
\and U. Y{\i}ld{\i}z\inst{\ref{inst32}}
\and the WISH team
          }

   \institute{Leiden Observatory, Leiden University, PO Box 9513, 2300 RA Leiden, The Netherlands\label{inst1}
              \email{ewine@strw.leidenuniv.nl}
         \and
         Max-Planck Institut f\"ur Extraterrestrische Physik (MPE), Giessenbachstr. 1, 85748 Garching, Germany\label{inst2}
         \and
         Niels Bohr Institute \& Centre for Star and Planet Formation,
         Copenhagen University, {\O}ster Voldgade 5--7, 1350
         Copenhagen K, Denmark\label{inst3}
         \and
         Max Planck Institute for Astronomy, K{\"o}nigstuhl 17, 69117, Heidelberg, Germany\label{inst4} 
         \and
         Institute for Particle Physics and Astrophysics, ETH Zurich, 8093 Zurich, Switzerland\label{inst5}
         \and
     Department of Astronomy, The University of Michigan, 1085 S. University Ave., Ann Arbor, MI 48109-1107, USA\label{inst6}
\and
Lab.\ d'astrophysique de Bordeaux, Univ. Bordeaux, CNRS, B18N, all{\'e}e
Geoffroy Saint-Hilaire, 33615 Pessac, France
\label{inst7}
\and
National Research Council Canada, Herzberg Astronomy and Astrophysics, 5071 West Saanich Rd, Victoria, BC, V9E 2E7, Canada\label{inst8}
\and
Department of Physics \& Astronomy, University of Victoria, Victoria, BC, V8P 1A1, Canada\label{inst9}
\and
Department of Space, Earth and Environment, Chalmers University of Technology, Onsala Space Observatory, 439 92 Onsala, Sweden\label{inst10}
\and
INAF - Osservatorio Astronomico di Roma, Via di Frascati 33, 00074, Monte Porzio Catone, Italy\label{inst11}
\and
Observatorio Astron\'{o}mico Nacional (OAN), Calle Alfonso XII, 3, 28014, Madrid, Spain\label{inst12}
     \and
     SRON Netherlands Institute for Space Research, PO Box 800, 9700 AV, Groningen, the Netherlands\label{inst13}
     \and
     Kapteyn Astronomical Institute, University of Groningen, PO Box 800, 9700 AV, Groningen, The Netherlands\label{inst14}
\and
Max-Planck-Institut f\"{u}r Radioastronomie, Auf dem H\"{u}gel 69, 53121 Bonn, Germany\label{inst15}
\and
INAF -- Istituto di Astrofisica e Planetologia Spaziali, via Fosso del
Cavaliere 100, 00133 Roma, Italy\label{inst16}
\and
Division of Geological and Planetary Sciences, California Institute of Technology, Pasadena, CA 91125, USA\label{inst17}
\and
LERMA \& UMR8112 du CNRS, Observatoire de Paris, PSL University, Sorbonne Universit\'es, F-75014, Paris, France\label{inst18}
\and
Instituto de Fisica Fundamental (IFF-CSIC), Calle Serrano 123, 28006, Madrid, Spain\label{inst19}
\and
Korean Astronomy and Space Science Institute, Daejeon 34055, Korea\label{inst20}
\and
Anton Pannekoek Institute for Astronomy, University of Amsterdam, Science Park 904, 1098XH, Amsterdam, The Netherlands\label{inst21}
\and
INAF, Osservatorio Astrofisico di Arcetri, Largo Enrico Fermi 5, 50125, Firenze, Italy\label{inst22}
\and
Department of Physics \& Astronomy, University of Waterloo, 200 University Avenue, Waterloo, ON, N2L 3G1, Canada\label{inst23}
\and
National Astronomical Observatory of Japan, 2-21-1 Osawa, Mitaka, Tokyo 181-8588, Japan\label{inst24}
\and
Kavli Institute for Astronomy and Astrophysics, Peking University, Yiheyuan Lu 5, Haidian Qu, 100871 Beijing, People's Republic of China\label{inst25}
\and
Institute of Astronomy, Faculty of Physics, Astronomy and Informatics, Nicolaus Copernicus University, Grudziadzka 5, 87-100 Torun, Poland\label{inst26}
\and
Department of Physics and Astronomy, San Jose State University, One Washington Square, San Jose, CA 95192-0106, USA\label{inst27}
\and
Center for Astrophysics | Harvard \& Smithsonian,
60 Garden Street, Cambridge, MA,
02138, USA\label{inst28}
\and
Leiden Institute of Chemistry, Gorleaus Laboratories, Leiden University, PO Box 9502, NL-2300 RA Leiden, the Netherlands\label{inst29}
\and
Department of Astronomy, Stockholm University, 106 91 Stockholm, Sweden\label{inst30}
\and
INAF - Osservatorio Astronomico di Cagliari, via della Scienza 5, 09047, Selargius, Italy\label{inst31}
\and
Jet Propulsion Laboratory, California Institute of Technology, 4800 Oak Grove Drive, Pasadena, CA 91109, USA\label{inst32}
\and
Department of Physics \& Astronomy, Johns Hopkins University, Baltimore, MD 21218, USA\label{inst33}
\and
SRON Netherlands Institute for Space Research, Sorbonnelaan 2, 3584 CA Utrecht, Netherlands\label{inst34}
\and
School of Physics and Astronomy, University of Leeds, Leeds LS2 9JT, UK\label{inst35}
\and
Center for Space and Habitability (CSH), University of Bern, Gesellschaftsstrasse 6, 3012, Bern, Switzerland\label{inst36}
}

   \date{Received 1 August 2020; Accepted 14 December 2020}

 
  \abstract
 {Water is a key molecule in the physics and chemistry of star and planet formation, but it is difficult to observe from Earth. The {\it Herschel} Space Observatory provided unprecedented sensitivity as well as spatial and spectral resolution to study water. The Water In Star-forming regions with {\it Herschel} (WISH) key program was designed to observe water in a wide range of environments and provide a legacy data set to address its physics and chemistry.} 
{The aim of WISH is to determine which physical components are traced by the gas-phase water lines observed with {\it Herschel} and to quantify the excitation conditions and water abundances in each of these components.
This then provides insight into how and where the bulk of the water is formed in space and how it is transported from clouds to disks, and ultimately comets and planets. }
{Data and results from WISH are summarized together with those from
  related open time programs. WISH targeted $\sim$80 sources along the two
  axes of luminosity and evolutionary stage: from low- to high-mass
  protostars (luminosities from $<$1 to $>10^5$ L$_\odot$) and from
  pre-stellar cores to protoplanetary disks. Lines of H$_2$O and its
  isotopologs, HDO, OH, CO, and [O I], were observed with the HIFI and
  PACS instruments, complemented by other chemically-related molecules
  that are probes of ultraviolet, X-ray, or grain chemistry. The
  analysis consists of coupling the physical structure of the sources
  with simple chemical networks and using non-LTE radiative transfer
  calculations to directly compare models and observations.}
{Most of the far-infrared water emission observed with {\it Herschel}
  in star-forming regions originates from warm outflowing and shocked
  gas at a high density and temperature ($>10^5$ cm$^{-3}$, 300--1000
  K, $v\sim 25$ km s$^{-1}$), heated by kinetic energy
  dissipation. This gas is not probed by single-dish low-$J$ CO lines,
  but only by CO lines with $J_{\rm up}>14$. The emission is compact,
  with at least two different types of velocity components seen. Water
  is a significant, but not dominant, coolant of warm gas in the
  earliest protostellar stages. The warm gas water abundance is
  universally low: orders of magnitude below the H$_2$O/H$_2$
  abundance of $4 \times 10^{-4}$ expected if all volatile oxygen is
  locked in water. In cold pre-stellar cores and outer protostellar
  envelopes, the water abundance structure is uniquely probed on
  scales much smaller than the beam through velocity-resolved line
  profiles. The inferred gaseous water abundance decreases with depth
  into the cloud with an enhanced layer at the edge due to
  photodesorption of water ice. All of these conclusions hold
  irrespective of protostellar luminosity.  For low-mass protostars, a
  constant gaseous HDO/H$_2$O ratio of $\sim$0.025 with position into
  the cold envelope is found. This value is representative of the
  outermost photodesorbed ice layers and cold gas-phase chemistry, and
  much higher than that of bulk ice. In contrast, the gas-phase NH$_3$
  abundance stays constant as a function of position in low-mass pre-
  and protostellar cores. Water abundances in the inner hot cores are
  high, but with variations from $5\times 10^{-6}$ to a few
  $\times 10^{-4}$ for low- and high-mass sources.  Water vapor
  emission from both young and mature disks is weak.  }
{The main chemical pathways of water at each of the star-formation stages have been identified and quantified. Low warm water abundances can be explained with shock models that include UV radiation to dissociate water and modify the shock structure. UV fields up to $10^2-10^3$ times the general interstellar radiation field are inferred in the outflow cavity walls on scales of the {\it Herschel} beam from various hydrides. Both high temperature chemistry and ice sputtering contribute to the gaseous water abundance at low velocities, with only gas-phase (re-)formation producing water at high velocities.  Combined analyses of water gas and ice show that up to 50\% of the oxygen budget may be missing. In cold clouds, an elegant solution is that this apparently missing oxygen is locked up in larger $\mu$m-sized grains that do not contribute to infrared ice absorption. The fact that even warm outflows and hot cores do not show H$_2$O at full oxygen abundance points to an unidentified refractory component, which is also found in diffuse clouds. The weak water vapor emission from disks indicates that water ice is locked up in larger pebbles early on in the embedded Class I stage and that these pebbles have settled and drifted inward by the Class II stage. Water is transported from clouds to disks mostly as ice, with no evidence for strong accretion shocks. Even at abundances that are somewhat lower than expected, many oceans of water are likely present in planet-forming regions. Based on the lessons for galactic protostars, the low-$J$ H$_2$O line emission ($E_{\rm up}<300$~K) observed in extragalactic sources is inferred to be predominantly collisionally excited and to originate mostly from compact regions of current star formation activity. Recommendations for future mid- to far-infrared missions are made.}


   \titlerunning{WISH: physics and chemistry from cloud
     to disks probed by {\it Herschel} spectroscopy}
   \authorrunning{E.F. van Dishoeck et al.}
   \maketitle
%

\newpage
   
\section{Introduction}
\label{sec:intro}

Of the more than 200 detected interstellar molecules
\citep{McGuire18}, water is special because it combines two of the
most abundant elements in the Universe and plays a key role in the
physics and chemistry of star- and planet-forming regions.  On
planets, water is widely acknowledged as essential for potential
habitability and the emergence of life \citep{Chyba05,Kaltenegger17}.
This makes the question of how much water is present in forming
planetary systems, and whether this amount depends on the location and
birth environment, highly relevant.
Water ice also plays a role in promoting the coagulation of small dust
grains to pebbles, rocks and ultimately planetesimals, the building
blocks of planets, by enhancing the mass of solids due to freeze
out. Such icy planetesimals (asteroids, comets), in turn, may have
delivered much of the water and organic molecules trapped in ices to
oceans on planets such as Earth that have formed inside the water iceline
\citep[see][for
reviews]{Morbidelli12,Morbidelli18,vanDishoeck14PPVI,Hartmann17,Altwegg19}.

In the early phases of star formation, water vapor is an exceptional
tool for studying warm interstellar gas and the physical processes
taking place during star formation. This diagnostic capability stems
from water's large abundance variations between warm gas, where it is
copiously produced \citep[e.g.,][]{Draine83,Kaufman96}, and cold gas,
where it is mostly frozen out \citep[e.g.,][]{Bergin02}.  Thus, water
vapor emission can be used as a ``beacon'' that signals where energy
is deposited into molecular clouds. This happens especially in the
deeply embedded stages when jets and winds from the protostars
interact with the surroundings, and when the (proto)stellar luminosity
heats envelopes and disks.  Water vapor lines are also particularly
sensitive to small motions inside clouds, such as those that are due
to gravitational collapse or expansion.  Water\footnote{Unless
  specified, the term ``water'' in this article can imply both water
  vapor and water ice. The term ``ice'' is used to indicate all
  volatiles in solid form, which includes -- but is not limited to --
  water ice.} is therefore highly complementary to other molecules
such as CO in probing the protostellar environment. Finally, water
actively contributes to the energy balance of warm gas as a gas
coolant, with its importance likely varying with protostellar
evolution \citep{Nisini02}.

The quest for understanding the water ``trail'' from clouds to
planet-forming disks is complicated by the fact that water in the
Earth's atmosphere prevents the direct observation of rotational lines of
water gas with ground-based telescopes, except for some high
excitation (masing) transitions and a few isotopolog lines. Indeed,
the first detection of interstellar water was made from the ground
through the 22 GHz water maser line in Orion
\citep{Cheung69}. Space-based observatories are needed to probe the
full spectrum of water vapor. As a light hydride, water vapor has its
principal rotational transitions at far-infrared rather than
millimeter wavelengths, for which development of appropriate detector
technology was required.

A related complication is the fact that star formation takes place
over many size scales \citep{Shu93,Beuther07}: pc-sized clouds can
fragment and contract to $\sim$0.1 pc pre-stellar cores, which can
subsequently collapse to form a protostar. Its envelope has a typical
size of a few thousand au and feeds a growing disk of a few hundred au
radius from which material can accrete onto the young star. Early in
the evolution, the protostar will develop a bipolar jet and wind,
which interact with their surroundings to create outflows on scales
ranging from $<0.01$ up to 0.5 pc. Even in the nearest star-forming
clouds at $\sim$150 pc distance, these sizes range from many arcmin to
$<$1 arcsec on the sky. A single-dish telescope in space with a
spatial resolution of typically 0.5$'$ cannot resolve protoplanetary
disks at far-infrared wavelengths, in contrast with what the Atacama
Large Millimeter/submillimeter Array (ALMA) now does routinely at
longer wavelengths. However, a single-dish telescope can image
protostellar envelopes and their outflows. Motions range from $<0.1$
km s$^{-1}$ to $>100$ km s$^{-1}$ so that very high spectral
resolution is needed to resolve line profiles which can typically only
be provided by heterodyne instruments. This high spectral resolution
allows to infer some spatial information for unresolved sources with
systematic motions, such as for gas in rotating disks.

Pioneering infrared and submillimeter space missions have provided
considerable insight into the water cycle in space. The {\it Infrared
  Space Observatory} (ISO) \citep{Kessler96} covered the full mid- and
far-infrared wavelength range observing both water gas and ice, but at
modest spectral and spatial resolution and sensitivity. The {\it
  Submillimeter Wave Astronomy Satellite} (SWAS) \citep{Melnick00}
observed water with heterodyne spectral resolution in a
$3.3'\times 4.5'$ beam, but only through the ground-state line of
ortho-H$_2$O and H$_2^{18}$O near 557 and 547 GHz, as did the
Swedish-led $Odin$ satellite \citep{Nordh03}. $Odin$ had a factor of
two higher angular (2$'$ circular beam) and spectral (0.5 MHz)
resolution than SWAS, with spectra from these two missions agreeing
very well when degraded to the same resolution
\citep{Larsson17}. Finally, the {\it Spitzer Space Telescope}
\citep{Werner04} covered the mid-infrared wavelength range at low
spatial and spectral resolution but with much higher sensitivity than
ISO, thus probing both water ice and highly excited water gas. These
missions, combined with the {\it Kuiper Airborne Observatory} and
ground-based observations of water ice, have demonstrated that water
is mostly frozen out as ice in cold clouds and that water vapor
becomes abundant at high temperatures such as associated with
outflows. Detailed results from these missions are summarized
elsewhere and will not be repeated here \citep[see
e.g.,][]{Melnick09,Hjalmarson03,vanDishoeck04,vanDishoeck13,Pontoppidan10}.

The {\it Herschel Space Observatory} \citep{Pilbratt10} covered the
55--672 $\mu$m range and improved on previous space missions in all
relevant observational parameters. With a 3.5m telescope, its
diffraction-limited beam ranged from $44''$ at the longest to $9''$ at
the shortest wavelengths, increasing the spatial resolution with respect
to $SWAS$ at the longest wavelengths by a factor of 5 and making it a
much better match to the sizes of protostellar sources. Its Heterodyne
Instrument for the Far-Infrared (HIFI) \citep{deGraauw10} provided
spectra at very high spectral resolution ($R=\lambda/\Delta \lambda$ up to
$10^7$) at a single position over the 480--1250 GHz (600--240
$\mu$m) and 1410--1910 GHz (210--157 $\mu$m) ranges, thus covering
many water lines and fully resolving their profiles. The Photodetector
Array Camera and Spectrometer \citep[PACS,][]{Poglitsch10} obtained
spectra with moderate resolving power ($R=(1-5)\times 10^3$) in the 55-190
$\mu$m range at each pixel of a $5\times 5$ array allowing efficient
instantaneous imaging of water lines and full far-infrared spectral
scans. The sensitivity of both HIFI and PACS was a factor of
$>$10--100 better than previous missions due to improved detector
technology and increase in collecting area. The combination of these
two instruments made {\it Herschel} eminently suited to study both hot
and cold water in space. The Spectral and Photometric Imaging Receiver
(SPIRE) \citep{Griffin10} also covered some of the lower-frequency
water and CO lines at 194--318 and 294-672 $\mu$m but at much
lower spectral resolution than HIFI and was generally not used as part
of WISH.

\begin{figure*}[tb]
  \centering
    \includegraphics[width=14cm]{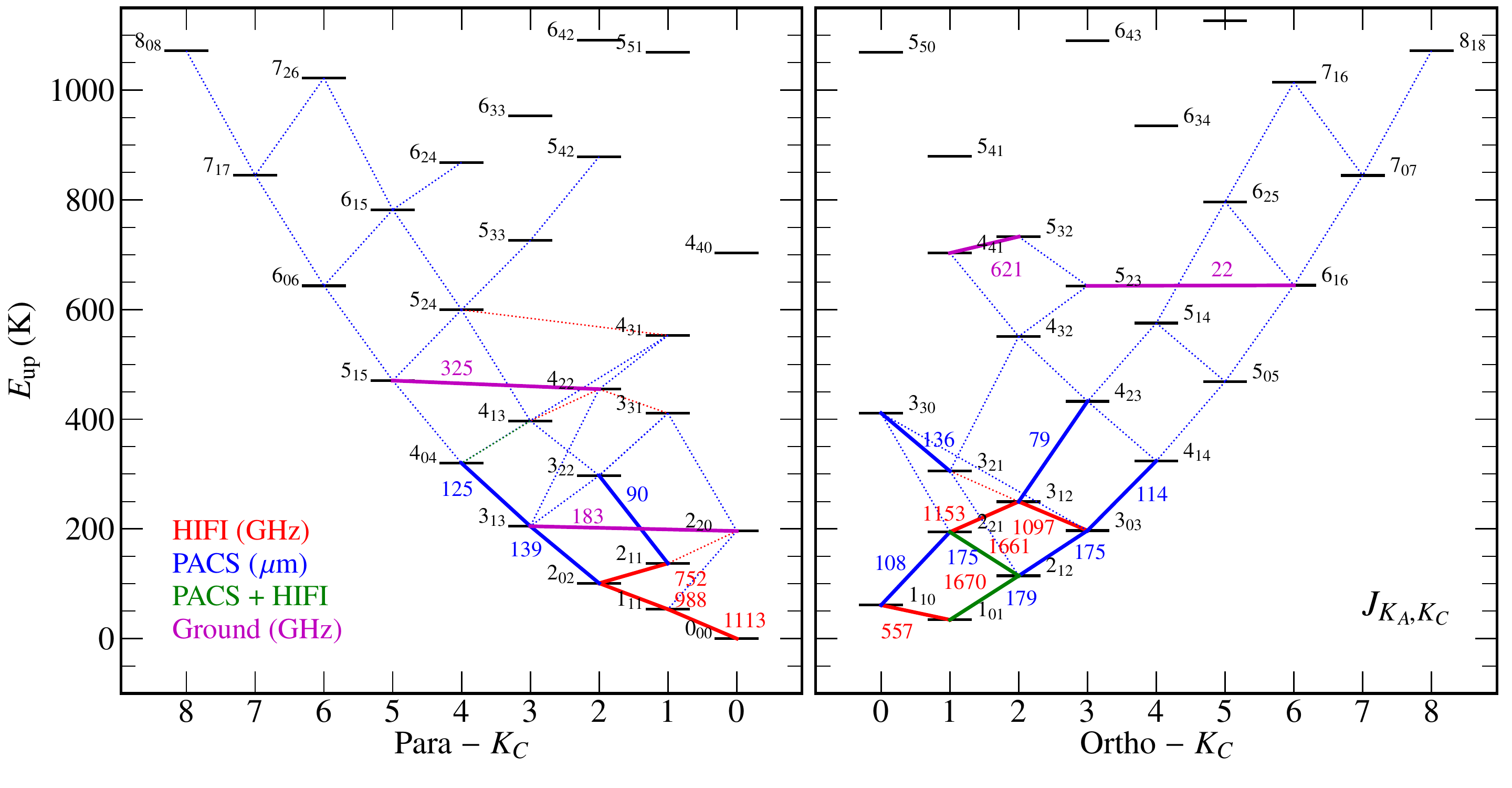}
    \caption{Water rotational energy levels within the ground
      vibrational level. Transitions targeted with HIFI are indicated
      in red with frequencies in GHz, those with PACS in blue with
      wavelengths in $\mu$m. Dashed lines indicate water transitions
      detected in PACS and HIFI spectral scans. The 179 $\mu$m line
      observed with both PACS and HIFI is indicated in green. Purple
      lines indicate water (isotopolog) lines that can be observed
      from the ground with frequencies in GHz; some of them (e.g., 621
      GHz) have also been observed with HIFI. The 22 GHz
      $6_{16}-5_{23}$ maser line observed at cm wavelengths is
      indicated.}
         \label{fig:waterlevels}
\end{figure*}

The goal of the Water In Star-forming regions with {\it Herschel}
(WISH) guaranteed time Key Program (KP) was to use water vapor as a
physical and chemical diagnostic and follow the water abundance
throughout the different phases of star- and planet formation.  A
sample of about 80 young stellar objects (YSOs) was targeted covering
two axes: mass and time (or equivalently, evolutionary stage). The
WISH sources have luminosities from $<1$ to $>10^5$ L$_\odot$, thus
including low (LM), intermediate (IM) and high mass (HM)
protostars. They also cover a large range of evolutionary stages, from
pre-stellar cores prior to collapse, through the embedded protostellar
stages, to the revealed phase when the envelope has dissipated and
-- for the case of low-mass sources -- the pre-main sequence stars become
optically visible but are still surrounded by gas-rich disks. The full
list of sources is included as Table~1 in \citet{vanDishoeck11}, which
also describes the WISH program and highlights initial results. The
gas-poor debris disk stage was not covered in WISH but was targeted
spectroscopically by other programs such as GASPS \citep{Dent13}.

The WISH program was organized along the mass and time axes (see
Table~\ref{tab:programs} in the Appendix) with slightly different
observing strategies for each of the subprograms.  For all embedded
protostars, a comprehensive set of water observations has been carried
out with HIFI and PACS (Fig.~\ref{fig:waterlevels}) with a mix of low-
and high-excitation lines of H$_2$O, H$_2^{18}$O, H$_2^{17}$O, and the
chemically related species O and OH, targeted. In addition, a number
of high-$J$ CO and its isotopologs lines are included in the spectral
settings, as are some hydrides and C$^+$ that are diagnostic of the
presence of X-rays or UV that are emitted either by the source itself
or at the star-disk interface \citep[see Table 2
in][]{vanDishoeck11}. Thus, the WISH data also cover the major
coolants of the gas. For the cold pre-stellar cores and protoplanetary
disks, at most a few water lines were observed but with very long
integrations. Later in the {\it Herschel} mission, when additional
guaranteed time became available, some HDO lines were added, since the
HDO/H$_2$O ratio is a particularly powerful tracer of the water
chemistry and its history. The $1_{11}-0_{00}$ line of HDO at 893 GHz
is a particularly powerful probe of cold HDO gas.

Since the WISH sample was limited in the number of sources in each
luminosity or evolutionary bin, several follow-up open time programs
have been pursued to validate the conclusions. For example, the {\it
  William Herschel Line Legacy} survey (WILL) doubled the number of
low-mass protostars in a much more unbiased way \citep{Mottram17}. A
program targeting all protostars in the Cygnus star-forming cloud
added greatly to the intermediate- and high-mass samples
\citep{SanJose15}. Very deep integrations on a number of
protoplanetary disks allowed more general conclusions for those
sources to be drawn \citep{Du17}. Similar extensions hold for many of
the WISH subprograms, the most important of which are summarized in
Table~\ref{tab:programs}. The collection of WISH and these programs is
indicated here as WISH+.

Complementary guaranteed and open time {\it Herschel} programs have
provided information on water in star-forming regions as well, most
notably the HIFI spectral surveys as part of the CHEmical Survey of
Star-forming regions (CHESS; \citealt{Ceccarelli10}) and Herschel
observations of EXtraOrdinary Sources (HEXOS;
\citealt{Bergin10hexos}) KPs. The Dust, Ice and Gas in Time (DIGIT;
\citealt{Green13}) and Herschel Orion Protostar Survey (HOPS;
\citealt{Manoj13}) KPs complemented WISH by carrying out full PACS
spectral scans for a larger sample of low-mass embedded YSOs. The
PRobing InterStellar Molecules with Absorption line Studies
(PRISMAS) targeted the chemistry of water and other hydrides in the
diffuse interstellar gas \citep{Gerin16}. At the other end of the
water trail, the Water and related chemistry in the Solar System
(HSSO; \citealt{Hartogh09}) probed water and its deuteration in a
variety of solar system objects, including comets.  Some of their
results will be put in the context of the WISH+ program.

The main aims of the WISH program are to determine:

\begin{enumerate}[label=(\roman*)]
\item
  How and where water is formed in space.

\item Which physical components of a star-forming cloud water traces,
  and what the water abundances are in each of these components. Also,
  what fraction of the total oxygen reservoir is accounted for by
  water, and what the importance is of water vapor as a coolant of
  warm gas.

\item
What the water trail is
from clouds to disks, and ultimately to comets and planets.

\end{enumerate}

Overall, WISH+ has published nearly 90 papers to date which are
summarized on the WISH website {\tt www.strw.leidenuniv.nl/WISH}. This
paper synthesizes these WISH+ results, together with related results
from the above mentioned {\it Herschel} programs. It focuses on
questions (i) and (iii) and the latter part of question (ii),
including previously unpublished data, new models and new analyses.
The physical components traced by water vapor have been addressed in a
number of synthesis papers, most notably \citet{SanJose16},
\citet{Herpin16}, \citet{vanderTak19}, \citet{Kristensen17b} and
\citet{Mottram17} for WISH+, \citet{Karska18} for WISH+WILL+DIGIT, and
\citet{Manoj16} for HOPS+DIGIT, with the latter paper limited to CO.

In terms of question (iii), this paper describes the evolution of
water from cores to disks.  The bigger picture of the delivery of water to
planets and Earth itself is summarized in other reviews
\citep[e.g.,][]{vanDishoeck14PPVI,Morbidelli18}.  Also, this paper is
centered around {\it Herschel}'s contributions to these
questions. Other facilities provide important complementary
information, most notably ground-based infrared telescopes on ices
\citep[e.g.,][]{Boogert15}, and millimeter interferometers such as ALMA
and the Northern Extended Millimeter Array (NOEMA) through high
resolution images of quiescent warm water in protostellar sources
\citep[e.g.,][]{vanderTak06,Persson14,Harsono20}, but they will be
mentioned only sparsely.

This paper is organized as follows. Section 2 summarizes the main
observing strategy, sample selection and data reduction. Section 3 and
Appendix~\ref{app:waterroutes} provide an overview of
the main water formation and destruction processes, especially those
that are relevant for star-forming regions.
Also, the WISH modeling approach is outlined. Section 4 summarizes the
main results of {\it Herschel} WISH+ and related programs on water,
especially its distribution, line profiles and excitation, and links
them to observations of the CO rotational ladder. The inferred
characteristics hold universally from low- to high-mass
protostars. Section 5 describes the surprisingly low water abundance
inferred for the different types of warm outflowing gas and shocks
associated with star-forming regions and the implications for shock
models. Comparison is made with chemically related molecules such as
OH observed with {\it Herschel} and shock tracers such as SiO and
grain surface products such as CH$_3$OH.

Sections 6--9 follow water from pre-stellar cores through protostars
to disks for low- and high-mass sources. Section 6 focuses on the cold
pre-stellar cores, whereas Section 7 describes similar results for the
outer cold protostellar envelopes using the high-resolution HIFI line
profiles to infer the water abundance profile as a function of position
in the envelope. New analyses of HDO and NH$_3$ (another grain surface
chemistry product) are presented. Section 8 highlights the puzzling
low water abundances in the inner hot core of several protostellar
envelopes where all ices should have sublimated. Section 9 focuses on
recent new results on water in young and mature disks around pre-main
sequence stars, setting the stage for planet formation.  Section 10
and Appendix~\ref{app:obudget} bring all the
information on oxygen reservoirs in star-forming regions together and
compare them with diffuse clouds and comets to address the puzzling
question of the missing oxygen. Section 11 concludes with a discussion
and overview of what we have learned from {\it Herschel} on the above
water questions, from low to high-mass protostars, and from clouds to
disks. Section 12 summarizes the main points in bullet form and looks
forward to future observations of warm water using ALMA, the {\it
  James Webb Space Telescope} (JWST) and Extremely Large
  Telescopes (ELTs) on the ground and in space, as well as surveys of water
ice using the SPHEREx mission. It also makes recommendations for
future far-infrared space mission concepts such as the {\it Space IR
  telescope for Cosmology and Astrophysics} (SPICA)
\citep{Roelfsema18} and the {\it Origins Space Telescope}
\citep{Origins19}. For deep observations of cold water vapor, the WISH+ {\it
  Herschel} data will remain unique for a long time.

\section{Observations}
\label{sec:obs}

\subsection{Observed lines}
\label{sec:obslines}

Water is an asymmetric rotor with energy levels characterized by the
quantum numbers $J_{K_A,K_C}$, grouped into ortho ($K_A+K_C$=odd) and
para ($K_A+K_C$=even) ladders because of the nuclear spin statistics
of the two hydrogen atoms. In contrast with CO, its energy level
structure is highly irregular, resulting in transitions scattered
across the far-infrared and submillimeter wavelength range.
Figure~\ref{fig:waterlevels} summarizes the H$_2^{16}$O lines in the
vibrational ground state that were targeted by WISH using HIFI and
PACS. Also, water lines serendipitously detected in full spectral
scans are indicated.  The full list of lines, including isotopologs
and related species, is summarized in Table 2 of
\citet{vanDishoeck11}. To this list, the HDO $1_{11}-0_{00}$ 893.639
GHz line has been added for a number of embedded protostars and the
H$_2$O $5_{24}-4_{31}$ line at 970.3150 GHz
($E_{\rm up}=599$~K)\footnote{Energies of levels are indicated in
  short as $E_{\rm up}$ in K units rather than $E_{\rm up}/k_B$ with
  $k_B$ the Boltzmann constant.} for high-mass sources.

\begin{figure}
  \centering
 \includegraphics[width=7.5cm]{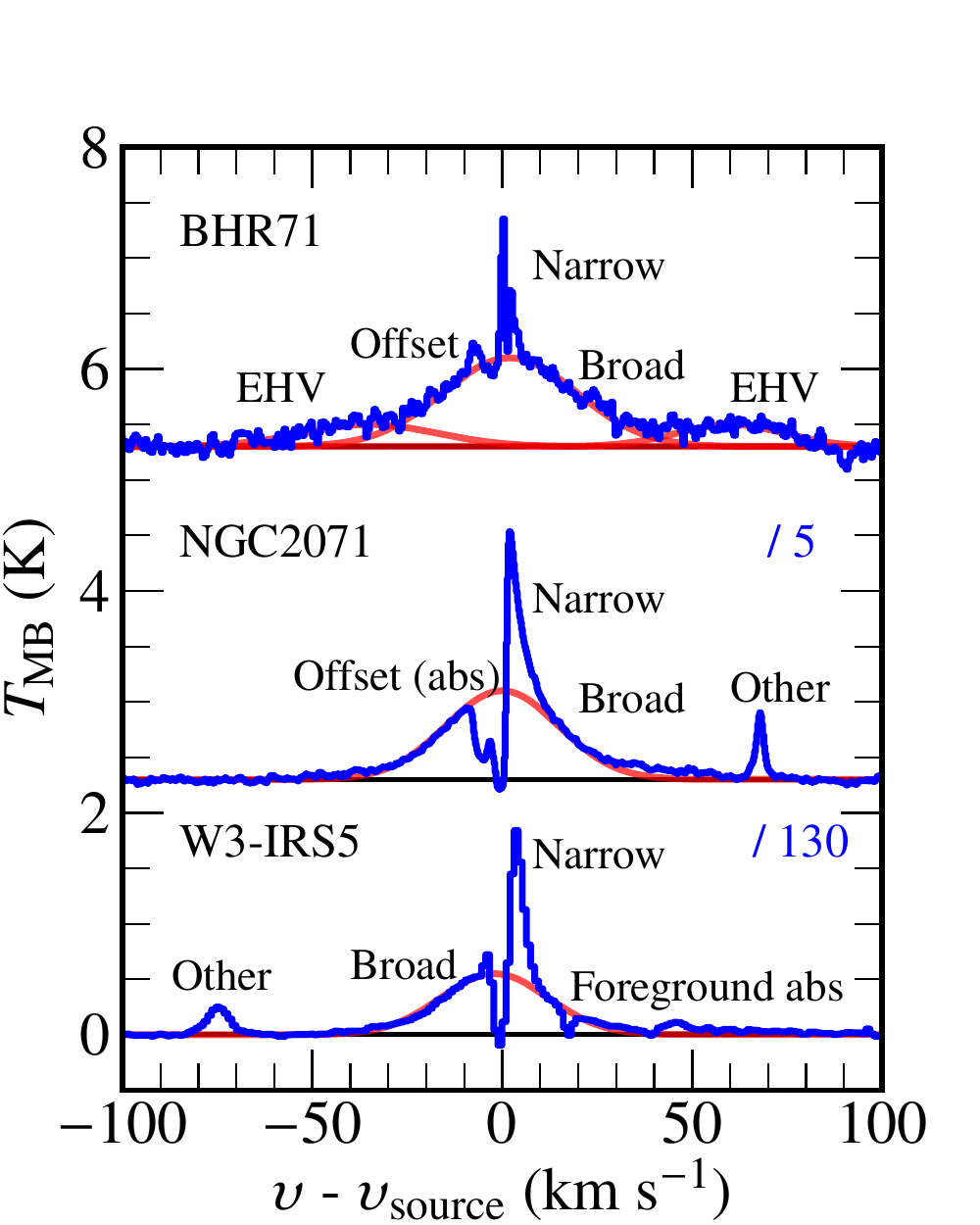}
 \caption{Water spectra of the $1_{10}-1_{01}$ 557 GHz line taken with
   {\it Herschel}-HIFI (38$''$ beam) toward a low-, intermediate- and
   high-mass YSOs (top to bottom). The spectra highlight the
   complexity of water profiles with multiple kinematic absorption and
   emission components (see text and Table~\ref{tab:profiles} for
   the interpretation of these components). Other refers to lines from
   molecules other than H$_2$O.}
         \label{fig:water_mass}
\end{figure}

The targeted water vapor lines were chosen to cover both low- and
high-energy levels as well as a range of opacities in three groups.
First, lines connected with the ground state levels of o- and p-H$_2$O
at 557, 1113, and 1670 GHz\footnote{Frequencies are abbreviated to
  values commonly used to designate these lines in the WISH
  papers. They may differ by 1 GHz from the properly rounded values.}
are prime diagnostics of the cold gas. They usually show strong
self-absorption or can even be purely in absorption against the strong
continuum provided by the source itself (Fig.~\ref{fig:water_mass}).

The second group, the `medium-$J$' lines originating from levels with
$E_{\rm up}$ around 100--250 K, probe the warm gas and are not (or
less) affected by absorption. The $2_{02}-1_{11}$ line at 988 GHz,
$3_{12}-3_{03}$ at 1097 GHz, and $3_{12}-2_{21}$ at 1153 GHz (for LM
and IM sources), are particularly useful; the latter two lines
originate from the same upper level but have Einstein $A$ coefficients
that differ by a factor of 6.3. Because of high optical depths of
H$_2^{16}$O, optically thin(ner) isotopic lines are crucial for the
interpretation. Thus, deep integrations on a number of ground-state
and medium-$J$ H$_2^{18}$O and H$_2^{17}$O lines are included in WISH
for embedded sources. They proved to be particularly useful for the
high-mass protostars.  For the low-mass subsample, H$_2^{18}$O
specific settings were only observed toward Class 0 sources which
have brighter lines than the Class I sources.
No water lines had been seen prior to {\it Herschel} for Class I
sources, so longer integrations were put on the $1_{10}-1_{01}$ 557
GHz transition to ensure detections.

\begin{figure}
  \centering
 \includegraphics[width=7.5cm]{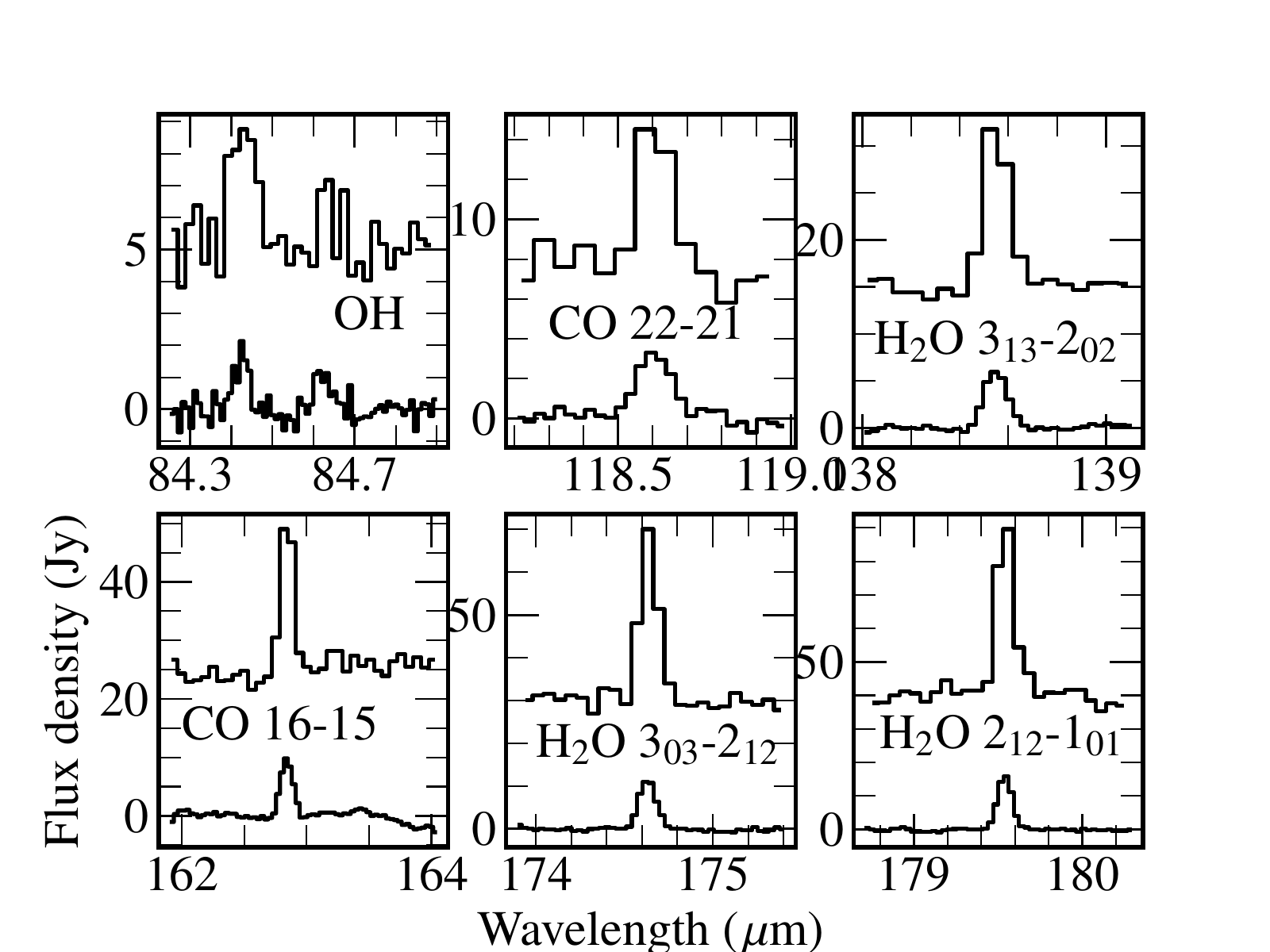}
 \caption{Examples of PACS spectra of NGC 1333 IRAS4B with 
   the same spectrum shown twice in each panel: (top) taken in the full
   spectral scan mode at Nyquist sampling; (bottom) taken in
   individual line scan mode, which has a factor 3 higher spectral
   sampling. }
         \label{fig:pacscomp}
\end{figure}

The third group are highly excited lines originating from levels
$>$300 K, which are only populated in high temperature gas and strong
shocks (e.g., higher-lying backbone lines). Most of these lines are
covered by PACS in its individual line scan mode and are thus not
velocity resolved. They have $E_{\rm up}$ up to 1324 K at the shortest
PACS wavelengths \citep{Herczeg12}, which is high in energy, but not
yet high enough to probe the hottest shocked gas seen at mid-infrared
wavelengths \citep[e.g.,][]{Melnick08}. The PACS spectra also cover
many OH lines and the two [O I] fine-structure lines. Full PACS
spectral scans were obtained for four low-mass WISH sources to provide
an unbiased view of all the far-infrared lines. Both modes have the
same spectral resolving power, but the spectral sampling for full
scans is much lower than that for individual line scans. Combined with
the shorter integration time per line, this affects the quality of
individual line spectra in full spectral scans as illustrated in
Figure~\ref{fig:pacscomp}.

CO lines are included in the HIFI settings up to the $J$=10--9
transition ($E_{\rm up}=304$~K), both of the main and minor CO isotopologs
\citep{Yildiz13,SanJose13}. As described in \S~4, these lines are not
yet high enough in energy to probe the same physical components as the
mid-$J$ water lines. The important CO $J$=16--15 ($E_{\rm up}=752$~K)
does so and was targeted in a separate program, COPS-HIFI
\citep{Kristensen17b}. PACS covers CO lines starting at $J$=14--13 up
to $J$=49-48 with $E_{\rm up}$=6725 K \citep{Herczeg12,Goicoechea12},
so there is a small overlap to connect PACS and HIFI for CO ladder
observations.

PACS provides small $5\times$5 maps over a $\sim 1'\times 1'$ region
for all sources. Additional, sometimes larger, maps were made with HIFI
and PACS of selected water and CO lines together with the continuum
for a few selected low-mass \citep[e.g.,][]{Nisini10,Bjerkeli12} and
high-mass protostars \citep[e.g.,][Kwon et al., unpublished]{Jacq16,Leurini17}.

Simple hydrides such as CH$^+$, OH$^+$, and H$_2$O$^+$, also have their
lowest transitions at far-infrared wavelengths that can be observed
with {\it Herschel}. Since these molecules are either part of the
water network, have formation pathways that involve energy barriers or
are dissociation products, they provide complementary information on
the chemistry, gas temperatures and the irradiation by far-UV or
X-rays and were therefore included in WISH. Moreover, HCO$^+$, an ion
whose main destroyer is water, is serendipitously covered through its
$J$=6--5 line in the WISH H$_2^{18}$O $1_{10}-1_{01}$ setting.

Even though WISH had a generous guaranteed time allocation of 425 hr,
a trade-off between number of sources, lines and integration times per
line had to be made. Integration times range from as little as 10--15
minutes per line for embedded protostars with bright water emission,
to 10--20 hr per setting to detect the very weak water lines from cold
pre-stellar cores and protoplanetary disks.

\subsection{Sample}
\label{sec:sample}

The WISH KP source sample of $\sim$80 sources is summarized in Table~1
of \citet{vanDishoeck11}, with updates on source distances and
luminosities listed in Table~1 of \citet{Kristensen12} for low-mass
and Table~1 of \citet{vanderTak19} for high-mass protostars.  Spectral
energy distributions (SEDs) of most sources, including updated
far-infrared fluxes from {\it Herschel}, are presented in
\citet{Kristensen12,Karska13,vanderTak13,Green13,Mottram17}.

WISH selected its source samples among well known archetypal sources,
based on a wide array of complementary data.  The program
distinguishes low (LM, $<10^2$ L$_\odot$), intermediate (IM,
$10^2-10^4$ L$_\odot$) and high-mass (HM, $>10^4$ L$_\odot$)
protostars. Also, for a subset of low-mass protostars, positions in
the outflow, well separated from the source, were targeted in a
separate subprogram. In terms of evolutionary stages, pre-stellar
cores, embedded protostars (both Class 0 and Class I, separated by
$T_{\rm bol} <$ and $>$ 70 K, \citealt{Evans09}), and pre-main
sequence stars with disks (Class II), are included. For high mass
protostars, the evolutionary stages range from Infrared Dark Clouds
(IRDC) to High-Mass Protostellar Objects (HMPO, both IR ``quiet'' and IR
bright) to UltraCompact H II regions (UC HII), although the latter two
stages have some overlap \citep{Beuther07,Motte18}.

The WISH sample favors luminous sources with particularly prominent
and extended outflows in the case of low-mass sources. The latter was
a requirement to be able to target separate on-source and off-source
positions in the outflow subprogram. As demonstrated by other {\it
  Herschel} programs such as WILL and DIGIT, such large extended
outflows are not representative of the general population of low-mass
protostars \citep{Mottram17,Karska18}. All WISH conclusions have,
however, been verified by these additional samples.

The number of sources per WISH (sub)category ranges from a few for the
cold line-poor sources (pre-stellar cores, disks) to 10--20 for warm
line-rich objects. The latter samples are large enough to allow
individual source peculiarities to be distinguished from general
trends. The radiation diagnostics subprogram observed hydrides other
than H$_2$O in a subsample of low- and high-mass protostars. As noted
above, the number of sources per subcategory has in many cases been
more than doubled thanks to the additional programs listed in
Table~\ref{tab:programs}.  Details of the observations and data
reduction are summarized in Appendix~\ref{app:reduction}.

\section{Water chemistry}
\label{sec:chemistry}

\subsection{Overall oxygen budget}
\label{sec:oxygenbudget}

For the interpretation of the {\it Herschel} data, it is important to
compare observed water abundances with the maximum water abundance
that can be expected based on the available elements in interstellar
gas.  The average abundance of elemental oxygen with respect to total
hydrogen nuclei in the interstellar medium is measured to be
[O]=$5.8 \times 10^{-4}$ \citep{Przybilla08}, with an uncertainty of
about 20\%. This value is close to the current best estimate of the
solar oxygen abundance of $4.9\times10^{-4}$
\citep{Asplund09,Grevesse10} and nearly identical to the solar system
value \citep{Lodders10}. The notation [X] indicates the abundance of
element X in all its forms. Some fraction of this oxygen,
$(0.9-1.4)\times 10^{-4}$ (16--24\%), is locked up in refractory
silicate material in the diffuse interstellar medium
\citep{Whittet10}. The abundance of volatile oxygen (that is, the
oxygen not tied up in some refractory form) is measured to be
$3.2\times 10^{-4}$ in diffuse clouds \citep{Meyer98}, so this is the
maximum amount of oxygen that can cycle between water vapor and ice in
dense clouds.

Counting up all the forms of detected oxygen in diffuse clouds, the
sum is less than the overall elemental oxygen abundance. This missing
fraction increases with density \citep{Jenkins09}.  Thus, a fraction
of oxygen is postulated to be in some yet unknown refractory form,
called Unidentified Depleted Oxygen (UDO), which may increase from
$\sim$20\% in diffuse clouds up to 50\% in dense star-forming regions
\citep{Whittet10,Draine20}. For comparison, the abundance of elemental
carbon is $3 \times 10^{-4}$, with about 50--70\% of the carbon
thought to be locked up in solid carbonaceous material
\citep{Henning98}. If CO (gas + ice) contains the bulk of the volatile
carbon, its fractional abundance should thus be about
$1\times 10^{-4}$ with respect to total hydrogen, thus accounting for
up to 30\% of the volatile oxygen. Indeed, direct observations of warm
CO and H$_2$ gas provide a maximum value of CO/[H]=$1.4\times 10^{-4}$
\citep{Lacy94}, that is, a little less than half the volatile
oxygen. Subtracting the amount of oxygen in CO from
$3.2\times 10^{-4}$ leaves $\sim 2\times 10^{-4}$ for other forms of
volatile oxygen in dense clouds.

Often water abundances are cited with respect to H$_2$ rather than
total hydrogen. Thus, the maximum expected abundance if most volatile
oxygen is driven into water is H$_2$O/H$_2$=$4 \times 10^{-4}$, with
H$_2$O/CO$\approx (1.4-2)$. This maximum H$_2$O abundance takes into
account the fraction of oxygen locked up in CO and silicates (at
24\%), as well as a minimal 20\% fraction of UDO such as found in
diffuse clouds. A more detailed discussion of the oxygen budget can be
found in \S~\ref{sec:obudget} and Appendix~\ref{app:obudget}.

\begin{figure}
  \centering
\includegraphics[width=9cm]{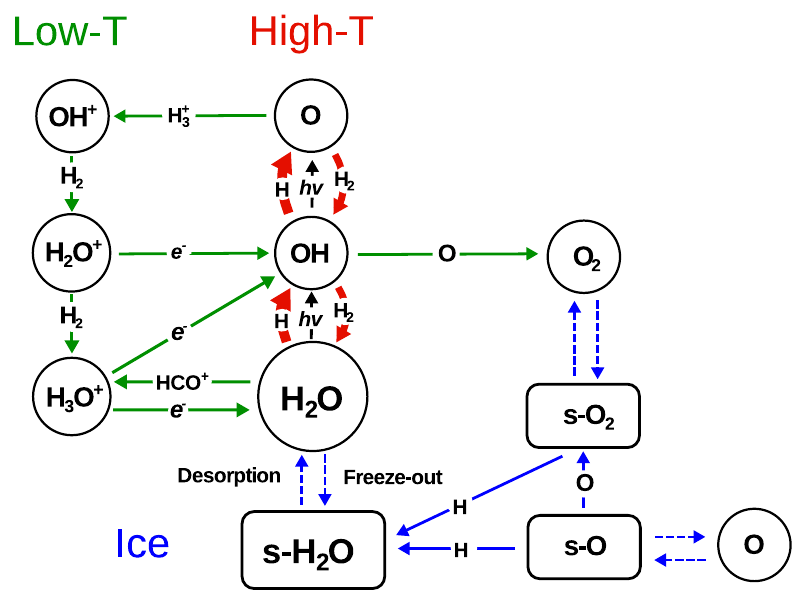}
\caption{Simplified gas-phase and solid-state reaction network leading
  to the formation and destruction of H$_2$O. s-X denotes species X on
  the ice surface. Three routes to water can be distinguished: (i)
  ion-molecule gas-phase chemistry which dominates H$_2$O formation at
  low temperatures and low densities (green); (ii) high-temperature
  neutral-neutral chemistry which is effective above $\sim$250 K when
  energy barriers can be overcome (red); and (iii) solid-state
  chemistry (blue). The full solid-state network is presented in
  \citet{Lamberts13}. Chemically related molecules discussed in this
  paper such as OH, O$_2$ and HCO$^+$ are shown as well.}
\label{fig:waternetwork}
\end{figure}

\subsection{Simplified water chemistry networks}
\label{sec:simplechemistry}

Detailed networks of the water chemistry in the gas and on the grains
are presented in \citet{vanDishoeck13} and \citet{Lamberts13}, and the
three main routes to water are described in
Appendix~\ref{app:waterroutes}: (i) cold ion-molecule chemistry; (ii)
high-temperature gas-phase chemistry; and (iii) ice chemistry.  In
addition, the link between the water network and that forming CO$_2$,
another potentially significant oxygen carrier, is described
there. The main reactions are illustrated in
Fig.~\ref{fig:waternetwork}

For coupling chemistry with detailed physical or hydrodynamical
models, often smaller chemical networks are preferred to make such
calculations practical and computationally feasible for large grids of
models. Moreover, these simple models often allow the key physical and
chemical processes to be identified.  Two independent simple water
chemistry models have been developed within WISH, that by
\citet{Keto14} for cold pre-stellar cores and the SWaN (Simple Water
Network) model for protostellar envelopes by \citet{Schmalzl14}. They
include only a few of the reactions shown in
Fig.~\ref{fig:waternetwork}:
(a) cycling of water from gas to ice through freeze-out and UV
  photodesorption; (b) photodissociation of gaseous water to OH and
  atomic O, and (c) grain-surface formation of water ice from
  atomic O.

The \citet{Keto14} model explicitly includes gaseous OH as an
intermediate channel, as well as the high temperature O + H$_2$ and OH
+ H$_2$ reactions, although the latter reactions never become
significant in cold cores. The SWaN model ignores all OH reactions,
but explicitly includes thermal desorption of water in the hot core
region above 100~K.

Both simplified models have been extensively tested against the full
water chemistry models of \citet{Hollenbach09} \citep[see Fig.\ 3 and
4 of][]{Keto14} and those of \citet{Visser11}, \citet{Walsh12} and
\citet{Albertsson14} \citep[see Fig.\ B.2 of][]{Schmalzl14}. While
differences of factors of a few in absolute abundances can readily
occur, the overall profiles are similar and robustly seen in all
models as long as the same physical and chemical parameters are
adopted.

Multilayer ice models such as developed by \citet{Taquet14} and
\citet{Furuya16} predict similar abundance profiles of water as the
two-phase simple SWaN models do, but differences in absolute water
abundances can be introduced because multilayer interstellar ices are
inhomogenous.

To compare models with observations, not just the water chemistry but
also the water excitation and radiative transfer need to be treated
correctly. The various methods and modeling approaches adopted by WISH
are described in \S~\ref{sec:waterexcitation} and
Appendix~\ref{app:modeling}.

\subsection{Water deuteration}
\label{sec:deuteration}

\begin{figure}
  \centering
\includegraphics[width=9.5cm]{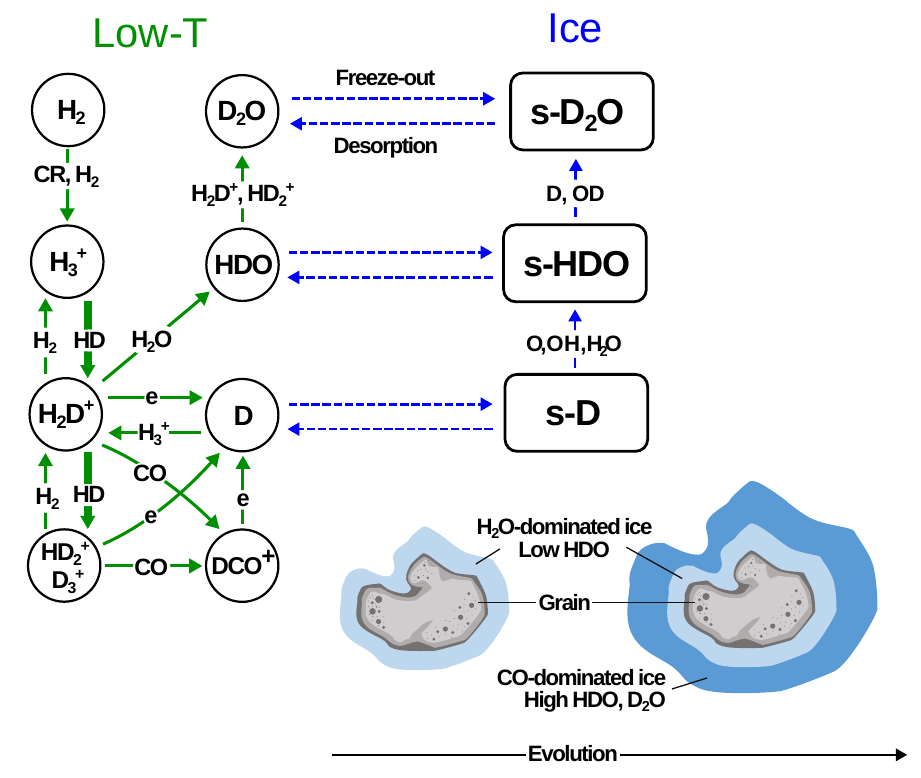}
\caption{Simplified reaction network illustrating the important
  reactions in the deuteration of water and other molecules. The
  left-hand side illustrates the cold gas-phase chemistry, leading to
  high fractionation of gaseous H$_2$D$^+$ and atomic D, and
  ultimately gaseous HDO and D$_2$O.  The right-hand side illustrates
  how this enhanced D ends up on the ice and leads to enhanced solid
  HDO and D$_2$O. The bottom cartoon shows the different ice layers on
  a grain: the H$_2$O-dominated layer formed early in the cloud
  evolution with low HDO/H$_2$O $< 10^{-3}$, and the CO-dominated
  layer formed at higher densities which is rich in deuterated water,
  with D$_2$O/HDO$>>$HDO/H$_2$O \citep{Furuya15}.}
\label{fig:hdonetwork}
\end{figure}

Deuterated water, HDO and D$_2$O, is formed through the same processes
as shown in Fig.~\ref{fig:waternetwork} but there are a number of
chemical reactions that can enhance the HDO/H$_2$O and D$_2$O/H$_2$O
ratios by orders of magnitude compared with the overall [D]/[H] ratio
of $2.0 \times 10^{-5}$ found in the local interstellar medium
\citep{Prodanovic10}. Details can be found in \citet{Taquet14} and
\citet{Furuya16}, and in \citet{Ceccarelli14} for deuterium
fractionation in general (see also \S~\ref{sec:waterhdocold}). The
processes are illustrated in Fig.~\ref{fig:hdonetwork}. The most
effective water fractionation occurs on grains, due to the fact that
the relative number of D atoms landing from the gas on the grain is
enhanced compared with that of H atoms. In other words, atomic D/H in
cold gas is much higher than the overall [D]/[H] ratio
\citep{Tielens83,Roberts03}. This high D/H ratio landing on grains
naturally leads to enhanced formation of OD, HDO and D$_2$O ice
according to the grain-surface formation routes.

The high atomic D/H ratio in the gas, in turn, arises from the
well-known fractionation reactions initiated by the H$_3^+$ + HD $\to$
H$_2$D$^+$ + H$_2$ reaction, which is exoergic by $\sim$ 230 K and is
thus very effective at low temperatures $\leq 25$ K
\citep[e.g.,][]{Watson76,Aikawa99,Stark99,Sipila15}. The H$_2$D$^+$ abundance
is further enhanced when the ortho-H$_2$ abundance drops (preventing
the back reaction) and when the main H$_3^+$ and H$_2$D$^+$ destroyer,
CO, freezes out on the grains
\citep{Pagani92,Pagani09,Roberts03,Sipila10}. The latter
processes become more important as the cloud evolves from a lower
density to a higher density phase
\citep{Dartois03,Pagani13,Brunken14,Furuya15}. Dissociative
recombination of H$_2$D$^+$ and other ions such as HD$_2^+$, D$_3^+$ and
DCO$^+$ with electrons then produces enhanced atomic D which gets
incorporated in the ices (Fig.~\ref{fig:hdonetwork}). As a result, the
outer ice layers, which are produced when the cloud is denser and colder,
have higher HDO/H$_2$O ratios --- by orders of magnitude --- than the
inner layers and the bulk of the ice.  The enhanced H$_2$D$^+$ also
leads to enhanced H$_2$DO$^+$ and thus HDO in cold gas, which may play
a role at the lower density edge of the cloud.

Another characteristic of this layered ice chemistry is that the
D$_2$O/HDO ratio is much higher than the HDO/H$_2$O ratio
\citep{Furuya16}. Moreover, the deuteration of organic molecules
formed through hydrogenation of CO ice in the later cloud stages is
generally much higher than that of water
\citep{Cazaux11,Taquet14,Furuya16}.

\begin{figure*}[t]
  \centering
    \includegraphics[width=14cm]{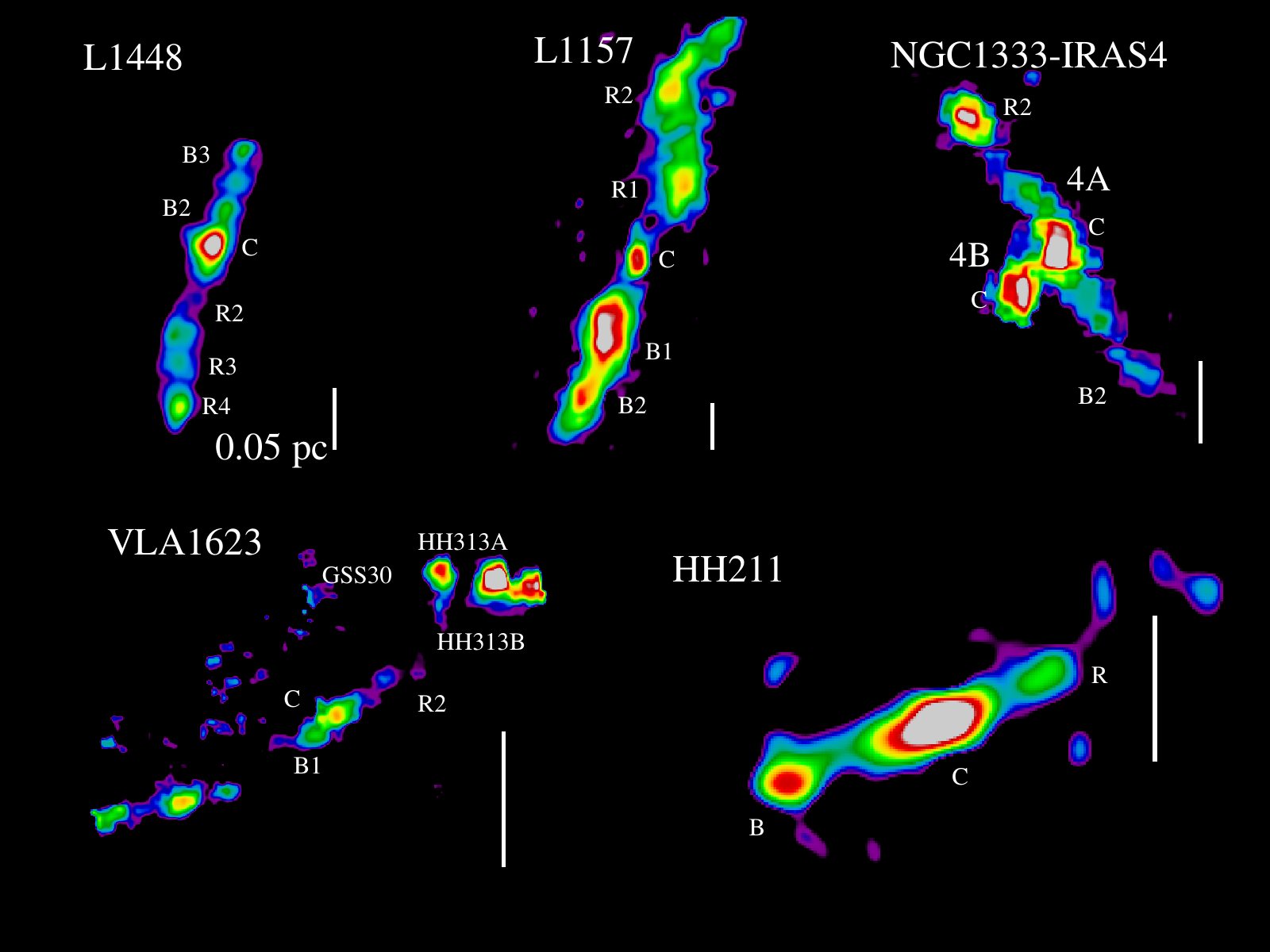}
    \caption{Gallery of water maps in the $2_{12}-1_{01}$ 179 $\mu$m
      line made with {\it Herschel}-PACS of the low-mass outflow
      sources L1448-MM ($d$=235 pc), L1157 (325 pc), NGC 1333 IRAS4
      (235 pc), VLA1623 (140 pc) and HH211 (250 pc). The white bars
      indicate a 0.05 pc scale.
      The central sources are indicated by C, with coordinates and
      details of data given by
   \citet{Nisini10,Nisini13,Bjerkeli12,Santangelo14a,Santangelo14b,Dionatos18}.
    }
    \label{fig:maps}
\end{figure*}

The desorption processes -- photodesorption at low ice temperatures
\citep{Oberg09h2o,Arasa15,CruzDiaz18} and thermal desorption at high
ice temperatures -- have a small to negligible effect on the deuterium
fractionation. In other words, the gaseous HDO/H$_2$O and
D$_2$O/H$_2$O ratios should reflect the ice ratios following
desorption if no subsequent gas-phase reactions are involved. It is
important to note that photodesorption is only effective in the outer
few layers of the ice, whereas thermal desorption removes the bulk ice
mantle. This selective formation and removal of ice layers turns out
to be important in the interpretation of HIFI observations of
HDO/H$_2$O ratios in cold versus warm gas.

In warm gas, the exchange reaction D + OH $\to$ H + OD
is likely barrierless and can be effective in enhancing OD, especially
since the reverse reaction seems to have a barrier of around 800 K
\citep{Thi10dh}. Photodissociation of HDO can also enhance OD
compared with OH by a factor of 2--3, which could be a route to
further fractionation. Finally, in high temperature gas in disks the
exchange reaction H$_2$O + HD $\leftrightarrow$ HDO + H$_2$ is often
included.  Similarly, there are reactions inside the ices that can
both enhance and reduce the water fractionation
\citep{Lamberts15,Lamberts16} but are not considered here \citep[see
discussion in][]{Furuya16}.

\begin{figure*}
  \centering
 \includegraphics[width=12cm]{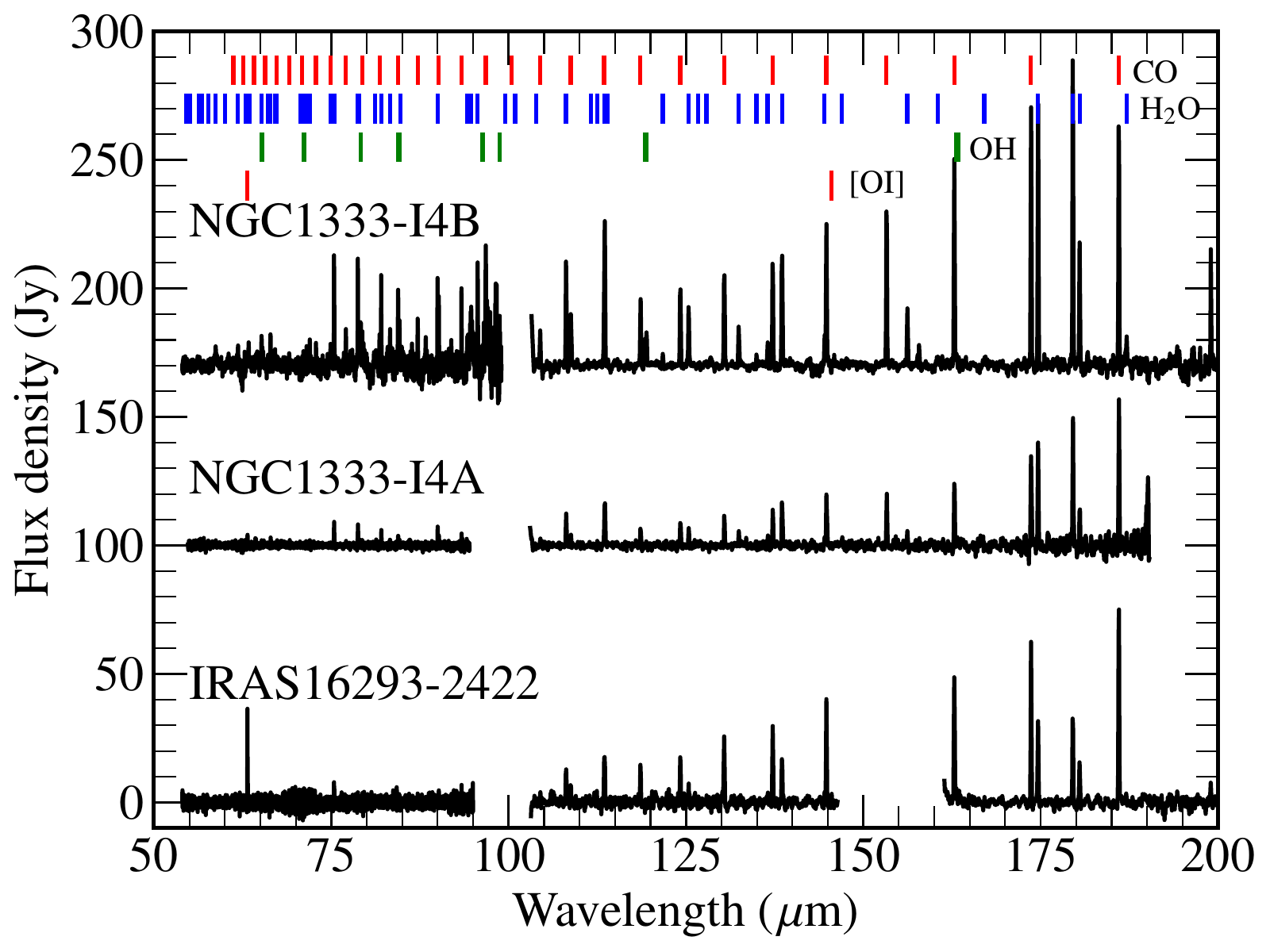}
 \caption{Full PACS spectral scans of the low-mass protostars NGC 1333
   IRAS4B \citep{Herczeg12} and IRAS 1333 IRAS4A, both taken as part
   of WISH, compared with that of IRAS16293 -2422, taken as part of
   CHESS. The IRAS4B spectrum, which is extracted at the
   offset and localized shock position just south of the protostar, is
   much richer in lines than the other two sources, likely because of
   less extinction at far-infrared wavelengths in the blue outflow
   lobe.}
         \label{fig:pacs_scans}
\end{figure*}

\section{Herschel water spectra and maps}
\label{sec:waterdata}

In this section, we briefly summarize the main features characterizing
water emission observed with {\it Herschel}, largely based on WISH+
programs. Figure~\ref{fig:water_mass} illustrates the complexity of
individual water profiles for protostars observed with HIFI over more
than $\pm$50 km s$^{-1}$ whereas Fig.~\ref{fig:maps} shows the spatial
distribution imaged with PACS. Figure~\ref{fig:pacs_scans} presents
examples of full spectral scans with PACS with many lines
detected. The main conclusion from the combined analysis of the water
maps, the broad water line profiles, and its excitation, is that most of
the observed gaseous water is universally associated with warm
  outflowing and shocked gas of several hundred K. All
water lines observed by {\it Herschel} within WISH show thermal
emission, so nonmasing, in contrast with the 22 GHz maser often
associated with star-forming regions. At least two different types of
kinematic components are involved, with water being a significant (but
not necessarily dominant) coolant. In contrast, water emission is not
associated with the slower, colder entrained outflow gas traced by
low-$J$ CO line wings.

Cold quiescent water vapor is also detected but is primarily seen in
absorption at the protostellar position; only very few cloud positions
show weak narrow (FWHM $\lesssim$few km s$^{-1}$) water emission lines.
These are further discussed in \S~\ref{sec:protocold}.  Recall that {\it
  Herschel} does not observe the bulk of water in cold clouds
directly, since this is locked up in ice.

\subsection{Water maps of outflows} 
\label{sec:maps}

\begin{figure*}[bt]
  \centering
\includegraphics[width=12cm]{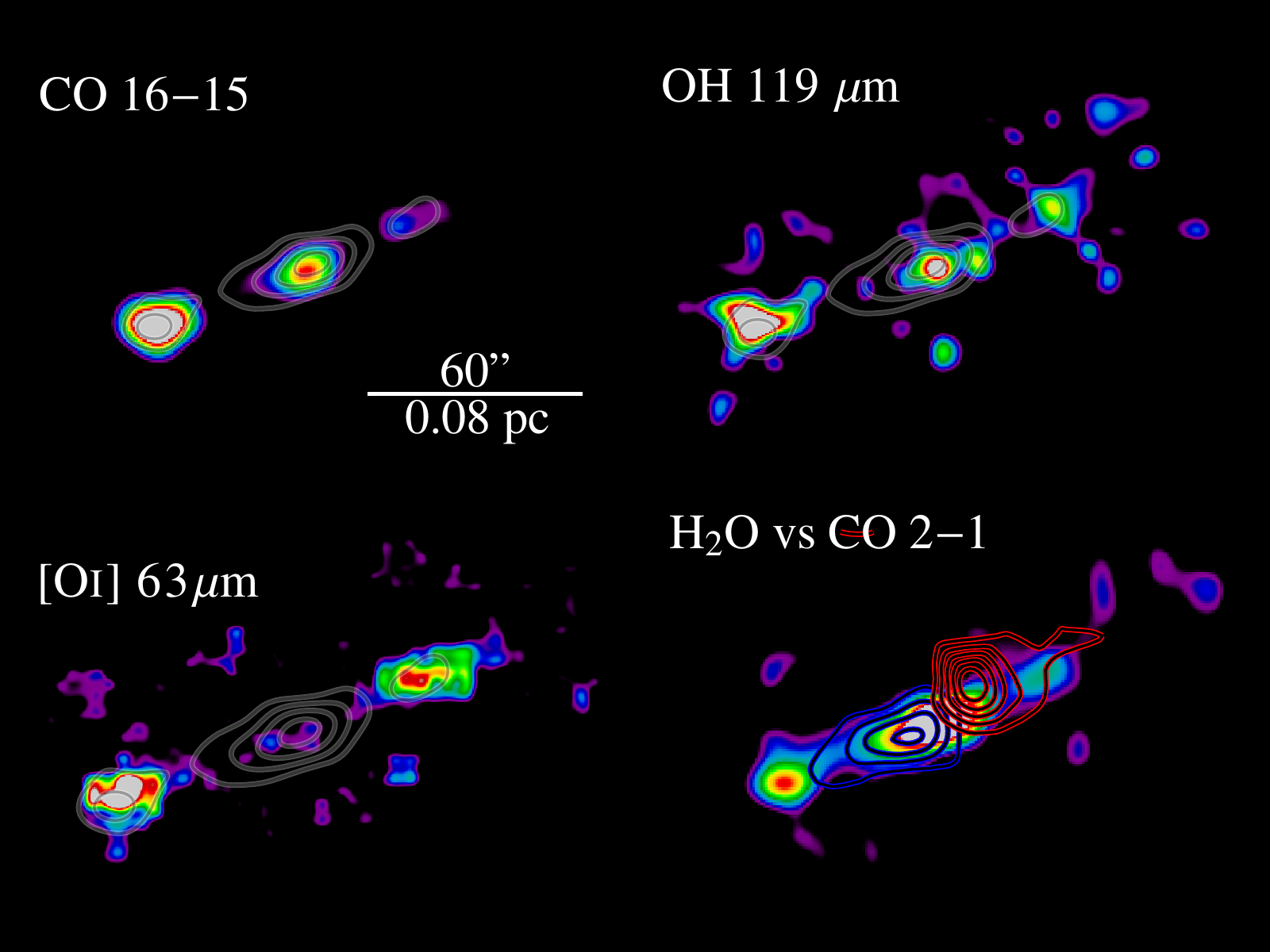}
\caption{Map of the water 179 $\mu$m emission in the blue outflow lobe
  of the HH 211 low-mass Class 0 source
  (L$_{\rm bol}\approx 3.6 L_\odot$, $d$=250 pc) compared with that of
  other lines observed with {\it Herschel}-PACS. The color
  maps present, from left to right: a high-$J$ CO line, the OH 119
  $\mu$m doublet, the [O I] 63 $\mu$m line and the H$_2$O 179 $\mu$m
  line overlaid with a low-$J$ CO ($J$=2-1) line. The H$_2$O map is
  overlaid in light gray contours on the other panels. Note the
  similarity between the H$_2$O and CO high-$J$ maps, but not with CO
  low-$J$. Data from \cite{Tafalla13,Dionatos18}.}
\label{fig:hh211}
\end{figure*}

Figure~\ref{fig:maps} presents maps in the H$_2$O 179 $\mu$m line
imaged over several arcmin scales at $\sim 13''$ resolution with {\it
  Herschel}-PACS for a number of low-mass sources
\citep{Nisini10,Nisini13,Santangelo14b,Dionatos18}.  Water vapor
emission is clearly associated with the powerful large scale outflows
from these sources.  Close inspection shows, however, that the water
emission is systematically shifted from that of low-$J$ CO emission
commonly used to trace outflows \citep[e.g.,][]{Tafalla13}: water only
spatially coincides with maps of the higher-$J$ CO lines with
$J_{\rm up}>14$ and with the H$_2$ mid-IR lines S(1)--S(4)
\citep[e.g.,][]{Nisini13,Bjerkeli12,Santangelo14a,Santangelo14b,Neufeld14}
(Fig.~\ref{fig:hh211}).  Thus, water and high-$J$ CO probe a
fundamentally different gas component than that commonly studied with
low-$J$ CO lines. This also limits the use of low-$J$ CO lines to
determine, for example, abundances in shocked gas
(\S~\ref{sec:shocks}).

Bright compact water emission is seen at the central protostar
position, in contrast with the thermal emission from other molecules
that are associated with outflows such as SiO and CH$_3$OH. This
suggests an additional production route of water in the warm inner
protostellar envelope beyond thermal desorption of ices.  Along the
outflow, the water emission is clumpy, with unresolved individual
peaks (at the 13$"$ resolution of the 179 $\mu$m line) corresponding
to shock spots.  In addition, weaker more extended water emission is
observed.

The [O I] 63 $\mu$m line is commonly associated with jets powering the
outflows. Indeed, {\it Herschel}-PACS maps show strong [O I] emission
along the outflow direction (Fig.~\ref{fig:hh211}). In some cases
($\sim$10\% of sources) even velocity resolved PACS spectra have been
seen, indicating speeds of more than 90 km s$^{-1}$ with respect to
that of the source, characteristic of jets
\citep[e.g.,][]{vanKempen10,Nisini15,Dionatos17,Karska18}.  Comparison
of mass flux rates from [O I] and CO suggests that this atomic gas is not
the dominant driver in the earliest stages: the jets in the Class 0
phase are mostly molecular. However, toward the Class I stage, the
jets become primarily atomic, and ultimately ionized in the Class II
stage.

The HH 211 outflow shown in Fig.~\ref{fig:hh211} is an example of a
source where the atomic jet has just enough momentum to power the
molecular jets and the large scale outflow \citep{Dionatos18}. Maps in
the chemically related OH molecule exist for only a few sources and
usually show more compact emission than that of water. In the case of
HH 211, OH peaks primarily on source and at the bow-shock position.
It is important to note that HH 211 is the only protostellar source
which was mapped completely with {\it Herschel}-PACS full spectral
scans. All other outflows were only mapped in selected lines.

The sources included in Figure~\ref{fig:maps} represent those with
well-known extended outflows; the WISH low-mass protostar sample
contains a relatively large fraction of them. However, this is not
representative of the low-mass population. As noted in
\S~\ref{sec:sample}, the bulk of the low-mass Class 0 and I sources in
the full WISH+WILL+DIGIT sample show little or no extended emission
beyond the central PACS 13$''$ spaxel: out of 90 sources only 18 show
extended water emission in the 47$''$ PACS footprint
\citep{Karska13,Karska18}. This means that the currently active
mechanism exciting water in the bulk of the sources is limited to a
$\sim$1000 au radius from the central source.  The same conclusion
holds for higher mass sources \citep[e.g.,][]{Leurini17}. Even in
Orion-KL, the bulk of the water emission is relatively compact and
associated with the high-mass protostars and their outflows on scales
of $<$25000 au ($<$0.1 pc) \citep[e.g.,][]{Goicoechea15}. For more
distant high-mass sources at a few kpc, this emitting region would
correspond to a single PACS spaxel.  Indeed, HIFI maps of high-mass
sources in the 988 GHz $2_{02}-1_{11}$ line show emission confined to
$\lesssim 20''$ \citep{Jacq16,vanderTak19}.

\subsection{Cooling budget}
\label{sec:cooling}

{\it Herschel}-PACS full spectral scans from the WISH, WILL and DIGIT
programs (see Fig.~\ref{fig:pacs_scans} for examples) have directly
measured the total far-infrared line cooling, $L_{\rm FIRL}$ of the
warm gas \citep{Karska14,Green16,Karska18}. For those sources for which only
selected individual line scans have been observed, corrections for the
missing lines have been made. For the low-mass source sample
consisting of 90 targets, $L_{\rm FIRL}$ does not change
significantly with evolution from the Class 0 to the Class I stage:
median values are $4.5\times 10^{-3}$ L$_\odot$ for Class 0 and
$3.7\times10^{-3}$ L$_\odot$ for Class I sources, respectively
\citep{Karska18}.

\begin{figure}
  \centering
 \includegraphics[width=9.5cm]{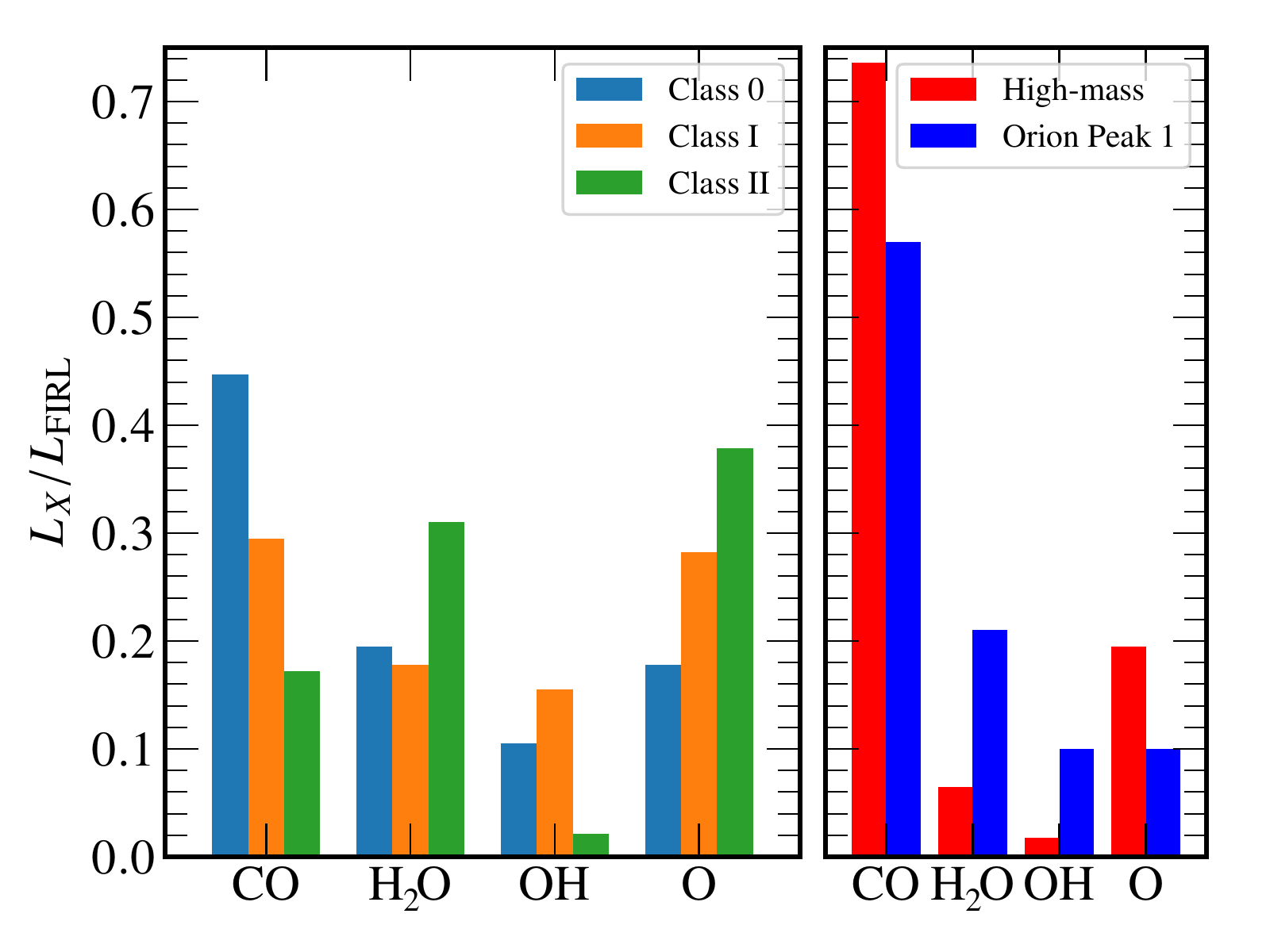}
 \caption{Fractions of gas cooling contributed by far-infrared lines
   of CO, H$_2$O, OH and [O I] to the total far-infrared line cooling,
   as function of evolutionary stage for low-mass sources (left) and
   for high mass sources (right). The Class II sample is small and
   differences with Class I are not statistically significant.  
   [O I] is seen to become relatively more important with evolution, whereas
   H$_2$O and OH contribute little for high-mass sources.}
         \label{fig:water_cooling}
\end{figure}

The relative contribution of individual coolants does change with
evolution and between low- and high-mass sources.
Figure~\ref{fig:water_cooling} (left) summarizes the far-infrared line
emission of CO, H$_2$O, OH and [O I] as a fraction of $L_{\rm FIRL}$.
The LM Class 0 and I numbers are taken from \citet{Karska18}, and this
sample is large enough that statistical uncertainties are small.  The
Class II values are from \citet{Karska13} based on the sample from
\citet{Podio12}. This sample is biased toward Class II sources with
strong optical jets which may be more appropriate for comparison with
the Class 0 and I outflows than disk-only sources. The Class II sample
is too small to conclude whether differences between Class I and II
are significant. The HM data are taken from \citet{Karska14} and for
Orion Peak 1 from \citet{Goicoechea15}.

In the earliest stages of LM sources, CO and H$_2$O are the dominant
coolants, with [O I] becoming relatively more important in the
Class I and II stages. Interestingly, the fractional cooling of CO decreases by
a factor of 2 whereas that of H$_2$O stays roughly constant from Class
0 to I. This conclusion differs from that of \citet{Nisini02}, who
found a significant decrease in water cooling based on earlier ISO
data.  The absolute [O I] line cooling is similar from Class 0 to
Class II, but its fraction increases as the jet changes from being
mostly molecular to being primarily atomic \citep{Nisini15}.

High-mass sources are found to have a smaller fraction of H$_2$O
cooling compared with low-mass sources: their far-infrared line
emission is dominated by CO. The main reason for this difference is
that several H$_2$O and some OH lines are found in absorption rather
than emission at the high-mass protostellar source positions,
suppressing their contributions to the cooling. However, while
globally the energy released in the water lines is low, the central
emission is higher with the absorption transferring energy to the
cooler layers.  To assess the effect of this on-source absorption on
the cooling budget, the Peak 1 outflow position in Orion has been
added to the figure. This comparison demonstrates that the conclusion
that CO dominates the cooling also holds off source where there is no
absorption in the water lines.

Consistent with \citet{Nisini02}, the total far-infrared line cooling
is found to strongly correlate with $L_{\rm bol}$, see Fig.~14 in
\citet{Karska14} and Fig.~17 in \citet{Karska18}. The ratio of line
cooling over dust emission, $L_{\rm FIRL}$/$L_{\rm bol}$, decreases
from $\sim 10^{-3}$ for low- and intermediate-mass sources to
$\sim 10^{-4}$ for high-mass sources. Nevertheless, this implies
that total far-infrared line luminosity can be used as a direct
measure of the mechanical luminosity deposited by the jet and wind into
the protostellar surroundings. These data also provide estimates of
the mass loss rates \citep{Maret09,Mottram17,Karska18}.

\subsection{Water emission: Two universal components}
\label{sec:profiles}

The water lines imaged by PACS reveal the location of warm water vapor
emission, but are spectrally unresolved.  In contrast, HIFI spectra
reveal the full kinematic structure of the gas but are available
mostly at a single position. Figure~\ref{fig:water_mass} illustrates
that the observed water line profiles are universally broad toward all
types of protostars, from low to high mass, and show dynamical
components not previously seen from the ground in low-$J$ CO or other
molecules
\citep[e.g.,][]{Kristensen10,Kristensen11,Kristensen12,Kristensen13,Johnstone10,Herpin12,vanderTak13,Mottram14,SanJose16,Conrad20}.
They are complex, with emission out to $\pm$ 100 km s$^{-1}$, and
often have narrow absorption superposed on the ground-state lines at
the source position.  Nevertheless, comparing different water
transitions, a maximum of only four different Gaussian components can
be distinguished, each of them associated with a different physical
component (Fig.~\ref{fig:water_mass}, Table~\ref{tab:profiles}, see
also Fig.~\ref{fig:averagespectra} in Appendix):

\begin{table*}[t]
  \caption[]{Physical components seen in water and high-$J$ CO emission lines profiles$^{a}$}
         \begin{tabular}{l c c c c}
            \hline 
            \hline
            \noalign{\smallskip}
   Source type &  Velocity & FWHM$^a$ & CO Excitation & Possible origin$^b$  \\
            & Characteristics	& (km s$^{-1}$)	& (K) 	\\
            \noalign{\smallskip}
            \hline
            \noalign{\smallskip}
       Pre-stellar & narrow & $<5$ & $<$10  & quiescent, infall  \\[7pt]
       Low mass Class 0 & broad,  & 24 & 300   
        & warm outflow (cavity shocks, disk wind, turbulent mixing layers)  \\
                     & medium/offset   & 18 & 700
                     & hot spot shock  \\
                     & medium/EHV  &  & 700
                     & hot EHV bullets  \\
                     & narrow &  3.5 & $<$100 
                     & quiescent, infall, expansion   \\[7pt]
       Low mass Class I & broad & 15 & 300   
        & warm outflow (cavity shocks, disk wind, turbulent mixing layers)  \\
                     & narrow &  2.6 & $<$100
                     & quiescent, infall, expansion   \\[7pt]
       Intermediate mass    & broad & 32 & 300   
        & warm outflow (cavity shocks, disk wind, turbulent mixing layers)  \\
                     & narrow &  4.6 & $<$100 
                     & quiescent, infall, expansion   \\ [7pt]
       High mass     & broad & 24 & 300   
       & warm outflow (cavity shocks, disk wind, turbulent mixing layers)  \\
                     & narrow &  5.6 & $<$100 
                     & quiescent, infall, expansion   \\ [7pt]
       Disks & narrow & $<4$ & 20--300  & rotating disk   \\
            \noalign{\smallskip}
            \hline
         \end{tabular} 

$^{a}$ Median width (FWHM) of water lines at the source
         position, taken from \citet{SanJose16} and Table 5.4 of
         \citet{SanJose15}. Values refer to the median of the WISH
         sample; those including other samples, such as the WILL
         sample, are similar within the uncertainties of a few km s$^{-1}$.
         $^b$ Based on Table
         1 of \citet{Kristensen17b}.
         \label{tab:profiles}
\end{table*}

\begin{enumerate}

\item A broad component (typical FWHM$>$15 km s$^{-1}$) centered at
  the source velocity and heated by kinetic energy dissipation. One
  scenario is that this water emission is produced in non-dissociative
  $C-$type shocks along the outflow cavity, called ``cavity shocks''
  \citep{Mottram14}.  Alternative explanations for the observed line
  profiles and gas heating include MHD ``disk winds'' heated by
  ion-neutral drift \citep{Yvart16}, and ``turbulent mixing layers''
  forming between the protostellar wide-angle wind and infalling
  envelope \citep{Liang20}.

\item A medium-broad offset component (typical FWHM=5--15 km s$^{-1}$)
  slightly blue-shifted from the source velocity. Based on its
  velocity offset and distinct chemistry, this component is likely
  tracing dissociative $J-$type shocks, called ``spot shocks'', at
  the base of the outflow cavity.

\item A medium-broad component with similar FWHM as 2 but
  significantly offset from the source velocity, previously called
  Extremely High Velocity (EHV) gas or ``bullets''
  \citep{Bachiller91}. Although separately listed here, types 2 and 3
  form a continuous sequence of velocity offset features in the
  profiles and are therefore all considered $J$-type spot shocks
  \citep{Mottram14}.

\item A narrow component (FWHM$<$5 km s$^{-1}$) at the source
  velocity, seen primarily in ground-state water lines in absorption
  and arising in quiescent cloud or envelope material. The absorption
  can also be part of a P-Cygni type profile, consisting of strong
  emission lines with blue-shifted absorption pointing to an
  expanding shell of gas. Also, inverse P-Cygni profiles are seen with
  red-shifted absorption indicating infall.

\end{enumerate}

Components 1--3 involve heating by dissipation of kinetic energy
through one or more of the proposed mechanisms. In contrast, component
4 involves radiative heating, either by the bulk luminosity of the
protostar heating the dust in the surrounding envelope which then
transfers its heat to the gas through collisions (a `passively' heated
envelope), or by UV irradiation of the young star-disk system
impinging on the gas and dust in outflow cavity walls or disk surfaces
and heating it through the photoelectric effect.

Roughly 70-80\% of the integrated H$_2$O emission comes from the broad
part of the profile (component 1), whereas the medium-broad offset
parts contribute 20-30\% (components 2+3) for low-mass sources. For
intermediate and high-mass sources, the quiescent component 4
contributes an increasingly higher fraction, up to 40\% for the
highest luminosity cases \citep{SanJose16}.

The first two components are also seen in CO $J$=16--15
($E_{\rm up}= 752$ K, $n_{\rm crit}\approx 4 \times 10^6$ cm$^{-3}$)
profiles observed with HIFI for low-mass protostars
\citep{Kristensen17b}. Extremely high velocity features are
occasionally present as well in the 16--15 lines. The fact that they
are not yet seen in the CO 10--9 profiles emphasizes the need for high
spectral resolution observations at 1.5 THz and above. The broad and
medium-broad velocity components can be associated physically with the
two components universally seen in CO rotational diagrams of the same
sources, as the $T_{\rm rot}\sim 300$~K (broad) and $\sim$700~K
(medium-broad) component, respectively (Table~\ref{tab:profiles}; see
also below).

At the outflow positions, well offset from the LM and IM source positions by
$>$1000 au, the water lines are also broad, with line widths of 10--40
km s$^{-1}$ on either the blue or red side of the flow
\citep{Vasta12,Tafalla13}.
The wings at the outflow positions generally follow those seen at the
source position in shape and extent although the level of agreement
depends on whether a low or high excitation line is chosen for
comparison \citep[see below and examples
in][]{Bjerkeli12,Santangelo14b}.  Similar to the source position, the
outflow positions show evidence for two physical components, at
intermediate and high velocity offsets, which have different
distributions and source sizes \citep{Santangelo14b}. In addition, EHV
gas is seen in a few cases, most notably at various positions in the
L1448-MM outflow \citep{Nisini13} (see Fig.~\ref{fig:water_mass} for
BHR71). The EHV component is actually more prominent in water lines
than in the low-$J$ CO lines in which it was originally detected
\citep{Bachiller90,Bachiller91}.

\subsubsection{Profiles across evolutionary stages}

\begin{figure}[t]
\centering
  \includegraphics[width=8.5cm]{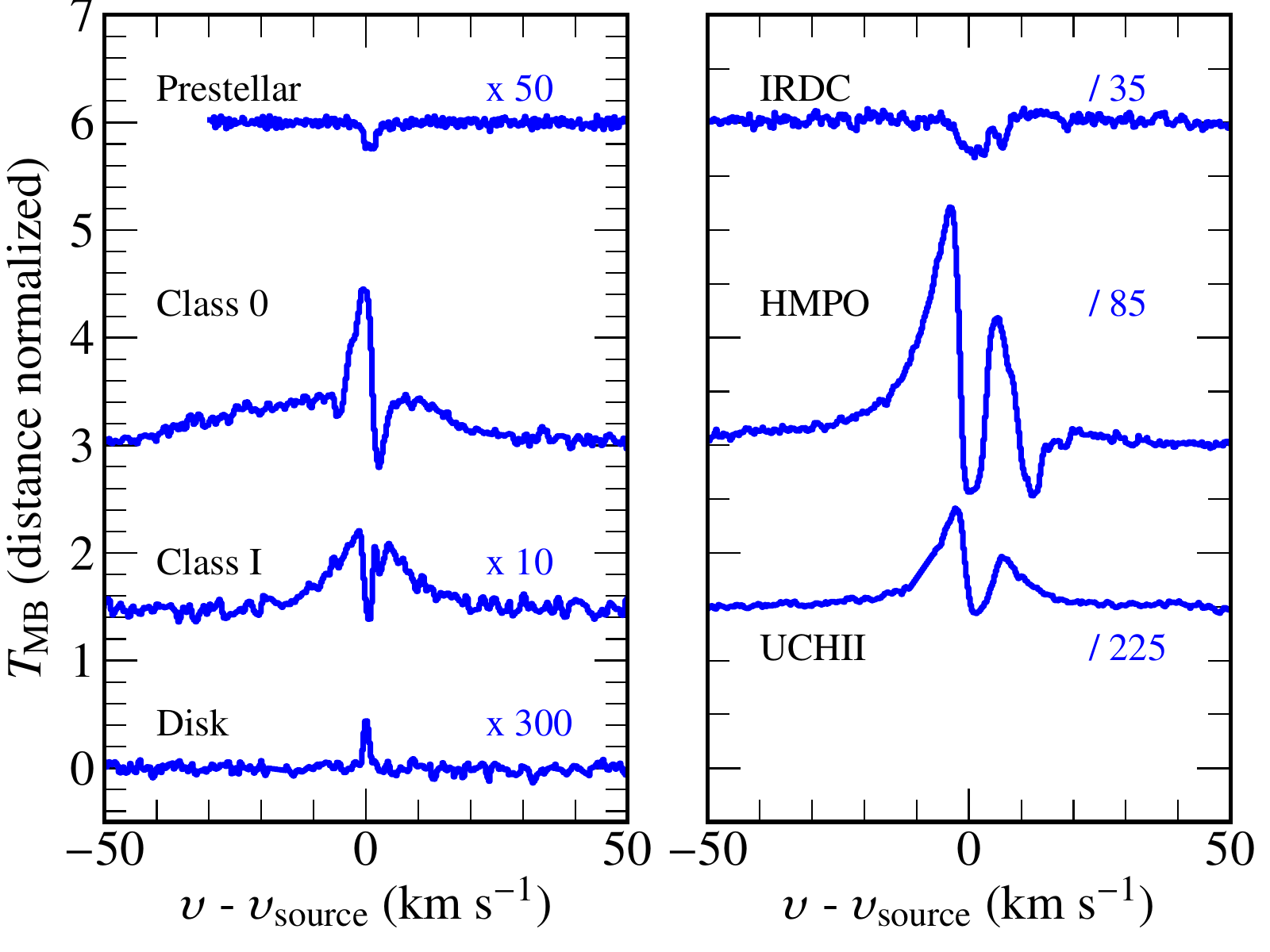}
 \caption{Water $1_{10}-1_{01}$ 557 GHz spectra at various
   evolutionary stages, from pre-stellar cores to protostars and
   disks. Left: Low-mass YSOs, showing from top to bottom the starless
   core Oph H-MM1, the Class 0 source Ser SMM4, the Class I source
   Elias 29 and the disk TW Hya. Right: high-mass YSOs, showing from
   top to bottom the starless core G11-NH$_3$, the HMPO DR21(OH), and
   the UC HII region NGC 7538 IRS1.}
         \label{fig:h2o_masstime}
\end{figure}

Figure~\ref{fig:h2o_masstime} (left) compares the complexity and number
of components seen in velocity-resolved water line profiles with
evolutionary stage across low-mass sources, whereas
Fig.~\ref{fig:h2o_masstime} (right) does so for high-mass objects.

Water emission from quiescent starless or pre-stellar cores
prior to star formation is very weak or not detected, both for
low-mass \citep{Caselli10,Caselli12} and higher mass cores
\citep{Shipman14}. If any feature is seen, it is usually in absorption
against the weak central continuum (Fig.~\ref{fig:h2o_masstime}).

As soon as the protostar is formed, complex water line profiles
appear, especially in the earliest most deeply embedded stages. The
complexity decreases with time: very few medium/offset components
attributed to $J-$type spot shocks are seen in low-mass Class I
sources. The FWHM and Full Width at Zero Intensity (FWZI) of the
profiles also decrease from Class 0 to Class I, suggesting that the
outflows become less powerful as the source evolves
(Table~\ref{tab:profiles}). These results hold irrespective of source
sample used to determine the mean or median width within each
Class. Figure~\ref{fig:h2o_masstime} (right) shows that within the
high-mass stages from HMPO to UC HII, no significant differences in
line widths are seen \citep{SanJose16,SanJose15}.

For low-mass Class II sources without an envelope or molecular
outflow, water emission is extremely weak and remains undetected for
the bulk of the protoplanetary disks observed either with HIFI
\citep{Bergin10,Du17} or PACS
\citep{Fedele13,Dent13,Riviere15,Alonso17}.

Overall, comparing the different evolutionary stages in
Fig.~\ref{fig:h2o_masstime}, it is clear that water emission `turns'
on only when there is star formation activity in the cloud and it
diminishes quickly once sources exit the embedded phase.

Water vapor emission is generally not detected from molecular clouds
away from protostars. The only exception are Photon Dominated Regions
(PDRs), that is, clouds exposed to enhanced UV radiation where water
has been observed to have a single narrow emission line
component. Examples are the Orion Bar PDR \citep{Choi14,Putaud19}, the
Orion Molecular Ridge \citep{Melnick20} and the extended $\rho$ Oph
cloud \citep[see Fig.~1 in][]{Bjerkeli12}.

\subsubsection{Profiles across mass range}

Nearby intermediate mass sources show very similar features to those
of low-mass Class 0 sources, with all four components present
\citep{Johnstone10,vanKempen16} (see NGC 2071 in
Fig.~\ref{fig:water_mass}). More distant intermediate mass sources and
high-mass sources generally show only two components: the broad
profile (component 1) and the quiescent envelope
absorption and/or emission (component 4) \citep{SanJose16}
(Fig.~\ref{fig:h2o_masstime}). Either these sources are more evolved,
such as the low-mass Class I sources, or the $J-$type spot shocks are too
beam diluted due to their larger distance to be detected.

The FWHM and FWZI of the water profiles change little from low- to
high-mass protostars, suggesting a similar underlying launching
mechanism of the jet. This point is illustrated by the average
profiles for three water transitions for each of the types of sources
presented in \citet{SanJose16} (see Fig.~\ref{fig:averagespectra} in
Appendix).  Also, the maximum velocity $v_{\rm max}$ reached by gas
seen in CO is similar to that of water and independent of the CO line
used, even though the line profiles change greatly from low- to
high-$J$ and their FWHM are generally smaller than that of water
\citep{Kristensen17b,SanJose16}.  At the outflow positions,
$v_{\rm max}$ is also similar among tracers such as H$_2$O, CO and SiO
\citep{Santangelo12,Bjerkeli12}.

The narrow emission component originating in the quiescent envelope
becomes more prominent for intermediate and high-mass sources due to
their higher envelope mass, up to 40\% of the emission seen in the
low-lying H$_2^{16}$O lines.  This component is better probed through
the excited lines of the less abundant isotopologs H$_2^{18}$O and
even H$_2^{17}$O: indeed, their line profiles show a narrow(er)
emission component with FWHM $<$6 km s$^{-1}$ that lacks the broad
outflow wings
\citep{Johnstone10,Chavarria10,Marseille10,Choi15,Herpin16}.

Interestingly, for high-mass sources, the H$_2^{18}$O $1_{11}-0_{00}$
line profiles at 1101 GHz seen in absorption are remarkably similar to
the difference between the H$_2$O $2_{02}-1_{11}$ 988 GHz and
$2_{11}-2_{02}$ 752 GHz profiles, whose upper energy levels lie at 101
and 137 K respectively. This suggests that the narrow H$_2^{18}$O
absorption originates in the envelope just inside the 100 K radius
where water sublimates from the grains; it does not arise in the cold
envelope gas (see also \S~\ref{sec:protocold} and \ref{sec:hotcore})
\citep{Jacq16,vanderTak19}.

\begin{figure}[t]
  \centering
    \includegraphics[width=2.8cm]{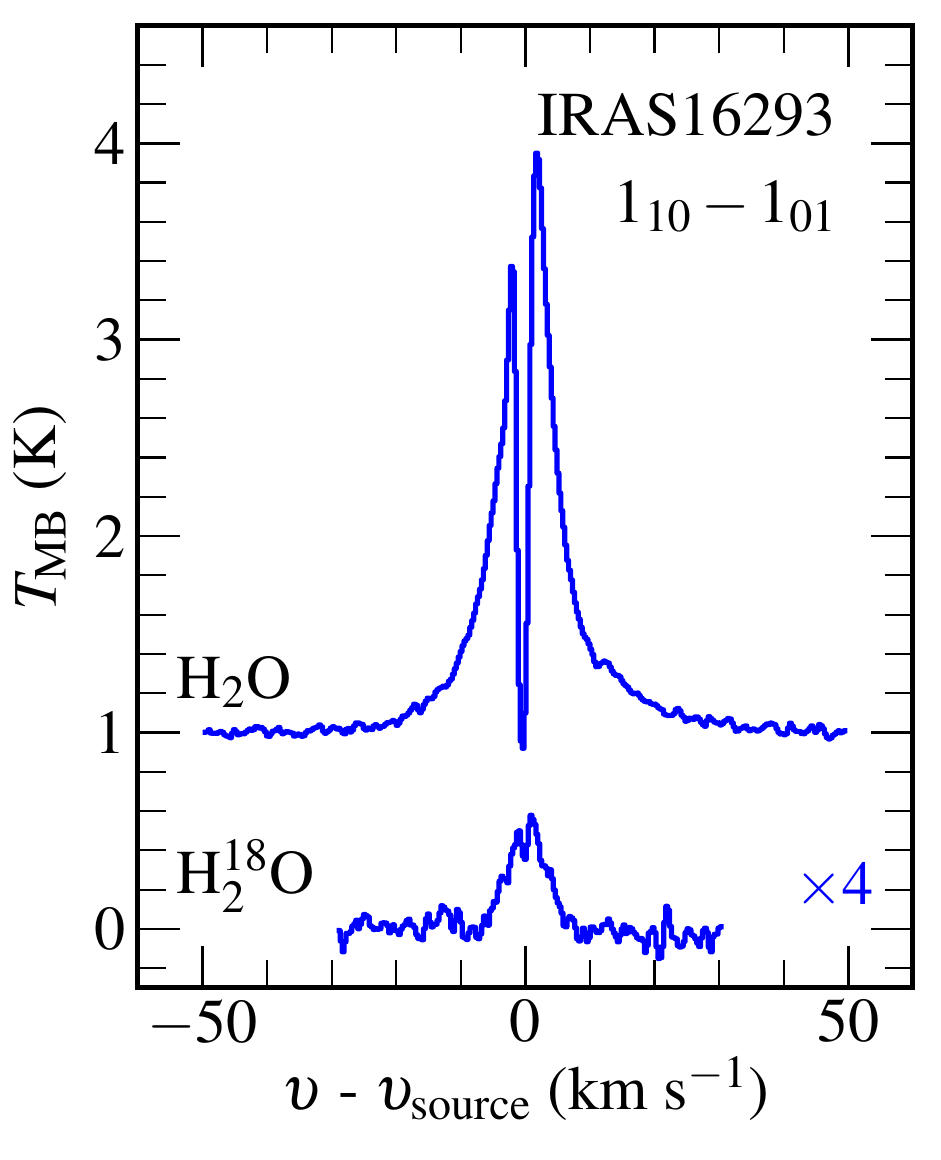}
    \includegraphics[width=3.0cm]{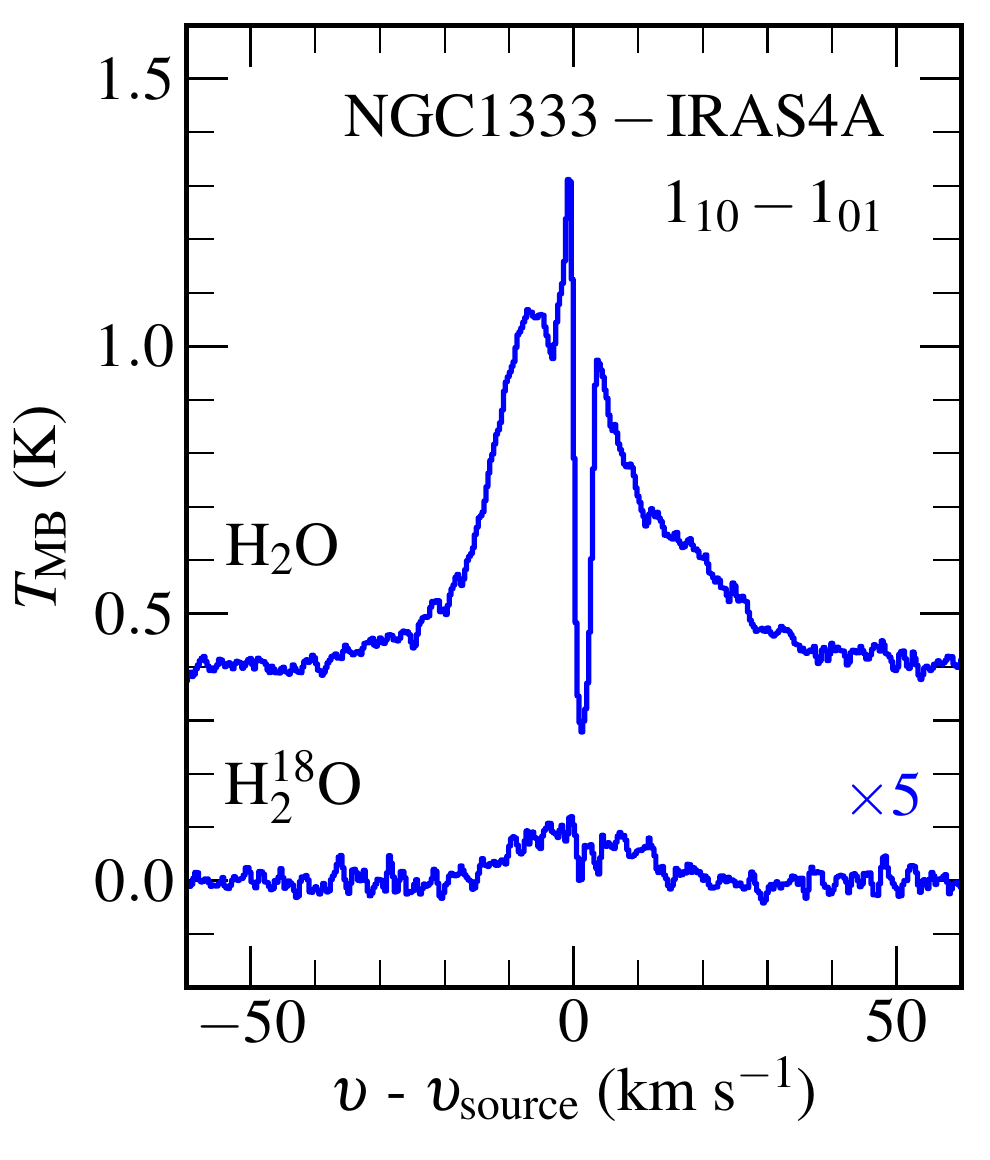}
    \includegraphics[width=3.0cm]{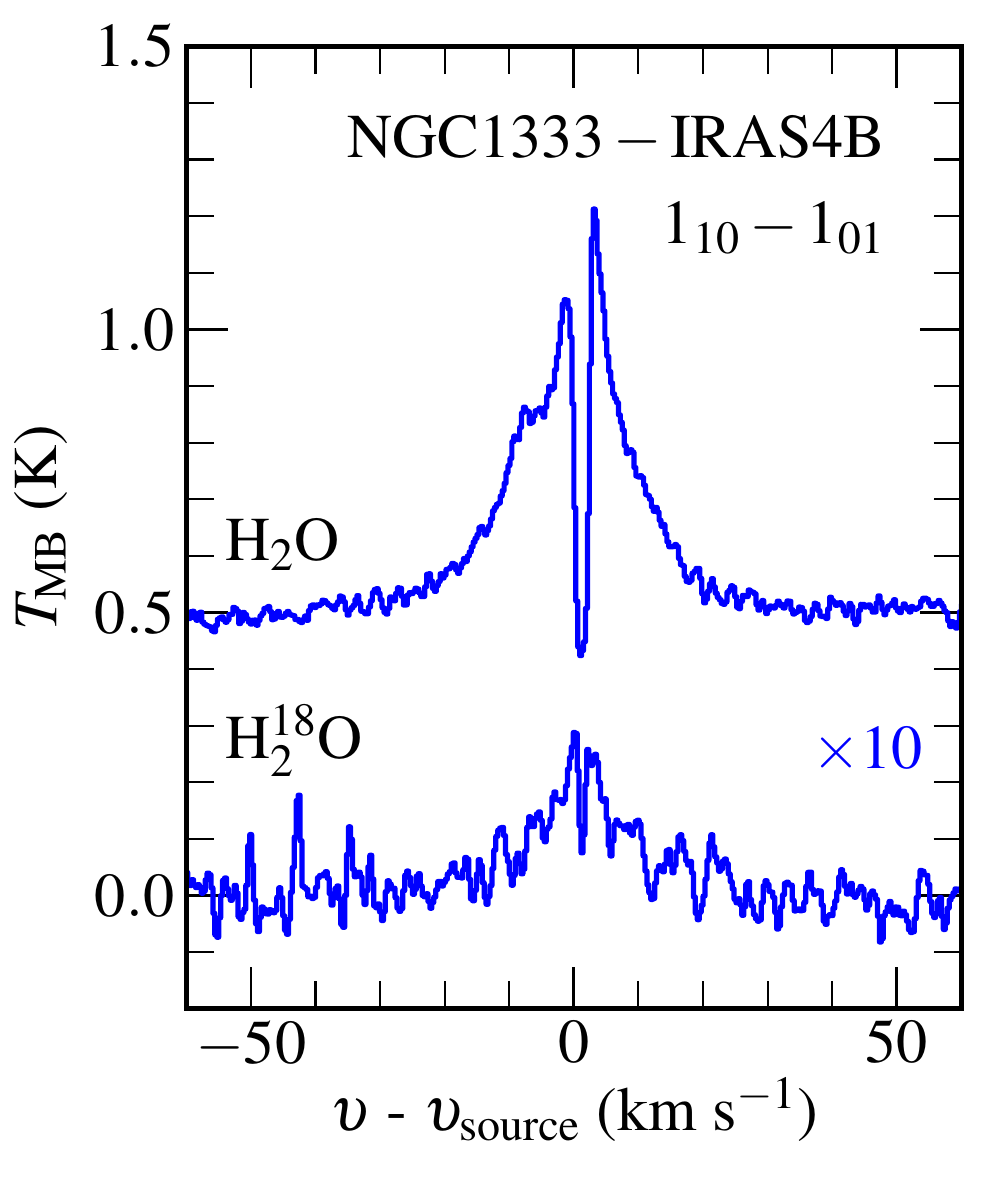}
    \caption{Water $1_{10}-1_{01}$ 557 GHz and H$_2^{18}$O 548 GHz
      lines for three low-mass sources.  The H$_2^{18}$O line for
      IRAS16293-2422 is seen to be narrower compared with that of the NGC 1333
      sources. Data from \citet{Coutens12} and \citet{Mottram14}.}
         \label{fig:h218o}
\end{figure}

These H$_2^{18}$O features are not seen for low-mass sources: in spite
of low noise levels, very few low mass sources show detections in the
WISH and WILL samples, even in the ground-state lines
\citep{Mottram14,Mottram17}. Also, when detected
(Fig.~\ref{fig:h218o}), the H$_2^{18}$O lines are in emission and
broad, in contrast with narrow excited $^{13}$CO and C$^{18}$O lines
\citep{Kristensen10,Yildiz13}. The spatially extended broad
H$_2^{18}$O outflow component may actually block any narrower emission
line arising closer to the protostar if it is optically thick. One
exception may be the molecule-rich low-mass protostar IRAS16293-2422,
part of the CHESS survey \citep{Coutens12}, where both H$_2^{18}$O and
H$_2^{17}$O lines have been detected with widths that are clearly
narrower than those of H$_2$O (Fig.~\ref{fig:h218o}).

Is quiescent warm H$_2^{18}$O emission completely absent in low-mass
sources? A dedicated deep (up to 5 hr integration per line) open time
program has revealed narrow features superposed on the broad outflow
profile in the excited water $3_{12}-3_{03}$ lines near 1097 GHz for
all three water isotopologs, but only in a few sources
\citep{Visser13} (see \S~\ref{sec:hotcore} and
Fig.~\ref{fig:visserspectra} in Appendix).

Taken together, the broad and medium/offset components of the water
line profiles can be used to study its chemistry in warm outflowing
and shocked gas, from low- to high-mass sources. For intermediate and
high-mass sources, the narrow H$_2^{18}$O emission profiles probe
the water abundance in the quiescent envelope. For low-mass sources, only
the narrow part of the H$_2^{16}$O (inverse) P-Cygni profiles can be
used to determine the quiescent water abundance, since their
H$_2^{18}$O profiles are still broad and optically thick, hiding any
narrow emission component. As will be shown in \S~\ref{sec:protocold},
these complex (inverse) P-Cygni profiles are actually a remarkably
powerful tool to determine the water abundance profile throughout the
cold envelope and cloud.

\subsection{Water excitation}
\label{sec:waterexcitation}

\subsubsection{Comparing observed fluxes and line profiles}

\begin{figure}[t]
  \centering
    \includegraphics[width=7.5cm]{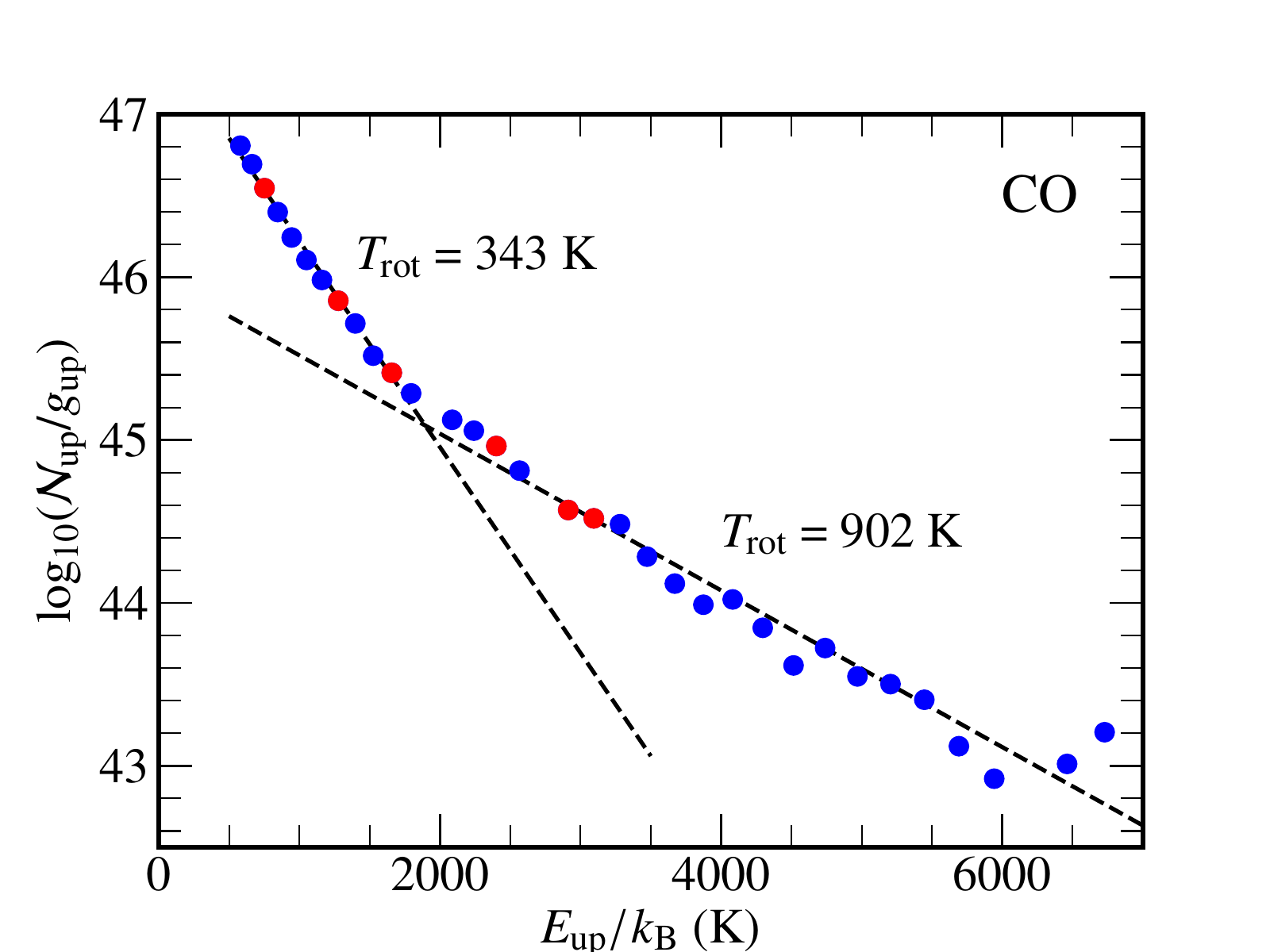}
    \includegraphics[width=7.5cm]{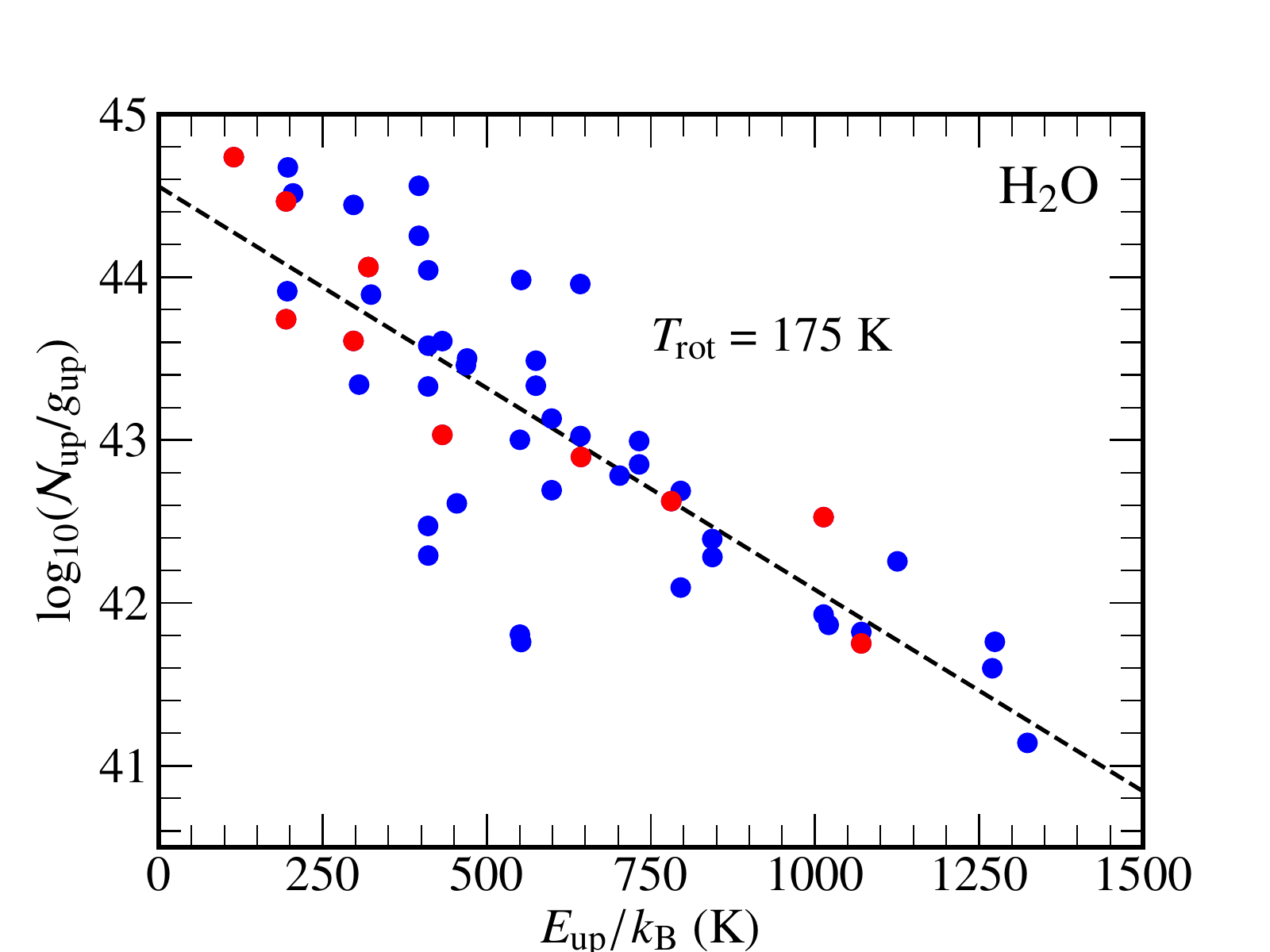}
    \includegraphics[width=7.5cm]{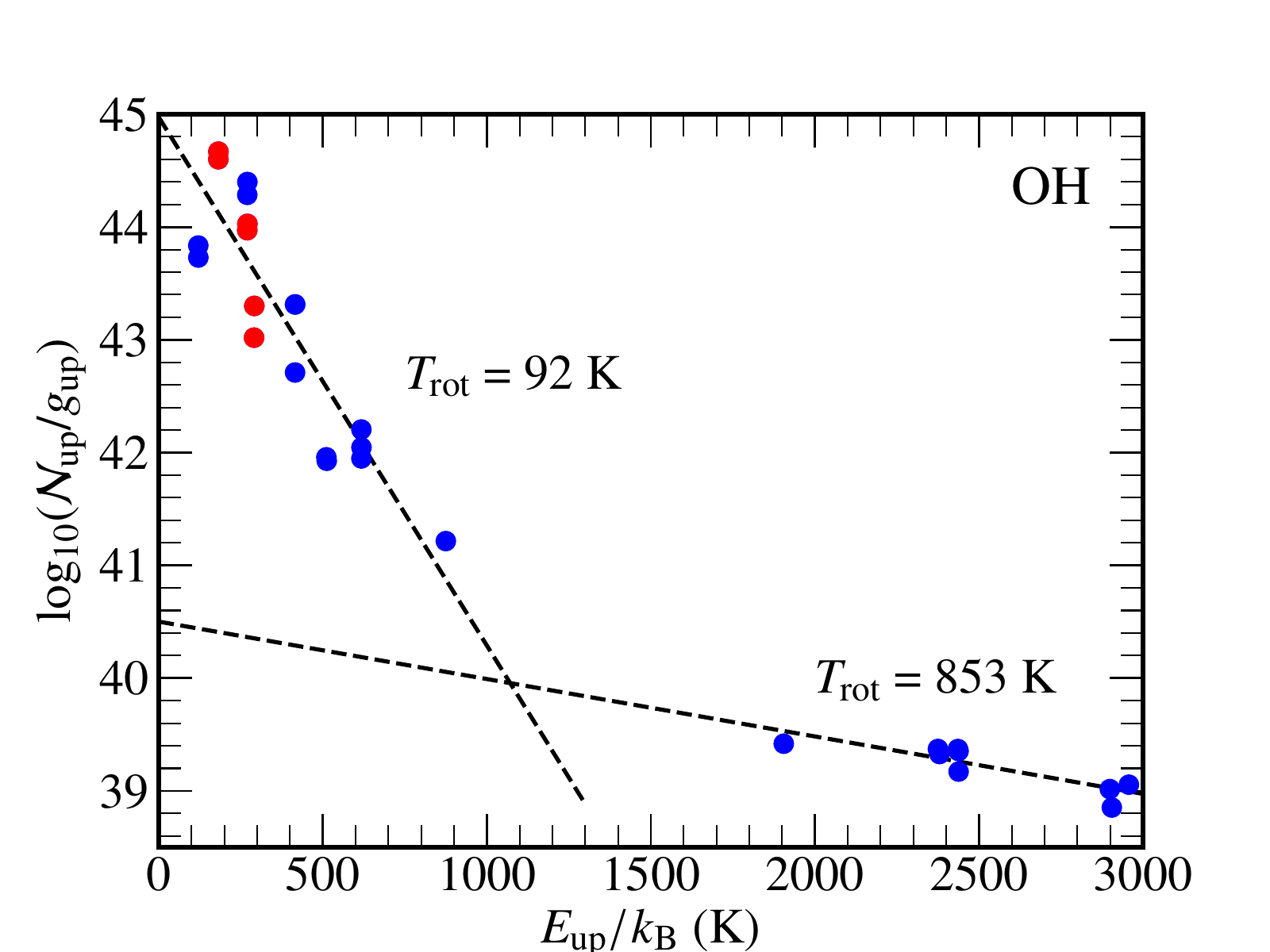}
    \caption{Comparison of the rotational diagram of CO (top), H$_2$O
      (middle) and OH (bottom) for the low-mass YSO NGC 1333 IRAS4B
      obtained from the full PACS spectral scan. The red points
      indicate those lines targeted in WISH and WILL for sources for
      which no full spectral scans have been obtained. The scatter in
      the water diagram is dominated by the difference in optical
      depth of the lines. Different energy ranges are covered for
      each of the species.}
         \label{fig:n1333rotdiag}
\end{figure}

The integrated intensities and luminosities of the various water vapor
lines obtained with HIFI and PACS are well correlated with each other
at the central protostellar positions, both for low- and high-mass
sources \citep[e.g.,][]{Karska14,Mottram14,SanJose16}. Rotational
diagrams using mostly PACS lines (Fig.~\ref{fig:n1333rotdiag}, middle)
indicate median water excitation temperatures of $\sim$140 K for
low-mass sources \citep{Herczeg12,Goicoechea12,Karska18} increasing up
to $\sim$250 K for high-mass sources \citep{Karska14}, with some
spread around the mean. The DIGIT survey finds an average value of
$\sim$190 K \citep{Green13}. Because water is subthermally excited,
these temperatures are lower limits to the kinetic temperature(s) of
the emitting gas. Also, because of high but varying optical depths of
the lines, these water excitation temperatures have a limited meaning.

\begin{figure}[bt]
  \centering
 \includegraphics[width=8cm]{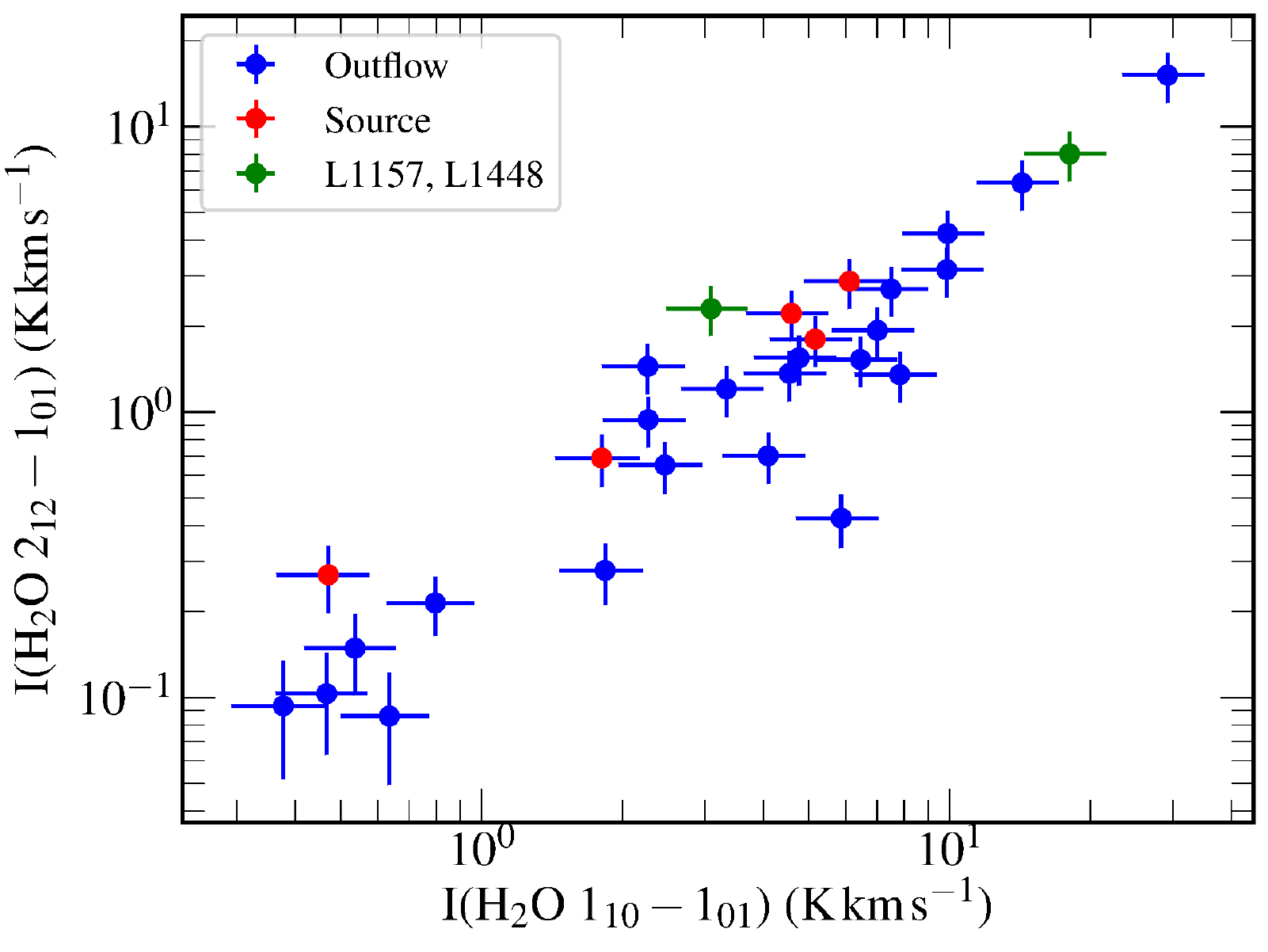}
 \caption{Correlation of the water $1_{10}-1_{01}$ 557 GHz (HIFI) and
   $2_{12}-1_{01}$ 1670 GHz/179 $\mu$m (PACS) line intensities convolved to
   a common angular resolution toward various outflow and source
   positions. The outflow data points (blue) are from
   \citet{Tafalla13}, whereas the source positions (red) use data from
   \citet{Kristensen12}. This figure demonstrates similar line ratios
   and thus similar conditions at the on and off source
   positions. Also, the two sources with strong EHV components, L1157
   and L1448-MM (green), do not stand out in this figure.}
         \label{fig:tafalla}
\end{figure}

Do these conclusions also hold at positions off source? For low-mass
sources, water vapor maps have been made only in two of the ground
state transitions (557 GHz and 1670 GHz/179 $\mu$m). The flux ratio of
these two lines is again very constant across the entire outflow
\citep{Tafalla13}. Moreover, as Figure~\ref{fig:tafalla} shows, the
protostellar positions do not stand out in this correlation,
suggesting similar conditions on and off source.

This good correlation between water lines holds not only for
integrated fluxes but even as a function of velocity.  HIFI
velocity-resolved observations of the different water transitions
with $E_{\rm up}$ ranging from 50 to 250 K reveal surprisingly
similar line profiles at the low-mass protostellar positions,
irrespective of excitation and beam size \citep{Mottram14}. The same
is often true for the intermediate and high-mass protostars for those lines
whose profiles are not affected by absorption
\citep{Herpin16,SanJose16} (Fig.~\ref{fig:averagespectra} in Appendix).

\subsubsection{Comparison with CO excitation}

\begin{figure}[bth]
  \centering
    \includegraphics[width=8cm,angle=0]{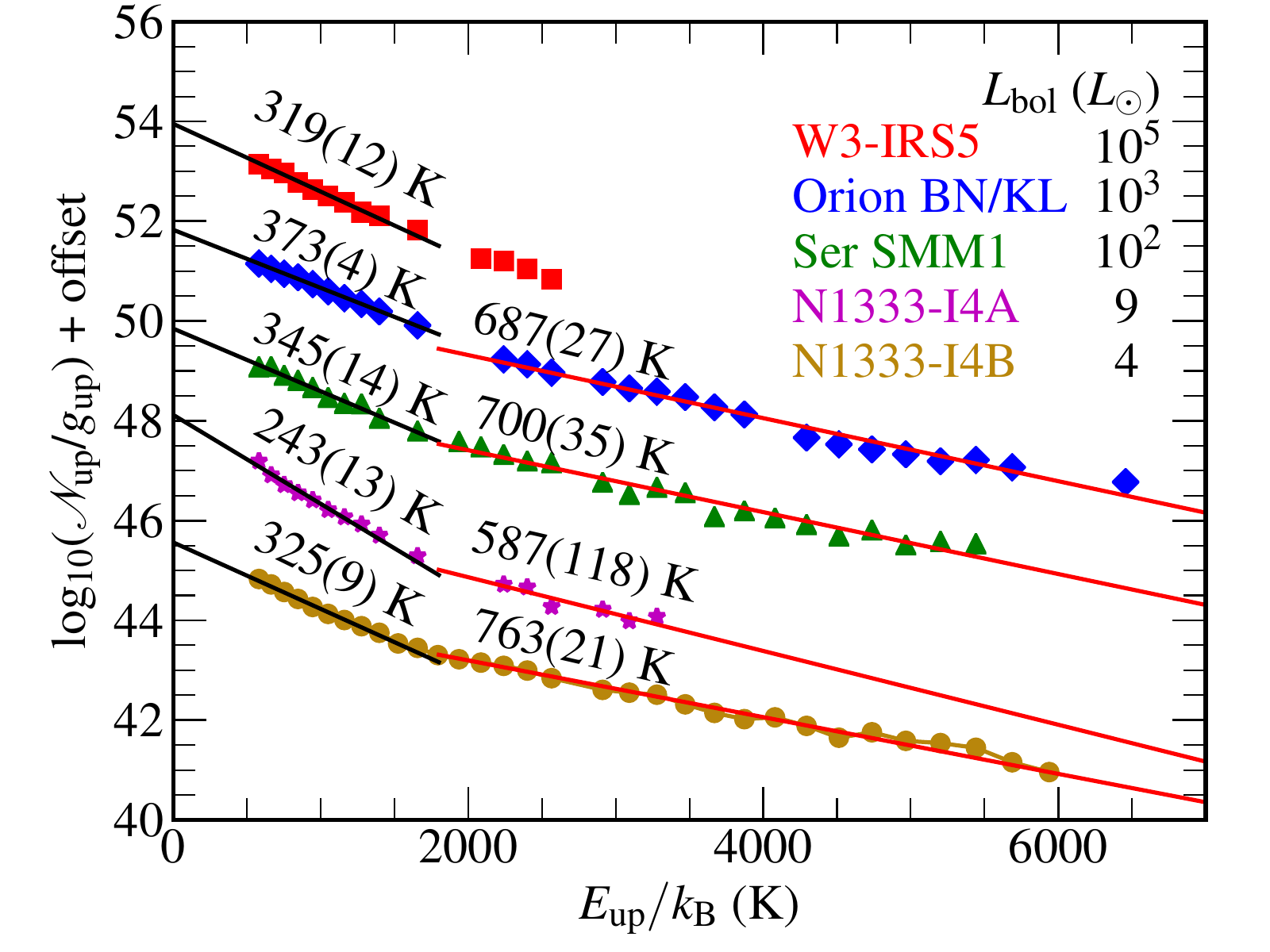}
    \caption{CO excitation diagrams from low to high-mass protostars
      based on (velocity-unresolved) PACS data. Two
      excitation temperature components of $T_{\rm rot}\sim$300 and
      $\sim$700 K are seen across the mass
      range. The Orion rotation diagram refers to Peak 1. Data from
      \citet{Herczeg12,Karska14,Goicoechea12,Goicoechea15}. }
          \label{fig:COrotdiag}
\end{figure}

Surveys of the CO rotational ladder for many protostars with PACS have
shown that its excitation for $J_{\rm up}>14$ can be universally
fitted by two rotational temperatures: $T_{\rm rot}\sim 300$~K (warm)
and $\sim$700~K (hot) (Fig.~\ref{fig:n1333rotdiag} top,
Fig.~\ref{fig:COrotdiag}), respectively, with about 80\% of the CO
flux originating in the warm component and 20\% in the hot component
\citep[e.g.,][]{Goicoechea12,Karska13,Green13,Green16,Manoj13,Lee13,Dionatos17,Karska18,Yang18}. This
constancy holds both across evolutionary stages as well as across mass
or luminosity \citep{Karska14,Matuszak15,Karska18}. The temperature of
the hot CO component shows a larger scatter between sources than the warm
component but always has $T > 500$~K.

A key step forward in the interpretation of these two CO rotational
temperature components was made by \citet{Kristensen17b} through
velocity resolved CO $J$=16--15 profiles, a transition also covered by PACS.
This CO line has very similar line profiles to those of water. The
broad component was shown to correspond to the
$T_{\rm rot}\sim$300 K component, and the medium-offset spot shocks
to the $\sim$700 K component.

Taken together, {\it Herschel} has shown that three ranges of CO lines
can be identified \citep{Goicoechea12,Yildiz13,Karska18}: (i) low-$J$
($J_{\rm up}<14$) probing warm quiescent and entrained outflow gas;
(ii) mid-$J$ ($J_{\rm up}$=14--24) probing the 300 K broad component;
and (iii) high-$J$ ($J_{\rm up}>24$) probing the hot 700 K shock
component.

\subsubsection{CO and water excitation analysis: inferred physical conditions}

\paragraph{On-source position.} To infer the physical conditions of the gas
from these observed rotational temperatures, non-LTE excitation
calculations need to be performed using programs such as RADEX
\citep{vanderTak07}. Such analyses have been primarily carried out for
LM and IM sources. For CO, it has been demonstrated that both
components require high densities ($>10^6$ cm$^{-3}$), with kinetic
temperatures $T_k$ of $\sim$300--400 and 700--1000 K, respectively
\citep{Goicoechea12,Karska18}. In some sources with particularly
bright CO lines extending to mid-infrared wavelengths, an even hotter
component with $T_k\approx 2500$ K may be needed.
A single hot (up to 4000 K) component at very subthermal excitation
conditions, $T_{\rm rot} << T_k$, can provide good solutions to the
entire CO excitation ladder as well \citep{Neufeld12}, but this
solution is not consistent with the multiple velocity components seen
for high-$J$ CO and H$_2$O line profiles \citep{Kristensen17b} and
will therefore not be considered further here.

Do these same conditions also reproduce the water excitation?
Analyses of the HIFI lines using RADEX for a single slab model
demonstrate that water is indeed collisionally and sub-thermally
excited. Best fits are obtained for densities in the range of
$10^5-10^8$ cm$^{-3}$, both for the broad component and the
medium-broad or offset spot shocks. Kinetic temperature is not well
constrained due to the limited range of energy levels covered with
HIFI but is typically a few hundred K. Moreover, to fit the absolute
fluxes, the size of the emitting area has to be small compared with the
beam size, only of order ten to a few hundred au \citep{Mottram14}.

PACS observed much higher excitation lines of water.  An early
analysis for the line-rich low-mass protostar NGC 1333 IRAS4B
(Fig.~\ref{fig:pacs_scans}, \ref{fig:COrotdiag}) resulted in a
single component fit with $T_k\approx 1500$ K, $n\approx 3\times 10^6$
cm$^{-3}$ and an emitting radius of $\sim$100 au ($<1''$) centered in
the blue outflow lobe located slightly off source \citep{Herczeg12}. Thermal
excitation at a much lower temperature of 170 K and very high
densities $n>10^{11}$ cm$^{-3}$ on source, as suggested based on
mid-infrared {\it Spitzer} observations \citep{Watson07}, was excluded
in this case. Similarly, modeling of the PACS spectral scan of the
Serpens SMM1 intermediate mass source gave $T_k\approx 800$~K,
$n\approx 5\times 10^6$ cm$^{-3}$ with an emitting region of radius
$\sim$500 au \citep{Goicoechea12}. Comparable conditions are found for
L1448-MM \citep{Lee13}.

In summary, all detailed analyses for low- and intermediate-mass
sources agree that the on-source CO and H$_2$O emission originates
from high density $n>10^6$ cm$^{-3}$, warm ($T_k \gtrsim 300$~K up to
1000 K) gas with a small emitting area. There is little evidence for a
very hot component of several thousand K based on far-infrared
data. For the more distant high-mass sources, the analysis is
complicated by the fact that shock and envelope emission are more
intermixed. However, if the RADEX analysis is limited to the water
wing emission, the conditions are inferred to be similar, albeit with somewhat
larger emitting areas up to 5000 au \citep[see \S 3.6
in][]{SanJose16}.

\paragraph{Outflow positions.} In contrast with the protostellar positions,
the velocity-resolved profiles of water lines at outflow spots in
low-mass sources can be markedly different from each other. 
Low excitation lines often show excess emission at high velocity compared
to the higher excitation lines. This can be interpreted in terms of
density variations, with the gas at low velocity being denser than
that at high velocity \citep{Santangelo12,Vasta12}.  Indeed, when the
water line intensities are consistently fitted together with spectrally
resolved high$-J$ CO and mid-IR H$_2$ observations, the physical
conditions of the two water components are as follows
\citep{Santangelo14b}: a warm component at $T_k \sim 300-1000$ K, $n
\approx 10^6-10^8$ cm$^{-3}$, and size $10-15''$; and a hot component
at $T_k > 1000$ K, $n \approx 10^4-10^5$ cm$^{-3}$, and size $1-5''$,
for sources at $\sim$200 pc (see also \citealt{Busquet14}).

These numbers are consistent with the analysis of the water excitation
based on the 557/1670 GHz line ratio at the outflow positions
(Fig.~\ref{fig:tafalla}), which gives pressures
$nT\approx 4\times 10^9$ cm$^{-3}$ K \citep{Tafalla13}. Such pressures
are about 4 orders of magnitude higher than the surrounding quiescent
gas. For $T_k<4000$ K, which is the temperature above which molecules
are collisionally dissociated, this implies $n>10^6$ cm$^{-3}$,
consistent with the warm component. The association with H$_2$
mid-infrared emission gives $T_k>300$~K as a lower limit.

Thus, while the physical parameters at the outflow positions are
similar to those inferred at the protostellar positions (warm and
dense), the emitting areas are much larger off source, with sizes up
to a few thousand au for the warm component. Also, temperatures can be
higher in the offset hot component. This compact hot component is
suggested to be associated with the jet impacting the surrounding
material in a bow shock, whereas the extended warm and dense component
originates from the weaker shocks further downstream.
Since shocks at off-source positions can expand in more directions
than close to the protostellar base, the larger emitting area and
somewhat lower densities found at the outflow positions are not in
contradiction.

\subsection{Water as tracer of envelope infall and expansion}
\label{sec:infall}

The absorption and narrow emission features probe the quiescent
envelope material. Their line profiles are asymmetric in a fraction of
sources (see, for example, Fig.~\ref{fig:water_mass} and
\ref{fig:h2o_masstime}), and indicate either infall motions (inverse
P-Cygni) or expansion (regular P-Cygni) on envelope scales (few
thousand au).  The incidence of infall profiles is higher in low-mass
Class 0 than Class I sources \citep{Kristensen12,Mottram17}. Still,
the majority of the Class 0 sources surprisingly do not show infall
motions in any molecular feature \citep{Mottram17}. The incidence of
expansion is similar for Class 0 and I. Taken together, the WISH +
WILL samples show infall in 7+6=13 sources and expansion in 6+2=8
sources, out of a total of 29+49=78 sources. The small sample of
intermediate mass sources shows primarily P-Cygni profiles indicative
of expansion.

The majority of the high-mass sources, about 2/3, show infall whereas
the remainder show expansion \citep{Herpin16,vanderTak19}. Infall
is seen not just in the outer cold envelope gas but also in warm dense
gas close to the protostar through H$_2^{18}$O absorption
\citep{Jacq16,vanderTak19}. High mass sources often have many more
narrow absorption lines offset from the source velocity due to low
density diffuse clouds along the line of sight, some of which may be
close by and infalling onto the high-mass core
\citep{Marseille10,vanderWiel10,Gerin16}. 

Quantitative modeling of the (outflow subtracted) water line profiles
provides mass infall rates from cloud to envelope of order
$10^{-5}-10^{-4}$ M$_\odot$ yr$^{-1}$ for low-mass protostars
\citep{Mottram13}, to $10^{-4}-10^{-2}$ M$_\odot$ yr$^{-1}$ for
high-mass protostars \citep{vanderTak19}. The latter values are
comparable to those derived from inverse P-Cygni profiles observed in
NH$_3$, either with HIFI or with the Stratospheric Observatory for
Infrared Astronomy (SOFIA) \citep{Wyrowski16,Hajigholi16}.

\subsection{Summary of physical components}

Table~\ref{tab:profiles} summarizes the profile characteristics and
their proposed origin.  The key point is that the bulk of the broad
water and high-$J$ CO ($J_{\rm up}>14$) emission go together and arise
in two physical components associated with outflows and shocks, seen
as $T_{\rm rot}$=300 and 700 K in the CO data and as the broad and
medium-broad offset components. The implied physical conditions for
both components are high densities ($n>10^5-10^8$ cm$^{-3}$) and with
warm ($T_k \approx 400$~K) and hot ($T_k \approx 1000$~K)
temperatures, respectively. Emitting sizes are small, only $\sim$100
au for low-mass sources up to 5000 au for distant high-mass sources.

The narrow emission and absorption lines probe the more quiescent envelope
gas.  Overall, the infall and expansion signatures highlight water’s
ability to pick up small motions along the line of sight on envelope
scales. This will prove to be extremely useful to constrain the
physical structure and abundance profiles of water and related
molecules in the cold outer parts of envelopes
(\S~\ref{sec:protocold}). Water is in this aspect more effective than
well-known envelope tracers such as low-$J$ C$^{18}$O or HCO$^+$.

\subsection{Ortho/para ratio}
\label{sec:orthopara}

The ortho/para ratio of water, or equivalently its ``spin'' temperature,
is in principle also a diagnostic of its physical and chemical
history. At low temperatures ($< 10$~K) the ortho/para ratio tends to
0 whereas at $T>50$ K, the equilibrium ortho/para ratio would be
3. Thus, low ortho/para ratios have been invoked to trace a low
formation temperature of H$_2$O on cold grain surfaces
\citep{Mumma11}. However, the ortho/para ratio is controlled by a
complex combination of processes
\citep[e.g.,][]{Tielens13,vanDishoeck13}, including possible changes
in the ortho/para ratio upon ice desorption \citep{Hama18}, and will
not be summarized here. Overall, water spin temperatures may tell
astronomers less about the water formation location than previously
thought.

Within the WISH project, the data are consistent with an ortho/para of
3 for warm water \citep[e.g.,][]{Herczeg12,Mottram14,Herpin16}. There
have been some claims of ortho/para ratios lower than 3 in other
sources \citep[e.g.,][]{Hogerheijde11,Choi14,Dionatos18} but
differences in optical depth of the ortho and para lines can also
mimic such low ratios if not properly accounted for
\citep{Salinas16}. In fact, a more detailed analysis of many water
lines observed with HIFI in the Orion Bar shows values close to 3
\citep{Putaud19}.

In summary, there is no convincing evidence for water ortho-to-para ratios
significantly below the LTE value of 3 to be present in the ISM.

\section{Shocks}
\label{sec:shocks}

Assuming that the bulk of the water emission observed by {\it Herschel}
originates from high temperature shocked gas, WISH+ allows tests of
chemical models of different shock types.  Key questions to address
are (i) whether the basic prediction that all volatile oxygen not
locked up in CO is driven into water at high temperatures holds, in
which case values of H$_2$O/H$_2 \approx 4\times 10^{-4}$ and
H$_2$O/CO$\approx (1.4-2)$ are expected (\S~\ref{sec:oxygenbudget}),
(ii) what the relative contributions of ice sputtering and gas-phase
(re)formation are to the total gaseous water abundance.

\begin{table*}
\caption{Summary of H$_2$O/H$_2$ abundance determinations using {\it Herschel} data in warm outflows and shocks}
\begin{tabular}{l l l c l l}
\hline 
\hline
Source & Type & Instrument & H$_2$O/H$_2$ & H$_2$ from & Reference \\
       &       &     & (10$^{-4}$) & \\
\hline

{\bf On source} \\ [5pt]
Several & LM & HIFI & 0.02 (LV+HV) & CO 16-15$^a$ & \citet{Kristensen17b} \\
AFGL 2591 & HM & HIFI & 0.001  & CO 16-15 & Kristensen, unpubl. \\
NGC1333 IRAS4B & LM & PACS & 1.0 & high-$J$ CO & \citet{Herczeg12}\\
Serpens SMM1 & LM/IM & PACS & 0.4 (hot) & high-$J$ CO & \citet{Goicoechea12} \\
NGC 7129 IRS & IM & HIFI/PACS & 0.2--0.3 & high-$J$ CO&
   \citet{Johnstone10} \\
             & & & & & \citet{Fich10} \\
NGC 6334I & HM & HIFI & 0.4 & mid-$J$ CO & \citet{Emprechtinger10} \\
DR21(OH) & HM & HIFI & 0.32$^b$ & high-$J$ $^{13}$CO
   & \citet{vanderTak10} \\ [5pt]

{\bf Outflow} \\ [5pt]
L1157 B1 & LM & HIFI & 0.008(LV) & mid-$J$ CO & \citet{Lefloch10} \\
     &     & & 0.8(HV) & & \\
L1157 B2/R & LM-OF & HIFI/PACS & 0.01 & warm H$_2$ & \citet{Vasta12} \\
L1448 B2 & LM-OF & HIFI/PACS & 0.03 (warm) & warm H$_2$ &\citet{Santangelo12,Santangelo13} \\
   &  &  & 0.03-0.1 (hot) & hot H$_2$ & \\
L1448  & LM & HIFI/PACS & 0.005-0.01  & warm H$_2$& \citet{Nisini13} \\
VLA1623 & LM-OF & HIFI & $<$0.01$^c$ & warm H$_2$ & \citet{Bjerkeli12} \\
Several B/R & LM/IM-OF & HIFI/PACS & 0.003 & warm H$_2$ & \citet{Tafalla13} \\
HH 54 & OF & HIFI & $<$0.14 (B) & warm H$_2$ & \citet{Santangelo14a} \\
L1157 B1 & LM-OF & HIFI/PACS & 0.007-0.02 (warm) & warm H$_2$ & \citet{Busquet14} \\
   &  & HIFI/PACS & 1--3 (hot) & hot H$_2$ &  \\
NGC 1333 I4A R2 & LM-OF & HIFI/PACS & 0.007-0.01 (warm) & CO 16--15$^a$ & \citet{Santangelo14b} \\
   &  &  & 0.3--0.7 (hot) &  &  \\
NGC 2071 B/R& IM-OF & PACS & 0.3-0.8 & warm H$_2$+high$-J$ CO 
    & \citet{Neufeld14} \\
Orion-KL & HM-OF & PACS & $\leq$0.01 (warm,hot) & high-$J$ CO$^a$ & \citet{Goicoechea15} \\
        &  &  & $>$2 (very hot) & very high-$J$ CO &  \\
\hline
\end{tabular}
Note: uncertainties claimed by the authors range are typically a factor of 2 to a few.
HV=high-velocity; LV=low; B=Broad component
velocity. In most cases, the abundance refers to the outflow at the
central source position. B and R indicated blue- and red-shifted outflow spots
(OF) offset from the source. \\
$^a$ assuming CO/H$_2 \approx 10^{-4}$; if CO/H$_2$=$2\times
10^{-4}$ advocated in \S~\ref{sec:oxygenbudget} is used, the water
abundances are increased by a factor of two. \\
$^b$ from para-H$_2$O assuming ortho/para=3 \\
$^c$ ortho-H$_2$O abundance. \\
\\
\label{tab:shocks}
\end{table*}

\subsection{Inferred water abundances in shocks}

\subsubsection{On-source positions}

Table~\ref{tab:shocks} summarizes measurements of the water abundance
in warm outflowing and shocked gas made by {\it Herschel}. A list of
pre-{\it Herschel} results, mostly using $SWAS$ or ISO, can be found
in Table~4 of \citet{vanDishoeck13}. {\it Herschel} has advanced the
field in several ways. First, thanks to HIFI, the different velocity
components (see \S~\ref{sec:profiles}) can now be quantified
separately. Although the precise origin of the broad component is
still under discussion (see Table~\ref{tab:profiles}), this does not
matter for the observational derivation of abundances, and this
component will therefore be denoted in this section as shock. A second
key advantage is the simultaneous measurement of (spectrally-resolved)
high$-J$ CO originating in the same gas. Moreover, {\it Spitzer}
observations of the mid-IR H$_2$ lines are available for several
off-source shock positions providing a direct measure of the warm
H$_2$ (S(1)-S(5), few hundred K) or hot H$_2$ (S(5)-S(9), $>$1000 K)
column densities. At the source position such data are alas not
available since the mid-IR lines are too extincted
\citep{Maret09}. The availability of high-$J$ CO or H$_2$ data removes
one of the largest uncertainty in the
$X$(H$_2$O)=$N$(H$_2$O)/$N$(H$_2$) abundance determinations, namely
the H$_2$ column (the denominator).  Early measurements often used the
line wings of low$-J$ CO lines to estimate the H$_2$ column, which
{\it Herschel} has now shown not to be appropriate
(\S~\ref{sec:waterdata}). Hence, any literature values using low-$J$
CO as reference are not included in Table~\ref{tab:shocks}.

The most reliable determination of the water abundance as a function
of velocity comes from the detailed study of 24 low-mass protostars by
\citet{Kristensen17b} using HIFI CO 16--15 spectra as the reference. A
remarkably constant H$_2$O/CO abundance ratio of 0.02 is found,
independent of velocity. This constant abundance ratio even holds for
the EHV ``bullet'' component such as seen in L1448-MM. Assuming
CO/H$_2$=$10^{-4}$, both the warm broad and hot spot shock components
are found to have surprisingly low water abundances, with
H$_2$O/H$_2$=$2\times 10^{-6}$.  Only very hot gas may have a higher
water abundance \citep{Franklin08}: using the H$_2$O and CO PACS lines
in line-rich low-mass sources, H$_2$O abundances up to $\sim 10^{-4}$
have been found for the hottest gas \citep{Goicoechea12}.

For high-mass protostars, few observations of CO 16--15 are available
as reference. For the case of AFGL 2591, the observed line intensity
ratio of H$_2$O 557 GHz / CO 16--15 is $\sim$ 0.1 in the outflow wing.
This translates to an abundance ratio H$_2$O/CO $\sim 10^{-3}$ when
taking beam dilution into account and following the same analysis as
for the low-mass sources (Kristensen, unpublished). An early analysis
of the DR 21 shock using $^{13}$CO 10--9 as reference found an
abundance ratio H$_2$O/CO $\sim 7\times 10^{-3}$ in the shock. Thus,
the H$_2$O abundance in high-mass outflows may be even lower than
that inferred for low-mass protostars, although not as low as found
by \citet{Choi15}. Indeed, \citet{vanderTak19} conclude based on
H$_2^{18}$O absorption in line wings toward high-mass protostars that
water abundances in outflows are an order of magnitude higher than in
the cold envelopes, where values of $10^{-9} - 10^{-8}$ have been
found \citep[][and \S~\ref{sec:protocold} and
\ref{sec:hotcore}]{Boonman03}. Taken together, the H$_2$O/H$_2$
abundances for high mass outflows are found to be $10^{-7}-10^{-6}$.

\begin{figure}
\centering
 \includegraphics[width=7.5cm]{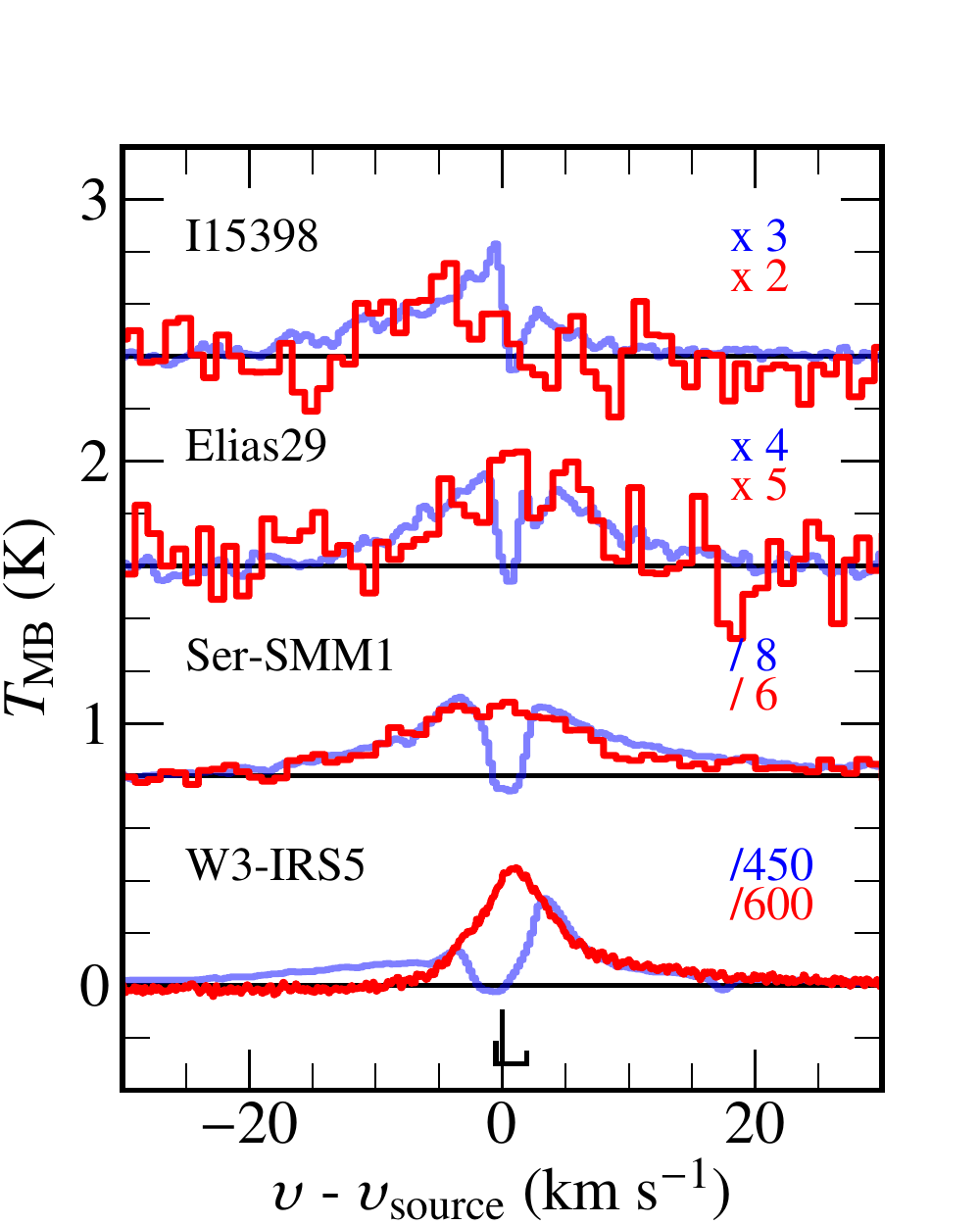}
 \caption{Comparison of HIFI spectra of the H$_2$O $1_{10}-1_{01}$ 557
   GHz line (blue) and the OH 1834 GHz triplet (red) toward a number
   of low and high-mass regions.  The splitting of the three OH lines
   in the triplet and their relative strength are indicated at the
   bottom of the figure. Data from Kristensen \& Wampfler
   (unpublished).}
\label{fig:ohh2o}
\end{figure}

\subsubsection{Outflow positions}

At the outflow positions, where direct measurements of the warm H$_2$
columns are available, the largest uncertainty stems from the
uncertainty in the physical parameters used to determine the H$_2$O
column densities, since the same data can be fitted with different
water column densities and filling factors.  Nevertheless, if the
H$_2$O, high-$J$ CO and H$_2$ data are fitted simultaneously, the
abundance of the warm H$_2$O component is better constrained.  The derived
abundance is universally found to be low, ranging between
$10^{-7}-10^{-6}$
\citep{Santangelo12,Santangelo13,Bjerkeli12,Nisini13,Tafalla13,Busquet14,Santangelo14a}. There
are suggestions that the abundance in the hot component ($>1000$~K)
component is higher, $> 5 \times 10^{-6}$, but this is more difficult to
constrain with the available data
\citep{Santangelo13,Santangelo14b}. Abundances close to $10^{-4}$ have
been found only for the very hot gas seen in the highest-$J$ H$_2$O
and CO lines but the lack of H$_2$O lines originating from very high
energy levels ($>$1000 K) in the {\it Herschel} data make it more
difficult to constrain this component
\citep{Lefloch10,Busquet14,Neufeld14,Goicoechea15}.

\citet{Melnick08} combined mid-infrared observations of pure
rotational H$_2$O and H$_2$ lines to infer water abundances
H$_2$O/H$_2$ in the range $ (0.2-6) \times 10^{-4}$ for shock
positions in the intermediate mass source NGC 2071 offset from the
source position. These mid-infrared data probe higher excitation
H$_2$O levels than the {\it Herschel} data and are thus better suited
to test the water predictions for $>$1000~K gas. The uncertainty in
the inferred water abundances reflects the uncertainty in the gas
density in the shock. \citet{Leurini14b} derived a fractional column
density ratio H$_2$O/H$_2$ $\approx (1.2-2.5)\times 10^{-5}$ for the
IRAS 17233–3606 outflow by modeling consistently both {\it Herschel}
H$_2$O and H$_2$ 2 $\mu$m infrared data, finding a maximum fractional
water abundance in the shock layer of $10^{-4}$.

Independent evidence that H$_2$O does not contain all of the oxygen
comes from mid-infrared absorption line observations with ISO and
SOFIA of hot water and CO along the lines of sight to massive
protostars, many of them the same sources as included in the WISH
sample \citep{vanDishoeck96,Boonman03h2o}. In particular, high
spectral resolution absorption line observations at 6 $\mu$m using
SOFIA-EXES give $N$(H$_2$O)/$N$(CO)$\approx 0.7$ for gas at $\sim
1000$~K, a factor of 2--3 less than expected if all oxygen is in water
\citep{Indriolo15exes}.

In conclusion, the water abundances observed by {\it Herschel} are
two--three orders of magnitude lower than predicted by standard
high temperature (shock) models that drive all volatile oxygen into
water. This conclusion holds in both velocity components on-source,
and at off-source positions as well. Water abundances as high as
$\sim 10^{-4}$ are perhaps only attained in the very hot ($> 1000$ K)
gas that is not well probed by {\it Herschel}. Even then, there is no
firm observational evidence yet for any shock associated with outflows
in which H$_2$O/H$_2$ is as high as $4\times 10^{-4}$, the value
expected if all volatile oxygen is in H$_2$O. See \S~\ref{sec:obudget}
and Appendix \ref{app:obudget} for further discussion.

\subsection{H$_2$O vs OH}
\label{sec:oh}

A key intermediate species in the O $\rightarrow$ H$_2$O reaction is
OH, so given the fact that the H$_2$O abundance is found to be low,
does this imply that the OH abundance is high?  WISH targeted a number
of OH transitions with PACS, particularly the low-lying transitions at
119, 84 and 163 $\mu$m (Fig.~\ref{fig:pacscomp}), and covered low-,
intermediate-, and high-mass sources. The PACS instrument did not
resolve the hyperfine components of the OH transitions (the splitting
is typically $<$ 5 km s$^{-1}$), but it did resolve the $\Lambda$
doubling.  Strong indications for an enhanced OH abundance come from
the observed OH/H$_2$O flux ratio in a large number of low-mass
protostars, which is found to be much higher than expected based on models of
shocks propagating into dark molecular clouds
\citep{Karska14Perseus,Karska18}. Strong OH emission is also detected
with PACS for high-mass protostars \citep{Karska14}.

Apart from these velocity-unresolved PACS observations, the 1834 and
1837 GHz OH triplets, which make up the 163 $\mu$m lines, were
observed with HIFI and detected toward W3 IRS5 and Ser SMM1
\citep{Wampfler11,Kristensen13}. The 1837 GHz triplet is completely
resolved toward W3 IRS5 and consists of narrow ($\sim$ 5 km s$^{-1}$)
hyperfine components superposed on top of a broad ($\sim$ 20 km
s$^{-1}$) outflow component. Toward Ser SMM1, the 1834 GHz transition
consists of a single broad component, which appears blue-shifted
($\sim$ 10 km s$^{-1}$) (Fig.~\ref{fig:ohh2o}).  The hyperfine
splitting is actually smaller for this triplet than for the 1837
triplet. The 1837 GHz line was also observed toward a number of other
low-mass sources within WISH but not detected. Comparing their limits
with the PACS 163 $\mu$m detections implies that the lines for these
sources must also be broad (FWHM at least 10-15 km s$^{-1}$)
\citep{Wampfler10}.

As part of WISH+, the COPS-HIFI program observed the 1834 GHz triplet
together with CO 16--15 for 24 sources \citep[][Kristensen \& Wampfler
unpublished]{Kristensen17b} (Figure~\ref{fig:ohh2o}). Where detected,
the OH line is broad and follows H$_2$O. The OH abundance with respect
to H$_2$O in the broad outflow component has been constrained to $>$
0.03 \citep[][Kristensen \& Wampfler, unpublished]{Wampfler11}.  For
comparison, OH/H$_2$O abundance ratios $>1$ have been found in
classical PDRs such as the Orion Bar where UV photodissociation
controls the chemistry \citep{Goicoechea11}.

\begin{figure}
\centering
 \includegraphics[width=9.0cm]{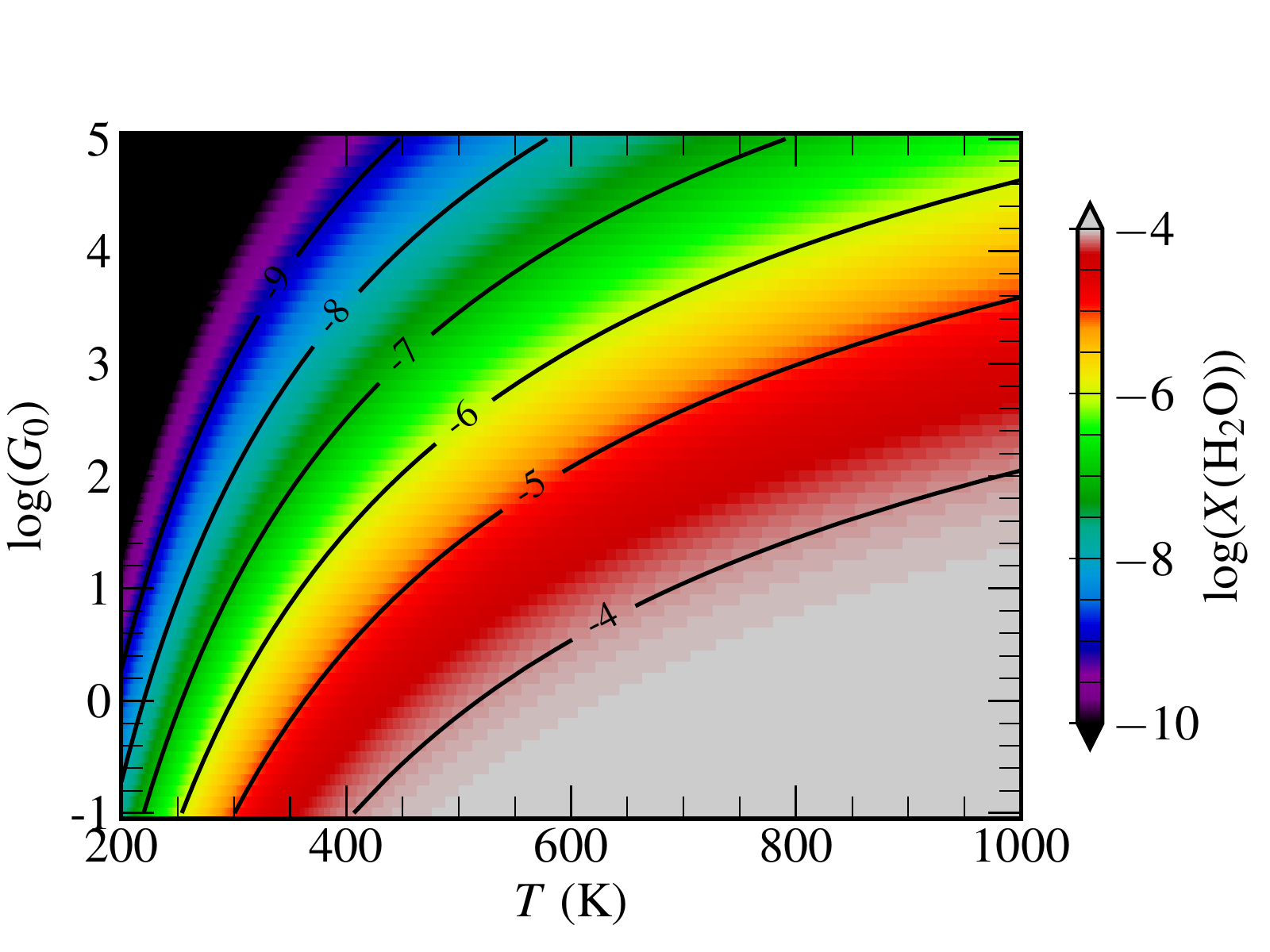}
 \caption{ Water abundance with respect to H$_2$  as function of
   temperature and UV radiation field using a simplified
   high-temperature chemistry model. The assumed density is $10^6$
   cm$^{-3}$; for higher or lower densities the abundance curves shift
   up or down, respectively, but the trend is the same. Based on
   \citet{Kristensen17b}.}
\label{fig:water_TUV}
\end{figure}

The OH line luminosity increases from Class 0 to Class I low-mass
protostars when compared to the total FIR luminosity
\citep{Wampfler13,Karska18} (Fig.~\ref{fig:water_cooling}). To first
order, this suggests that the OH abundance increases with evolutionary
stage.  As the sources evolve, the UV radiation from the accreting
protostar is able to penetrate deeper into the outflow, adding to
photodissociation of H$_2$O and thus increasing the abundance of
OH. However, in no case has OH been found to lock up a significant
fraction of the oxygen, perhaps implying that gas temperatures remain
below the activation threshold of the critical reactions (see also
\S~\ref{sec:UVshocks}, Fig.~\ref{fig:water_TUV}).

\subsection{Hydrides: Evidence for UV-irradiated outflow cavity walls}
\label{sec:hydrides}

The cold outer protostellar envelopes are punctured by outflow
cavities on scales of a few thousand au. Besides shocks stirring up
the gas, UV radiation can escape through the cavities and impinge on
the walls thereby heating them up \citep{Spaans95}. Thus, outflow cavity
walls are a source of bright far-infrared line radiation and simple
hydrides such as CH$^+$, OH$^+$ and H$_2$O$^+$ turn out to be a good
probe of them. These hydrides are readily detected in the WISH data
for many sources, often in absorption, with their velocity offsets
pointing to a spot shock or a more quiescent cavity wall origin.  The
results of the WISH hydride subprogram are summarized in
\citet{Benz16} with the analysis building on detailed modeling of the
effects of UV and X-rays on protostellar envelopes by
\citet{Stauber05} and \citet{Bruderer09b,Bruderer10mod}.

In brief, the observed hydride fluxes and flux ratios (e.g.,
CH$^+$/OH$^+$) can be reproduced by 2D models of UV illuminated outflow
cavity walls on scales of the {\it Herschel} beam. The implied UV
fluxes are up to a few$\times 10^2$ times the interstellar radiation
field, at distances of $\sim$2000 au and $\sim$0.1 pc from low- and
high-mass sources, respectively.  If the FUV flux required for
low-mass objects originates at the central protostar, a substantial
FUV luminosity, up to 1.5 L$_\odot$, is required.  Alternatively, some
of the UV can be produced by the high-velocity shocks themselves. For
high-mass regions, the FUV flux required to produce the observed
molecular ratios is smaller than the unattenuated flux expected from
the central object(s) at that radius, implying some extinction in the
outflow cavity or, alternatively, bloating of the protostar
\citep{Hosokawa10}. Another important conclusion is that there is no
molecular evidence for X-ray induced chemistry in low-mass objects on
the observed scales of a few$\times$1000 au, even though X-rays may be
important in destroying water on smaller scales (see
\S~\ref{sec:hotcore}).

\subsection{Implications of low water abundances in shocks}

\subsubsection{Previous model predictions}

Early shock models under dark cloud conditions find complete
conversion of all volatile oxygen into water for any shock that has
temperatures greater than $\sim$250--400 K. Above this temperature,
the O + H$_2$ and OH + H$_2$ reactions drive rapid water formation
\citep{Draine83,Kaufman96,Flower10} (Fig.~\ref{fig:waternetwork}). The
detection of high water abundances in the Orion shock by ISO was cited
as a confirmation of this prediction \citep{Harwit98}. For a
non-dissociative shock, the temperature just behind the shock is
typically $T_s = 375 b{_M}^{-0.36} [v_s/10\ {\rm km \ s^{-1}}]^{1.35}$
with the magnetic field strength $b_M=B/(n_{\rm H}$/cm$^{-3}$)$^{1/2}$
$\mu$G usually taken to be standard $b_M=1$ \citep{Neufeld06}. This
then implies shock velocities $v_s=10-15$ km s$^{-1}$ to obtain the
temperatures of a few hundred K required to drive the reactions.

If instead most oxygen is locked up in water ice in the pre-shocked
gas, higher shock velocities are needed to sputter water ice,
typically $\gtrsim 15$ km s$^{-1}$. Such shocks have peak temperatures
of $T_s\approx 650$~K \citep{Draine83,Jimenez08,Gusdorf11}.  Standard
theoretical models for non-dissociative shocks predict the complete
vaporization of ice mantles resulting in H$_2$O/CO$\sim (1.4-2)$
(\S~\ref{sec:oxygenbudget}).  Even if the ice comes off as atomic O or
OH, it should be quickly converted to H$_2$O at these temperatures and
densities, if shielded from dissociating radiation.

In dissociative $J$-type shocks much higher temperatures $> 10^4$ K
can be reached at shock velocities $\gtrsim 25$ km s$^{-1}$
\citep{McKee80}, which are sufficiently high to collisionally
dissociate H$_2$ and H$_2$O. These molecules, together with CO,
subsequently reform downstream in the cooling shock gas resulting in
similar abundances.

In conclusion, the low water abundances and low H$_2$O/OH intensity
ratios found by WISH+ are clearly in conflict with the standard
shock model predictions.

\subsubsection{UV-irradiated shocks}
\label{sec:UVshocks}

The above findings have led to the development of a new class of UV
irradiated shock models \citep{Melnick15,Godard19}. In such models,
the H$_2$O abundance is lowered by photodissociation into OH and
O. The UV radiation can either come from the shock itself, if fast
enough \citep{Neufeld89b}, or it can be external, for example from the
disk-star accretion boundary layer \citep{Spaans95}.

Lowering the water abundance in the shocked gas by photodissociation
is the simplest effect of UV radiation.  If the pre-shocked gas is
also UV irradiated and some of its H$_2$ dissociated, the entire shock
structure changes: the UV photons increase the atomic and ionization
fraction in the pre-shocked gas, resulting in a shock layer that is
more compressed (smaller) and hotter for a given shock velocity and
density. The UV radiation can also unlock more oxygen from water ice
in the pre-shocked gas and thus increase the amount of atomic oxygen
that can be converted into water.

Results from a limited set of irradiated shock models by M.\ Kaufman
(unpublished) for UV fields of $G_0$=0.1--10 times the standard
interstellar radiation field and pre-shock densities of $10^4-10^6$
cm$^{-3}$ indeed show much better agreement with {\it Herschel} data
than the older models when $G_0>1$ \citep{Karska18}. Such UV fluxes
are plausible. In particular, enhanced UV fields with
$G_0$=$10^2-10^3$ close to the protostar on scales of $\sim$1000 au
illuminating the outflow cavity walls have been independently inferred
from bright narrow extended mid-$J$ $^{13}$CO emission lines
\citep{Spaans95,Yildiz12,Visser12,Lee13,Yildiz15}. Moreover, as noted
above (\S~\ref{sec:hydrides}), the presence of several hydrides such
as CH$^+$ and OH$^+$ for both low- and high-mass protostars obtained
as part of WISH points to UV-irradiated outflow walls with similar
$G_0$ values \citep{Bruderer09b,Bruderer10mod,Benz10,Benz16}.

Figure~\ref{fig:water_TUV} illustrates the results of a simple
chemical model balancing the high temperature formation of water
through the reactions of O and OH with H$_2$ with photodissociation of
H$_2$O, where the UV radiation field is given by its enhancement $G_0$
with respect to the standard interstellar radiation field
\citep[see][for details]{Kristensen17b}.  For $G_0$ values of
$10^2-10^3$, the expected water abundance is indeed in the
$10^{-7}-10^{-6}$ range at temperatures of several hundred K.
Temperatures $>$1000 K are needed to drive the bulk of the oxygen into
water and reach water abundances as high as $10^{-4}$ if UV radiation
is present.

\subsection{High $T$ chemistry vs sputtering: SiO, CH$_3$OH, NH$_3$}
\label{sec:sputtering}

\begin{figure}
  \centering
    \includegraphics[width=8cm]{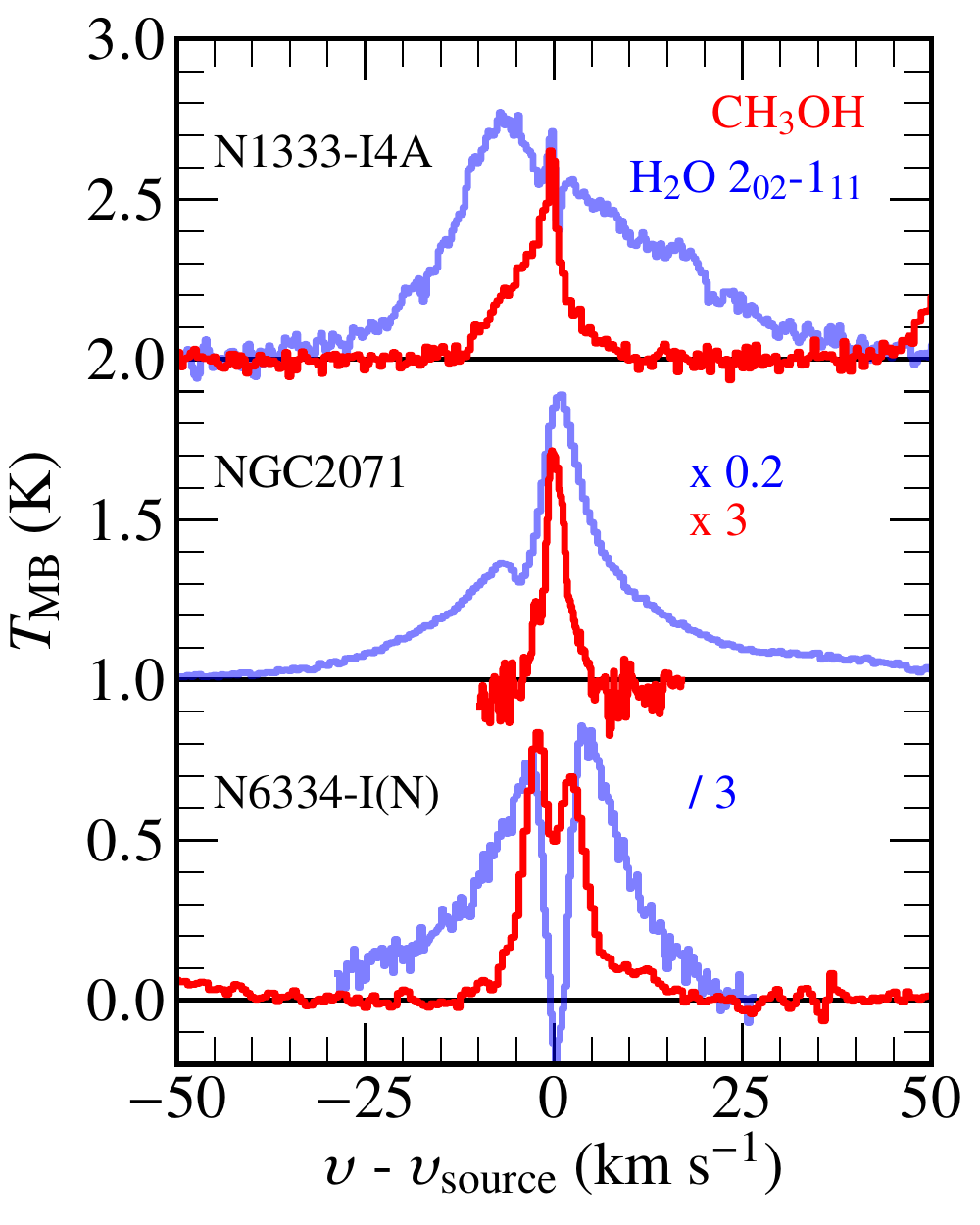}
    \caption{Comparison of the H$_2$O $2_{02}-1_{11}$ 988 GHz line
      with that of an ice species, CH$_3$OH, for a low, intermediate
      and high-mass source. In all cases the CH$_3$OH line
      is seen to be much narrower than that of H$_2$O. For NGC 1333 IRAS4A, the
      $7_0-6_0$ line ($E_{\rm up}$=78 K) at 338.4 GHz observed with
      the JCMT (15$''$ beam) is shown; for NGC 2071 the $6_2-5_1$ line
      ($E_{\rm up}$=86 K) at 766.7 GHz observed with HIFI is included
      \citep{vanKempen14} and for NGC 6334-I(N) the $5_1-4_0$ line
      ($E_{\rm up}$=56 K) at 538.5 GHz observed with HIFI is
      presented.}
         \label{fig:water_methanol}
\end{figure}

\begin{figure}
   \centering
    \includegraphics[width=7cm]{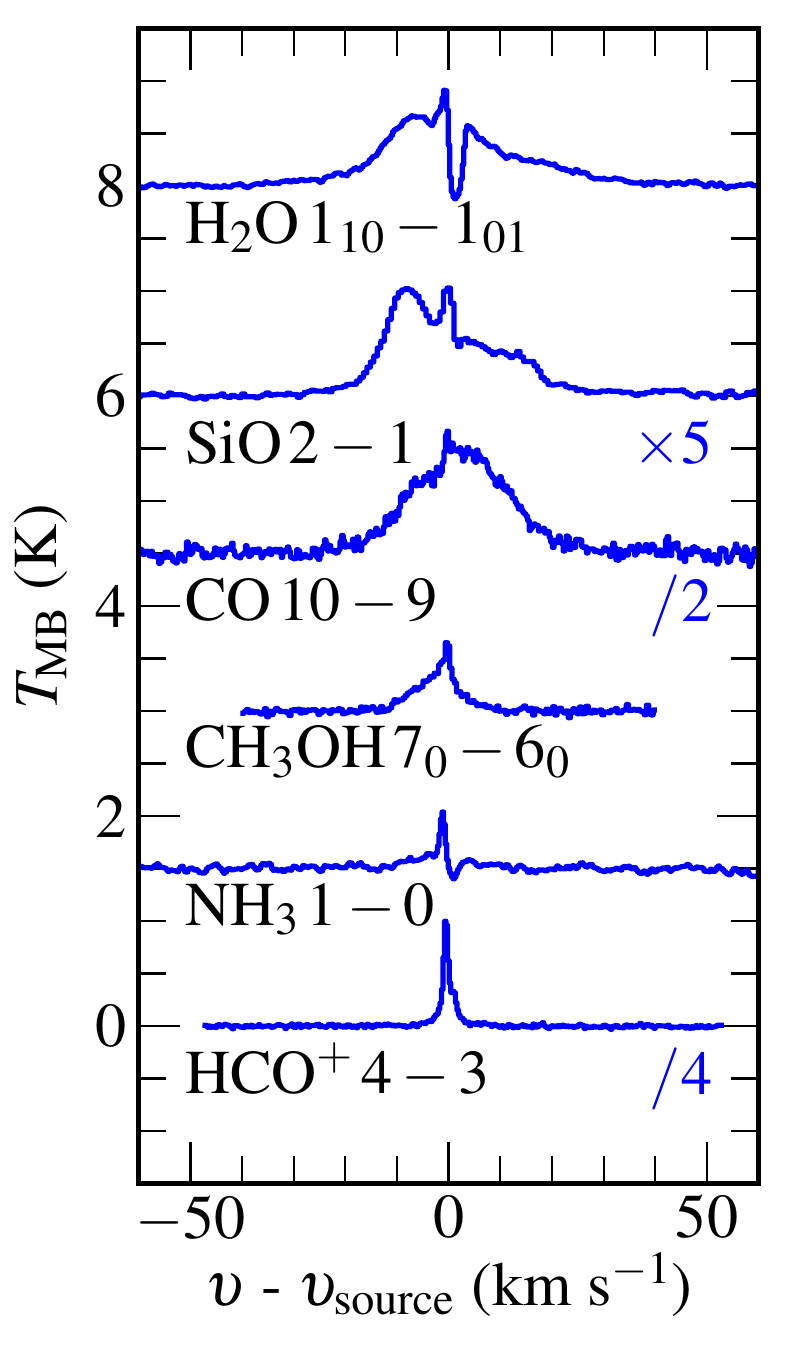}
    \caption{Comparison of the line profile of water with those of
      other chemically or physically related molecules for the
      low-mass protostar NGC 1333 IRAS4A. SiO and CO 10--9 trace shock
      and outflow; NH$_3$ and CH$_3$OH are both ice mantle products,
      similar to H$_2$O. The fact that their profiles are narrower suggests
      that H$_2$O is not just sputtered from ices but also produced by
      high temperature chemistry in the gas. The anticorrelation of
      the HCO$^+$ abundance with that of water is illustrated by its
      lack of broad line profiles.  }
         \label{fig:water_other}
\end{figure}

To what extent is the water seen in shocks produced by high
temperature gas-phase chemistry, as assumed in
Fig.~\ref{fig:water_TUV}, versus ice sputtering?  Shocks with
velocities of 15 km s$^{-1}$ needed for water ice sputtering are
observed for most embedded sources, except perhaps for the low-mass
Class I objects. To better distinguish the two mechanisms and isolate
the contribution due to sputtering, it is useful to compare the H$_2$O
line profiles with those of other abundant ice mantle constituents,
most notably CH$_3$OH and NH$_3$.

Figure~\ref{fig:water_methanol} compares H$_2$O and CH$_3$OH line
profiles for a low, intermediate and high mass protostar, observed
with {\it Herschel}-HIFI or from the ground in a similar beam
\citep{Suutarinen14,vanKempen14,Herpin16}. The H$_2$O profiles clearly
extend to much higher velocities than those of CH$_3$OH, with the
latter profiles dropping to zero beyond $\pm$15 km s$^{-1}$ from
source velocity, independent of the source luminosity. A quantitative
analysis shows one order of magnitude decrease in the CH$_3$OH/H$_2$O
column density ratio as the velocity increases to 15 km s$^{-1}$
\citep{Suutarinen14}. A similar decrease in the CH$_3$OH/CO ratio is
found, demonstrating that CH$_3$OH is destroyed in higher velocity
shocks, and that not just H$_2$O increased at larger
velocities. A similar lack of CH$_3$OH (and H$_2$CO) emission compared
with H$_2$O at higher velocities has been seen at shock positions off
source \citep[e.g.,][]{Codella10,Busquet14,Vasta12}.

These data suggest that sputtering of ices does indeed occur but that
the process either shuts off at velocities above 15 km s$^{-1}$, or
that any CH$_3$OH that is sputtered at higher velocities is
destroyed. Destruction of molecules can occur either during sputtering
or by reactions with atomic H in the shock. The fact that H$_2$O is
seen up to much higher velocities implies that either H$_2$O is not
destroyed (see below) or that gas-phase formation of H$_2$O must be
active above $\pm$10 km s$^{-1}$ to re-form any water destroyed during
sputtering or by other processes such as UV photodissociation
(Fig.~\ref{fig:water_TUV}). The data also suggest that sputtering
takes place at somewhat lower velocities than the expected 15 km
s$^{-1}$, or that the sputtered molecules have been slowed down
since being released into the gas.

For NH$_3$, the same phenomenon is observed
(Fig.~\ref{fig:water_other}). NH$_3$ is also found to emit only at low
velocities up to $\pm 15$ km s$^{-1}$ at shock positions off source
\citep[e.g.,][]{GomezRuiz16}. Since NH$_3$ has similar excitation
requirements as H$_2$O, this must be a chemical effect.  Similar to
the case for CH$_3$OH, destruction by atomic H in the shock is
proposed \citep{Viti11}. However, the barrier for the H + NH$_3$
reaction is about 5000 K, much higher than that for H + CH$_3$OH of
about 2200--3000 K.  The barrier for H + H$_2$O destruction is even
higher, $\sim 10^4$ K, allowing H$_2$O to survive following
sputtering. If only temperature-sensitive destruction would play a
role, the relative abundances of these three molecules could provide a
thermometer for the higher velocity gas. More likely, however,
high-temperature gas-phase formation of H$_2$O in the high velocity
gas controls the difference between these three species.

Shocks are also expected to sputter silicates from grain mantles
and cores for shock velocities above 20--25 km s$^{-1}$, with SiO
long known to be one of the best tracers of shocks
\citep[e.g.,][]{MartinPintado92,Caselli97,Schilke97,Jimenez08}.
How do the SiO and H$_2$O profiles compare? 
Figure~\ref{fig:water_other} presents the case of NGC 1333 IRAS4A showing
similarly broad profiles for the two molecules, as also found in some
other sources \citep[e.g.,][]{Leurini14b}.
In other sources such as L1448-MM, SiO is only found in the EHV
bullets but not in the broad component \citep{Nisini13}. In yet other
cases, SiO is found at intermediate velocities
\citep[e.g.,][]{Vasta12,Busquet14}. One explanation for L1448-MM is
that the jet gas is rich in atomic Si originating from its dust-free
launch position in the inner disk.  This then leads to SiO formation
in the jet by gas-phase processes, a mechanism proposed originally by
\citet{Glassgold91} and revisited theoretically by \citet{Tabone20},
not because of sputtering of ambient dust grains.  Whether SiO is also
seen in the broad component then depends on the details of the
wind-cloud interaction.  An alternative option is that these
differences reflect the time evolution of the SiO profiles going from
high to low velocities as the shocked gas slows down, as modeled by
\citet{Jimenez09}. Spatial evidence for SiO evolution from high to
lower velocities with distance from the source (or equivalently, time)
is found in ALMA images of outflows \citep{Tychoniec19}. Clearly,
H$_2$O line profiles are less sensitive to such time evolution
effects.

\subsection{H$_2$O vs HCO$^+$}

As Figure~\ref{fig:waternetwork} shows, HCO$^+$ is effectively
destroyed by reactions with H$_2$O so that an anticorrelation between
these two species is expected, both in position and velocity
\citep[e.g.,][]{Phillips92,Bergin98}. This anticorrelation has
recently been demonstrated observationally through mm interferometric
images of H$_2^{18}$O (or CH$_3$OH as a proxy) and H$^{13}$CO$^+$
\citep{Jorgensen13,vantHoff18hcop}.  These lines trace quiescent warm
envelope gas, however, not the shocked gas.

The broad shocked H$_2$O gas line profiles observed with {\it
  Herschel}-HIFI can be compared with HIFI HCO$^+$ $J$=6--5 profiles,
covered serendipitously in the 557 GHz setting, or with JCMT HCO$^+$
$J$=4--3 profiles from low- to high-mass sources convolved to the {\it
  Herschel} beam \citep{Carney16}. Figure~\ref{fig:water_other} shows
that HCO$^+$ generally avoids the high velocities and is limited to
$\pm$5 km s$^{-1}$ around source velocity, although in some sources a
weak underlying component out to $\pm$10 km s$^{-1}$ is seen
\citep[e.g.,][]{Kristensen10,SanJose15,Carney16,Benz16,Mottram17}.
HCO$^+$ thus seems to trace primarily parts of the envelope that are
away from locations in the outflow where water is enhanced, consistent
with the chemical expectations.

In summary, the combination of H$_2$O data with those of other species
confirm the chemical schemes outlined in Fig.~\ref{fig:waternetwork},
including the anticorrelation with HCO$^+$. Both high temperature
chemistry and ice sputtering contribute to the water abundance, with
ice sputtering limited to the low velocities.  UV radiation reduces
the water abundance in outflows to orders of magnitude below the
expected abundance of H$_2$O/H$_2$=$4\times 10^{-4}$.

\section{Cold dense pre-stellar cores}
\label{sec:prestellar}

{\it Herschel}-HIFI was unique in its ability to obtain
velocity-resolved water line profiles in cold dense clouds, and
will remain so in the coming decades. Such data allowed a much deeper
analysis than just deriving beam-integrated column densities, as is
normally done from spectra taken at a single position.  In fact, the
detailed line profiles combined with water's ability to probe motions
at a fraction of a km s$^{-1}$ make it possible to reconstruct the
actual water abundance profile as a function of position in the
core, in spite of not spatially resolving or mapping the cloud. This
allows stringent testing of the primary chemical processes controlling
the gaseous water abundance.
At the same time, constraints on physical parameters that are
difficult to determine otherwise are obtained, most notably the
external and internal UV radiation fields that control 
photodesorption and photodissociation \citep{Caselli12,Schmalzl14}.

A similar analysis can be carried out for other species, in particular
NH$_3$ observed with {\it Herschel-}HIFI. Since NH$_3$ is also
abundant in ices, comparison with H$_2$O can be insightful into how,
where and why the nitrogen chemistry in dark clouds is so different
from that of other species \citep{HilyBlant17,Caselli17}.

This section describes the results of the water abundance profile analysis
for dense cores prior to star formation combined with our new analysis
of the NH$_3$ profile; the next section does so for the cold outer
parts of protostellar envelopes which have an internal heat source.
In both cases, simplified chemistry networks as described in
\S~\ref{sec:simplechemistry} are coupled with a physical structure of
the source. These chemistry models can be run either in steady-state
or in a time-dependent mode at each position in the cloud.  For
pre-stellar cores at densities $>10^6$ cm$^{-3}$ the chemistry is
faster than the free-fall time of 0.03 Myr, so the chemistry reaches
steady-state before the physical conditions change \citep{Keto14}. In
the time-dependent case, abundances at $t=0$ need to be assumed, with
results depending on whether oxygen is initially in atomic or
molecular form.

\subsection{Observations and column densities}
\subsubsection{Low-mass cores}

Within the WISH program, only one low-mass pre-stellar core, L1544,
was observed deep enough with HIFI in the ground-state o-H$_2$O line
for detection and detailed analysis (13.6 hr integration)
\citep{Caselli12}.  The line is very weak but shows an inverse P-Cygni
profile. Assuming an ortho/para ratio of 3, the inferred H$_2$O column
for L1544 integrated over the entire line of sight is
$> 1 \times 10^{13}$ cm$^{-2}$ from the absorption part of the
feature. This corresponds to a fractional gaseous water abundance
H$_2$O/H$_2$ $> 1.4\times 10^{-10}$, with the lower limit stemming
from the moderate optical depth of the line. The fact that water is
seen in emission in the blue-shifted part of the line profile implies
high central densities, about $10^7$ cm$^{-3}$. The same deep spectrum
also includes the o-NH$_3$ $1_0-0_0$ transition, showing clear
emission lines and resolving the hyperfine structure for the first
time in space \citep{Caselli17}.

A second starless core observed in WISH, B68, has a slightly less
stringent upper limit of $N$(o-H$_2$O)$< 2.5\times 10^{13}$ cm$^{-2}$,
or a fractional abundance H$_2$O/H$_2 <2\times 10^{-9}$. Both of these
results re-inforce the original conclusions from $SWAS$ and $Odin$
\citep{Bergin02,Klotz08} that the gaseous water abundance is very low
in cold clouds and that most of the water must be locked up as water
ice.

Two additional very deep integrations (15 hr each) were obtained in an
OT2 program (PI: P.\ Caselli; Caselli et al., unpublished) on the
low-mass cores L183 and Oph H-MM1. Both show weak absorption features,
with that for Oph H-MM1 included in Fig.~\ref{fig:h2o_masstime}.  As
for B68, their lower central densities compared with L1544 do not
favor water emission but their central continuum emission is strong
enough to absorb against. Upper limits on the gaseous H$_2$O abundance
integrated along the line of sight are again low, $<10^{-9}$.

\subsubsection{High-mass cores}

Four high mass cores that were thought to have no associated star
formation prior to {\it Herschel} launch were targeted within
WISH. The low-lying H$_2$O lines are detected in all four sources
\citep{Shipman14}. Three of the four sources show outflow wings,
however, demonstrating that they are not truly
pre-stellar. Nevertheless, after subtraction of the outflow component,
their inverse P-Cygni profiles greatly resemble those of L1544 and of
the low-mass protostars discussed below, but over a wider range of
velocities. The fourth source, G11.11-0.12-NH$_3$, shows no signs of
star formation. It has narrow  (FWHM$\approx 2$ km s$^{-1}$)
water absorption lines (Fig.~\ref{fig:h2o_masstime}), even in H$_2^{18}$O,
that allow an accurate water column density determination. Assuming
ortho/para=3, the inferred abundance is again low, $4\times 10^{-8}$.

\begin{figure}
  \centering
\includegraphics[width=8.5cm]{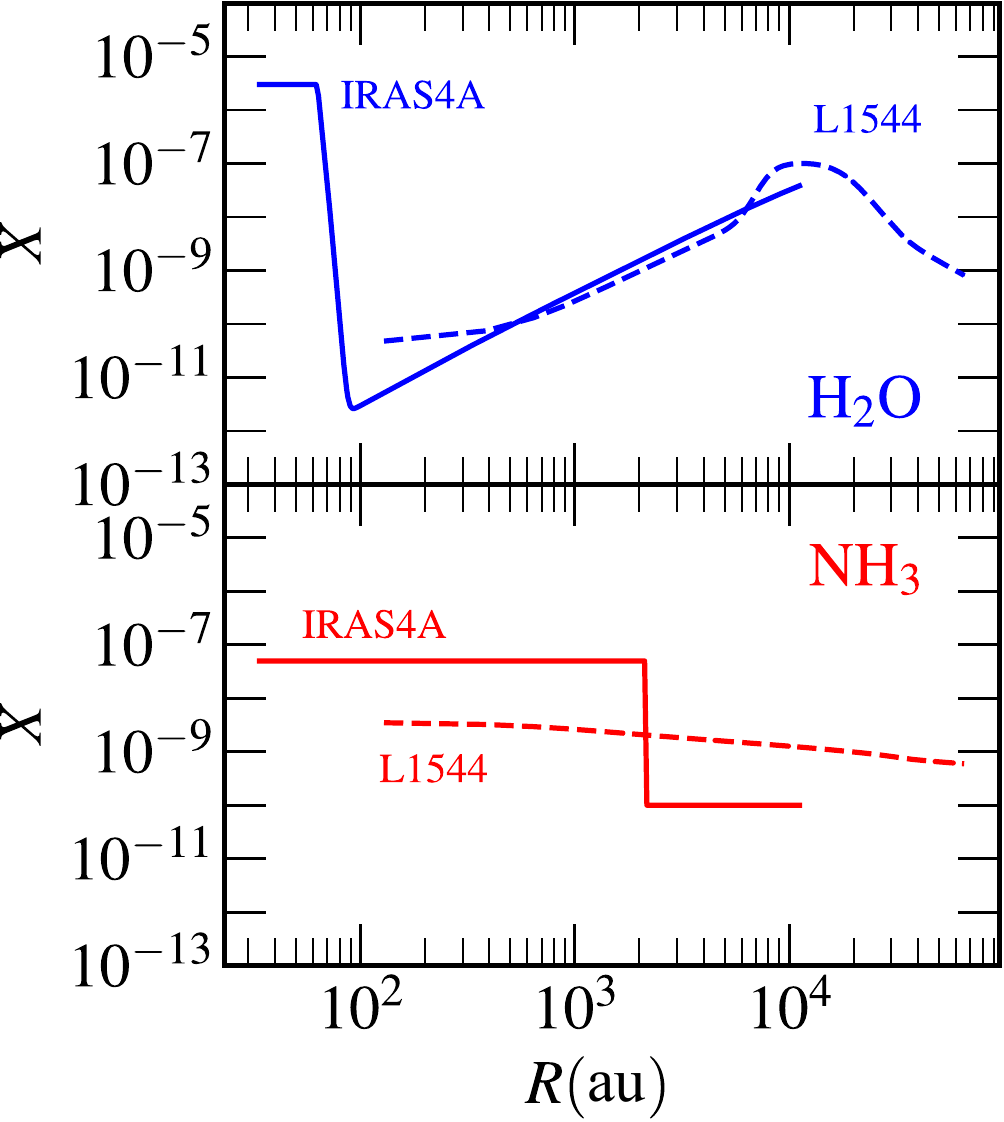}
\caption{Best-fit H$_2$O (top) and NH$_3$ (bottom) abundance profiles
  (with respect to H$_2$) as functions of position in the NGC 1333
  IRAS4A protostellar envelope (full lines) and the L1544 pre-stellar
  core (dashed lines) based on analysis of HIFI spectra. Note the
  similarity of the outer water abundance profiles for the pre-stellar
  core and outer protostellar envelope, and the very different
  chemical behavior of H$_2$O versus NH$_3$. Based on results from
  \citet{Caselli12,Mottram13,Caselli17} and new analyses (see
  Fig.~\ref{fig:N1333_hdo_nh3}). The inner hot core H$_2$O abundance
  for IRAS4A is taken from \citet{Persson16}.}
\label{fig:N1333_L1544}
\end{figure}

\subsection{Inferred H$_2$O and NH$_3$ abundance profiles}

{\it Herschel}-HIFI confirmed the overall picture of low gas-phase
water and high ice abundances in cold clouds, but could do much more
by inferring the water abundance profiles across the cores. These
profiles can then be compared with models such as those by
\citet{Hollenbach09} through coupling of the chemistry with a physical
model of the source to infer critical parameters \citep{Caselli12}.

L1544 is an excellent example, since it has gradients in temperature,
density and velocity through the core that are well constrained by
other observations.  The observed inverse P-Cygni profile allows the
blue-shifted emission part to be connected with the dense central part
of the core, and the red-shifted absorption with the infalling
lower-density outer part. The strong absorption indeed points to a
relatively high water abundance of $\sim 10^{-7}$ in the outer layer
as expected due to photodesorption of water ice by the external UV
radiation field (Fig.~\ref{fig:N1333_L1544},
Appendix~\ref{app:waterroutes}). The central emission requires a
significantly higher water abundance than the original value of
$10^{-12}$ in the Hollenbach et al.\ model. Adding cosmic-ray induced
UV photodesorption of water ice to the model raises the central
abundance to $\sim 10^{-10}-10^{-9}$, sufficient to explain the
emission. The best fitting water abundance profile for L1544 is
included in Fig.~\ref{fig:N1333_L1544}.

If grain surface formation followed by photodesorption would dominate
the production of gas-phase NH$_3$ as well, its abundance structure
should follow that of H$_2$O. However, such an abundance structure
gives a poor fit to the data \citep{Caselli17}. A much better fit is
obtained with a constant or slightly increasing (rather than
decreasing) abundance structure toward the center. Such an abundance
profile was previously inferred from radio observations with the Very
Large Array (VLA) of p-NH$_3$ by \citet{Crapsi07}. Thus, gaseous
NH$_3$ and H$_2$O have very different distributions in dense cores,
with a much higher contribution of cold gas-phase chemistry than ice
chemistry for NH$_3$ compared with H$_2$O \citep{LeGal14,Sipila19}.

\section{Protostellar envelopes: Cold outer part}
\label{sec:protocold}

Is this behavior for H$_2$O and NH$_3$ also seen for protostellar
sources?  The cold outer parts of protostellar envelopes are in many
aspects similar to pre-stellar cores, but protostars have a more
strongly centrally concentrated density structure as well as an
internal heating source, facilitating both absorption in the outer
layers and emission from the inner part of their envelopes. As
discussed in \S~\ref{sec:profiles} and ~\ref{sec:infall}, the
low-lying H$_2$O lines toward protostars --from low to high mass--
indeed show narrow (inverse) P-Cygni profiles superposed on the
outflow components in some fraction of sources (Fig.~\ref{fig:h218o}),
indicating either infall or expansion. When the outflow components are
subtracted, the residual line profiles can be analyzed in a similar
way as those for the pre-stellar cores.

\begin{figure*}[t]
  \centering
  \includegraphics[width=18cm]{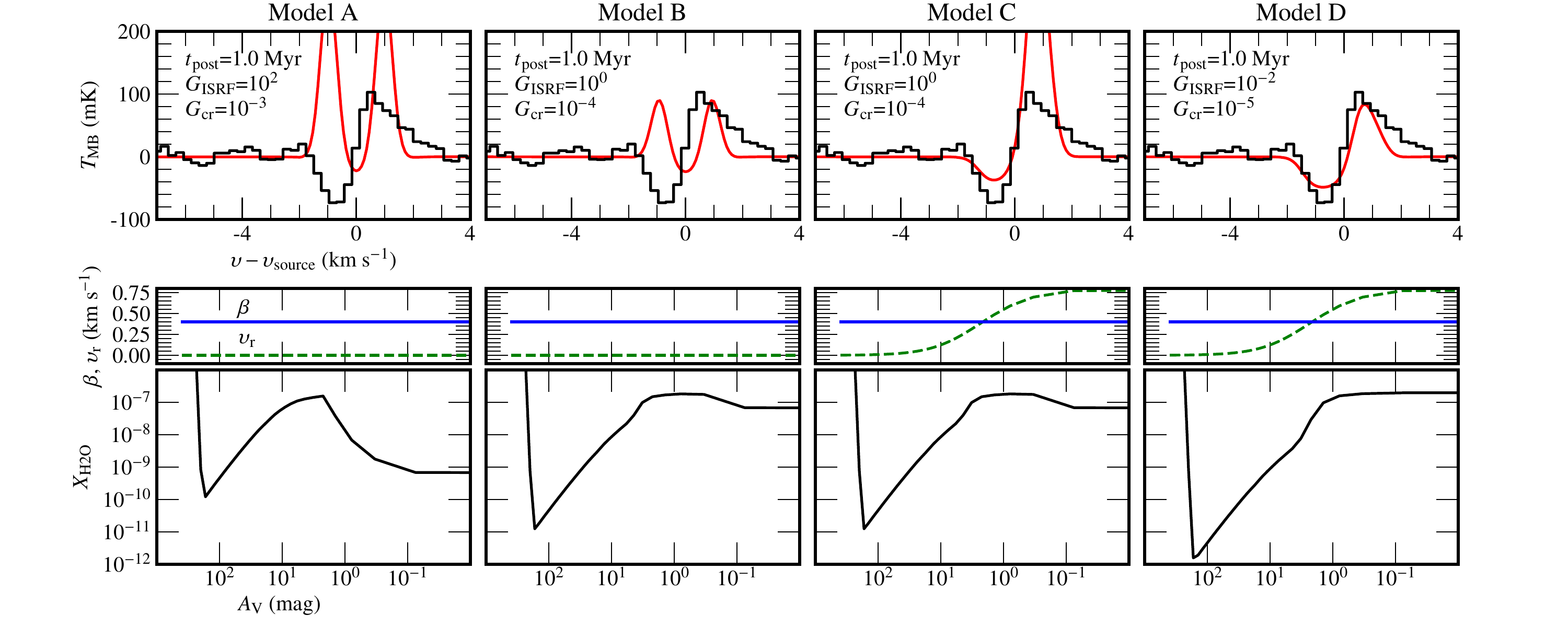}
  \caption{Illustration how various physical and chemical parameters
    can be constrained using the SWaN network. H$_2$O $1_{01}-1_{01}$
    557 GHz observations of the Class I source L1551-IRS5 are compared
    with SWaN models assuming different parameters for the FUV fluxes
    and velocity profile.  The top row shows the observations (black)
    together with the synthetic spectra (red), whereas the bottom row
    shows the adopted velocity profile (green) and water abundance
    profile (with respect to H$_2$) (black).  A constant Doppler
    broadening of $\beta$=0.4 km s$^{-1}$ is assumed, whereas the
    radial velocity profile is either taken to be constant at 0
    (Models A and B) or to have an expansion profile (Models C and
    D). The radiation fields $G_{\rm ISRF}$ and $G_{\rm CR}$ strongly
    decrease from model A to model D. A pre-stellar stage duration of
    0.1 Myr is used. Based on \citet{Schmalzl14}.}
         \label{fig:SWaN}
\end{figure*}

\subsection{Inferred abundance profiles}
\label{sec:abundanceprofiles}

\subsubsection{Simple chemistry analysis using SWaN network}

To analyze the water data, the temperature and density structure of
protostellar envelopes has been coupled with the (simplified) SWaN
water network (\S~\ref{sec:simplechemistry}) \citep{Schmalzl14}. An
important difference with pre-stellar cores is the temperature
structure: as soon as a protostar has turned on in the center, a
temperature gradient is established throughout the envelope, with
temperature decreasing with radius $R$ as roughly $R^{-0.4}$ in the
region where the far-infrared dust emission is optically thin. The
actual dust temperature structure can be computed with full continuum
radiative transfer calculations given a luminosity and density
structure of the core, both fitted to the SED and extent of the
submillimeter continuum emission assuming spherical symmetry
\citep[e.g.,][]{Jorgensen02,Kristensen12}. Gas and dust temperatures
are taken to be coupled, which is a valid assumption at these high
densities. The chemistry is run in time-dependent mode.  The initial
abundances for the protostellar phase are taken from a model of a
cold, constant density ($n_{\rm H}=2\times 10^4$ cm$^{-3}$) pre-stellar
cloud at an age of $\sim$1 Myr, starting the chemistry with oxygen in
atomic form. Thus, protostellar models consist of a two-stage
approach: a pre-stellar and protostellar stage, with the duration of
each of these stages as free parameters.

Figure~\ref{fig:N1333_L1544} shows the resulting gaseous water
abundance with radius of a typical low-mass source (NGC 1333 IRAS 4A)
\citep{Mottram13,Schmalzl14} in comparison with that inferred for
L1544 \citep{Keto14}.  In the coldest outer part of the cloud with
$T<100$~K, the abundance structure is remarkably similar to that of L1544,
with a water peak abundance of $\sim 10^{-7}$ due to ice
photodesorption at $A_V\approx$ few mag and then a rapid decrease in
gaseous water abundance going inward. This steep decrease is largely
due to the steeply increasing H$_2$ density, which is the denominator in
H$_2$O/H$_2$, coupled with H$_2$O freeze-out on grain mantles. Once
the threshold for thermal desorption is reached, the gaseous water
abundance quickly returns to the overall oxygen abundance in the model
of $\sim 10^{-4}$.

This general behavior has been demonstrated in detail for a number of
low-mass protostars, both Class 0 \citep{Mottram14} and Class 0-I
\citep{Schmalzl14}.  The main parameters that control the shape of the
line profiles are the external UV radiation $G_{\rm ISRF}$ and the
internal cosmic ray field $G_{\rm CR}$ (all in units of the
\citealt{Habing68} field\footnote{To convert to units of
  \citet{Draine78} field, divide $G_0$ by a factor of 1.7}), together
with the details of the velocity profile (infall or expansion). Best
fit models generally find an internal cosmic ray induced UV field that
is slightly below the normal value ($G_{\rm CR} \lesssim 10^{-4}$),
with $10^{-4}$ the standard number for a cosmic-ray ionization rate of
$\sim 10^{-17}$ s$^{-1}$ \citep{Shen04}. More precisely, the amount of
water vapor produced by photodesorption scales with the product
$n_{\rm gr} \sigma_{\rm gr} G_{\rm CR}$ with $n_{\rm gr}$ the number
of grains and $\sigma_{\rm gr}$ its geometrical cross section. Thus,
the results scale with the inverse of grain radius $a$, and somewhat
lower values of this product can also imply dust growth to
micron-sized particles deep inside the core, as has been found
observationally \citep{Pagani10}.

The results are insensitive to the timescale of the protostellar phase
$t_{\rm proto}$ over the 0.1--1 Myr range. The absolute values are,
however, highly sensitive to the duration of the pre-stellar stage. In
fact, the observations of both water gas and ice for the same line of
sight can only be reconciled within this modeling framework if a
rather short pre-stellar phase is assumed of $\sim$0.1 Myr to avoid
turning all oxygen into water ice (see discussion in
\S~\ref{sec:watergasice}).

Figure~\ref{fig:SWaN} illustrates the power of the simple SWaN network
to constrain these parameters for the case of the L1551-IRS5 Class I
source. From left to right, the external interstellar radiation field,
the velocity profile, and the internal cosmic-ray induced radiation
fields are systematically adjusted to get the best fit to the observed
H$_2$O $1_{01}-1_{01}$ 557 GHz line profiles. For this source, both a
low external and internal radiation field are required, together with
a radially expanding velocity field to get the P-Cygni line
profile. The best-fitting H$_2$O abundance profile in model D is
similar to that found in Fig.~\ref{fig:N1333_L1544} for IRAS 4A.

More generally, the inferred external UV field is very low for all
sources analyzed, $G_{\rm ISRF}\approx 0.01$, suggesting that the
outer protostellar envelopes on scales of $\sim$10000 au are shielded
from UV radiation by a water-free column with $A_V=2-3$ mag in
extent. For a few sources an elevated ISRF with $G_{\rm ISRF}>1$ is
found, most notably for sources in Corona Australis where the young
B-star R CrA sets up a PDR across the cloud \citep{Schmalzl14}.

\subsubsection{Step-function analysis}
\label{sec:step}

The alternative approach to analyze water data is to use a
simple step-function profile with an outer (freeze-out at $T<100$~K)
and inner (sublimation $>$100 K) abundance. A slightly more
sophisticated method is a ``drop abundance'' profile, with the latter
including an outer ice photodesorption layer 
\citep{Coutens12}. As shown in \citet[][their Fig.~4 and
14]{Mottram13}, the main features of the water chemistry are captured
by the drop profile, but the outer abundances have a limited
meaning. For example, for NGC 1333 IRAS4A the best fit drop abundance
profile has an outer abundance of $3\times 10^{-10}$ and a
photodesorption layer abundance of $3\times 10^{-7}$, whereas
Figure~\ref{fig:N1333_L1544} shows a steadily decreasing abundance
profile between $10^{-7}$ and $10^{-11}$.

\begin{figure}
  \centering
\includegraphics[width=8cm]{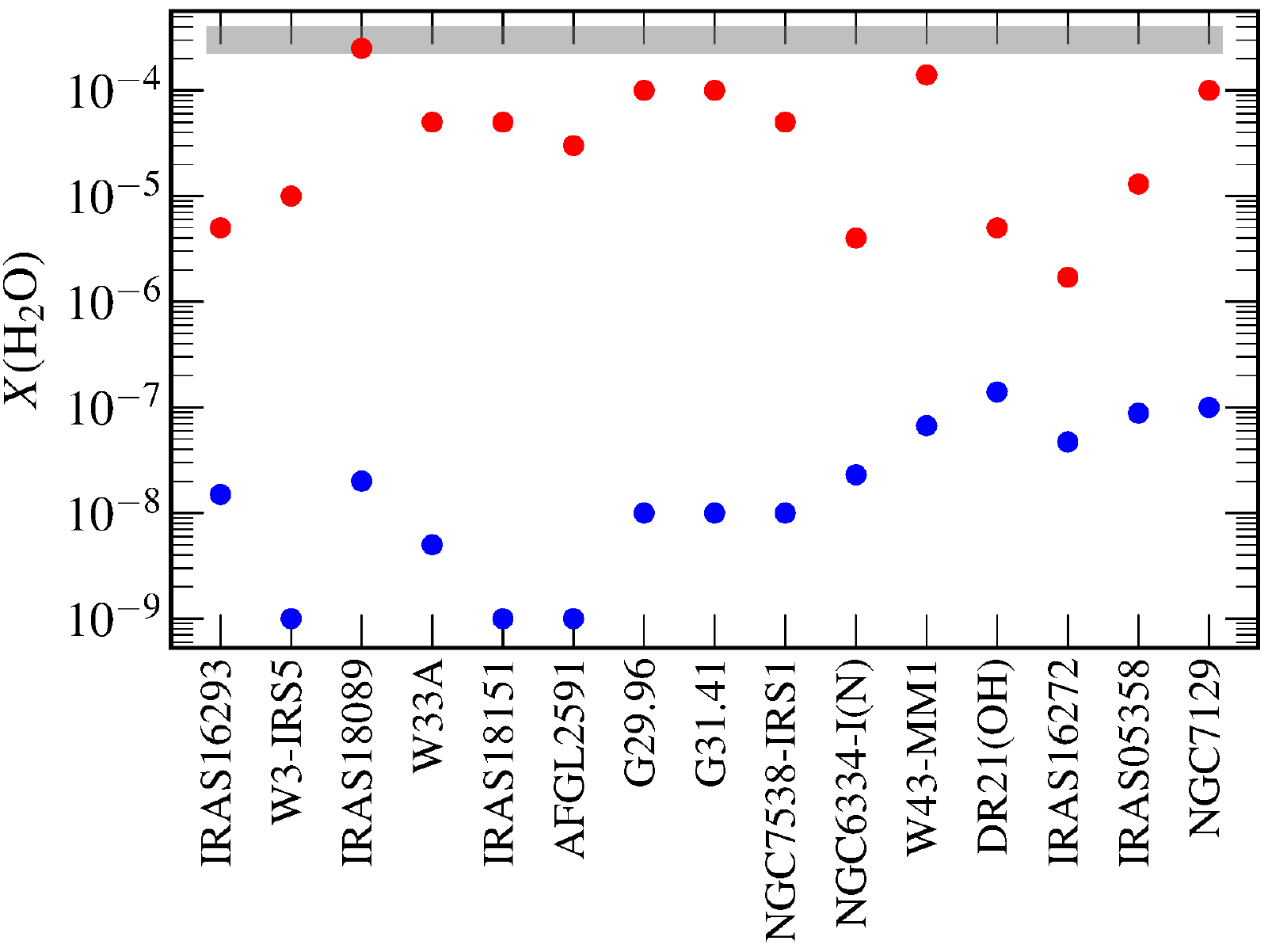}
\caption{Inner hot core (red circles) and outer cold envelope (blue
  circles) water abundances with respect to H$_2$ in high mass
  protostars, derived using a step-function analysis. Inferred
  abundances for a low mass (IRAS16293-2422) and an intermediate mass
  (NGC 7129) source are shown for comparison. Uncertainties are
  typically factors of a few, but see text for the limited meaning of
  outer abundances. The gray bar at the top shows the warm water
  abundance expected if all oxygen is driven into water. Based on
  \citet{Herpin16,Choi15thesis,Coutens12,Johnstone10}.}
         \label{fig:inout}
\end{figure}

One intermediate (NGC 7129) and thirteen high-mass protostars have
been analyzed using the step-function profiles
\citep{Johnstone10,Marseille10,Chavarria10,Herpin16,Choi15thesis}. The
inferred outer abundances $X_{\rm out}$ (with respect to H$_2$) range
from $10^{-9}$ to $10^{-7}$ and are plotted in
Figure~\ref{fig:inout}. There is no obvious trend with luminosity or
envelope mass, which is not surprising given its sensitivity to
external UV and the discussion above on analysis methods.

\subsection{H$_2$O vs NH$_3$ abundance profile}
\label{sec:nh3profiles}

\begin{figure*}[t]
  \centering
    \includegraphics[width=5.5cm]{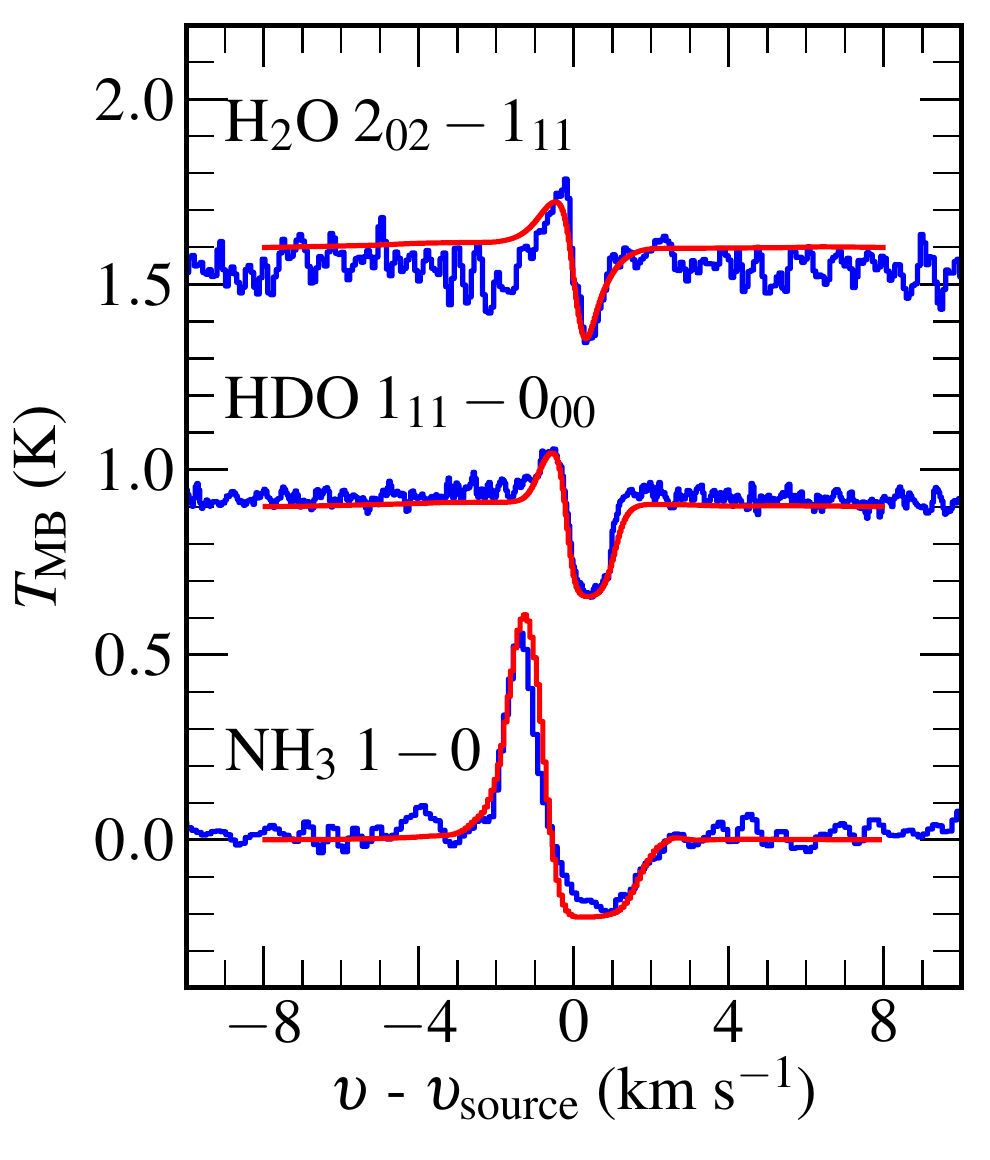}
    \includegraphics[width=5.5cm]{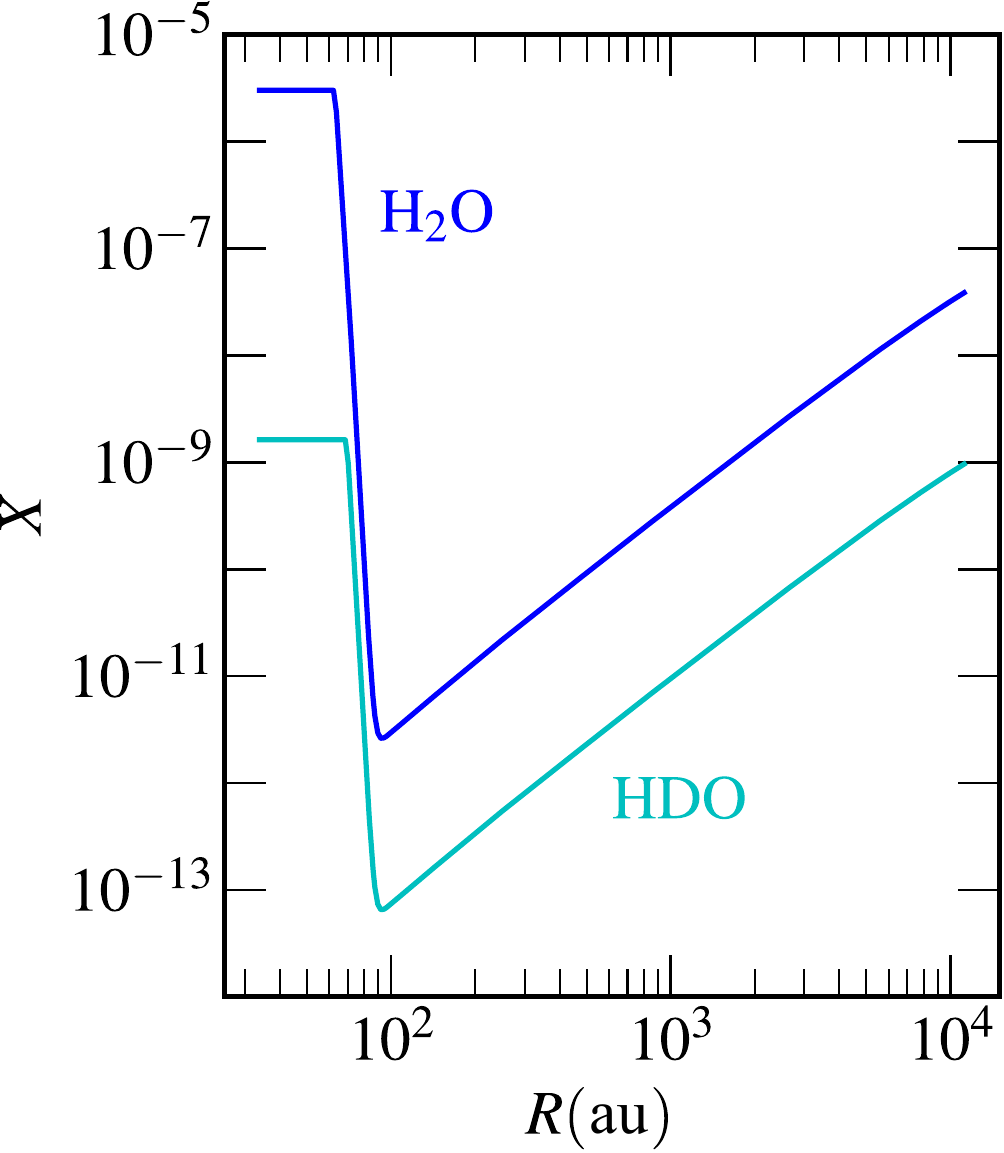}
    \includegraphics[width=5.5cm]{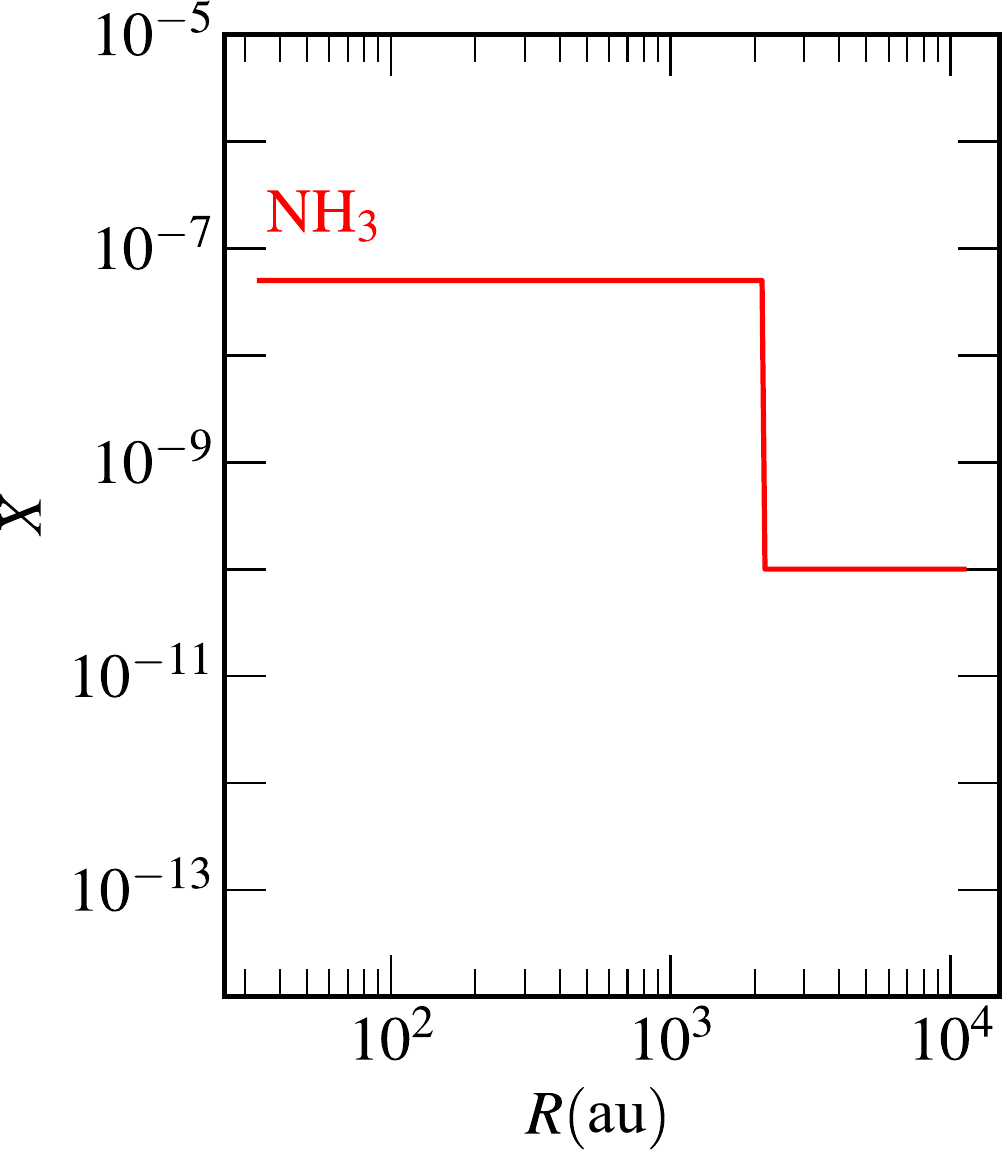}
    \caption{Left: Outflow-subtracted HIFI line profiles of H$_2$O,
      HDO and NH$_3$ (blue) toward the low-mass protostar NGC 1333
      IRAS4A overlaid with the best fit model line profiles
      (red). These model spectra correspond to the abundance
      structures presented in the middle and right panels. The inner
      H$_2$O and HDO abundances are set to those found by
      \citet{Persson14,Persson16}.}
         \label{fig:N1333_hdo_nh3}
\end{figure*}

HIFI observations of the NH$_3$ $1_0-0_0$ 572 GHz line have been taken
for a number of low-mass sources as part of an open time program (PI:
P.\ Hily-Blant). As for pre-stellar cores, the gas-phase NH$_3$
abundance in the outer envelope would be expected to follow that of
H$_2$O if both were produced primarily on the grains and returned to
the gas by photodesorption of ice. However, such an abundance profile
is again a poor fit to the NH$_3$ spectrum in the best studied case of
NGC 1333 IRAS4A. The best-fit NH$_3$ abundance is a constant abundance
of $7\times 10^{-8}$ over much of the envelope, with a slightly lower
value in the outermost photodissociation layer
(Fig.~\ref{fig:N1333_hdo_nh3}). Overall, the inferred NH$_3$ abundance
profile for this low-mass protostar is similar to that found for the
pre-stellar core L1544 (Fig.~\ref{fig:N1333_L1544}). The similarity of
the observed NH$_3$ spectra for other sources suggests that this
abundance structure is a common feature.

The inferred constant NH$_3$ abundance
is inconsistent with current chemical models \citep{Caselli17}.
Yet NH$_3$ ice is observed at an abundance of a few \% of that of
H$_2$O ice in a wide variety of sources
\citep[e.g.,][]{Bottinelli10,Oberg11,Boogert15}, so the abundance of
$s-$NH$_3$ is known to be about $10^{-6}$ with respect to hydrogen.
The bulk of the observed gaseous NH$_3$ with HIFI has an abundance of a few
$\times 10^{-8}$, which is a few \% of that of ice. This value is
higher than can be explained by photodesorption of ice. Thus, the
observed gas-phase NH$_3$ is likely formed primarily through gas-phase
reactions \citep{LeGal14} in a route that preserves a constant abundance with
increasing density and that is apparently not included, or with a too
low rate coefficient, in current gas-phase models \citep{Caselli17,Sipila19}.

\subsection{Cold water deuteration}
\label{sec:waterhdocold}

\begin{figure*}[t]
  \centering
  \begin{minipage}{0.23\textwidth}
    \includegraphics[width=3.5cm]{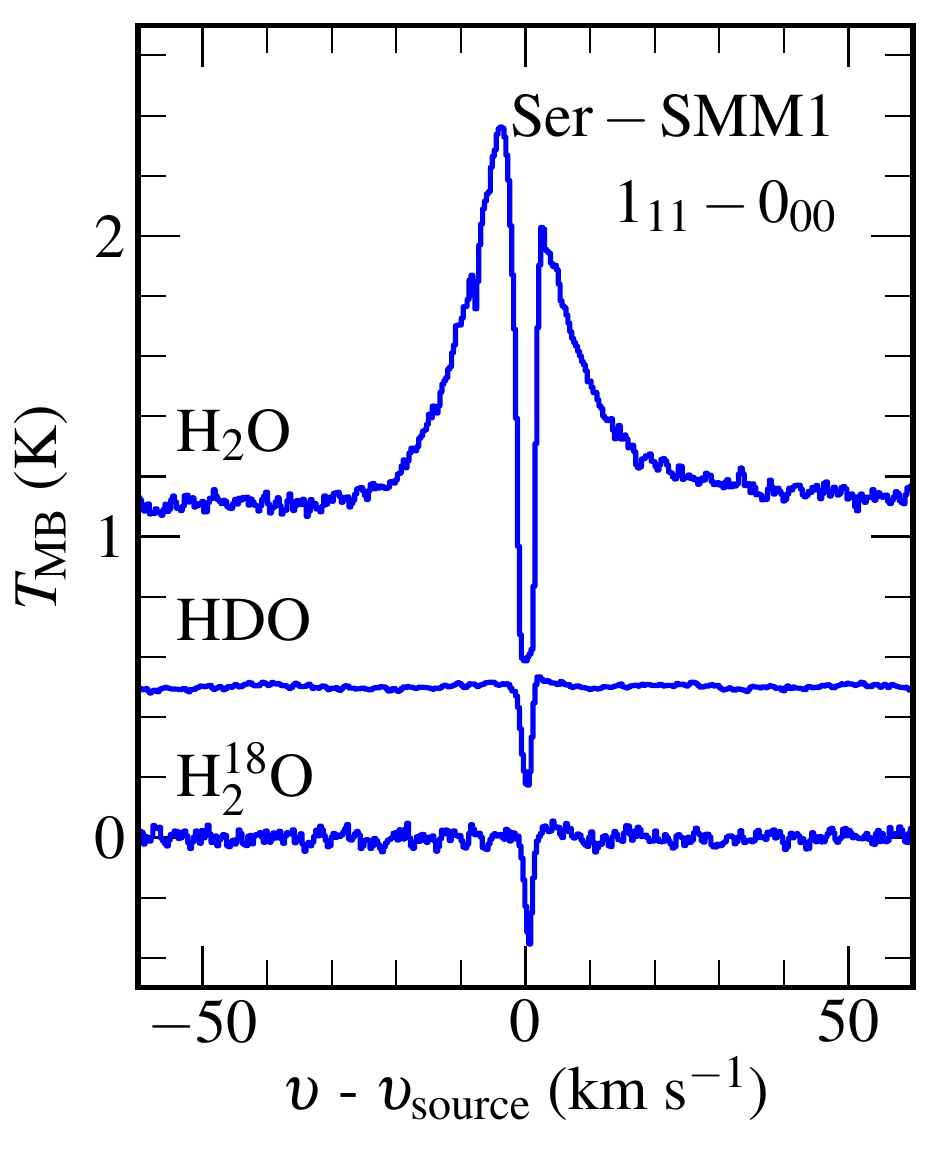}
  \end{minipage}
  \begin{minipage}{0.23\textwidth}
    \includegraphics[width=3.5cm]{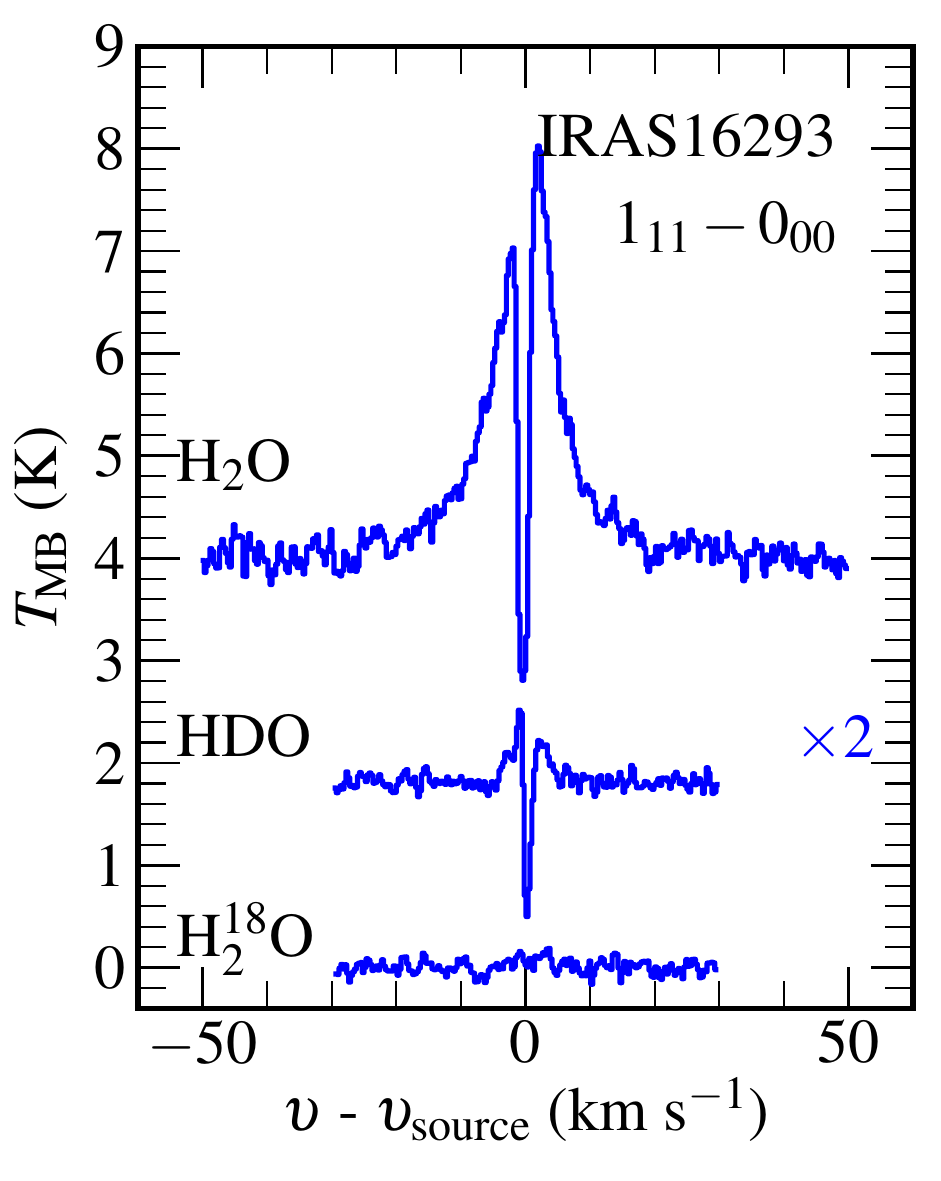}
 \end{minipage}
  \begin{minipage}{0.23\textwidth}
    \includegraphics[width=3.5cm]{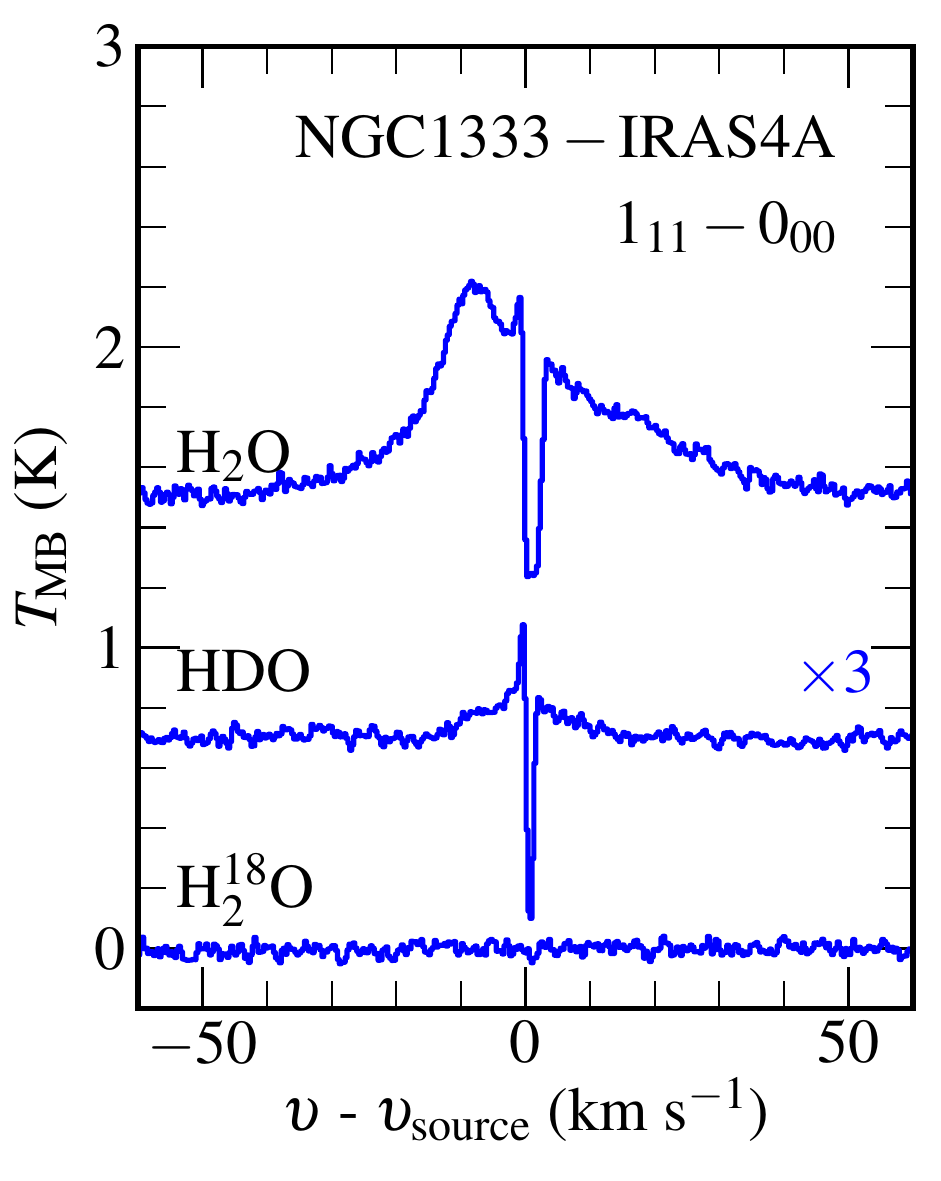}
 \end{minipage}
  \begin{minipage}{0.23\textwidth}
    \includegraphics[width=3.5cm]{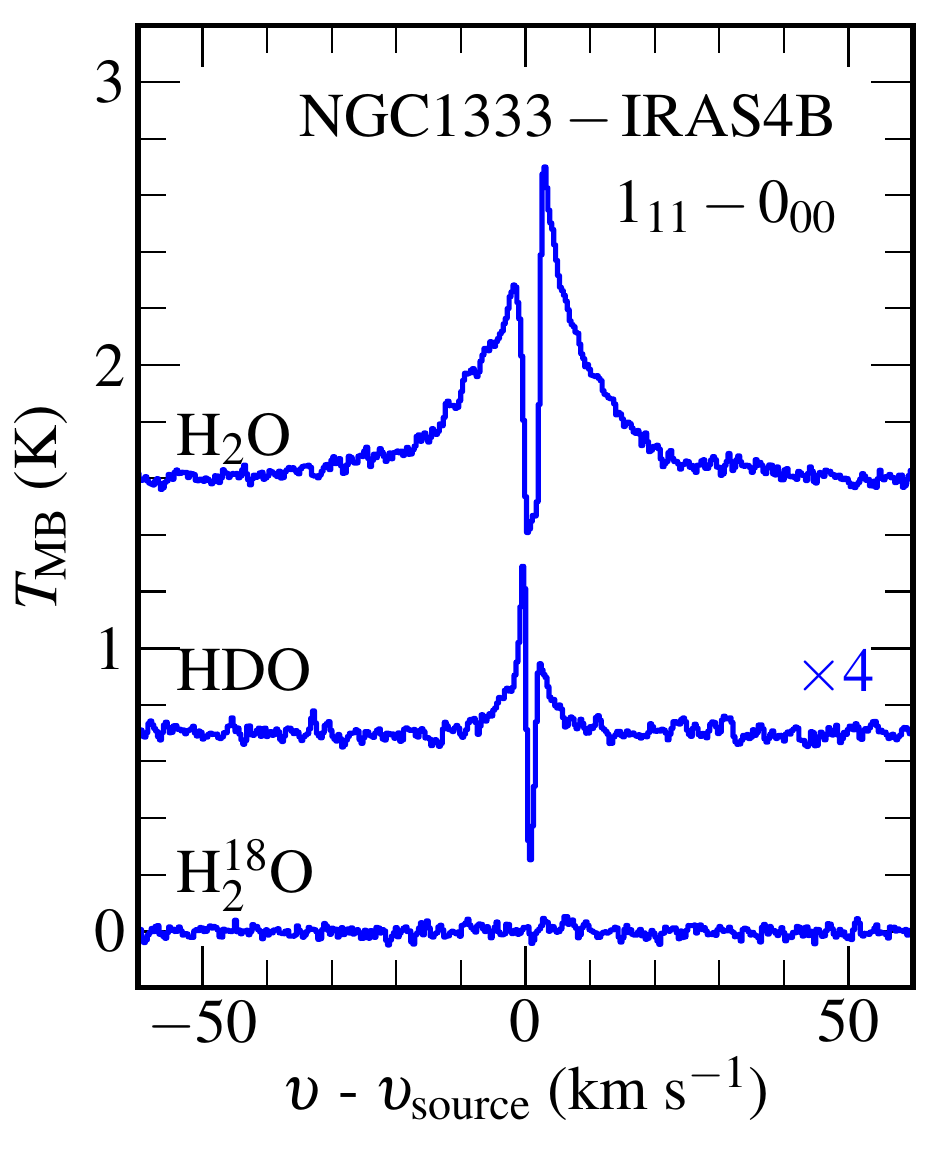}
 \end{minipage}
 \caption{HIFI observations of H$_2$O and H$_2^{18}$O $1_{10}-1_{01}$
   557 and 548 GHz lines together with the HDO $1_{11}-0_{00}$ 893 GHz line
   toward a number of low-mass protostars.}
         \label{fig:Gallery_hdo}
\end{figure*}

Lines of deuterated water, HDO, have been observed as part of the
(extended) WISH program, most notably the $1_{11}-0_{00}$ line at 893
GHz connecting with the HDO ground state. Figure~\ref{fig:Gallery_hdo}
provides examples of high quality HIFI $1_{11}-0_{00}$ spectra near 1
THz of both H$_2$O, H$_2^{18}$O and HDO, all obtained in a similar
beam of $\sim 20''$.

Similar to the H$_2$O lines, the 893 GHz line is very well-suited to
determine the HDO abundance profile, and thus the HDO/H$_2$O abundance
ratio, in the cold outer envelope. Higher-lying HDO lines such as
$2_{11}-2_{12}$ at 241 GHz and $3_{12}-2_{21}$ at 225 GHz, both of
which can be observed from the ground, are more sensitive to the inner
warm HDO abundance (see \S~\ref{sec:hotcore}).

The HDO abundance profile has been determined using the same procedure
as for H$_2$O and NH$_3$.  The simplest assumption is that HDO follows
H$_2$O but scaled by a constant factor.  Figure~\ref{fig:N1333_hdo_nh3}
shows that this assumption works very well for HDO: a constant
HDO/H$_2$O abundance ratio of 0.025 provides a very good fit to the
observed HDO and outflow-subtracted H$_2$O line profiles for the case
of the low-mass protostar NGC 1333 IRAS4A. A flat abundance profile,
such as for NH$_3$, does not fit well. Thus, HDO follows H$_2$O.

Since the observed gaseous H$_2$O results from photodesorption of
water ice, this result could imply that the same holds for HDO.
Detailed modeling of the H$_2$O and HDO photodesorption processes by
\citet{Arasa15} has shown that differences in efficiencies are very
small, so that no significant corrections to the observed numbers are
needed because of differences in desorption efficiencies. However,
since photodesorption only proceeds from the top few ice layers, these
data would then only probe HDO/H$_2$O in the outermost ice
  layers, not in the bulk of the ice (Fig.~\ref{fig:hdonetwork}). Moreover,
cold gas-phase chemistry can also contribute to the observed ratios
(\S~\ref{sec:deuteration} and below).

Using a step-function model, \citet{Coutens12,Coutens13} also infer
very high HDO/H$_2$O ratios of 0.05 for the outermost photodesorption
layers of the NGC 1333 IRAS4A and IRAS 16293 -2422 envelopes.
Averaged over the entire cold part of the envelopes, however, values
of 0.002--0.02 are found \citep{Coutens13}. For the Serpens SMM1
source, narrow unsaturated H$_2^{18}$O and HDO absorption lines are
detected (Fig.~\ref{fig:Gallery_hdo}) implying HDO/H$_2$O=0.002
integrated along the line of sight.
For one high-mass source, G34.26+0.15, \citet{Coutens14} obtain
HDO/H$_2$O=0.001-0.002 in the cold part of the envelope.

All observed HDO/H$_2$O ratios are much higher than the overall
[D]/[H] ratio of $2\times 10^{-5}$ in the gas, suggesting significant
fractionation of water in cold cores, with the upper range values
close to those found of some other highly-deuterated molecules such as
H$_2$CO or NH$_3$ \citep{Ceccarelli14}.  It is important to stress once more,
however, that the observed cold HDO/H$_2$O values do not reflect the
bulk HDO/H$_2$O ice ratios, which are typically
$(0.5-1)\times 10^{-3}$ as measured in warm hot cores where all ices
have just sublimated \citep{Persson14,Jensen19}.

As discussed in \S~\ref{sec:deuteration} and illustrated in
Fig.~\ref{fig:hdonetwork}, high levels of cold water deuteration are
well reproduced by various gas-grain chemical models
\citep{Cazaux11,Aikawa12,Taquet13,Albertsson14,Wakelam14}, especially
if the multilayer structure of the ice is taken into account
\citep{Taquet14,Furuya16}.  The bulk of the H$_2$O is formed early in
the cloud evolution, whereas most of the deuteration takes place in
the subsequent high density stage where the forward reaction producing
H$_2$D$^+$ receives a boost from CO freeze-out.
The outer layers of the ice are therefore expected to be richer in
deuterated water than the inner layers.

\begin{figure}
  \centering
\includegraphics[width=8cm]{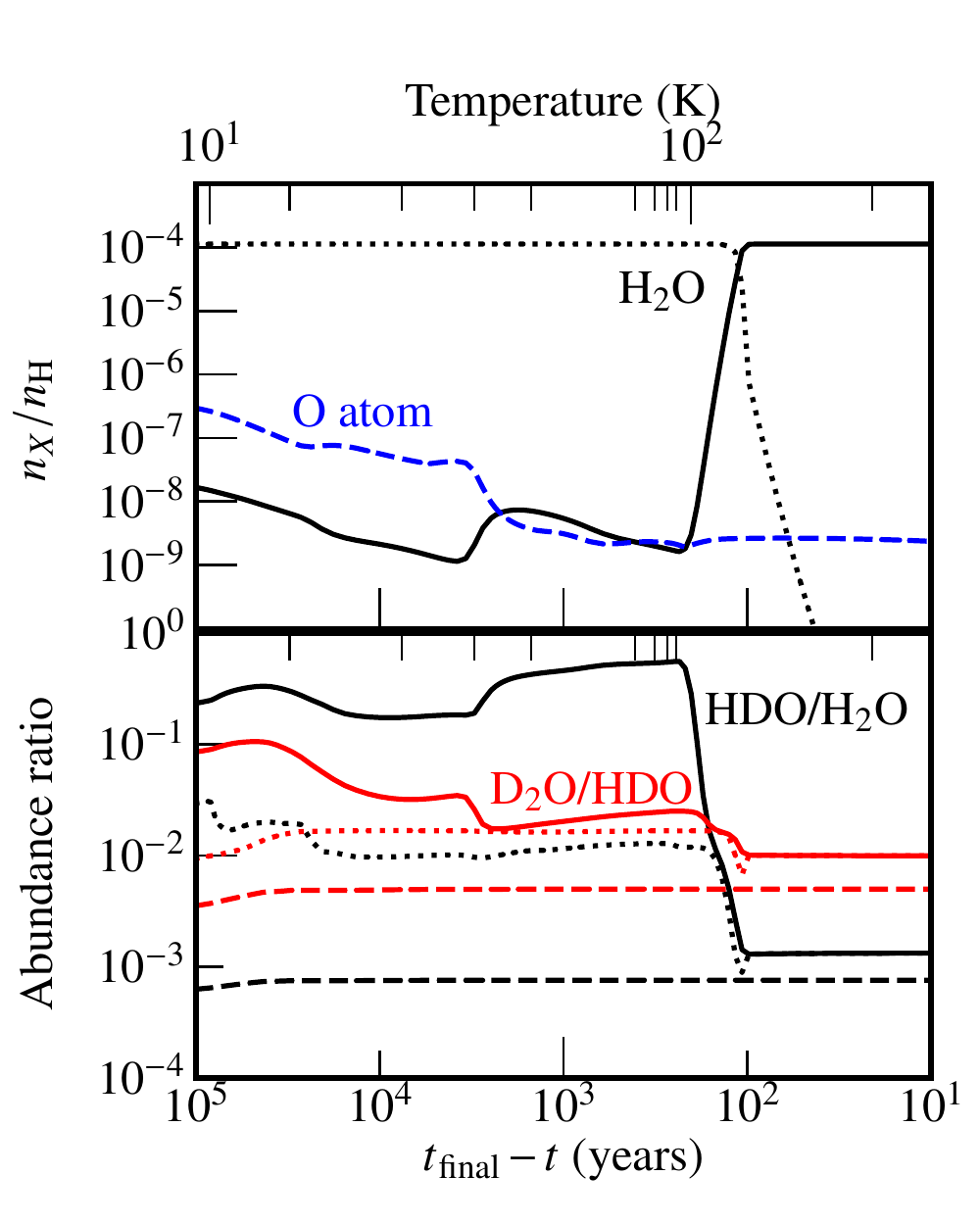}
\caption{H$_2$O, HDO and D$_2$O gas and ice abundances as function of
  time following an infalling trajectory from cold outer envelope (left) to
  inner hot core (right) using the full multilayer chemistry by
  \citet{Furuya16}. The top panel shows the water ice abundance (dotted line)
  in the cold outer part of the envelope, sublimating into gaseous water
  (solid line) when the parcel enters the $\sim 100$ K region.
  The bottom panel shows the computed HDO/H$_2$O (black) and
  D$_2$O/HDO (red) abundance ratios along the trajectory. Solid
  lines: gas phase, dotted lines: ice surface layers (top $\sim 5$\% of ice);
  dashed lines: bulk ice (bottom $\sim 95$\%).  The bulk
  D$_2$O/HDO ice $>>$ bulk HDO/H$_2$O ice, as also reflected in the
  warm gas ratios. The cold gaseous HDO/H$_2$O ratio is very high in
  this model due to cold gas-phase chemistry; the observed HDO/H$_2$O
  value for IRAS4A is HDO/H$_2$O=0.025.}
         \label{fig:hdo_models}
\end{figure}

Figure~\ref{fig:hdo_models} presents models of H$_2$O, HDO and D$_2$O by
\citet{Furuya16} that follow the chemistry along an infalling stream
line from the cold outer part to the warm inner envelope (see also
Fig.~3 in that paper). No free atomic oxygen is needed in the high
density phase to make deuterated water: the OH resulting from
photodissociated water ice can be used to produce HDO. The models of
\citet{Furuya16} and \citet{Taquet14} find that cold gas-phase
chemistry dominates over water ice photodesorption in setting the cold
gaseous HDO/H$_2$O value, and that both result in a roughly constant
HDO/H$_2$O ratio with depth into the envelope
(Fig.~\ref{fig:hdo_models}, bottom, full lines at $T<$100 K). The
model value of HDO/H$_2$O$>0.1$ in cold gas is however higher than the
HDO/H$_2$O ratio of 0.025 that has been observed for NGC1333 IRAS4A.

The main message is that the observed HDO/H$_2$O ratios in cold gas
can be understood using multilayer ice models but that they do not
reflect at all the bulk ice ratio.
To obtain the HDO/H$_2$O ratio for the bulk of the ice, for example
for comparison with cometary values, observations of hot cores are
needed where the entire ice mantle is thermally desorbed above
100 K, as mentioned above and further discussed in
\S~\ref{sec:hotcore}.

Alternatively, the bulk HDO/H$_2$O ice ratio could be measured by
infrared absorption. The deepest direct observational limits of
infrared ice features give HDO/H$_2$O$<$0.005--0.02
\citep{Dartois03,Parise03}. These limits are lower than the tentative
HDO ice detections of \citet{Aikawa12} at a level of HDO/H$_2$O of a
few \%. Deeper limits may be obtained with JWST toward ice-rich
sources although values down to $10^{-3}$ or lower, as expected for
bulk ice, will remain difficult to probe due to limitations on
feature-to-continuum ratios that can be achieved with the instruments.

\subsection{Water and the mystery of low O$_2$ abundances}
\label{sec:o2}

Standard gas-phase-only chemistry models of cold dark clouds predict
that most oxygen is transformed into O$_2$ with time
\citep[e.g.,][]{Goldsmith78,Millar97}. The low abundance of gaseous
O$_2$ in dense clouds measured by the $SWAS$ \citep{Goldsmith00} and
subsequently $Odin$ \citep{Larsson07} satellites therefore initially
came as a surprise.  They were readily explained by models that
include gas-grain chemistry which turn most oxygen into water ice
\citep{Bergin00}, a process now also demonstrated to proceed in
laboratory ice experiments at low temperatures
\citep{Ioppolo08,Miyauchi08,Oba09}. {\it Herschel}-HIFI has been able
to push the observational limits on O$_2$ even deeper, with gaseous
O$_2$ remaining undetected in most sources. O$_2$ has only firmly been
identified through multiple transitions in the Orion shock
\citep{Goldsmith11,Chen14} and one region in the Oph A cloud
\citep{Liseau12,Larsson17}, but not, for example, in the Orion PDR
\citep{Melnick12}.  Its abundance is only
O$_2$/H$_2$=$5\times 10^{-8}$ in the Oph A cloud.

It is clear that the stories of the O$_2$ and H$_2$O chemistries are
intimately linked, and that any model that explains water gas and ice
also needs to be consistent with the O$_2$ data
(Fig.~\ref{fig:waternetwork}). Unfortunately, deep upper limits on the
O$_2$ 487 GHz line exist for only one low-mass protostellar envelope,
the Class 0 source NGC 1333 IRAS4A \citep{Yildiz13o2}. Using a
drop-abundance profile, the inferred 3$\sigma$ limit is
O$_2$/H$_2 <6\times 10^{-9}$ in the cold outer part, the lowest O$_2$
limit so far. Even in the inner hot core, the observed O$_2$ abundance
cannot be more than $10^{-6}$, so O$_2$ is not locked up in ice at
high abundances and released at high temperatures in this particular
source. Recent ALMA interferometer data also imply a similarly low
gaseous O$_2$ abundance in the IRAS16293-2422 hot core
\citep{Taquet18}. Deep unpublished {\it Herschel} O$_2$ 487 GHz data
in a few high-mass protostars such as NGC 6334I and AFGL 2591 have
also failed to detect any line at the 3 mK km s$^{-1}$ (3$\sigma$)
level, pointing to similarly low O$_2$ abundances as for IRAS4A.

The low observed O$_2$ abundance of IRAS4A has been modeled by
\citet{Yildiz13o2} by coupling the temperature and density structure
of the source with a full gas-grain chemical model. Subsequent
radiative transfer models provide line intensities that can be compared
directly with the HIFI observations. The results indicate that only
models with a long and cold pre-stellar stage of (0.7--1) Myr, coupled
with a protostellar stage of about $10^5$ yr, are consistent with the
data. A long timescale at low temperatures is needed to transform all
O and O$_2$ into water ice, leaving no free oxygen to form O$_2$. 
These same models are also able to reproduce a very low O$_2$
abundance in the hot core. The long timescale seems in contradiction
with the conclusion in \S~\ref{sec:abundanceprofiles} of a short
pre-stellar timescale to avoid the overproduction of water ice. However,
no water ice observations exist for IRAS4A to test this model because
it is too faint at mid-infrared wavelengths. This conundrum of
reproducing both water gas, water ice and O$_2$ will be further
discussed \S~\ref{sec:obudget}.

A new twist to the O$_2$ puzzle has been provided by the detection of
abundant O$_2$ ice in comet 67P by the ROSINA instrument on board the
{\it Rosetta} mission \citep{Bieler15,Rubin19}. The inferred ratio is
O$_2$/H$_2$O $\approx 0.03$ in the ice, which would correspond to
O$_2$/H$_2 \approx 3\times 10^{-6}$ if released in the gas assuming
H$_2$O/H$_2$=$10^{-4}$. This is several orders of magnitude higher
than the limit for cold O$_2$ measured for IRAS4A and close to the
warm O$_2$ hot core limit \citep[see also][]{Taquet18}.

A detailed parameter study by \citet{Taquet16o2} shows that a
O$_2$/H$_2$O ice abundance as high as a few \% found for comet 67P can
only be achieved if the atomic H/O ratio in the gas is low, so that
water ice formation is suppressed and O$_2$ ice production promoted.
This, in turn, implies a rather specific narrow range of physical
parameters: a relatively high temperature, $\sim$20--25 K, high
density, $>10^5$ cm$^{-3}$, and low cosmic ray ionization rate,
$< 10^{-17}$ s$^{-1}$, for the pre-stellar cloud out of which our Solar
System formed. A key result of these models is that they are also
consistent with the low observed HO$_2$, H$_2$O$_2$ and O$_3$
abundances in 67P, in contrast with alternative models such as water
ice ``radiolysis'' (that is, processing of ice with ionizing photons)
\citep{Mousis16}. Similarly, other recent models that argue for a
primordial origin of the high O$_2$ in comet 67P by formation in a
cold pre-stellar cloud \citep[e.g.,][]{Rawlings19} do not attempt to
simultaneously reproduce these other species.

Such a relatively warm model could also explain the detection of O$_2$
in the SM1 core of the $\rho$ Oph A cloud, which is known to have
enhanced dust temperatures due to illumination by nearby massive
B-type star(s). A high gaseous O$_2$/H$_2$O abundance is only found at
an early chemical age \citep{Taquet16o2} which may not be unrealistic
given the measured infall speeds in this core \citep{Larsson17}. For
IRAS 4A, these same models are consistent with the low observed limits
because the pre-stellar cloud is much colder, 10 K.

\section{Hot cores: Dry or wet}
\label{sec:hotcore}

\subsection{Inferred abundances from observations}

In the warm inner envelopes close to the protostar, the dust
temperature becomes higher than 100 K, at which point all water ice
sublimates back into the gas. The size of this so-called ``hot core''
region scales roughly as
$R_{\rm T=100 K}\approx 2.3\times 10^{14} (\sqrt{L/L_{\odot}})$ cm in
a spherically symmetric envelope \citep{Bisschop07}. For $L$=1, 100
and $10^4$ $L_\odot$, the 100 K radius is at 15, 150 and 1500 au,
respectively. Thus, whether for low-mass sources at 200 pc or
high-mass sources at 2 kpc, the hot core region is $< 1''$ on the sky
and thus heavily beam diluted in the {\it Herschel} beams. The region
where $T>250$~K, the temperature at which all volatile oxygen would
be driven into water by gas-phase reactions, is even smaller.

Hot core water abundances have been derived by fitting the HIFI line
intensities of the higher-lying water lines, especially those of
H$_2^{18}$O or H$_2^{17}$O that are not affected by outflow emission
(see Figure~\ref{fig:visserspectra} in Appendix). Detailed radiative
transfer models using a step-abundance profile have been performed for
most of the WISH high-mass sources \citep{Herpin16,Choi15thesis} and
one intermediate mass source (NGC 7129) \citep{Johnstone10}. The
resulting inner abundances $X_{\rm in}$ (with respect to H$_2$) are included
in Figure~\ref{fig:inout}, together with a few results from the
literature using a similar approach: the low-mass source
IRAS16293-2422 \citep{Coutens12} and the high-mass source NGC 6334I
\citep{Emprechtinger13}, both from the CHESS program.  The narrow
H$_2^{18}$O absorption lines toward high-mass sources also indicate a
jump in water abundance of at least an order of magnitude at the
$\sim$100 K radius \citep{vanderTak19}.  For most low-mass sources, a
step-function analysis has not been possible due to the lack of narrow
HIFI H$_2^{18}$O lines (Fig.~\ref{fig:h218o}).

It is clear from Figure~\ref{fig:inout} that the inner water
abundances range from $5\times 10^{-6}$ to $> 10^{-4}$, with no trend
with luminosity or envelope mass, nor with the ratio $L/M_{\rm env}$,
which is thought to be an evolutionary indicator. A few sources have
inner abundances of $10^{-4}$ or higher, as expected from ice
sublimation and high temperature chemistry driving oxygen into water,
although none as high as the expected value of $4\times 10^{-4}$. The
only region with a water abundance that may be as high as
$6.5\times 10^{-4}$ is a small compact dense clump near the Orion hot
core found using HEXOS data \citet{Neill13}. Among a subset of WISH
high-mass sources, \citet{Herpin16} have found a possible trend of
higher inner water abundances with higher infall or expansion
velocities. This could suggest that sputtering of ice mantles
contributes to the water production in the inner region.

In an attempt to probe the inner warm water abundance for LM
protostars with {\it Herschel}, very deep (5 hr) HIFI observations of
the H$_2^{18}$O $3_{12}-3_{03}$ 1095 GHz ($E_{\rm up}=249$~K) lines
have been obtained for 6 sources \citep{Visser13}. This H$_2^{18}$O
line is detected in two low-mass sources, NGC 1333 IRAS2A and Serpens
SMM1 (Fig.~\ref{fig:visserspectra} in Appendix), and actually shows
narrow profiles with FWHM$\approx 3$ km s$^{-1}$, so the emission is
clearly not associated with the outflow. Deep limits are obtained for
the other sources (NGC 1333 IRAS4A, IRAS4B; GSS30, Elias 29); the
H$_2^{17}$O line covered in the same setting is not detected at the
same noise level (8 mK in 0.5 km s$^{-1}$ bin).

A detailed combined analysis of the IRAS2A H$_2^{18}$O HIFI 1095 GHz
spectrum and H$_2^{18}$O $3_{13}-2_{20}$ 203 GHz NOEMA data
reveals that even the HIFI water isotopolog line is likely optically
thick when coming from a hot core with a radius of 100 au. In
contrast, the 203 GHz emission is optically thin. It results from the
fact that the Einstein $A$ coefficient for the 1095 GHz line,
$1.65\times 10^{-2}$ s$^{-1}$, is a factor of 3400 larger than that of
the 203 GHz line, $4.8\times 10^{-6}$ s$^{-1}$. Moreover, the dust
continuum emission from the hot core starts to become optically thick
at 1 THz on 100 au scales, thus shielding some of the water
emission. Altogether, this means that only a lower limit can be placed
on the hot core water abundance of H$_2$O/CO$> 0.25$ or
H$_2$O/H$_2$$>2\times 10^{-5}$. Here the C$^{18}$O 10--9 ($E_{\rm
  up}=290$~K) line obtained in the same spectrum has been used as
reference for the warm gas assuming CO/H$_2$ $=8\times
10^{-5}$ inferred for this source \citep{Visser13,Yildiz12}.

Are most hot cores indeed ``dry'', or is this an artifact of the
analysis method?  For high-mass sources, the H$_2^{18}$O 203 GHz line
($E_{\rm up}=203$~K) has long been recognized to be an excellent
diagnostic of water in hot cores, with
inferred water abundances of $\sim 10^{-5}-10^{-4}$
\citep[e.g.,][]{Jacq88,vanderTak06,Wang12}. Millimeter interferometers
can now also image this line in low-mass hot cores with sub-arcsec
resolution \citep{Jorgensen10,Persson12,Jensen19}. These data show
narrow water lines (FWHM$\approx$1--2 km s$^{-1}$) located within a
$\sim 100$ au radius region. Inferred water abundances are low, only
$10^{-8}-5\times 10^{-6}$ (with respect to H$_2$) if analyzed within a
spherically symmetric model. However, on these scales much of the gas
and dust is likely in a flattened disk-like structure, even in the
Class 0 phase \citep{Murillo13,Tobin12,Tobin16}. The temperature in
the shielded midplane of the disk, where most of the mass is located,
is much lower than 100~K, so the amount of warm H$_2$ in the
denominator can be much smaller.

Modeling the small scale interferometer structure with a
nonspherically symmetric model consisting of a disk (compact source)
+ envelope indeed confirms that only a small fraction of the mass in
the inner region
is above 100~K \citep{Persson16}. This raises the fractional water
abundances with respect to H$_2$ by an order of magnitude to
$10^{-7} - 6\times 10^{-5}$, with the latter value applying to IRAS
2A. These water abundances are closer to, but still below, the
expected water abundance of $4\times 10^{-4}$.

In summary, when corrections for nonspherical symmetry and optical
depths are included, hot cores are not as ``dry'' as previously
thought, but they are not ``wet'' either, with water abundances still
a factor of a few to two orders of magnitude lower than expected
\citep[see also Fig.~11 in][]{Persson16}.  Overall, this comparison
highlights the fact that {\it Herschel} was not well suited to
determine hot core water abundances, due to the combination of severe
beam dilution, high optical depths in line and continuum, and
difficulties in isolating the hot core emission from that in the
outflows. Studying the chemistry of hot cores is squarely an
interferometer project, with NOEMA and now also ALMA opening up the
opportunity to image the 203 GHz line, and other H$_2^{18}$O lines
with small Einstein $A$ coefficients and low continuum optical depth,
for large numbers of sources. The challenge remains to constrain the
amount of warm gas on small scales to determine the water abundance,
that is, the denominator in the H$_2$O/H$_2$ ratio remains the main
uncertainty, rather than just the water column. This will require
unravelling the detailed disk-envelope physical structure on $<$100 au
scales.

\subsection{Water deuteration in hot cores}
\label{sec:waterhdohot}

There has been no shortage of determinations of the HDO/H$_2$O ratio
in warm gas, using both ground-based and {\it Herschel}-HIFI
observations. Single-dish ground-based determinations of high-mass
protostars, using typically the 225 and 241 GHz HDO lines to measure
HDO, typically find values as low as HDO/H$_2$O=$(2-8)\times 10^{-4}$
\citep[e.g.,][]{Jacq90,Gensheimer96,Helmich96,vanderTak06,Emprechtinger13,Coutens14},
with occasionally values up to $(2-5)\times 10^{-3}$
\citep{Neill13}. Interferometer observations of high-mass protostars
are still sparse but indicate values of $\sim 5\times 10^{-3}$ for the
disk-like structure of AFGL 2591 \citep{Wang12}. For low-mass
protostars in clustered regions, the interferometer data give
HDO/H$_2$O ratios in the range $(0.6-1)\times 10^{-3}$
\citep{Jorgensen10hdo,Persson14}. Interestingly, for isolated low-mass
protostars, the HDO/H$_2$O ratios are a factor of $\sim$2 higher
\citep{Jensen19}. Error bars on these ratios for nearby sources
observed with interferometers are small, of order 30\% or less, since
many assumptions about excitation temperature and emitting region for
the two species cancel out.

Even with this fairly large range of inferred values, it is clear the
HDO/H$_2$O ratios in warm gas are lower than those for the cold gas of
0.025 presented in \S~\ref{sec:waterhdocold} by an order of magnitude
or more. This is consistent with the models of \citet{Taquet14} and
\citet{Furuya16} discussed in that section and presented in
Fig.~\ref{fig:hdo_models} (bottom panel, compare full lines at $T$ $<$
and $>100$ K), namely that the bulk water ice sublimated in hot
cores has a lower deuteration fraction than the top ice layers and
cold chemistry probed in the cold outer envelopes (see also
Fig.~\ref{fig:hdonetwork}).

\section{Protoplanetary disks}
\label{sec:disks}

To address the question of the water trail from clouds to planets, the
water content of disks at different evolutionary stages needs to be
addressed. This section discusses {\it Herschel}'s contribution to our
knowledge of water in forming disks in the embedded Class 0/I
protostellar stages as well as in mature disks in the optically
visible Class II stage.

\begin{figure}
     \centering
    \includegraphics[width=8cm]{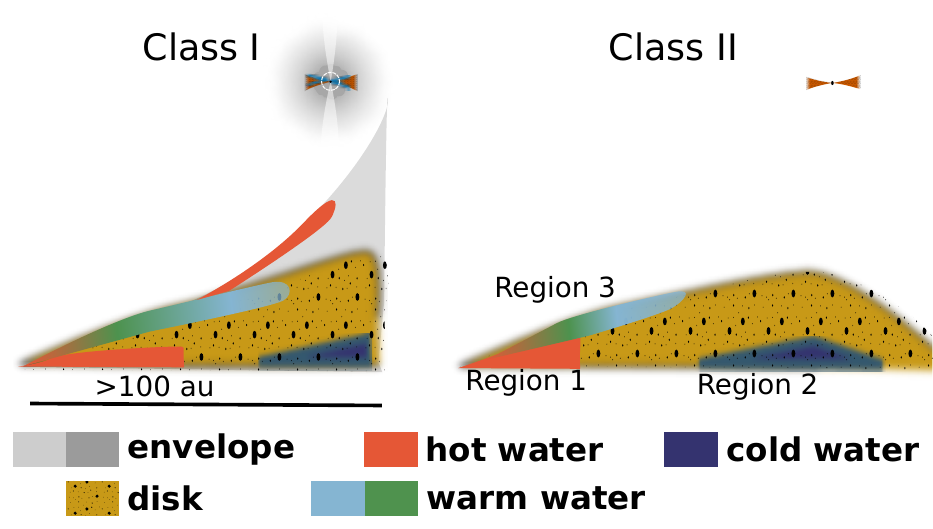}
    \caption{Water vapor reservoirs in young Class I systems and mature Class
      II disks.
      Region 1 lies inside the water ice line where high temperature
      chemistry drives all oxygen into water near the midplane
      ($\sim 10^{-4}$ abundance). Region 2 lies beyond the snow line
      and produces a small amount of water vapor ($10^{-11}$ up to
      $10^{-7}$) through photodesorption of water ice. Region 3
      concerns the warm surface layers of the disk out to large radii
      with moderate water vapor abundances limited by UV radiation
      ($\sim 10^{-6}$). Regions 1 and 3 extend further out
      into the disk for the warmer Class I sources. Also, hot water
      can be emitted in the hot core and in outflows or shocks along
      the outflow cavity in Class I sources. The small cartoons at the
      top illustrate the larger scale disk-envelope-outflow cavity
      system.}
         \label{fig:Harsonodisk}
\end{figure}

To put the {\it Herschel} observations in context and set the stage
for the discussion, it is useful to summarize the expected water vapor
reservoirs in disks based on chemical models
(Fig.~\ref{fig:Harsonodisk}).  Disks, whether young or mature, have
large gradients in temperatures and densities, both radially and
vertically. Their surface layers are exposed to intense UV radiation
from the young star. Thus, the chemistry varies strongly with position
in the disk, with the ``snow'' or ice-lines of different molecules
playing a key role \citep{Oberg11co}. Overall, models of protoplanetary
disk chemistry have identified three different water chemistry regimes
\citep[e.g.,][]{Woitke09,Gorti08,Walsh12,Albertsson14,Furuya13,vanDishoeck14PPVI,Walsh15}.
Region 1 lies inside the water ice line of $\sim$160 K and has a high
water abundance of at least 10$^{-4}$, with most oxygen contained in
water vapor by ice sublimation and high temperature chemistry. Region
2 lies in the outer disk well beyond the ice line, and has a low water
vapor abundance ranging from $10^{-11}$ up to $\sim 10^{-7}$. This
water vapor is produced at intermediate disk heights where UV
radiation can penetrate and water ice can be photodesorbed from grains
\citep{Dominik05,Bergin10}. Finally, in the warm surface layers of
disks (region 3), water vapor is produced by high temperature
chemistry in the ``warm finger'' which has also abundant OH. There the
water abundance is limited by photodissociation to a value of
$\sim$10$^{-6}$.

\subsection{Young disks}
\label{sec:youngdisks}

\begin{figure*}
\centering
\includegraphics[width=12cm]{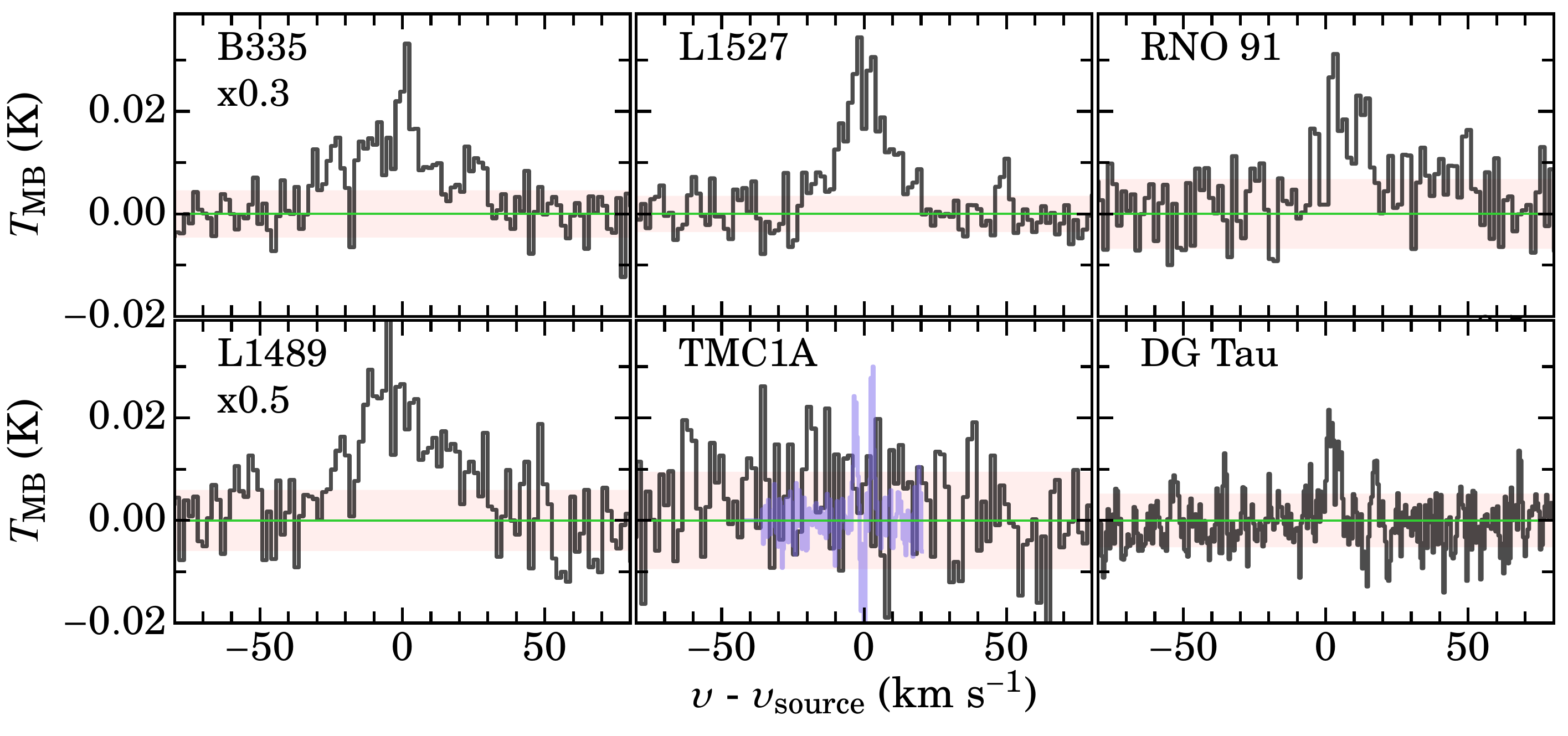}
\caption{H$_2$O $1_{10}-1_{01}$ spectra toward low-mass embedded
  protostars with and without disk-like structures. RNO91,
  TMC1A and L1489 are Class I sources with disks; L1527 is a Class
  0/I source with an embedded disk whereas B335 is a Class 0 source
  with no clear evidence for a disk down to 20 au.
  DG Tau  is a young Class II source with
  disk and jet. These are WBS spectra binned to 0.3 km s$^{-1}$ while
  B335 is binned to 0.6 km s$^{-1}$ velocity resolution.
  The horizontal green line indicates the baseline. For TMC1A, the
  blue line indicates the ALMA $^{13}$CO spectrum of the rotating
  disk to indicate its width \citep{Harsono18}.}
\label{fig:ClassIspectra}
\end{figure*}

The preceding sections have shown that most of the water emission from
low-mass protostars detected with {\it Herschel} arises in warm
outflows and shocks associated with jets and winds and their
interaction with the surrounding envelope, with the gas heated by
kinetic energy dissipation.  However, as the protostellar system
evolves from the deeply embedded Class 0 phase to the late Class I
phase, the envelope mass and outflow force decrease whereas the disk
grows in mass and size \citep{Hueso05}. Has {\it Herschel} detected
any water emission from these young disks?

Prior to the launch of {\it Herschel}, the presence of young disks
was heavily debated since numerical studies suggested
that strong magnetic fields inhibit the formation of disks
\citep{Galli93, ZYLi14}.
Recent millimeter interferometric observations of optically thin
molecular lines have however revealed flattened disk structures for
several Class 0 and I sources targeted by WISH+ that are characterized
by Keplerian motion
\citep[e.g.,][]{Lommen08,Tobin12,Murillo13,Harsono14,Yen17,Takakuwa18,vantHoff18,delaVillarmois19}.
Since these data clearly show that young disks are present, the disk's
contribution to the H$_2$O lines observed with {\it Herschel} needs to
be reassessed.

The sizes of the embedded disks are typically 50 to 100 au, so
$\sim 1 ''$ diameter at 140 pc \citep{Harsono14}.  Young disks in the
Class 0 and I phase are expected to be warmer than their Class II
counterparts with similar mass due to their higher accretion rate and
higher bolometric luminosity ($L \gtrsim 1$ L$_{\odot}$,
\citealt{Harsono15}).  Thus, warm young disks may have sufficient
amounts of water vapor that could contribute to the HIFI spectra.

\subsubsection{Cold water in young disks}
\label{sec:youngdiskscold}

A first, purely observational look is to compare the H$_2$O
$1_{10}-1_{01}$ 557 GHz spectra toward a number of Class 0/I
protostars with and without disks (Fig.~\ref{fig:ClassIspectra}).
ALMA observations of the Class 0 protostar B335 ($L = 3.3$
L$_{\odot}$) show no clear evidence for a Keplerian disk inside of
20 au \citep{Bjerkeli19}.  In contrast, the Class I protostar L1489
with a similar luminosity ($L=3.8$ L$_{\odot}$) is surrounded by a
very large rotating structure with a $\sim$700 au radius that may be
in transition to the Class II stage
\citep[e.g.,][]{Brinch07,Yen14}. L1527 is a borderline Class 0/I
source ($1.9-2.6$ L$_\odot$) with a $\sim 100$ au warm young disk seen
edge-on \citep{Tobin12,vantHoff18}. TMC1A (3.8 
L$_{\odot}$) is a Class I source with a bonafide disk
with 50--100 au radius \citep{Harsono14,Harsono18}.
Finally, the young Class II source DG Tau is included since
\citet{Podio13} suggest that their {\it Herschel}-HIFI observations of
the water ground-state lines at 557 GHz and 1113 GHz are emitted by
the disk ($M_{\rm disk}$=0.1 M$_{\odot}$, $L= 1$ L$_{\odot}$).  DG Tau
is an interesting case in that it also has a powerful optical jet
which emits strong UV and X-rays onto the disk surface
\citep{Guedel10,Guedel18}.

At first glance, all of these profiles look similar.  The B335
spectrum stands out because of its broad line profile with a FWHM of
40 km s$^{-1}$ typical of Class 0 sources while the water lines toward
the Class I protostars, when detected, have a typical FWHM of 15--20
km s$^{-1}$ (see \S~\ref{sec:profiles}).  Water is also detected
toward the Class 0 L1527 source with a FWHM of 20 km s$^{-1}$ but
its comparatively low value could be due to the outflow lying close to
the plane of the sky.  No water emission is detected from TMC1A, which
does have a large disk as well as a blue-shifted disk wind
\citep{Bjerkeli16}. Except for TMC1A, the integrated water line
intensities follow the observed correlation of $I$(H$_2$O) with
envelope mass $M_{\rm env}$ and measured outflow force established
for a much larger sample \citep{Kristensen12,Mottram17}, but there is
clearly no relation with $L_{\rm bol}$ for this subsample. The former
relation suggests that the line profiles largely reflect the warm 
gas associated with the outflow, which has weakened for the Class I
sources.  Association with the outflow is further strengthened by the
fact that spatially resolved H$_2$O emission has been seen for several
of these sources in the PACS data along the outflow direction, see
Figures~4, D.1 and D.2 in \citet{Karska13} for L1527, L1489 and TMC1A.

Could the ``double peaked'' profile reflect disk emission?
Figure~\ref{fig:ClassIspectra} includes the observed $^{13}$CO 2-1
line profile observed with ALMA from the inclined TMC1A disk and its
disk wind showing a typical FWHM of 8 km s$^{-1}$. This is close to
the maximum line width expected for outer disk rotation: even for
mature disks around Herbig stars ($M_*>1$ M$_{\odot}$), CO line
profiles do not span more than 10 km s$^{-1}$
\citep{Thi01,Dent05}. The comparison between L1489 and DG Tau  shows
that the DG Tau  line is narrower than most Class I sources, as
expected from a disk, although cloud or wind absorption can also
(asymmetrically) affect the profiles.

Are the observed line intensities consistent with those from a disk?
The simplest approach is to compare expected line intensities from
optically thick water emission within the ice line. For high accretion
rates, the midplane temperature increases and the water ice line
shifts outward, from radii of only a few au in mature, nonaccreting
disks to values as high as 60 au for accretion rates of $10^{-4}$
M$_\odot$ yr$^{-1}$ \citep{Harsono15}. There could also be water
emission from the remnant warm inner envelope
(\S~\ref{sec:hotcore}). The maximum water emission can then be
computed by considering optically thick line emission within a
$\sim$100 au radius at 100 K and then diluted by the {\it Herschel}
beam. This gives an integrated line intensity of $\sim$20 mK km
s$^{-1}$ for the 557 GHz line for disks at 150 pc, which is a factor
of 5 below the upper limit for TMC1A of $\sim 100$ mK km s$^{-1}$. For
low accretion rates, the model intensities drop steeply with the
reservoir of photodesorbed water (region 2) providing a minimum floor.

\begin{figure}
  \includegraphics[width=8cm]{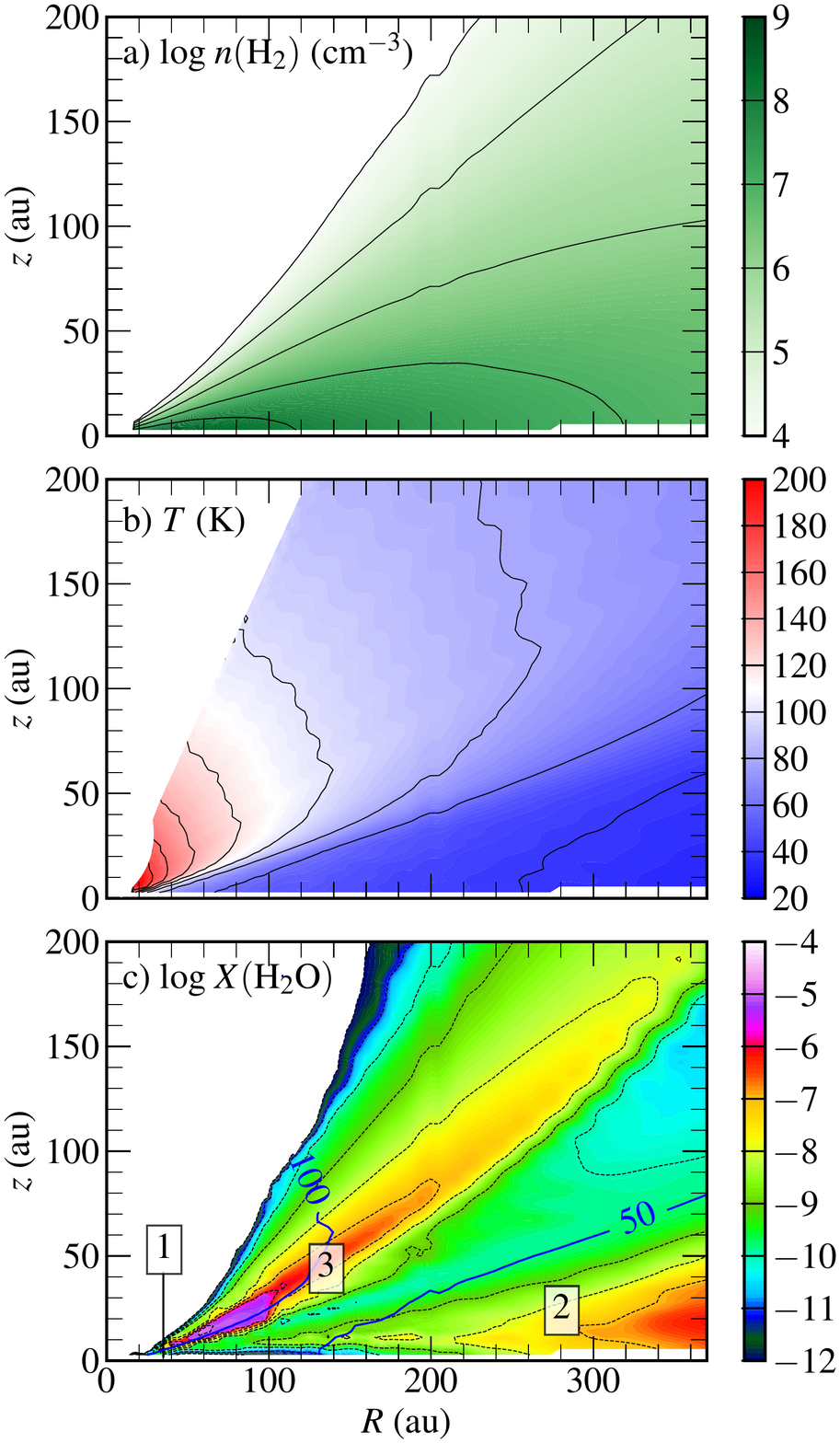}
\caption{Cross section of one quadrant of a model of a Herbig disk, tailored to that around HD 100546. (a) H$_2$ number density, (b) gas temperature (taken equal to the dust temperature), and (c) calculated H$_2$O abundance. Further details of the model can be found in the text and references. The H$_2$O abundance can be divided in three regions as indicated (see also Fig.~\ref{fig:Harsonodisk}).}
\label{fig:hd100model}
\end{figure}

In order to further quantify the young disk's water contribution to
the observed spectra, a good physical model of the envelope + disk
system inside of 100 au is required.  This limits such an analysis to
two Class I sources, TMC1A and L1489.  Here the case of TMC1A is
illustrated, for which the physical model is taken from
\citet{Harsono18}.
In these models, the water abundance in region 2 is assumed to vary
from $10^{-11} - 10^{-7}$. Region 3 is assumed to have an abundance of
$10^{-6}$ and is taken to lie at $T_{\rm gas} > 250$ K and $A_{\rm V}$
between 0.5 and 2 mag.  The UV radiation is typically $G_0>5000$ at
$A_{\rm V} < 0.5$ mag, so that water is easily photodissociated there.

The gas temperature is decoupled from the dust temperature in the
surface layers following \citet{Bruderer12} whereas an accretional
heating rate of 10$^{-7}$ $M_{\odot}$ yr$^{-1}$ is included in the
midplane.  The predicted H$_2$O $1_{10}-1_{01}$ peak line temperatures
from such a model are low, ranging from 0.15 to 2 mK for the highest
abundance in region 2, well below the noise level of 8--13 mK rms in
0.3 km s$^{-1}$ bin in the {\it Herschel} spectra.
In these models, most of the emission comes from region 3
($ T_{\rm gas} > 250$ K) due to the combined high excitation and
column density that extends up to 100 au. Region 2 only
contributes if the water abundance is as high as $10^{-7}$.

In conclusion, to detect cold water vapor emission from young disks
with {\it Herschel} the observations would need to have been much
deeper than the typical 15--30 min integration times used in WISH+
for Class I sources. For Class II disks, much longer integration times
of $>$10 hr have been adopted, but even then only few detections have
been found (see \S~\ref{sec:protoplanetary}). Given that Class 0 and I
disks are overall warmer and more massive than their Class II
counterparts with dust less evolved, they could have higher water
fluxes but future far-infrared space missions will be needed to test
this.

\subsubsection{Warm water in young disks}

Exploratory searches for warm water in young disks have recently been
undertaken with NOEMA and ALMA through the (nonmasing) H$_2^{18}$O
$3_{1,3}-2_{2,0}$ ($E_{\rm up} = 203$ K) 203 GHz and $4_{1,4}-3_{2,1}$
($E_{\rm up} = 322$ K) 390 GHz lines
\citep[e.g.,][]{Jorgensen10,Persson16,Harsono20}.  By targeting
H$_2^{18}$O, any outflow emission is minimized, as evidenced by the
narrow lines observed for Class 0 sources. In contrast with Class 0
sources, however, no H$_2^{18}$O lines are detected for Class I
sources. The current limits for four Class I sources suggest that any
warm water must be located inside of 10 au (region 1)
\citep{Harsono20}. No ALMA 203 GHz data have yet been taken toward
Class I sources, so future ALMA observations that are up to an order
of magnitude more sensitive and higher angular resolution than was
possible with NOEMA are needed.  For region 3, the H$_2^{18}$O column
is too low to contribute, in contrast with the case for the main water
isotopolog discussed in \S~\ref{sec:youngdiskscold}.

The nondetections of warm water with ALMA are significant for other
reasons as well.  It is well known that the infall of high velocity
envelope material onto the low velocity disk can cause a shock,
raising temperatures of gas and dust at the disk-envelope interface to
values much higher than those provided by heating by stellar photons
alone. These accretion shocks are widely found in models and
simulations \citep{Neufeld94,Li13} but have not yet been unambiguously
observed. In the early stages of disk formation, the accretion occurs
close to the star at high velocities, but that material ends up in the
star, not the disk. With time, the accretion quickly moves to larger
disk radii (tens of au) with typical shock speeds of $<$10 km s$^{-1}$
impacting gas with densities $> 10^7$ cm$^{-3}$. For these conditions,
molecules in the infalling gas largely survive the shock but gas
temperatures just behind the shock reach up to 4000 K. These high
temperatures drive most of the oxygen in OH and H$_2$O and should
result in bright far-infrared lines.  The dust temperature is also
raised, but only to about 50 K (assuming 0.1 $\mu$m grains), enough to
release weakly bound molecules but not strongly bound species such as
H$_2$O \citep{Visser09,Miura17}.  Some sputtering of ice mantles can
also occur.  The fact that no warm water vapor emission is detected
with ALMA near the young disk on 50 au scales for either L1527, Elias
29 or TMC1A \citep{Harsono20} limits the importance of any such
shocks.

Interestingly, warm water emission in the H$_2^{16}$O
$10_{2,9}-9_{3,6}$ ($E_{\rm up}$ = 1863 K) line at 321.2 GHz has been
observed with the SMA toward the Class I source HL Tau, known to have
a large disk structure seen in the continuum (100 au radius) with ALMA
\citep{ALMA15}. The nonmasing part of this line is blueshifted by 20
km s$^{-1}$, has a FWHM of 25 km s$^{-1}$, and extends over $3-4''$,
or $\sim$500 au. These characteristics suggest that the bulk of the
water emission originates in the protostellar jet or wind, not in the
rotating disk itself. Similarly, \citet{Watson07} found highly
  excited water lines in the {\it Spitzer} mid-infrared spectra of the
  Class 0 source NGC 1333 IRAS4B which they attributed to an accretion
  shock onto the young disk. However, \citet{Herczeg12} concluded
  based on {\it Herschel-}-PACS data (taken as part of WISH), that
  both the mid- and far-infrared water emission is consistent with an
  outflow origin.

For high-mass protostars, bonafide Keplerian disks of several hundred
au radius have been identified using the high spatial resolution of
ALMA \citep[e.g.,][]{Johnston15,Moscadelli19,Izquierdo18}.
The large beam of {\it Herschel} makes it difficult to isolate the
water coming from young disks for these objects, so they will not be
considered here.  Also, these disks are warm and are therefore better
targets for ALMA. Indeed, vibrationally excited water lines ($v_2$=1,
$5_{5,0}-6_{4,3}$, $E_{\rm up} = 3462$ K) with Keplerian rotation have
been detected toward disks around proto-O and B stars on $<$1000 au
scales \citep{Ginsburg18,Maud19}, but no water abundance has yet been
quantified.  Nevertheless, further spatially resolved water
observations of rotational transitions in vibrational states toward
disks around high-mass protostars will be interesting to understand
the water evolution during disk formation in different environments.

\begin{figure}
\includegraphics[width=9cm]{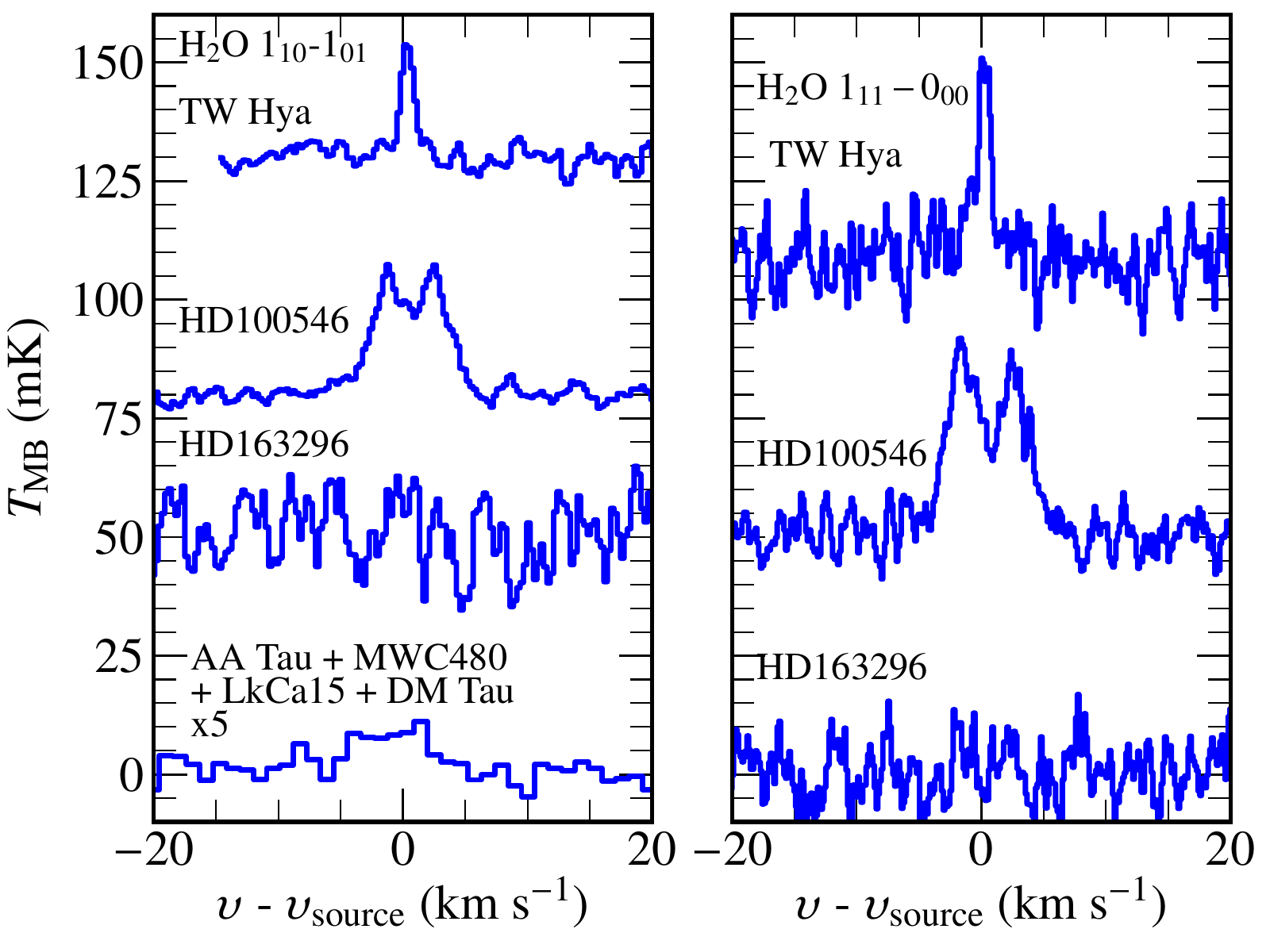}
\caption{Overview of detected emission of H$_2$O $1_{10}$--$1_{01}$
  (left) and $1_{11}$--$0_{00}$ (right) with \emph{Herschel}-HIFI to
  TW Hya, HD~100456, HD~163296 and the stacked results of AA~Tau,
  MWC~480, LkCa~15, and DM~Tau. Data from \citet{Hogerheijde11,Du17},
  {\it Herschel} archive and Hogerheijde et al.\ (unpublished).  }
\label{fig:diskspectra}
\end{figure}

\subsection{Protoplanetary disks}
\label{sec:protoplanetary}

\subsubsection{Cold water in mature disks}
\label{sec:protoplanetarycold}

\paragraph{Background.} \emph{Herschel}-HIFI was unique in probing the cold
water vapor reservoir in planet-forming disks, through
velocity-resolved observations of the two ground-state rotational
transitions of ortho-H$_2$O $1_{10}-1_{01}$ line at 557 GHz and the
para-H$_2$O $1_{11}-0_{00}$ line at 1113 GHz. With upper-level
energies of 53 and 61 K, and critical densities of $\sim 1\times 10^8$
and $\sim 7\times 10^8$ cm$^{-3}$, respectively, emission can be
expected from across the disk. This includes the cold ($<50$ K) outer
disk regions, although sub-thermal excitation conditions need to be
taken into account when interpreting the observations. It is these
outer disk regions that will dominate the signal in the large
\emph{Herschel} beams of $38''$ and $19''$ at the respective line
frequencies if the water vapor abundance is sufficiently high. Since
the beam sizes exceed those of disks (typically diameters of no more
than a few arcsec), the expected main-beam antenna temperatures are
small. Although \emph{Herschel} cannot spatially resolve the emission,
HIFI's high spectral resolution ($<$0.1 km~s$^{-1}$) allows the
emission lines from suitably inclined disks to be velocity-resolved,
thus providing constraints on the radial origin of the emission if
standard Kepler orbital speeds are assumed.

As discussed above, theoretical considerations predict water vapor to
occur in three distinct regions in disks, as illustrated schematically in
Figure~\ref{fig:Harsonodisk} and in Figure~\ref{fig:hd100model} for
one specific Herbig disk
\citep[e.g.,][]{Dominik05,Woitke09,Cleeves14}.
The small radial extent of the region 1 results in a negligible
contribution of the expected flux in the {\it Herschel} beam in the
ground-state lines ($<10\%$ of total emission).  In contrast with
young disks, region 3 also contributes little to the emission in the
ground-state lines in the {\it Herschel} beam ($<30\%$), because of
the low column density and the low H$_2$ density that precludes
efficient excitation of the lines. Outside these two regions, water is
generally expected to be frozen out onto cold ($<$150 K) dust grains,
as evidenced by the handful of detections of water ice in planet
forming disks \citep[e.g.,][]{Chiang01,McClure15,Min16,Honda16}.
Region 2 consists of the cold water vapor reservoir produced by UV
photodesorption of water ice into the gas phase
\cite[e.g,][]{Andersson08,Oberg09h2o}, with water vapor abundances of
up to $10^{-7}$ relative to H$_2$ predicted at intermediate disk scale
heights out to large radii. In this region, densities of a few times
$10^7$ cm$^{-3}$ and temperatures of 30--45 K are sufficient to
generate detectable emission in the \emph{Herschel} beam.

\paragraph{HIFI results and models.} The main observational result
from \emph{Herschel} on the ground-state rotational water lines from
mature disks, is that the emission is even weaker than expected on
theoretical grounds.  \emph{Herschel}-HIFI observed the ortho-H$_2$O
1$_{10}$--1$_{01}$ line toward ten disks around T~Tauri stars and four
disks around Herbig Ae stars as part of the WISH+ programs
(Table~\ref{tab:programs}). A subset of these targets were also
observed in the para-H$_2$O $1_{11}$--$0_{00}$ line with similar
results and are not further discussed here. The observations and their
results are presented in \citet{Hogerheijde11} and Hogerheijde et al.\
(unpublished), \citet{Salinas16}, and \citet{Du17}.

\begin{figure}
\includegraphics[width=9cm]{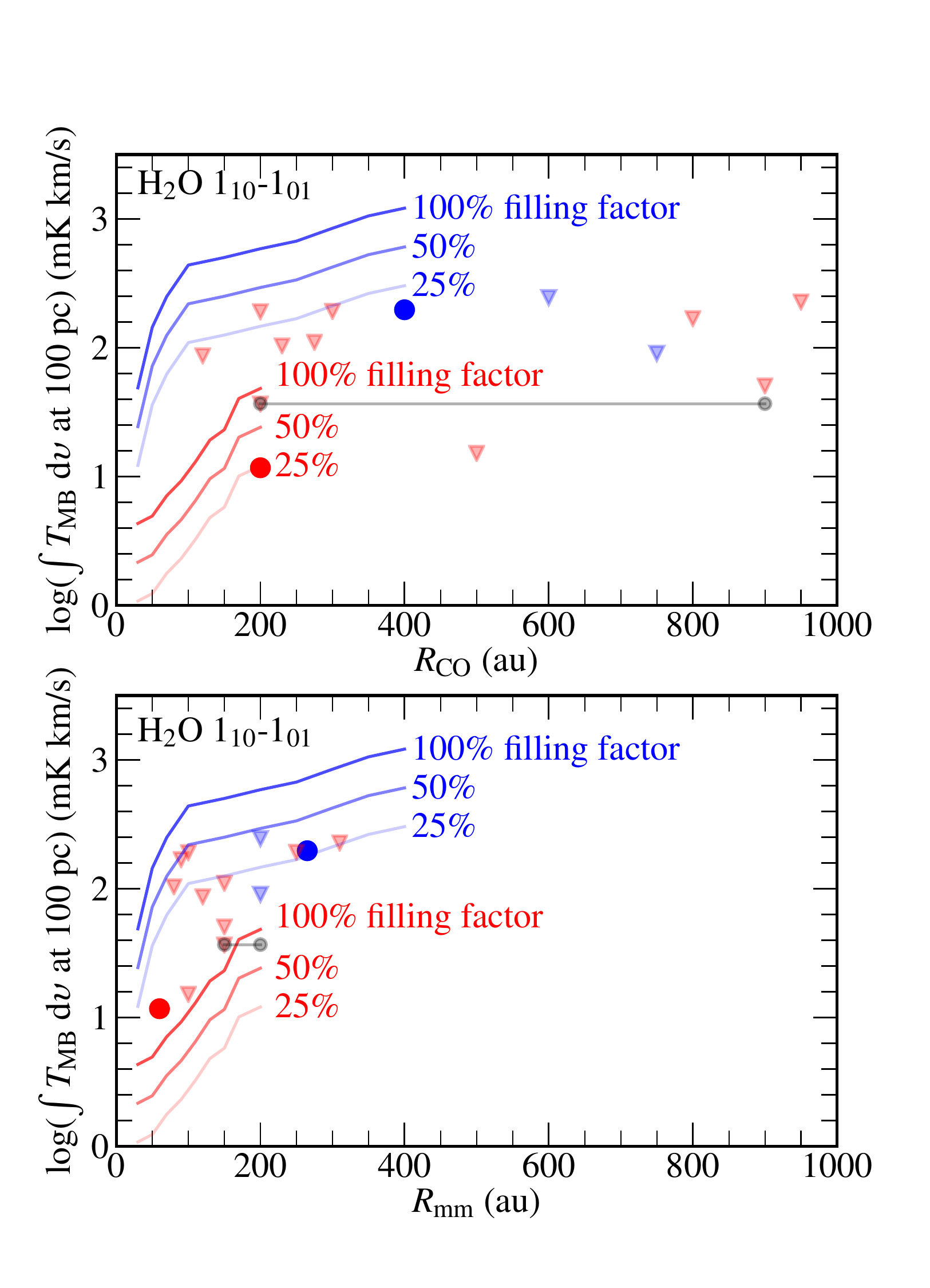}
\caption{Line luminosities of models and observations of ortho-H$_2$O $1_{10}$--$1_{01}$ for planet-forming disks observed with \emph{Herschel}-HIFI.
  Solid blue and red lines show predicted line luminosities for a Herbig disk model (blue) and a T Tauri disk model (red); see text for details. Note the logarithmic scale. The effects of non-unity filling factors are illustrated by shaded lines at 50\% and 25\%. The observational data are from \citet{Hogerheijde11,Du17} and Hogerheijde et al.\ (unpublished).  Filled circles show detections, blue for the Herbig star HD100546 and red for the T Tauri star TW Hya. The gray circles connected by the solid gray line show the detected line luminosity of the stacked detection of AA Tau, LkCa15, MWC~480 and DM~Tau. Down-pointing triangles show upper limits (blue: Herbig disks; red: T Tauri disks). Data are plotted versus the observed size of the disk as measured through CO emission (upper panel) and mm-wavelength continuum emission (lower panel); for the stacked detection (gray line) the range of disk sizes is plotted. 
Observed disk sizes are taken from \citet{Walsh14, Andrews12, Loomis17, Bergin16, Kudo18, Jin19, Qi03, Liu19, Fedele17, Huang16, Dutrey03, Dutrey16, Hughes09, Macias18, Isella16, Boehler18, Huelamo15, Cleeves16}.
Line luminosities and disk sizes are scaled to the latest GAIA
distance estimates.}
\label{fig:disklineluminosity}
\end{figure}

In summary, out of the fourteen observed sources, only two yielded
detected lines (TW Hya and HD100546; Fig.~\ref{fig:diskspectra}) in
spite of long integrations and rms levels down to 1.2--2.0 mK in 0.27
km~s$^{-1}$ channels. A stacked spectrum of AA Tau, DM Tau, LkCa~15
and MWC~480 also yielded a significant detection of the line. These
detections correspond to the data with the longest integration times
(typically $>$10 hr) and thus lowest noise levels. For all other sources,
observed to various rms levels, only upper limits were
obtained. Figure~\ref{fig:disklineluminosity} shows the detections and
upper limits in terms of velocity-integrated line luminosities, using
the most up-to-date distances of the sources obtained from GAIA.

Figure~\ref{fig:disklineluminosity}
plots the observed line luminosities and the upper limits as function
of the gas and dust outer radii of the disk as measured in $^{12}$CO
emission and through the mm-continuum emission. It also
compares the observations to the expected emission for a ``typical''
T~Tauri disk (red) and a ``typical'' Herbig Ae disk (blue). For the
T~Tauri disk, the TW~Hya model of \citet{Hogerheijde11} and
\citet{Salinas16} is used: a disk mass of 0.04 $M_\odot$,  disk outer
radius of 200 au, and a disk inclination of $7^\circ$. For the Herbig
disk, a model of the HD 100546 disk is adopted (Hogerheijde et al.\
unpublished): a disk mass of 0.01 $M_\odot$, a disk outer radius of 400
au, and a disk inclination of $42^\circ$. In both cases, the disk
temperature is calculated self-consistently and the disk chemistry
includes photodesorption due to stellar UV photons.
The H$_2$O excitation and emission line luminosity is calculated using
LIME version 1.9.5\footnote{https://github.com/lime-rt/lime}
\citep{Brinch10} and H$_2$O collisional rates from \citet{Dubernet09}
(taken from the LAMDA database \citealt{Schoier05}).
While the T~Tauri and Herbig disk models have a fixed outer radius,
the effect of confining the emission from water to a smaller region is
mimicked by artificially setting the water abundance to zero outside a
certain radius. Figure~\ref{fig:disklineluminosity} plots the
predicted emission from these models against this radius. Predicted
fluxes for Herbig disks are higher than for T Tauri disks because of
the higher temperature of the former; due to the subthermal excitation
of the water molecule, a factor of two difference in temperature
strongly affects the excitation.

Comparison of the model curves with the two detected disks (TW Hya and
HD 100546) and many of the upper limits confirms the earlier reported
results, namely that the H$_2$O ground-state rotational emission from disks
is weaker than expected by factors of 5--10.  \citet{Bergin10} and
\citet{Hogerheijde11} suggested that this indicates a reduction of
water ice at intermediate disk heights where stellar ultraviolet photons can
still penetrate. These authors hypothesize that settling of (icy)
grains can, over time, remove significant fractions of the water ice
reservoir \citep[see also][]{Kama16,Krijt16,Krijt20}.  A similar overall reduction of
volatile material was inferred by \citet{Du17} for the full sample.

\citet{Salinas16} put forward a different scenario for TW Hya, where
the observed fluxes are reproduced when the icy grains are taken to
spatially coincide only with the millimeter-sized grains as probed by
the submm continuum emission. \citet{Andrews12} find that in
the TW Hya disk, these grains are confined to $\sim 60$ au, much
smaller than the 200 au found for the gaseous outer disk radius as
seen in CO emission. Figure~\ref{fig:disklineluminosity} shows that
much better agreement between the TW Hya detection and the model curve
is found when the former is plotted at the detected radius of the
mm emission. Although in this case the model underpredicts the
line luminosity of TW Hya, it should be remembered that these model
curves are produced by setting the water abundance to zero outside the
plotted radius, and do not include transporting the water ice from
larger radii to smaller radii. No exact agreement would therefore be
expected in this simple minded comparison. In contrast to TW Hya, for
HD 100546 reducing the emitting area to that of the millimeter-sized grains
\citep[$\sim 265$ au,][]{Walsh14} 
does not sufficiently reduce the emission. 

Instead of reducing the size of the emitting region radially,
introducing a non-unity filling factor can also produce a reduction of
the line luminosity. Figure~\ref{fig:disklineluminosity} shows that
scenarios where as little as 25\% of the disk out to the CO outer
radius is filled with H$_2$O emitting material, reproduces the
observed line fluxes and many of the upper limits. Such filling
factors may be appropriate for several of the disks in the sample, if
most of the water ice is spatially coincident with the
mm-continuum emitting grains. High-resolution ALMA studies
show that such grains are confined in many (bright) disks to rings or
bands covering only small fractions of the disks
\citep[e.g.,][]{Andrews18,Long19,vanderMarel15}. For the particular
case of TW Hya, the disk is found to be relatively uniformly filled
with mm-continuum emitting grains out to $\sim$60 au, while
for HD100546 the disk's mm emission is dominated by a 15--45
au ring \citep{Pineda19}.  While such a ring takes up only 1\% of the
total disk area (out to 400 au), it does leave open the possibility
that the H$_2$O emission originates from only limited radial ranges
inside the disk.

The three scenarios outlined above can all explain the weak H$_2$O
emission found in the spatially unresolved \emph{Herschel}-HIFI
observations: (i) overall reduction of volatile material in the
vertical regions of the disk subjected to UV radiation; (ii) a uniform
radial confinement to the disk area rich in mm-continuum
emitting grains; and (iii) H$_2$O emission regions carved up in rings
or bands reminiscent of the resolved submm substructures seen
in many bright disks.

For one disk, it is possible to go one step further. The detected
spectrally resolved line toward the suitably inclined disk of HD
100546 (Fig.~\ref{fig:diskspectra}) holds information on the radial
origin of the emission. A simple model with a radial power-law
dependency for the line emission suggests that the H$_2$O emission
originates from radii between 40 au out to 250--300 au (Hogerheijde et
al.\ unpublished). While the outer radius is not very well determined
due to the adopted radial power-law drop off of the emission, the
inner radius is robust and indicates that the emission originates
outside the bright ring of mm continuum emission seen with ALMA
\citep{Pineda19}. Likely, this dust ring is optically thick at the
frequencies of the observed water lines (557 and 1113 GHz), obscuring
the water emission. The inferred region of water vapor emission
actually overlaps with the 40--100 au region where water ice
absorption has been detected \citep{Honda16}.

\paragraph{Summary and future prospects.} In summary, the
\emph{Herschel}-HIFI observations of the water rotational ground state
lines toward fourteen planet-forming disks show that water vapor is
present in at least five of these for which the lowest noise levels
were obtained. Theoretical models suggest that this water vapor likely
originates from photodesorption of icy grains, but that the reservoir
of available water ice is smaller than expected. Radial confinement in
a single region or several radial rings or bands, or vertical
confinement at heights below where UV photons can penetrate, can all
explain this. Whichever mechanism is at work, a close relation with
the settling, growth and radial transport of dust grains seems
implicated. The low gas-phase abundance of cold water vapor is also in
line with the inferred low gas-phase oxygen and carbon abundances in
the TW~Hya outer disk \citep{Favre13, Du15, Du17, Bergin16,
  Bergner19}. If low oxygen and carbon abundances go together, then
the low volatile carbon abundances inferred from weak CO emission in
many other mature disks \citep[e.g.,][]{Miotello17,Long17} suggests
that this is a common feature.

The prospects of further investigating the distribution of cold water
vapor across disks requires spatially resolved observations. ALMA
observations of lines of water isotopologs provide a possible
avenue. However, available H$_2^{18}$O transitions such as the 203 GHz
line (\S~\ref{sec:hotcore}) have lower-level energies exceeding 200 K,
and do not trace the cold reservoir of water ice ($<150$ K). Several
low-lying HDO transitions are observable with ALMA in Bands 8 and 10,
but their interpretation is complicated by the required knowledge of
the deuteration fractionation.

A highly promising avenue to constrain the radial distribution of
gas-phase water was pioneered by \citet{Zhang13}, who analyzed
multiple transitions of water from HIFI and PACS in TW~Hya to derive a
radial abundance profile and snowline location by adopting a
temperature structure for the disk. SPICA and the {\it Origins Space
  Telescope} (25--588 $\mu$m) offer the possibility to extend such
studies to a statistically significant sample of disks.

\subsubsection{Warm water in mature disks}
\label{sec:innerdisks}

 The {\it Spitzer}-IRS detected a wealth of highly-excited pure
  rotational lines of warm water at 10–30 $\mu$m in disks around a
  significant fraction of T Tauri stars
  \citep{Carr04,Carr08,Salyk08,Pontoppidan10,Salyk11,Carr11,Salyk15}, with
  line profiles consistent with a disk origin
  \citep{Pontoppidan10visir,Salyk19}. Typical water excitation
  temperatures are $T_{\rm ex}\approx 450$~K. Spectrally resolved
  groundbased near-IR vibration-rotation lines around 3 $\mu$m show
  that in some sources the hot water originates in both a disk and a slow
  disk wind \citep{Salyk08,Mandell12}.

\begin{figure}
  \centering
    \includegraphics[width=7.5cm]{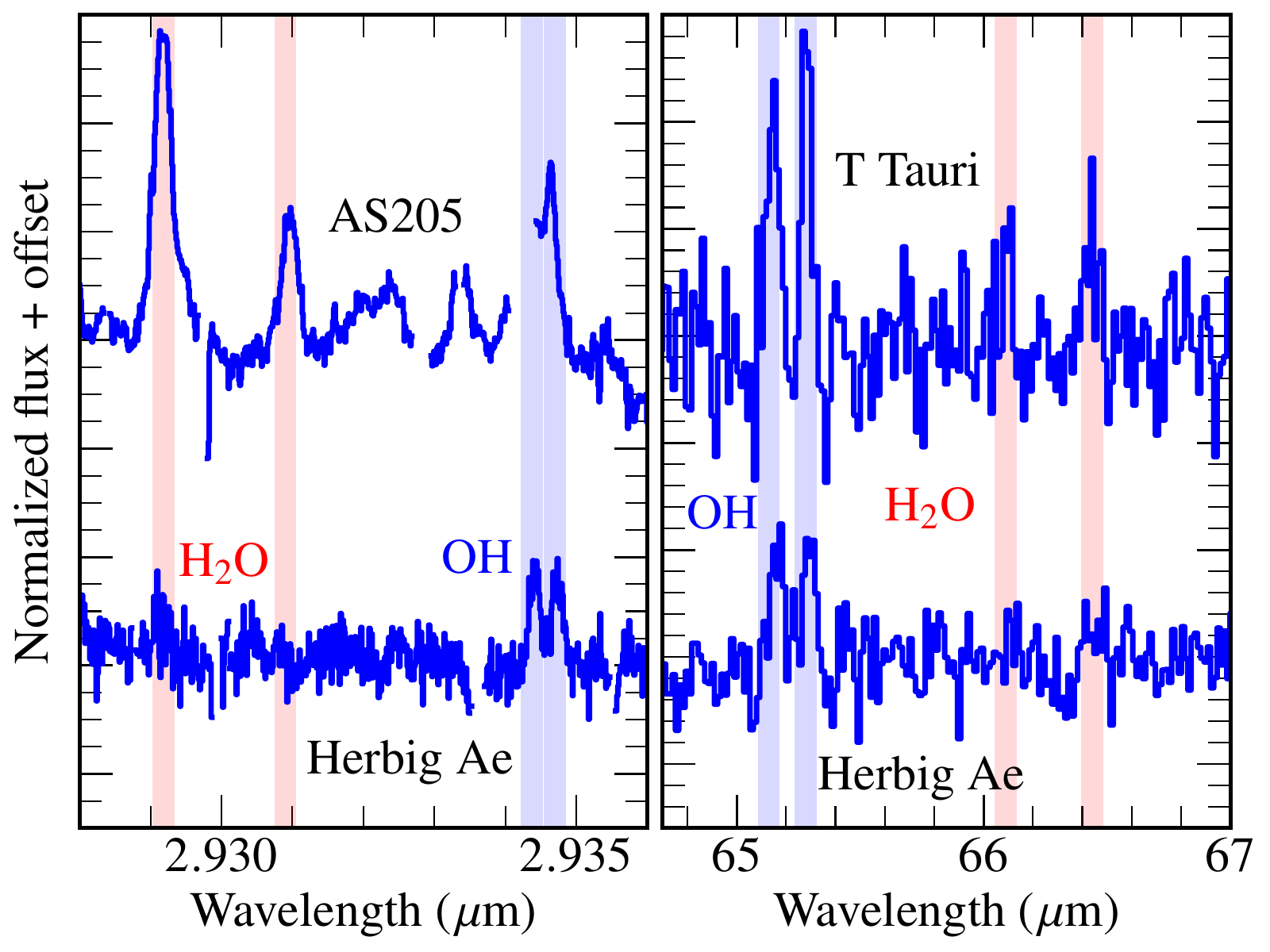}
    \caption{Water observations of protoplanetary disks at
      near-infrared wavelength with VLT-CRIRES and far-infrared wavelengths by
      {\it Herschel}-PACS in typical T Tauri and Herbig disks. The
      VLT-CRIRES spectra are for individual disks, AS 205 and HD
      250550, whereas the PACS spectra are obtained by stacking
      multiple spectra. Data from \citet{Fedele11,Fedele13}.}
               \label{fig:diskPACS}
\end{figure}

Abundance ratios extracted from the {\it Spitzer} observations are
uncertain because the lines are highly saturated and spectrally
unresolved. Nevertheless, within the more than an order of magnitude
uncertainty, H$_2$O/CO$\approx 1 - 10$ has been inferred for emitting
radii up to a few au \citep{Salyk11,Mandell12}.  This indicates that
the inner disks of those sources where water is prominently detected
have high water abundances of order 10$^{-4}$ and are thus not dry, at
least not in the surface layers down to where the dust becomes
optically thick at mid-IR wavelengths.

In contrast with the case for young disks in the Class 0 and I stages,
no NOEMA or ALMA observations of warm H$_2^{18}$O lines in Class II
disks have yet been published.  Emission of more highly excited lines,
however, has been detected using \emph{Herschel}-PACS
\citep[e.g.,][]{Fedele12,Fedele13,Riviere12} and these transitions,
when combined with longer wavelength HIFI and shorter wavelength
  {\it Spitzer}/VLT/Keck infrared lines,
provide insight into the gas-phase water abundance inside the
water snowline (Fig.~\ref{fig:diskPACS}).

Interestingly, the HD 100546 Herbig disk, which has prominent HIFI
lines pointing to cold water in the outer disk
(Fig.~\ref{fig:diskspectra}), does not show any detection of warm
water lines from the inner disk with PACS or VLT. In contrast, the HD
163296 disk has no detected HIFI lines, but does show a (stacked)
detection of warm water with PACS \citep{Fedele13}. These two stars
and their luminosities are similar, so the global disk thermal
structure cannot be the explanation. Instead, disk substructures in
gas and dust likely are. One possibility is that water ice is trapped
in the outer disk of HD 100546 by the bright dust ring at 40 au
\citep{Pineda19} (see discussion in \S~\ref{sec:protoplanetarycold}),
whereas the icy pebbles have drifted inward for HD 163296. This would
be consistent with the high CO abundances detected inside the CO
snowline in the HD 163296 disk \citep[e.g.,][]{Booth19}.

Another illustrative case of the importance of dust traps in
regulating the inner warm water reservoir is provided by the TW Hya
disk, the only other disk for which HIFI ground-state cold water lines
are clearly detected. Based on {\it Spitzer} spectra of warm water
combined with H$_2$, the inner few au of the TW Hya disk are found to
be oxygen poor by a factor of $\sim$50. This points to an efficient
water ice dust trap just outside the snowline around 2.4 au
\citep{Bosman19TWHya}.

More generally, \citet{Banzatti17} find a correlation between water
line fluxes and size of an inner disk gas cavity, as measured from CO
ro-vibrational line profiles for a set of T Tauri and Herbig
disks. Water emission first disappears at near-infrared wavelengths
(hot water) and then at mid-infrared wavelengths (warm water) as the
radius of gas emission expands out to the water ice line. This
suggests that infrared water spectra are a good tracer of inside-out
water depletion in regions 1 and 3 from within to outside the snow
line.  \citet{Banzatti20} also find an anticorrelation between the
water line fluxes in {\it Spitzer} data and the radius of the mm-sized
dust disk, suggesting that small disks where icy dust grains have
drifted inward have higher warm water abundances.

\section{Water and the oxygen budget}
\label{sec:obudget}

Sections 5--9 have quantified the abundances of water vapor and other
oxygen-containing molecules such as OH and O$_2$ in different components
of protostellar sources: outflows and shocks, pre-stellar cores, cold
and warm envelopes around protostars, and in young and mature
disks. This section looks at the question whether we have now
accounted for all of the oxygen.

The broader issue of the oxygen budget from the diffuse ISM to dense
clouds and comets was introduced in \S~\ref{sec:oxygenbudget} and is
summarized in much more detail in Appendix \ref{app:obudget}. The
overview Table~\ref{tab:obudget} and Figure~\ref{fig:obudget} are
included here in the main text of the paper. As discussed in
\S~\ref{sec:oxygenbudget}, some fraction of oxygen appears to be
locked up in an unknown form called UDO (Unidentified Depleted Oxygen)
even in the diffuse ISM, assuming a total oxygen abundance of
[O]=$5.75\times 10^{-4}$ or 575 ppm.

 \begin{figure*}
  \centering
\includegraphics[width=13cm]{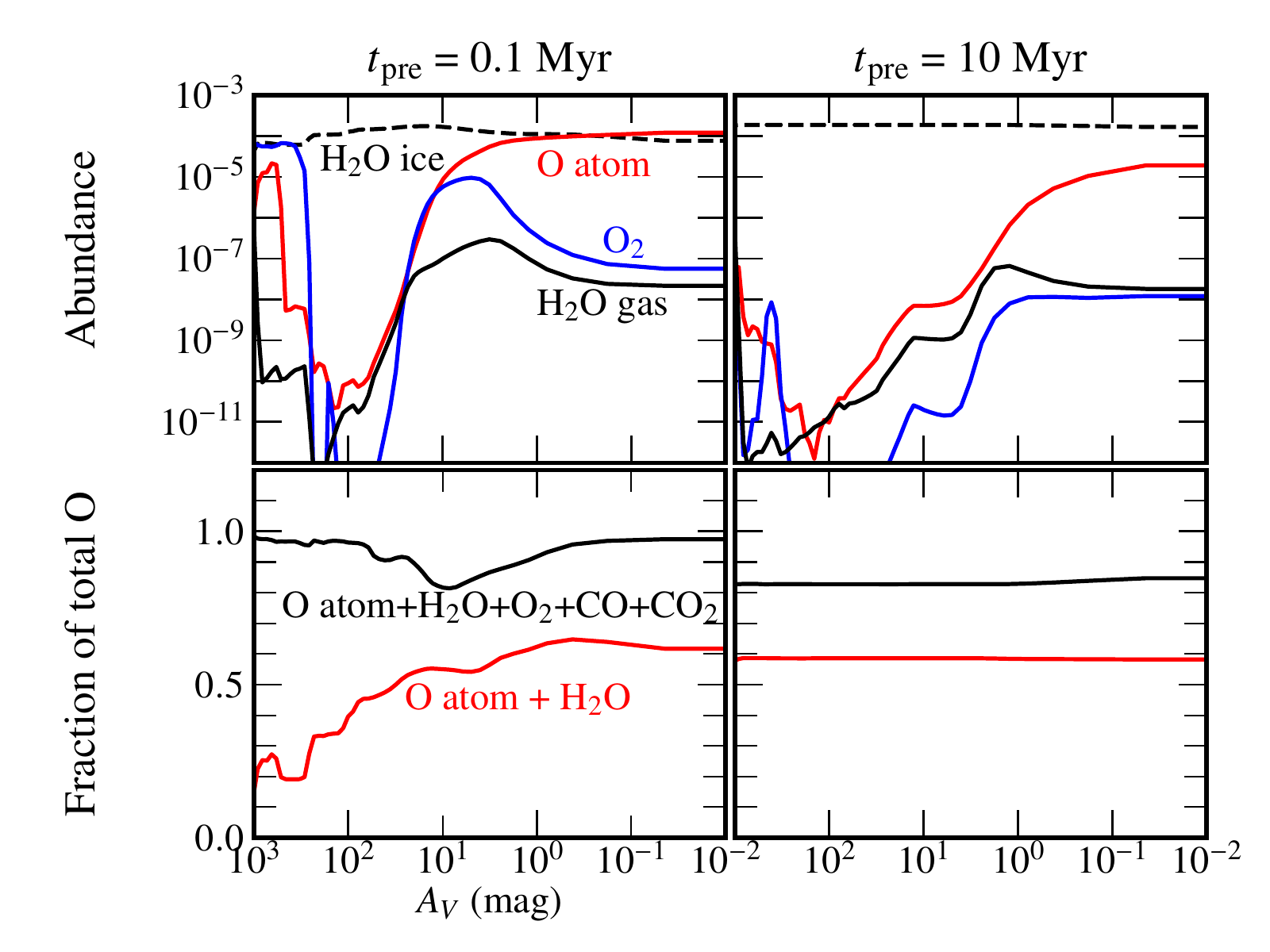}
\caption{Full multilayer gas-grain model results for water and related
  chemical species for the temperature and density structure of the
  NGC 1333 IRAS4A envelope model using the \citet{Furuya16} chemical
  network, for two different timescales of the pre-stellar phase: 0.1
  and 10 Myr.  Top panel: gaseous O, O$_2$ and H$_2$O (full lines) and
  H$_2$O ice (dashed lines). O$_2$ is seen to be strongly reduced in
  the 10 Myr case. The bottom figure includes the sum of the
  abundances of other major O-bearing species in the models in both gas +
  ice. The remaining oxygen is in H$_2$CO, CH$_3$OH and other larger
  organic species. }
         \label{fig:iras4amodel}
\end{figure*}

\subsection{Water gas + ice abundance in protostellar envelopes: the oxygen budget mystery}
\label{sec:watergasice}

The HIFI lines have allowed a detailed gaseous water abundance profile
through the envelope to be derived (\S~\ref{sec:protocold}), but this
chemical analysis ultimately has to be consistent with observations of
other oxygen carrying species, most notably water ice and O$_2$
gas (\S~\ref{sec:o2}). There are only a limited number of sources for
which both water ice and gas have been observed along the same line of
sight, since most deeply embedded Class 0 sources were too faint at
mid-infrared wavelengths for ice absorption studies with {\it
  Spitzer}.  In fact, there is no source for which all three species
--- water gas, water ice and deep O$_2$ gas limits --- are available.

Nevertheless, surveys of water ice exist for large samples of
infrared-bright low- and high-mass protostars, revealing typical water
ice column density ratios $N_{\rm s-H2O}$/$N_{\rm H} \sim 5 \times
10^{-5}$
\citep[e.g.,][]{Gibb04,Pontoppidan04,Boogert08,Oberg11,Whittet13,Boogert15}.
Similar values of $(2-6)\times 10^{-5}$ are measured toward stars
behind dark quiescent clouds \citep{Boogert13}.
The column density of hydrogen nuclei, $N_{\rm H} \approx 2 N$(H$_2$),
is inferred either from the silicate optical depth, or from the color
excess toward the star, or from a combination of the two.  Observed values
show a range of a factor of two around this mean.

The important implication of these water gas + ice observations is
that water gas contains a negligible fraction of oxygen and that even
water ice locks up only a moderate fraction, $\lesssim 20\%$, of the
volatile interstellar oxygen abundance with respect to total hydrogen
of $3.2 \times 10^{-4}$,  that is, the oxygen that is not locked up in
silicates (\S~\ref{sec:oxygenbudget}). In other words, not all
volatile oxygen observed in diffuse clouds ends up as water ice in the
dense cloud phase. A similar conclusion was reached by comparing gas
and ice for high-mass protostellar envelopes by \citet{Boonman03}.  As
shown in \S~\ref{sec:o2}, O$_2$ contains at most a few \%.

There are four possible explanations for this conundrum: (i) water ice
formation is suppressed, with most oxygen locked up in other volatile
species; (ii) water ice or gas is destroyed after formation, with
oxygen driven into other species; (iii) the water ice abundance
inferred from observations is underestimated; and (iv) the fraction of
oxygen locked up in refractory UDO increases from diffuse to dense
clouds. The following analysis focuses on low-mass protostars, but the
same arguments hold for high-mass protostellar envelopes.

\subsubsection{Suppressing water ice formation}

Regarding option (i), our modeling of both the water gas and ice using
the SWaN network demonstrates that the only way to get a low water ice
column is to start with a low water ice abundance in the pre-stellar
stage \citep{Schmalzl14}. Water ice cannot be effectively destroyed in
protostellar envelopes at temperatures lower than 100 K, but some
additional water ice can be made in the coldest outer parts where the
dust temperature is below $\lesssim 15$~K, which could increase the
ice column by up to a factor of $\sim$2 \citep[see Fig.~4
in][]{Schmalzl14}. This limiting temperature of $\sim$15~K is set by
the binding energies of atomic and molecular oxygen, either of which
is needed to make water ice. Once the temperature is above this
critical temperature, which holds for the bulk of the protostellar
envelope, no new water ice can form and the water ice abundance is
`frozen in' at the value given at the start of the protostellar phase,
$t=0$. The Schmalzl et al.\ models use $E_b$=800 K for atomic O, but the more
recent higher values of $E_b$=1660 K on water ice \citep{He15} only
exacerbate the problem by increasing the amount of water ice that can
be formed in the protostellar stage. 

How can the water ice abundance in the pre-stellar stage be limited?
Starting from oxygen in atomic form, as would be appropriate for an
earlier more diffuse cloud phase, water ice formation is fast
\citep{Cuppen07}. Current models thus require a short pre-stellar
phase, typically only 0.1 Myr at densities $>10^4$ cm$^{-3}$, to limit
water ice formation \citep{Schmalzl14}.
This timescale is shorter than the estimated duration of $\sim$0.5 Myr
from pre-stellar core statistics \citep{Enoch08}.

Full gas-grain models can test how much oxygen is processed into other
species than O, O$_2$ and H$_2$O. Figure~\ref{fig:iras4amodel}
presents the abundances in gas and ice of major oxygen-bearing species
as functions of position in the IRAS4A envelope, together with their
sum, using the three-phase chemical network of
\citet{Furuya16,Furuya17}. Initial abundances are identical to those
in \citet{Schmalzl14}, with [O$_{\rm vol}$]=$3.2\times 10^{-4}$. The
temperature and density structure of the source is taken to be the
same as that used to analyze the H$_2$O and HDO spectra in
\S~\ref{sec:abundanceprofiles} (Fig.~\ref{fig:N1333_hdo_nh3}).

After a short cold pre-stellar phase of 0.1 Myr, most oxygen is still
in atomic form with an O abundance of 1.6$\times 10^{-4}$
(Fig.~\ref{fig:iras4amodel}, left panels). The water ice abundance
varies with depth, but its integrated abundance along the line of
sight is only 0.5$\times 10^{-4}$ with respect to total H, consistent
with the ice observations.  In contrast, for a long pre-stellar phase of
10 Myr (right panels), all oxygen is in water ice at the start of the
protostellar phase with no change with depth. Velocity resolved
observations of cold atomic [O I] in emission or absorption could in
principle distinguish between these scenarios, but may be difficult to
interpret because of large optical depths.

In what form is the remaining oxygen in these models?  As expected,
the O$_2$ abundance is indeed strongly affected by the duration of the
pre-stellar phase: at 10 Myr, all the oxygen is in water ice 
with very little gaseous O$_2$, in contrast with the 0.1 Myr case.  As
the bottom panels of Figure~\ref{fig:iras4amodel} show, some fraction
is in CO and CO$_2$ gas or ice. At most 20\% of the volatile oxygen,
about 60-70 ppm, is in other species in either model, mostly H$_2$CO,
CH$_3$OH and minor oxygen-containing species. All of these results are
consistent with those presented in \citet{Schmalzl14}, which used a
much simpler chemical model.

The models presented in Fig.~\ref{fig:iras4amodel} assume a static
physical structure with time. Figure~\ref{fig:hdo_models} (top) shows
the result of the water chemistry in a collapse model such as
described by \citet{Furuya16} in one dimension and \citet{Visser09} in
two dimensions. There a streamline is followed from the outer edge of
the envelope inward, assuming a pre-stellar phase of 10
Myr. Consistent with the static case, the H$_2$O ice abundance stays
high even though different conditions are experienced ``en route''. In
conclusion, the only way to suppress H$_2$O ice formation is 
through a short pre-stellar stage.

\subsubsection{Water destruction after formation}

For option (ii), episodic accretion and heating during the
protostellar phase has been proposed as a possible way to decrease the
water ice abundance. It is now well established that most protostars
undergo multiple luminosity outbursts during the embedded phase of
star formation, increasing the luminosity by up to a factor of 100 for
a short period of time, typically $<$100 yr
\citep[e.g.,][]{Evans09,Dunham12,Audard14}. The enhanced luminosity
heats the envelopes to higher temperatures, sublimating water ice out
to larger radii and re-freezing it once the envelope has cooled back
down again, but with some delay due to the fact that the freeze-out
time is longer than the cooling time \citep{Visser12b,Frimann17}. If
water gas could be driven into other oxygen-containing molecules
during a luminosity burst before re-freezing, this could suppress the
water ice abundance compared with the pre-stellar stage and lessen the
requirement on its short time scale. However, detailed chemical models
have failed to make this scenario work in practice under protostellar
conditions \citep[e.g.,][]{Taquet16o2,Eistrup18o2}.

\begin{table*}[t]
\caption[]{Oxygen budget of various species in ppm assuming overall [O]=575 ppm}
\begin{tabular}{l l r r r r r}
\hline
\hline
\noalign{\smallskip}
Form & Species &  Diffuse clouds & Low-mass YSO & High-mass YSO & Shocks & Comets \\
\noalign{\smallskip}
\hline
Dust & Refr.\ Dust      & 140 & 140 & 140 & 140 & 140\\
     & Refr.\ Organics & $-$ & $-$ & $-$ & $-$ &  50\\
\noalign{\smallskip}
Gas & O               & 320  &  ? &  ?     & 30 & $-$ \\
    & H$_2$O          & $<1$ &  40  & $\leq 62$  & 80 & $-$ \\
    & CO              &   1  & 100   & 100       & 100 & $-$ \\
    & CO$_2$          & $-$  & $<1$  & $<1$      & $<1$ & $-$ \\
    & O-other         & $-$  & $-$   & $-$    & $-$ & $-$ \\
\noalign{\smallskip}
Ice & H$_2$O         & $<2$ & 38 & 31 & $-$ & 323 \\
    & CO             & $<1$ & 10 & 3  & $-$ & 10 \\
    & CO$_2$         & $<1$ & 24 & 8 & $-$ & 30 \\
    & CH$_3$OH       & $<1$ & 3 &  4 & $-$ & 1 \\
    & O-other        & $<1$ & 1 & 1 & $-$ & 21 \\
\noalign{\smallskip}
UDO &                & 115 &  219 & 226 & 225 & $-$ \\
\noalign{\smallskip}
\hline
\end{tabular}

See Appendix~\ref{app:obudget} for references to the adopted numbers
in this table and their uncertainties. The ice abundances for
protostars are as observed, 
without hiding ices in large grains. \\
\label{tab:obudget}
\end{table*}

\subsubsection{Ice columns underestimated due to grain growth}

Another solution (iii) is that the water ice absorption measurements
underestimate the true water ice column. This could be the case if
some fraction of grains along the line of sight are large enough
(typically a few $\mu$m) that they do not show any absorption
feature. This situation has been proposed for the diffuse $\zeta$ Oph
cloud where also some fraction of oxygen is unaccounted for
\citep{Poteet13}. For our much denser protostellar sources, only a
small fraction of the grains needs to be large.  As a toy-model
example, assume two dust populations: a standard one with grain size
$a_s$=0.1 $\mu$m, and a large-grain population with $a_\ell$=10
$\mu$m. The latter population would not show water ice absorption even
if covered with ice. The relative abundances are adjusted so that the
overall grain abundance is 6.5$\times 10^{-13}$ with respect to
hydrogen if all grains have $a=0.1$ $\mu$m (corresponding to a
gas/dust mass ratio of $\sim 100$).  The parameter to vary is the
ratio of large to small grains $f_{\rm large}$. The equations in the
SWaN network by \citet{Schmalzl14} have been adjusted for oxygen
freeze-out onto the two different grain sizes. The fraction of
traceable water ice by infrared absorption is then taken to be the
amount of ice on small grains relative to the total amount of water
ice on both small and large grains.  This analysis shows that a
fraction $f_{\rm large}=0.01\%$ of the grain population in 10
$\mu$m-sized grains (by numbers) would catch 50\% of the atomic oxygen
during freeze-out and make it invisible.  Assuming $a_\ell$=1 $\mu$m
would move the 50\% point to $f_{\rm large}=$1\%.

This scenario is highly plausible, given the substantial observational
evidence for grain growth to $\mu$m size in dense clouds
\citep[e.g.,][]{Pagani10,Boogert13,Steinacker15} and protostellar
envelopes \citep[e.g.,][]{Chiang12,Miotello14classi}. These
$\mu$m-sized grains likely coagulated from ice-coated small particles
(since they carry the largest surface area), thereby natively locking
up water ice within larger grains. This would have the added benefit
that there does not need to be a major fraction in atomic oxygen but
there is still room for some fraction of UDO.  For dark quiescent
clouds prior to star formation, which also show low water ice
abundances, the observed low water ice abundances could be an early
evolutionary effect with water ice still being formed, although even
there grain growth could start in the denser parts.

\subsubsection{Increasing UDO from diffuse to dense clouds}

The fourth possibility is that oxygen is locked up in increasing
amounts in some refractory form called UDO as the density increases
(Table~\ref{tab:obudget}) \citep{Jenkins09,Whittet10}. This scenario
is further discussed in \S~\ref{sec:obudget4}.

In summary, the combined observations of gaseous and solid water in
protostellar envelopes allow the water gas and ice abundances to
be constrained quantitatively as functions of radius.
The initial pre-stellar ice abundance is a crucial parameter that sets
the ice abundance profile in protostellar envelopes. Unless the
pre-stellar stage is very short, it is difficult to prevent all
volatile oxygen being turned into water ice. Alternatively, the
pre-stellar stage could be longer, and the observed low ice abundances
are then most easily explained by some fraction of grains having grown
to at least $\mu$m size. The final option is that the fraction of
oxygen in UDO has increased from diffuse to dense clouds.

\subsection{Hot cores}

The above models and discussion concerned the outer part of the
protostellar envelope where water is frozen out.
Section~\ref{sec:hotcore} shows that the gaseous water abundance in
the inner hot core region inside the water ice line is also generally lower
than the expected value of H$_2$O/H$_2$=$4\times 10^{-4}$.
If most hot cores are indeed dry, where is the remaining oxygen?  Any
water ice locked up in large grains, as in option (iii), should also
have sublimated.  Photodissociation by UV radiation from the young
star-disk accreting boundary layer (LM) or the stars themselves (HM)
can rapidly destroy water, but only in a narrow region of visual
extinctions or in outflow cavities through which UV photons can escape
\citep{Stauber04,Visser12}. X-rays may be more effective in destroying
water, but only if the gas temperature is less than $\sim$250 K
\citep[][Notsu et al. unpublished]{Stauber06}. Once the temperature is
higher, water persists thanks to the rapid gas-phase reactions that
drive oxygen back into water in regions shielded from high energy
radiation (see also Fig.~\ref{fig:water_TUV}).

\begin{figure*}
  \centering
 \includegraphics[width=14cm]{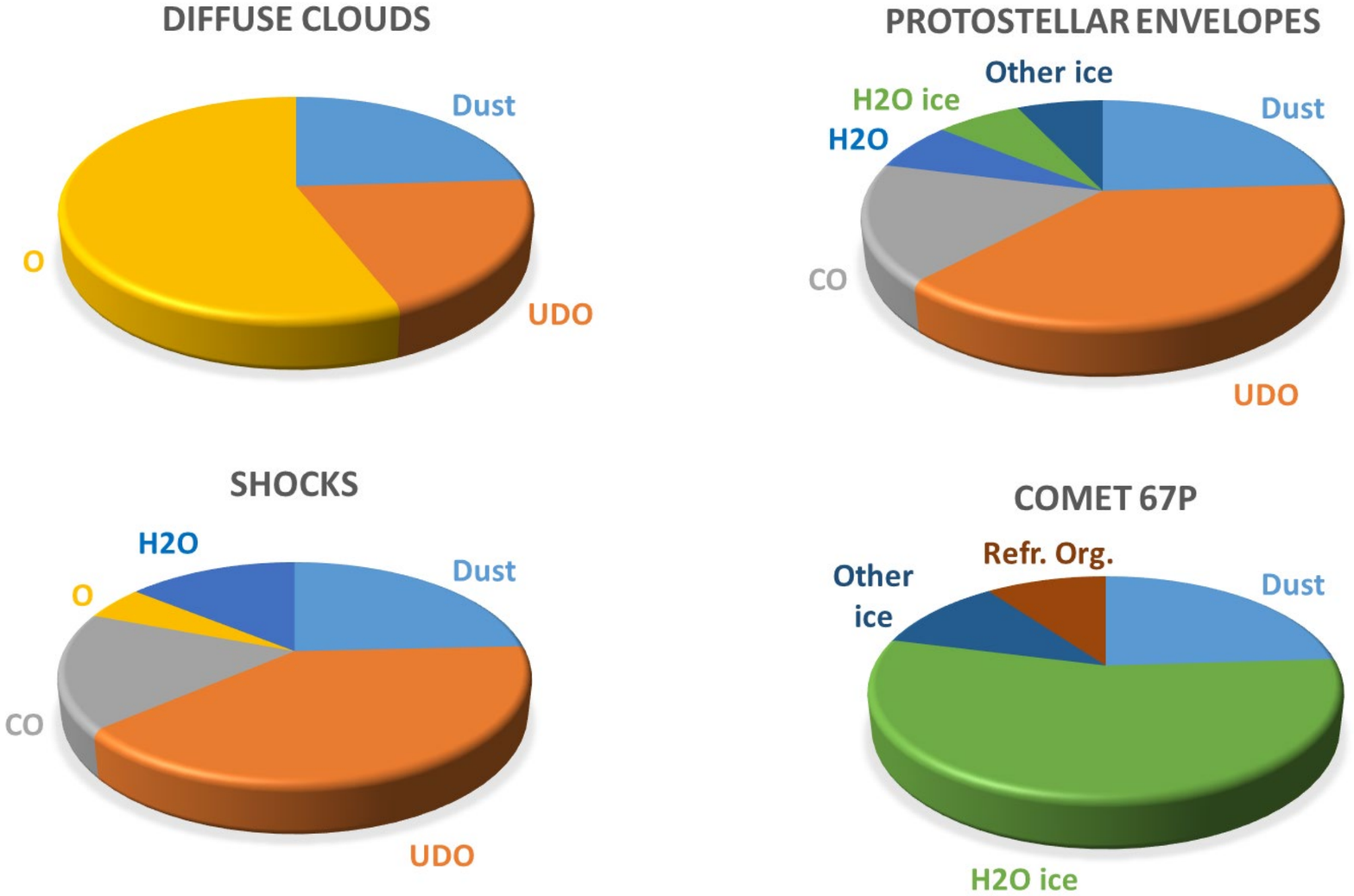}
 \vspace{-1.6cm}
 \caption{Partitioning of oxygen in different forms of gas, ice and
   dust in (a) diffuse clouds; (b) protostellar envelopes (low- and
   high-mass); (c) shocks; and (d) comet 67P. The amount of oxygen in
   refractory dust is kept constant at 140 ppm in all cases.}
         \label{fig:obudget}
\end{figure*}

Which oxygen species are left as reservoirs?  The lack of {\it
  Herschel}-HIFI O$_2$ detections from the hot core region in low- and
high-mass protostars limits its contribution to less than a few
\%. CO$_2$ gas is also known to be a minor contributor, at least for
high-mass protostars \citep{vanDishoeck98fd,Boonman03co2}. This leaves
atomic O as the major unknown in hot cores, even though models expect
any atomic O to be rapidly transformed into either H$_2$O or O$_2$
under dense warm conditions. There are currently no observations that
can constrain the amount of atomic O in hot cores. The only
alternative would be to have a large fraction of oxygen locked up in
UDO which is apparently not vaporized or atomized above 100~K.

\subsection{Shocks}

Section~\ref{sec:shocks} and Table~\ref{tab:shocks} have shown that
the abundances of H$_2$O in the different velocity components probed
by {\it Herschel} are well below the expected value of
H$_2$O/H$_2$=$4\times 10^{-4}$. Even if enhanced due to UV radiation,
OH is also not a major oxygen reservoir (\S~\ref{sec:oh}). Is the bulk
of the oxygen in warm outflows and shocks in atomic oxygen, if not in
H$_2$O or OH? Irradiated shock models indeed predict that atomic
oxygen may be a primary reservoir, although published results are for
pre-shock densities of only $10^4$ cm$^{-3}$ \citep{Godard19}. There
are numerous low resolution spectra of the [O I] 63 and 145 $\mu$m
lines taken toward protostars
\citep[e.g.,][]{Green13,Watson16,Mottram17}, but those data are
generally used to determine mass outflow rates assuming an oxygen
abundance with emission dominated by the jets
\citep{Nisini15}. Without spectrally and/or spatially resolved data,
no independent atomic oxygen abundance can be derived.

Velocity-resolved observations of the [O I] 63 $\mu$m
line with the GREAT instrument on SOFIA show that most of the [O I]
emission in star-forming regions indeed originates in shocks
\citep{Leurini15,Kristensen17a,Gusdorf17}. However, for the NGC 1333
IRAS4A shock position R1, atomic oxygen accounts for only 15\% of the
oxygen budget in the high velocity hot gas. CO appears to be the
dominant oxygen carrier at an abundance of
CO/H$_2 \approx 2\times 10^{-4}$; H$_2$O is a minor component, as is
OH at this position \citep{Kristensen17a}. For the distant high-mass
source G5.89-0.39, \citet{Leurini15} use spectrally resolved [O I]
data to derive atomic oxygen column densities which could be as large
as those of CO, or even larger depending on assumptions, but no
hydrogen columns are available to quantify abundances. H$_2$O and OH
are again minor components.

Thus, even in high temperature (shocked) gas, some significant
fraction of oxygen may still be unaccounted for since it is also not
observed in [O I].  Only in the very hottest gas could the H$_2$O
abundance approach the value required to contain most of the oxygen
\citep{Melnick08,Neill13}.  The important conclusion from this and
related work (see Appendix~\ref{app:obudget}) is that this fraction of
UDO must be locked up in a highly refractory form (such as refractory
organic material) that is not vaporized or atomized even for shock
velocities up to 50 km s$^{-1}$ and temperatures up to several hundred
K.

\subsection{Oxygen budget from clouds to comets}
\label{sec:obudget4}

Figure~\ref{fig:obudget} summarizes the partionizing of oxygen between
dust, gas and ice in the various regions discussed above. The adopted
numbers are summarized in Table~\ref{tab:obudget} with detailed
references and motivation for choices presented in
Appendix~\ref{app:obudget}. Uncertainties in each entry are difficult
to quantify and are discussed in the Appendix, but they are such that they
do not change the overall picture.

Interestingly, the amount of unaccounted oxygen in the form of UDO is
comparable between low-mass protostars, high-mass protostars and
shocks at $\sim$225 ppm, even though each individual UDO number
  is uncertain by $\sim$50\% and very different techniques and
instruments have been used in all three cases, with a mix of emission
and absorption lines involved. It is roughly double the amount of UDO
compared with diffuse clouds, consistent with the finding that the
amount of UDO increases with density \citep{Jenkins09,Whittet10}. The
absolute amount of UDO depends in all cases strongly on the adopted
overall [O] budget at 575 ppm; if the lower solar abundance of 490 ppm
is used, the fraction of UDO may be halved but it does not
disappear. In fact, \cite{Draine20} argue for a higher ISM
[O] abundance of 682 ppm with 66 ppm in UDO unaccounted for in diffuse
  clouds.

Table~\ref{tab:obudget} and Figure~\ref{fig:obudget} include the
oxygen reservoirs in comet 67P/Churyumov-Gerasimenko, which has been
studied in exquisite detail by the {\it Rosetta}
mission. \citet{Rubin19} provide an overview of the abundances in
volatiles measured by the ROSINA instrument, and those in refractory
solids (dust) measured primarily by the COSIMA instrument. This table
assumes an ice:dust mass ratio of 1:1. By definition, there is no room
for UDO in these measurements. However, some fraction of oxygen is
found to be in refractory organic material, about 9\% or 50 ppm.

There has been some speculation that UDO could consist mostly of
refractory organic material that is unobserved in diffuse and dense
clouds \citep{Whittet10}. The measurements for comet 67P suggest that
this component is not enough to explain all of the UDO in protostars
and shocks. However, its amount is
quite close to that found locked up in complex organic molecules in
large chemical models of protostellar envelopes
(\S~\ref{sec:watergasice}). This suggests that the chemistry may
proceed from the more volatile small organic molecules to larger more
refractory species as material is exposed to higher temperatures and
UV radiation ``en route'' from cloud to the comet-forming zones of
disks \citep[e.g.,][]{Greenberg90,Drozdovskaya16}.

For diffuse clouds, the measurement uncertainties are large enough
that refractory organics of the type seen in comet 67P could be the
major UDO reservoir, especially if the diffuse cloud numbers of
\citet{Draine20} are adopted.

\begin{figure*}
  \centering
    \includegraphics[width=14cm]{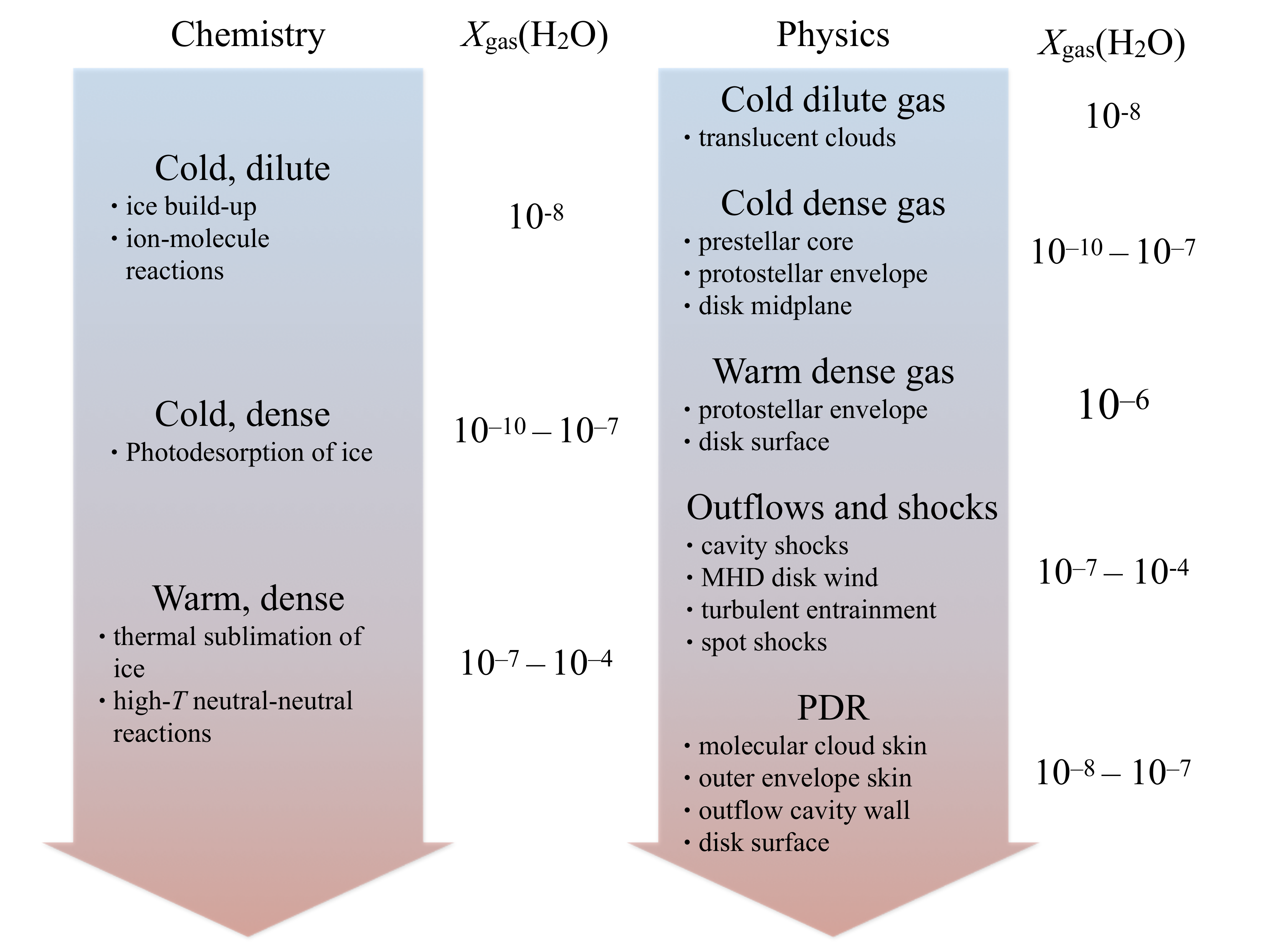}
    \caption{Physical components dominating water emission with
      evolutionary stage (right) and principal chemical processes at
      each temperature regime (left).  Typical water abundances with
      respect to H$_2$ associated with each evolutionary or
      temperature regime are indicated.}
         \label{fig:Lars}
   \end{figure*}

In summary, the observations of protostars and shocks discussed in \S
5--9 indicate that a significant fraction of oxygen, of order 100-200
ppm, is not accounted for. This section provides possible solutions to
this conundrum. In all cases, some small fraction of order 50 ppm may
be in refractory organics, assuming that this fraction is similar in
the ISM as in comets. For cold protostellar envelopes, an elegant
solution for the remaining missing oxygen is that ice columns are much
larger than assumed because grains have grown to larger sizes
suppressing the ice features. For hot cores and shocks, this solution
does not work since the water ice in large grains should have been
sublimated or sputtered so the situation is more puzzling. If gaseous
atomic oxygen is indeed negligible, then the oxygen must be in some
refractory form that does not vaporize or atomize, even in strong
shocks up to 1000 K.

Two sets of observations are urgently needed. First, more measurements
of velocity resolved atomic oxygen lines (both [O I] 63 and 145
$\mu$m) in protostars and shocks can quantify its contribution in more
sources. The current numbers hinge on just a couple of heterodyne [O
I] spectra. Only the upGREAT instrument on SOFIA is capable of such
measurements \citep{Risacher18}.

Second, deep observations of mid-infrared lines of H$_2$O and H$_2$
probing the hottest gas in shocks are warranted to determine whether
the UDO material is vaporized at $T>1000$~K and the H$_2$O abundance
starts to approach H$_2$O/H$_2$=$4\times 10^{-4}$. So far, only the
data for NGC 2071 and Orion-KL point in this direction
(\S~\ref{sec:shocks}). This will be possible with the JWST-MIRI
instrument \citep{Wright15} at shock positions offset from the
protostar to avoid large extinctions. The simultaneous measurement of
H$_2$ and other atomic and molecular lines should provide
better constraints on shock density, which is one of the largest
uncertainties in the analysis.

\section{Discussion}
\label{sec:discussion}

Based on the detailed analyses for individual physical components and
evolutionary stages, we can now come back to the three questions that
WISH aimed to address (\S~\ref{sec:intro}), as well as the two axes of
WISH: from low- to high mass, including implications for extragalactic
water observations, and the evolution of water from cloud to disk. The
focus is on {\it Herschel}'s contributions to these questions.
Finally some current and future opportunities for observing
water will be summarized together with lessons for future far-infrared
missions.

\subsection{How and where water is formed in space}
\label{sec:question1}

{\it Herschel} observations, in particular the velocity-resolved HIFI
data, have allowed the main chemical processes leading to water in
star- and planet forming regions to be identified and quantified
(Fig.~\ref{fig:Lars}).
Of the three routes illustrated in Fig.~\ref{fig:waternetwork}, the
water chemistry in star-forming regions is clearly controlled by a
close coupling between the gas and ice phases.  Water ice builds up in
the cold dilute phase with some gaseous water vapor produced by
ion-molecule reactions and destroyed by UV photodissociation. In cold
dense cores, the bulk of the water ice has formed, with a small
fraction released into the gas by nonthermal processes such as
photodesorption (see \S~\ref{sec:prestellar} and \ref{sec:protocold}
for more details). The new {\it Herschel} observations have
highlighted the importance of cosmic-ray induced photodesorption deep
inside clouds. Also, the multilayer nature of ice formation is key to
interpreting data on deuterated water.

Once the protostar turns on, heating occurs both actively through
shocks and winds associated with the outflows and passively by warming
up the dust and gas in the surrounding envelope through the
protostellar luminosity. Shocks can sputter ices and produce copious
water vapor through high-temperature gas-phase reactions: the {\it
  Herschel} data provide evidence that both processes are taking place
with high temperature gas-phase chemistry dominating at the high
velocities. Envelope heating results in water ice sublimation at its
iceline around 100 K. Once the star has formed, it emits strong UV
radiation setting up a PDR on the
neighboring cloud, envelope or disk surface. X-rays may also play a
role in destroying water in the inner envelope and hot core.
Figure~\ref{fig:Lars} (left) summarizes the chemical processes, whereas
\S~\ref{sec:question2b} discusses the abundances.

   \begin{figure*}
     \centering
    \includegraphics[width=12cm]{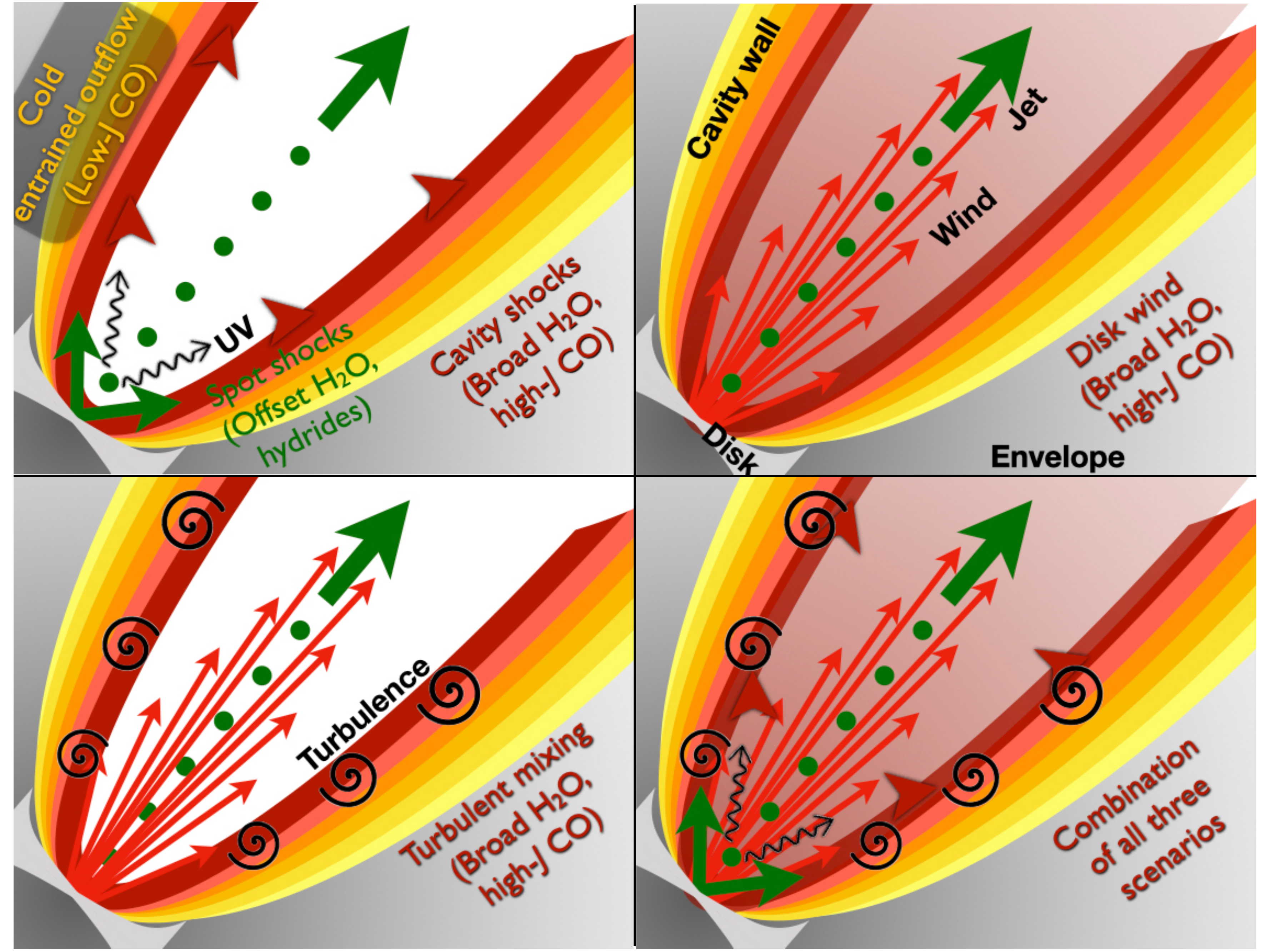}
    \caption{Cartoon illustrating in which physical components
        most of the H$_2$O, CO and other molecular line emission
        originates for the different scenarios presented in the
        text. These include the outflow cavity shocks and spot shocks
        (top left), the disk wind (top right), the turbulent
        entrainment (bottom left), or a combination of the three
        (bottom right). Each panel shows the jet with bullets (green),
        the outflow cavity wall (red/yellow gradient), disk and
        envelope (gray-scale). Arrows are meant to indicate gas
        motion.}
         \label{fig:Larscartoon}
   \end{figure*}

\subsection{Physical components traced by water vapor}
\label{sec:question2}

\subsubsection{Water}

Table~\ref{tab:profiles}, Fig.~\ref{fig:water_mass} and
\ref{fig:h2o_masstime} summarize the main water vapor line
characteristics at each of the physical stages, for low and high-mass
sources.  Water vapor in cold low-density gas is primarily seen in
narrow ($\sim 1$ km s$^{-1}$) absorption lines against a bright
far-infrared continuum. Narrow emission lines are seen only in cold
high density quiescent gas such as those found in pre-stellar cores,
in PDRs, and in the outer parts of protoplanetary disks near the
midplane. Relatively narrow water emission lines (a few km s$^{-1}$)
originate in warm, dense gas in protostellar envelopes (isotopologs,
away from outflows).

The water vapor emission seen by {\it Herschel} is, in contrast,
mostly in broad lines associated with various types of kinetic heating
(cavity shocks, disk winds, turbulent entrainment) and medium-broad
offset lines (spot shocks). Regions of strong water vapor emission are
generally compact, with an extent comparable to or less than one {\it
  Herschel} beam ($\lesssim$20$''$, \S~\ref{sec:maps}). Even within
this beam, the emitting sizes are generally small, of order 100 au for
low-mass protostars.

Three possible heating mechanisms are suggested for the broad warm
component involving kinetic energy dissipation (component 1,
\S~\ref{sec:profiles}, Fig.~\ref{fig:Larscartoon}).  First, cavity
shocks heat and compress the envelope along the cavity walls via
non-dissociative, 15--20 km s$^{-1}$, $C$-type shocks
\citep{Mottram14}.  Second, dusty MHD disk winds \citep{Yvart16}
launch dense gas from the inner disk at 0.2 out to 5--20 au, with heating
taking place via ion-neutral drift in the accelerating flow. This results in a
broad range of outflow velocities, spanning a few to 80 km s$^{-1}$.
Finally, turbulent entrainment heats gas within a mixing layer flowing
between the wide-angle protostellar wind and the infalling envelope
\citep{Liang20} and also provides a natural explanation for broad line
profiles peaking at systemic velocity.  In addition, the mixing-layer
model allows for a static outflow cavity that does not expand beyond
observed sizes over the typical Class 0 phase duration.
It is possible that all three mechanisms operate
simultaneously. Spatially resolved observations of the warm dense gas
together with mid-infrared searches for shock tracers (e.g., atomic
lines of Fe, S or Si) will be required to determine which dominates.

Overall, the conclusion from the {\it Herschel} data is that water
emission is a highly sensitive probe of the physics of different
star-formation phases with unique absorption or emission line
characteristics that are not easily seen by any other molecule. Viewed
from a more global perspective: strong water emission with large line
luminosities clearly points to active star-formation sites.

\subsubsection{Other molecules}

{\it Herschel} has also provided insight into the physical components
traced by other molecules and their lines, with a summary provided in
Figure~\ref{fig:Larscartoon} (see also Fig.~8 in \citealt{SanJose16}
and Fig.~5 in \citealt{Mottram14}). High-$J$ CO lines with
$J_{\rm up}\geq 14$ trace the same regions as water. In contrast, the
low-$J$ CO lines probe the entrained outflow gas physically separated
from the actively heated and shocked material.

OH shows broad line profiles consistent with those of H$_2$O and thus
probes the same components (\S~\ref{sec:oh}). Its
abundance is enhanced because the warm gas is exposed to UV radiation
dissociating H$_2$O into OH and changing the overall shock
structure. Hydrides such as OH$^+$, H$_2$O$^+$ and CH$^+$ have
narrower lines (but not as narrow as those of CH) and originate
primarily in the UV irradiated outflow cavity walls and spot shocks
\citep{Benz16}.

Other molecules that are abundant in ices, such as NH$_3$ and CH$_3$OH,
mostly trace the hot core region and the low-velocity part of the
shock. HCO$^+$, whose abundance anticorrelates with that of H$_2$O,
largely avoids the outflow region. SiO can trace the fast EHV part of
the shock similar to H$_2$O, but its abundance and line profiles show a more
rapid evolution across sources than that of H$_2$O.

\subsubsection{Cooling budget}

Water lines are a significant coolant of gas in warm gas, but not
dominant. For low-mass Class 0 and I sources, water contributes
typically at the 20\% level, whereas CO does so at 30-45\% of the
total far-infrared line cooling. The remainder of the far-infrared
cooling budget is provided by OH and [O I]. For high mass sources, CO
becomes more important, up to 70\% of total, whereas the contribution
from water lines drops because more water lines occur in absorption
rather than emission.

Significant gas cooling is also provided by the H$_2$ mid-infrared
lines \citep{Maret09}. These are not considered here but will be
further quantified by future JWST observations.

\subsection{Typical water abundances in each physical
  component}
\label{sec:question2b}

Figure~\ref{fig:Lars} summarizes the typical water vapor abundances
inferred for each of the physical components. Only average order of
magnitude values for each type of source are given, with details
described in \S 5--9.  For individual sources, abundances are accurate
to typically a factor of a few or better.

This figure illustrates that the gaseous water abundance varies
significantly and is thus a good diagnostic of different physical
regimes. However, it never reaches the maximum abundance of
H$_2$O/H$_2$ $\approx 4\times 10^{-4}$ expected if all volatile oxygen
is driven into water. Only the warmest gas has abundances approaching
$10^{-4}$, but in most cases the gas components probed by {\it
  Herschel} have abundances of $10^{-7}-10^{-6}$ with respect to
hydrogen.

In cold gas, water is mostly frozen out as ice so water vapor
abundances are very low, down to $10^{-10}$.  Interestingly, the most
accurate constraints on water abundance profiles have been made for
cold cores or outer protostellar envelopes, where the velocity profile
permits reconstruction of the water abundance as a function of position
(see \S~\ref{sec:abundanceprofiles}). At the edge of the core, H$_2$O
gas is produced by photodesorption of water ice at an abundance of
$\sim 10^{-7}$; deeper into the core it drops to less than
$10^{-10}$. The same holds for protoplanetary disks, but then in the
vertical direction: water vapor is UV photodesorbed in layers at
intermediate heights, dropping to very low abundances near the
midplane.

\subsection{Water: From low to high mass}
\label{sec:lowtohigh}

Section~\ref{sec:waterdata}, together with
Figures~\ref{fig:water_mass} and \ref{fig:h2o_masstime} and Table~1,
demonstrate the large similarity in water line profiles from low- to
high-mass YSOs. As summarized in \citet{SanJose16}, the profiles can
be decomposed in similar physical components associated with
kinetically-heated gas (warm outflows and shocks) and warm dense
envelopes, but with the relative contribution from the envelope
increasing from low- to high mass (see
Appendix~\ref{app:additionalspectra}). The same physical components
can also be seen in high-$J$ CO lines, but only for $J_{\rm up}
>14$. Even then, they are less prominent than in water lines.  The
similarity in profiles means that common physical mechanisms are at
work in the outflows of sources across more than five orders of
magnitude in luminosity and that the gas cooling structure appears
universal.  The water lines probed by {\it Herschel} and the high-$J$
CO emission are always dominated by material at similar conditions:
temperatures of a few hundred K and densities from $10^5-10^8$
cm$^{-3}$.

The water line luminosities show a strong, near-linear correlation
with bolometric luminosities of the sources, $L_{\rm bol}$. For
optically thick, but effectively optically thin lines such as those of
water, this empirical relation simply implies that there is more gas
at the required conditions for more luminous sources: material is
heated further from the source and deeper into the cavity wall. This
is consistent with the larger emitting areas found for high mass
sources from the water radiative transfer analysis compared with that
for their lower mass counterparts (few thousand au versus 100 au).

Overall, the intensity ratios of the various water lines probed by
{\it Herschel}-HIFI are also very similar across the luminosity range
(and even as function of velocity, modulo absorption features),
pointing to the similar physical conditions inferred
above. Surprisingly, however, one line ratio stands out: the 752 GHz
$2_{11}-2_{02}$/988 GHz $2_{02}-1_{11}$ ratio is generally less than 1
for a wide range of sources, but lies above 1 for high-mass protostars
\citep{SanJose16}. One possible explanation is that radiative pumping
plays a more significant role for higher-mass protostars than for
their lower- and intermediate-mass counterparts.  Alternatively,
absorption in the 988 GHz line affects the line ratio in high-mass
sources \citep{vanderTak19}. In many aspects, the intermediate mass
sources look more similar to low-mass protostars, or an unresolved cluster
of low-mass protostars within the {\it Herschel} beam, than a
scaled-down version of high-mass sources.

The water abundances in the various physical components are also found
to be similar across the luminosity range. Inner and outer envelope
water abundances, when analyzed in a similar manner, show a comparable
range of values (Fig.~\ref{fig:inout}). Water abundances in warm
outflowing gas are low by at least two orders of magnitude due to UV
irradiation in all sources, although there is some debate to what
degree shocks associated with high-mass sources have even lower
abundances than their low-mass counterparts due to more intense UV
(Table~\ref{tab:shocks}).

\subsection{Implications for extragalactic water observations}
\label{sec:extragalactic}

\begin{figure}
     \centering
    \includegraphics[width=8cm]{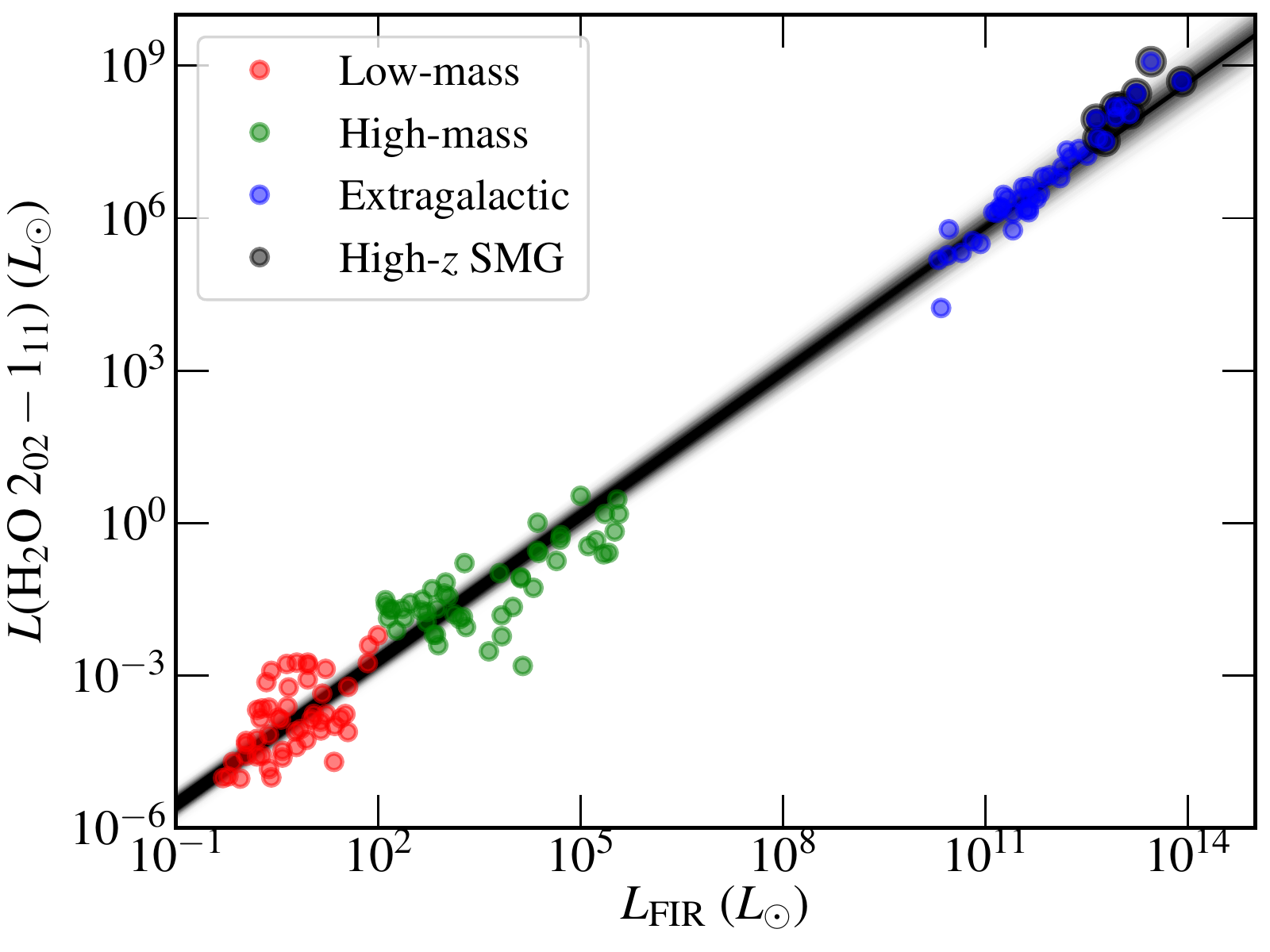}
    \caption{Line luminosity of the H$_2$O 2$_{02}$—1$_{11}$
      transition at 988 GHz versus the FIR luminosity for low- and
      high-mass sources and extragalactic sources. The gray line is
      the best-fit power-law function, with an index of
      0.95$\pm$0.02.}
         \label{fig:exgal}
   \end{figure}

The inferred linear relations of line luminosity with bolometric
luminosity for Galactic protostars hold over many more orders of
magnitude when observations of extragalactic sources from
\citet{Yang13} are included, both for water and for mid- to high-$J$
CO lines \citep{SanJose13,SanJose16}. Figure~\ref{fig:exgal} shows an
updated version of this correlation for the 988 GHz line including the
low-mass sources from the WILL survey \citep{Mottram17} and low- to
intermediate-mass sources in the Cygnus star-forming cloud
\citep{SanJose15}. This strong relation with FIR emission has been
interpreted differently by the two communities, however (Kristensen \&
Bergin, subm.). In Galactic sources, it is well known from the {\it
  Herschel} data that water and high-$J$ CO are associated with
shocked gas where the lines are collisionally excited
\citep[e.g.,][]{Mottram14}. The emission is spatially compact, not
extended over entire molecular cloud scales (\S~\ref{sec:maps}). In
fact, away from the immediate surroundings of protostars and their
outflows, the water emission and abundance drops steeply, within a
single {\it Herschel} beam for high-mass sources ($\lesssim$0.1 pc)
\citep{Jacq16,vanderTak19}. PDRs do not contribute significantly on
cloud scales \citep{Melnick20}.  The far-infrared luminosity in this
interpretation traces the material in which stars are currently
forming (the protostellar envelopes), but this infrared radiation does
not affect the water excitation: there is no direct causal relation.

In contrast, the extragalactic community uses the tight correlation
between the H$_2$O line luminosity and FIR luminosity to infer that
the water excitation is dominated by FIR pumping, as opposed to
collisional excitation in outflows. In the extragalactic scenario,
water emission thus originates in the entire molecular cloud in which
stars form, and is not directly associated and co-located with star
formation
\citep[e.g.,][]{vanderWerf11,Gonzalez14,Yang16,Liu17,Jarugula19}.

As shown by Kristensen \& Bergin (subm.), the low- and mid-$J$
water line ratios ($E_{\rm up} < 300$ K) do not change significantly
between Galactic and extragalactic sources.  Most notably, the
peculiar 752/988 GHz line ratio found for high-mass sources is not
found for extragalactic sources \citep{Yang13,SanJose16}. Thus, the
baseline interpretation is that the same mechanism that is responsible
for water emission in Galactic sources also holds for extragalactic
sources: water emission arises in outflows and shocks associated with
(clusters of) currently forming stars.  Thus, any water line that does
not suffer significant self-absorption is a good and clean tracer of
buried star formation activity.

By inference, the same interpretation would hold for mid- to high-$J$
CO lines. For high $J_{\rm up} \geq 14$, there is an observed close
relation between CO and water lines with both tracing currently active
outflow heating (\S~\ref{sec:maps} and \ref{sec:profiles}). For
mid-$J$ CO lines there can also be a contribution from the warm
envelope. For example, for $^{12}$CO $J = 10-9$, the fraction of
envelope emission is around 50\% for high-mass sources
\citep{SanJose16}. Since envelope emission also traces current star
formation, it does not necessarily affect the interpretation, albeit
that there could be extra scatter due to the variation in outflow
versus envelope contribution for each source.

Is all water emission associated with current star formation activity?
A subset of extragalactic sources are known to drive outflows on much
larger galaxy scales, often seen perpendicular to the galactic disk
where they can escape freely. The jets and flows are thought to be
driven by Active Galactive Nuclei (AGN) buried in the center of the
galaxy \citep[see][for a review]{Blandford19}. These large
galaxy-scale outflows are well traced by OH lines, not by H$_2$O
lines, observed by {\it Herschel}-PACS \citep{Sturm11,Gonzalez17}. OH
is much more abundant than H$_2$O due to the intense UV radiation in
the outflows and cavity walls away from the shielded regions. Thus,
H$_2$O is the better probe of star formation activity whereas OH may
be the better probe of AGN activity.

A big advantage of extragalactic observations is that even for small
redshifts, the water lines shift into frequency ranges that are not
obscured by the Earth's atmosphere and can be well observed from the
ground. ALMA is already providing a rich database on water lines in
extragalactic sources and will continue to do so in the coming
decades. The {\it Herschel} mission studying water in Galactic
star-forming regions was needed to properly
interpret these ALMA observations, thereby revealing the physical origin of
emission. The next step will be to calibrate water emission with measured
star-formation rates on galactic scales, such that water emission can be used
to quantify star-formation rates in distant high-redshift galaxies.

\subsection{Water: From cores to disks}
\label{sec:evol}

The third goal of the WISH+ program was to follow the trail of water
from its formation in molecular clouds to the sites where new planets
are formed. Addressing this question requires observations of both
water vapor and water ice using a wide range of instruments. The
broader picture of water from clouds to planets is reviewed elsewhere
\citep{vanDishoeck14PPVI,Hartmann17,Morbidelli18}. Here the unique
contributions that {\it Herschel} has made are highlighted, in concert
with recent millimeter interferometry (ALMA, NOEMA) results for warm
water.

\subsubsection{Molecular clouds: Water ice formation}

{\it Herschel}-HIFI data of cold clouds and outer protostellar
envelopes coupled with detailed chemistry and radiative transfer
modeling have demonstrated that most water is formed as ice in the
early stages of cloud evolution (\S~\ref{sec:prestellar} and
\ref{sec:protocold}). This is consistent with direct observations of
water ice at infrared wavelengths in quiescent and star-forming
clouds. {\it Herschel} studies, most notably those as part of WISH,
have demonstrated a good quantitative understanding of the cold water
chemistry and ice formation in the pre-stellar stage.  The amount of
water ice that can be produced at subsequent warmer protostellar
stages is only a small fraction of the total ice.

It is likely that a significant fraction of available volatile oxygen
(i.e, not locked up in silicates) is contained in water ice in cold
clouds. The fact that not all of this ice is detected in infrared
absorption spectra either points to a short pre-stellar stage, or to a
small fraction of grains that have grown to at least micrometer size
by the dense cloud stage. Some ill-determined fraction of oxygen is
contained in other forms, either in volatile O-containing species or
more refractory material called UDO (\S~\ref{sec:obudget}). While the
UDO fraction may change with evolutionary stage, it is not expected to
affect the bulk of the water trail.

How much water ice is made in dense clouds?  A typical 1 M$_\odot$
cloud forms about $6\times 10^{52}$ water molecules assuming a water
ice abundance of $10^{-4}$ with respect to H$_2$, over a period of
about $10^6$ yr. This amounts to more than 1 million oceans of water
available for new planetary systems that are formed in the cloud (1
Earth ocean = 1.5x10$^{24}$ gr of water = 5$\times 10^{46}$ water
molecules).  This large number of ``oceans'' implies that a
  significant water reservoir is available to be supplied.  However,
  there are a number of overall loss factors, both in the supply
  to the natal disk and then to a nascent planet (see below).  Thus
  this number should not be taken to imply a direct link to Earth’s water
  \citep[e.g.,][]{vanDishoeck14PPVI}.

\subsubsection{Protostars: Rapid water production in outflows and
  shocks but lost to space}

{\it Herschel} data have confirmed the rapid production of water in
warm kinetically heated gas associated with protostellar outflows and
winds and their impact on the surrounding envelope. Using the observed
mass outflow rates and water abundances, of order 1--5 oceans of water
vapor per year can be produced in a single low-mass outflow over the
period of the lifetime of the embedded phase, which is about $10^5$ yr
\citep{Kristensen18}. Thus, the amount of water molecules formed in
outflows per year is comparable to that in cold clouds. However, all
of this water is carried away by the outflows and thus lost to
space. There water is either photodissociated or frozen out back onto
the grains as ice \citep{Bergin98}.

\subsubsection{From clouds to disks: Inheritance versus reset}

The transport and survival of water ice from the collapsing core onto
the forming disk is a crucial step in the water trail. How and where
material falls into the disk and with what speed is still poorly
understood and ill constrained observationally.  Directly linked to
this question is the extent to which the chemical composition is
preserved from cloud to disk (``inheritance''), or whether it will be
modified en route (``reset'') due to the changing temperatures and
UV radiation along the infall paths
\citep[e.g.,][]{Visser09,Hincelin13,Drozdovskaya16}. In the most
extreme case of strong accretion shocks, reset implies
complete vaporization of all molecules and dust grains back to atoms,
with subsequent reformation; milder versions of reset include
sputtering of ices, and gas and ice chemistry modifying abundances.

To test the inheritance versus reset scenarios, key observational
diagnostics of these processes are needed.  There is some
observational evidence for chemical changes near the centrifugal
barrier in the form of strong SO and SO$_2$ emission
\citep[e.g.,][]{Sakai14,delaVillarmois19}, but implied
  temperatures are low, $\sim$50 K, and their origin and implications
for any shock are still unclear.

The results for young disks found in
\S~\ref{sec:youngdisks} argue for significant inheritance of water ice
since no water vapor, warm or cold, is observed to be associated with
young disks in the Class I phase, as would be expected when accretion
shocks onto the disk would sputter the ices or produce abundant water
gas.
Also, a water-vapor rich warm envelope is ruled out by the current
{\it Herschel} and ALMA+NOEMA data, although deeper and higher spatial
resolution ALMA data could put up to an order of magnitude stronger
constraints on any warm water produced by accretion shocks.
Even the disks themselves are cold enough by the Class I
stage that they have very low water vapor abundances in the outer disk
and surface layers, suggesting that the bulk of the water is locked up
as ices onto grains at an early stage of disk formation.

Models by \citet{Visser09} and \citet{Furuya17} that follow the water
chemistry along many infall trajectories have found that most water in
disks is indeed inherited as ice. A small fraction may be sublimated
during a luminosity burst of the young (proto)star moving the water
ice-line outward, as often happens in the Class 0 and I phases, but
this water ice quickly refreezes again without chemical alteration
\citep{Taquet16o2}. Only trajectories that come close to the outflow
cavity wall result in water dissociation due to exposure to UV
radiation or X-rays.

\subsubsection{Disk evolution and transport}
\label{sec:disktransport}

Figure~\ref{fig:Harsonodisk} provides a useful framework to illustrate
the different water reservoirs in disks. {\it Herschel} has shown that
gas and dust evolution as well as dynamics can alter this simple
picture, as discussed in \S~\ref{sec:protoplanetary}.  Once in the
disk, the icy grains can grow further and settle to the midplane
\citep{Krijt16}. There is observational evidence based on the spectral
index at mm wavelengths, as well the required dust opacity for hiding
lines in the inner disk, that grains have already grown to mm or cm
size in disks during the embedded phase
\citep[e.g.,][]{Kwon09,Testi14,Harsono18}.  These large icy grains
will drift inward or be trapped in pressure bumps where they can grow
further to water-rich planetesimals
\citep[e.g.,][]{Raymond17,Schoonenberg17}. The {\it Herschel}-HIFI
data of disks indicate that water vapor likely follows the large icy
dust grains that have settled and drifted inward.

Once the drifting icy pebbles cross the ice line, they will release
their water into the gas resulting in an inner disk that is rich in
water vapor, with abundances that can even exceed the canonical value
of $10^{-4}$ by more than two orders of magnitude within 1 Myr
\citep{Ciesla06,Bosman18b}.  Accretion onto the star will however
remove this oxygen-rich gas and there is some disagreement on the
relative importance of replenishment versus removal. If removal
dominates, it may drain the inner disk of water-rich gas in as little
as 0.1 Myr \citep{AliDib14}. Tracing the water vapor in this inner
disk requires mid-infrared observations. {\it Spitzer} data have
indeed provided evidence for both high and low water abundances in the
inner disk linked to dust disk size and structures
\citep[e.g.,][]{Banzatti17,Banzatti20} (see \S~\ref{sec:innerdisks}).

\begin{figure}[t]
  \centering
\includegraphics[width=9cm]{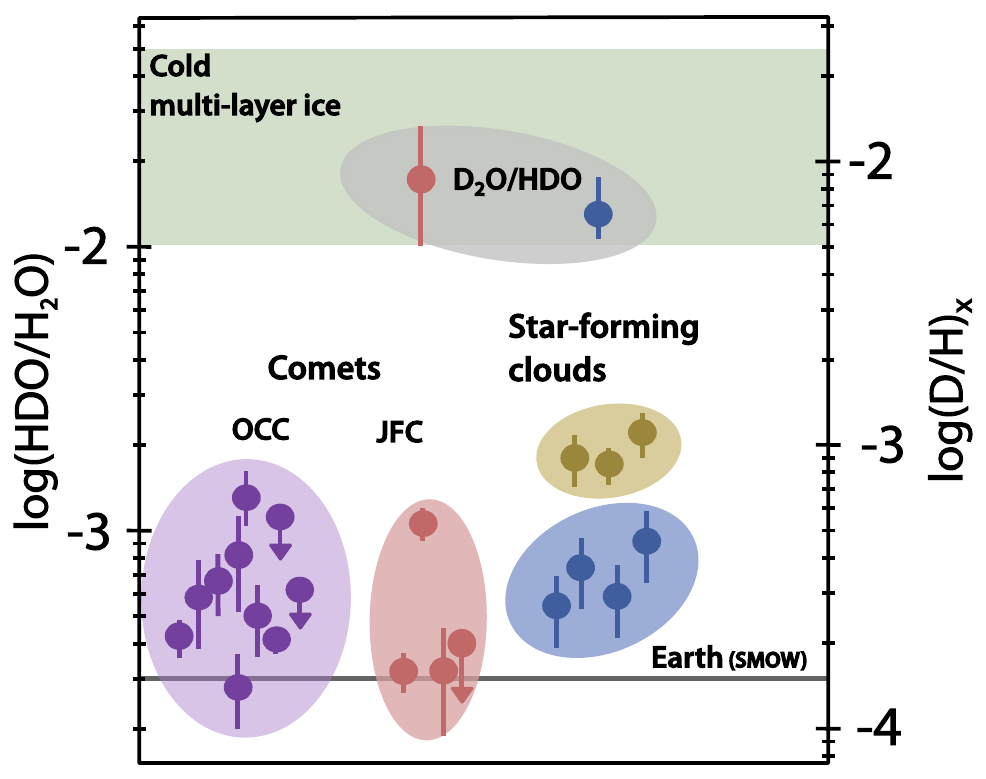}
\caption{HDO/H$_2$O (lower points) and D$_2$O/HDO (two top points)
  abundance ratios measured for warm gas in low-mass protostellar
  regions and in comets (JFC=Jupiter Family Comets; OCC=Oort Cloud
  Comets). The HDO/H$_2$O data are taken from the summary in
  \citet{Jensen19} (brown: isolated protostars; blue: protostars in
  clustered regions), the D$_2$O/HDO star-forming cloud ratio from
  \citet{Coutens14} and the HDO/H$_2$O and D$_2$O/HDO ratios in comet
  67P (JFC high points) from \citet{Altwegg17}. The much higher
  HDO/H$_2$O ratios found in the top layers of multilayer ice models
  are indicated with the light green zone and are similar to the
  HDO/H$_2$O values of 0.025 measured with {\it Herschel}-HIFI in cold
  outer envelope gas. }
         \label{fig:dhsummary}
\end{figure}

What do the HIFI data imply for the availability of water in new
planetary systems? The HIFI detection of gaseous H$_2$O in the TW Hya
disk implies a reservoir of water ice of a few thousand oceans, which
suggests plenty of water to build a habitable planet such as
Earth. However, it is unclear whether all of this water can reach the
habitable zone close to the star. The formation of a single
Jupiter-like planet in a disk could lock up the bulk of this available
oxygen reservoir. Also the flow of icy pebbles from outer to inner
disk can be stopped at a dust trap outside the ice line. Indeed, the
inner TW Hya disk is known to be poor in oxygen and carbon, likely due
to a dust trap outside of 20 au locking up CO and one at 2.4 au
locking up H$_2$O \citep{Zhang13,Schwarz16,Bosman19TWHya,McClure20}
(see also \S~\ref{sec:innerdisks}). In such cases, scattering of
ice-rich planetesimals from the outer to the inner disk could happen
only at later evolutionary stages when the gas has disappeared, and in
that way deliver water to forming terrestrial planets in the habitable
zone \citep{Morbidelli12}.

\subsubsection{Deuteration as a probe of water inheritance}

An alternative way to probe water ice processing en route from cloud
to disk is ice deuteration, which is particularly sensitive to
physical parameters \citep{Ceccarelli14}: if most water ice is
inherited, the HDO/H$_2$O ice ratio set in the cold pre-stellar stage
should be preserved in subsequent stages. The situation is however not
as simple as initially thought because the layered structure of ices
needs to be taken into account. Also, HDO has not yet been detected in
mature Class II disks.

Ice formation models that include a multilayer structure have shown
two stages: the initial cloud formation stage in which most of the
H$_2$O ice is formed, and the later dense core phase in which the bulk
of the HDO and D$_2$O ice are made, with only little additional H$_2$O
ice \citep{Taquet14,Furuya16} (Fig.~\ref{fig:hdonetwork}).
A typical grain of $\sim$0.1 $\mu$m has nearly 100 ice layers if all
oxygen is converted to ices. The bottom 75--95\% of those layers have
HDO/H$_2$O around $10^{-4}$ whereas the top 25--5\% have HDO/H$_2$O as
high as $10^{-2}-10^{-1}$ (Fig.~\ref{fig:hdo_models}). This results in
overall bulk HDO/H$_2$O ice ratios of $\sim 10^{-3}$ within a factor
of two \citep{Taquet14,Furuya16}. A characteristic feature of this
cold, layered ice chemistry is that doubly deuterated water is even
more enhanced: the bulk D$_2$O/HDO$>>$ bulk HDO/H$_2$O by about a
factor of 10 (Fig.~\ref{fig:hdo_models}).

What do the {\it Herschel} and new NOEMA+ALMA mm interferometry data
tell us?  Figure~\ref{fig:dhsummary} summarizes the observational
results for protostellar envelopes compared with comets, using data
summarized in \citet{Jensen19}. First, the derived high gaseous
HDO/H$_2$O of 0.025 in the cold outer core (\S~\ref{sec:waterhdocold})
from the HIFI data is entirely consistent with the multilayer ice
formation (green shaded area), and is much higher than the values
found in hot cores. The high value reflects both the fact that only
the top, deuterium-enriched ice layers are photodesorbed, with
low-temperature ion-molecule chemistry further enhancing the
deuteration (Fig.~\ref{fig:hdo_models}) \citep{Furuya16}.  Also,
\citet{Coutens13D2O} found D$_2$O/HDO $>$ HDO/H$_2$O in the cold outer
layers of the IRAS16293-2422 core from {\it Herschel}-HIFI D$_2$O, HDO
and H$_2^{18}$O lines consistent with the models.

Second, the ``hot core'' HDO/H$_2$O abundances of
$(0.5-1)\times 10^{-3}$ (\S~\ref{sec:hotcore}), which presumably
represent thermally sublimated ices, are consistent with the bulk ice
deuteration found in models of the cold pre-stellar phase
(Fig.~\ref{fig:hdo_models}). They also cover the same range as the
cometary data shown in Fig.~\ref{fig:dhsummary}. The three brown
HDO/H$_2$O ratios for star-forming clouds that lie somewhat higher in
this figure refer to isolated cores that are generally somewhat colder
than the clustered regions \citep{Jensen19}. Since our Solar System is
thought to be born in a cluster \citep{Adams10}, the lower blue points
are more appropriate for comparison with comets.

The ``smoking gun'' for the inheritance scenario comes from the
observed high abundance of D$_2$O, with D$_2$O/HDO $\approx 7$
HDO/H$_2$O for the NGC 1333 IRAS2A hot core
\citep{Coutens14}. Similarly, the detection of D$_2$O in comet
67P/Churyumov Gerasimenko provides an important clue
\citep{Altwegg17}. The inferred D$_2$O/HDO ratio for 67P is a factor
of 17 higher than the HDO/H$_2$O ratio.  Both values (high points in
Fig.~\ref{fig:dhsummary}) are consistent with pre-stellar multilayer
water ice formation and not with water reformed in the outer part of
the solar nebula disk.

Could the deuteration fractions actually be modified en route from
cloud to disks or be established within disks?  The delivery of
layered ice mantles from cloud to disk has been investigated by
\citet{Furuya17}.  These models confirm that the majority of
pre-stellar water ice is retained upon delivery to the disk without
significant UV processing and ice sublimation. The HDO/H$_2$O ratio is
somewhat lowered (by up to a factor of two) because of selective loss
of the upper ice layers but is largely retained when averaged over the
entire disk, as is D$_2$O/HDO. Locally, the HDO/H$_2$O can deviate
more from the original value, especially in the upper layers of the
outer disk.

How can this inheritance case be distinguished from the full reset
case in disks? \citet{Cleeves14} show that starting from the elemental
[D]/[H] ratio, mature Class II disks cannot produce enough HDO to
explain the measured level of water deuteration in comets of
HDO/H$_2$O$\approx (0.5-1)\times 10^{-3}$ \citep[see
also][]{Willacy09}, a conclusion that is strengthened if cosmic rays
are excluded from disks by stellar winds. There have been some claims
that disk models with vertical turbulent mixing can reach the required
HDO/H$_2$O levels \citep{Furuya13,Albertsson14}, but the recent
observational limits on turbulent mixing
\citep[e.g.,][]{Flaherty18,Flaherty20,Teague18} make those models less
realistic.

In any case, there would be two clear ways to distinguish
the inheritance and reset models: (i) in-situ disk chemistry models
predict an increasing HDO/H$_2$O ratio with radius due to the
decreasing temperature profile, whereas this is not always the case
for the inheritance models; (ii) reformed ices in the disk have
D$_2$O/HDO $\sim$0.1$\times$HDO/H$_2$O whereas the original pre-stellar
ices have $\sim 10 \times$ HDO/H$_2$O \citep{Furuya17}.

Although the results for D$_2$O/HDO hinge on only one protostar and
one comet observed so far, the HDO/H$_2$O ratios as well as the models
discussed above point to the same conclusion. Further observations of
D$_2$O in protostellar sources should have high priority, and new ALMA
data indeed strengthen the inheritance case (Jensen et al.,
unpublished).

In conclusion, several lines of observational and modeling evidence on
deuterated water, together with the results for young disks
(\S~\ref{sec:youngdisks}), point toward the bulk of the water being
formed in the cold pre-stellar stage and incorporated largely
unaltered into disks and comets, where they are locked up quickly into
larger grains and planetesimals.

\subsection{Lessons for future missions}

A general lesson from WISH+ is that most of the new science comes from
the deepest and longest integrations, either to detect weak lines or
to get high signal-to-noise on strong line profiles, pushing the
instruments to the limits. For WISH, such long integrations were
possible to plan because of the freedom that the GTO time offers. In
contrast, time allocation committees, in particular on Galactic
science, are often reluctant to grant significant time per
source. Future missions should allow for a mechanism for deep
integrations on Galactic sources early in the mission.

The importance of future mid- and far-infrared observations and
millimeter interferometry to trace water has been highlighted near the
end of each of the sections, \S 5--10.  Summarized per facility or
instrument, each has its strength in providing (part of the) answers
on the water trail from cloud to disk. In fact, it is clear from the
discussions in \S 5--10 that the full picture can only be pieced
together by the combination of many different observations, with the
far-infrared wavelength range providing a central role. Roughly in
order of when instruments become available, the opportunities for
adding to our understanding of water are as follows.

\begin{itemize}

\item Millimeter interferometers, most notably ALMA, will be very
  powerful to image warm water on subarcsec scales ($<$30 au radius in
  nearby star-forming clouds) in large numbers of hot cores and
  (young) disks through the H$_2^{18}$O 203 and 390 GHz and various
  HDO and D$_2$O lines. Mm arrays can observe optically thin water
  isotopologs at low frequencies at which the dust continuum emission
  is optically thin, in contrast with space-based data at high
  frequencies. So far, ALMA has observed just a few sources; large
  surveys are needed.  ALMA can also put constraints on water released
  or produced at the accretion shock. By observing HDO and D$_2$O in
  the same beam, direct and model-independent probes of HDO/H$_2$O and
  D$_2$O/HDO in warm gas are obtained. For outflow shocks, higher
  excitation (nonmasing) H$_2^{16}$O lines that are not obscured by
  the atmosphere can be imaged.
  
\item The VLT-CRIRES+ and SOFIA-EXES high-resolution $R\approx 10^5$
  near- to mid-infrared spectrometers are well suited to observe hot
  water absorption lines at 3 and 6 $\mu$m arising in the inner
  envelope or outflow toward protostars. By spectrally resolving the
  lines, both the origin as well as accurate optical depths and thus
  column densities can be determined.

\item VLT-CRIRES+ and Keck-NIRSPEC, and in the future ELT-METIS
  (planned $\sim$2028), can also spectrally resolve H$_2$O
  vibration-rotation emission lines at 3 $\mu$m in disks at
  $R\approx 10^5$ and constrain their location through systematic
  velocity patterns. Moreover, ELT-METIS can spatially resolve the
  lines down to a few au and distinguish a disk surface layer versus
  disk wind origin. These 3 $\mu$m vibration-rotation lines are also
  bright in comets, and provide an accurate way to determine cometary
  HDO/H$_2$O when combined with near-simultaneous observations of HDO.

  Individual mid-infrared lines in the 10 $\mu$m region can be
  targeted with VLT-VISIR at $R\approx 10^4$ in bright sources; higher
  resolution instruments such as TEXES on Gemini are only offered
  sporadically. ELT-METIS will offer high spatial resolution at
  N-band, but with a spectral resolving power of only a few hundred,
  limiting the line-to-continuum ratio.

\item The SOFIA-GREAT instrument is unique in observations of
  spectrally resolved lines above 1 THz, most notably of the [O I] 63
  and 145 $\mu$m lines. Such observations will be of crucial
  importance to nail down the oxygen budget in shocks and other warm
  regions. OH, NH$_3$ and high-$J$ CO can also be observed, albeit
  with reduced sensitivity compared with {\it Herschel}. SOFIA-4GREAT
  now also allows observations of the H$_2^{18}$O ortho-ground state
  line in bright sources, including comets. Also, velocity resolved [C
  II] line mapping as a tracer of UV radiation on molecular cloud
  scale is possible with SOFIA-UPGREAT.

\item The JWST-MIRI instrument will probe hot and warm water
  abundances in shocks and inner disk surface layers, through the 6
  $\mu$m vibration-rotation band and the pure rotational lines at
  $>$10 $\mu$m. For disks, this combination of lines will allow more
  accurate assessments whether the inner regions inside the ice line
  are enhanced or depleted in water. For shocks, the combination of
  observed H$_2$O and H$_2$ lines will be key to determine whether
  water accounts for the full oxygen budget in hot gas, or whether any
  UDO is still required. JWST-NIRSPEC can probe the hottest water
  emission at 3 $\mu$m, as well as the primary H$_2$O, HDO and CO$_2$
  ice absorption bands at 3--5 $\mu$m, even spatially resolved on
  10-20 au scales.

\item The approved SPHEREx mission with a launch date around 2024 will
  obtain low resolution absorption spectra at 0.75--5 $\mu$m toward
  $> 10^6$ lines of sight with strong background sources and thus
  measure H$_2$O ice and other oxygen-containing ices such as CO$_2$,
  OCN$^-$, CO, and OCS along the lines of sight, further testing and
  quantifying ice formation and ice chemistry models.

\item Future far-infrared missions such as the proposed {\it Origins
    Space Telescope} and the proposed but recently cancelled SPICA
  mission are particularly well suited to trace the water trail by
  observing a very wide range of mid- and far-infrared water lines
  (Fig.~\ref{fig:waterlevels}), covering both hot (1000~K), warm
  ($\sim$200 K), and cold ($\sim$30 K) gas with a resolving power of
  up to $3\times 10^5$ for {\it Origins}. This includes the critical
  179 $\mu$m line connecting with the ground state. For protostellar
  envelopes and shocks, the analysis would have to take into account
  that there are likely multiple physical components (as found by
  WISH) that are not spectrally resolved. Also, the flux of
  low-lying water lines and [O I] may be significantly affected by
  absorption.
  
  For disks, far-infrared missions would be a very powerful tool to
  locate the water snowline  and quantify the water vapor
  in the different disk water reservoirs in statistically significant
  samples of disks. Indeed, models show that the largest sensitivity
  to the location of the snow line is provided by lines in the 40--60
  $\mu$m region, which is exactly the wavelength range without recent
  observational facilities. Moreover, the 44 and 62 $\mu$m
  (crystalline) water ice features can be observed in emission,
  allowing to quantify the bulk of the water and oxygen in the
  emitting layer at intermediate disk heights. In the current design
  {\it Origins} is four orders of magnitude more sensitive to the mass of
  cold water vapor emission in disks compared to {\it
    Herschel}. Observing a thousand disk systems across the range of
  stellar masses will enable meaningful comparison to the exoplanet
  composition and inventory.

\end{itemize}

None of these planned space missions include a THz heterodyne
capability.  Only the Russian-led {\it Millimetron} proposal has such
an instrument in its baseline plan, whereas the {\it Origins}-HERO
instrument is an upscope option. This lack of heterodyne facility will
strongly limit studies of cold quiescent water gas in clouds and outer
envelopes as well as in protoplanetary disks for which $R>10^6$ is
needed to retrieve abundance profiles, whether from absorption or
emission line profiles. For example, relating the water abundance to
substructures in disks, such as appears the case for the HD 100546
disk (\S~\ref{sec:protoplanetary}), cannot be done without full
velocity resolution.  All three water lines connecting with the
ground-state -- 557, 1113, 1670 GHz (179.5 $\mu$m) --- are important,
as are their isotopolog lines. ALMA cannot do such studies due to the
blocking of these lines by the Earth's atmosphere.
To determine HDO/H$_2$O in cold gas, the 893 GHz $1_{11}-0_{00}$ HDO
line is particularly useful. For warmer regions such as protostars and
shocks, spectrally resolved profiles of the 988, 1097 and 1153 GHz
lines should be added to retrieve physical parameters of individual
velocity components.

\section{Conclusions}

\label{sec:conclusions}

Here we summarize in abbreviated form the main conclusions that water
observations with {\it Herschel} have taught us on the physics and
chemistry in star-forming regions, and their evolution from cloud to
disk.  We also reiterate some of the broader implications of the WISH+
program, for example for the interpretation of extragalactic data on
water.  Key points are that {\it Herschel} has revealed new physical
components, has tested and confirmed basic chemical networks, and has
pointed the way for the interpretation of subsequent (ground-based)
data of other molecules, including the various transitions of CO
itself, both in Galactic and extragalactic sources.

\subsection{Water line profiles and the physical components they
  trace}

\begin{itemize}

\item {\bf Water traces active star formation sites:} the bulk of the
  observed gaseous water emission with {\it Herschel} in star-forming
  regions is associated with warm kinetically heated gas in outflows
  and shocks that trace gas of several hundred K. This same gas is
  traced by CO lines with $J_{\rm up}>14$, but not by lower-$J$ CO
  lines.

\item {\bf Water spectral profiles are complex:} at least two
  different physical components are universally seen in water and
  high-$J$ CO lines for low-mass sources: a broad component (cavity
  shock, disk wind, turbulent entrainment) with
  $T_{\rm rot}$(CO)$\approx 300$ K, and a medium-broad offset
  (slightly blue-shifted) dissociative $J-$type `spot shock' with
  $T_{\rm rot}$(CO)$\approx 700$ K. The latter category includes
  extremely high velocity features with velocities out to $\pm$100 km
  s$^{-1}$ seen in a small fraction of sources. For higher-mass
  sources, only the broad component is seen.  Very few cloud positions
  show narrow (FWHM$<$few km s$^{-1}$) water emission lines.

\item {\bf Water profiles probe small motions:} inverse P-Cygni
  profiles due to infall are seen in a fraction of protostellar
  sources, indicating mass infall rates from cloud onto envelope of
  $10^{-5}-10^{-4}$ M$_\odot$ yr$^{-1}$ for low-mass protostars, and
  increasing up to $10^{-4}-10^{-2}$ M$_\odot$ yr$^{-1}$ for high-mass
  sources. A slightly smaller fraction of sources show
  regular P-Cygni profiles indicating expansion.
  
\item {\bf Water emission is compact:} for most low-mass sources, the
  water emission, and thus the mechanisms that produce it, is limited
  to $\sim$1000 au radius from the central source, with emitting areas
  even smaller. Even for high-mass sources such as Orion, bright water
  emission is compact and limited to $<$25000 au ($<$0.1 pc)
  scale. For cases where water emission is observed off source, the
  emitting areas are larger as the warm gas can expand in more
  directions.
  
\item {\bf Water traces warm, dense gas:} the inferred physical
  conditions in the water emitting gas are high density ($>10^5$
  cm$^{-3}$) and kinetic temperature (300--1000 K) with small emitting
  areas on source, of order 100 au for low-mass protostars and 1000 au
  for high-mass protostars. {\it Herschel} was not sensitive to a
  possible very hot component of several thousand K.

\item {\bf Water points to UV-irradiated outflow cavities and shocks:}
  the data on H$_2$O, OH and other hydrides indicate UV fields up to
  $10^2-10^3$ times the general interstellar radiation field in both
  low- and high-mass sources in outflow cavity walls on scales of the
  {\it Herschel} beam ($\sim 20''$). A new class of UV-irradiated
  outflows or shocks is required to explain the data.

\item {\bf Water is a significant but not dominant coolant:} Far-infrared
  line cooling in low-mass protostars is dominated by CO and H$_2$O in
  the earliest low-mass Class 0 and I stages, with [O I] becoming
  relatively more important in the later stage. The total H$_2$O line
  cooling does not change from Class 0 to Class I, whereas that of CO
  decreases by a factor of 2.  The absolute [O I] cooling is similar
  from Class 0 to Class II, but its fraction increases as the jet
  changes from being mostly molecular to being primarily atomic. For
  high-mass sources, line cooling is dominated by CO, with a much
  lower fraction of H$_2$O and OH cooling than for low-mass sources.

\item {\bf Water in extragalactic sources traces buried star formation
  activity}, originating from scales much smaller than entire molecular
  clouds.

\end{itemize}

\subsection{Water abundances in different components}

\begin{itemize}

\item {\bf Water abundance in outflows and shocks is low:} The water
  abundance in warm outflows and shocks is universally found to be
  low, only $\sim 10^{-6}$, much less than the H$_2$O/H$_2$ abundance
  of $4 \times 10^{-4}$ expected if all volatile oxygen is driven into
  water. Only very hot gas ($>1000$~K) may have water abundances close
  to $10^{-4}$. Simplified chemical models confirm that such high
  temperatures are needed to drive the bulk of the oxygen into water
  in dense gas with high UV fields.

\item {\bf High temperature routes to water confirmed:} comparison
  with line profiles of other ice species such as CH$_3$OH and NH$_3$
  shows that ice sputtering is only significant at low velocities (out
  to $\pm$ 10 km s$^{-1}$). High temperature gas-phase formation
  dominates water production at high velocities.

\item {\bf Water abundance profiles in cold cores retrieved:} velocity
  resolved HIFI line profiles have allowed the water abundance to be derived
  as a function of position across pre-stellar cores and 
  protostellar envelopes, even if the sources are not spatially
  resolved or mapped.
  A relatively high gaseous water
  abundance of $\sim 10^{-7}$ is found in the outer layers,
  decreasing roughly inversely with density deeper into the core.

\item {\bf Simplified water chemistry networks explain data:} small gas-grain
  chemical models including freeze-out of O, formation of H$_2$O ice,
  (photo-)desorption of ice, and photodissociation of gaseous H$_2$O,
  work well to explain water profiles in a wide range of low-mass
  sources. The best fits imply a range of the external radiation field
  $G_{\rm ISRF}$, cosmic-ray induced field $G_{\rm CR}$ and
  pre-stellar core lifetimes.

\item {\bf Water ice forms early:} Water ice is primarily formed in the cold
  pre-stellar stage. Some small enhancement (but no destruction) is
  possible in the cold outer envelopes up to $\sim 15-20$~K.

\item {\bf Bulk HDO/H$_2$O ice is lower than cold HDO/H$_2$O gas:}
  HDO ground-state line profiles obtained with {\it Herschel} are well
  fitted by a constant HDO/H$_2$O abundance ratio of $\sim$0.025 in
  cold protostellar envelopes. This value reflects the high HDO/H$_2$O
  abundance ratio in cold gas and in the outermost layers of the water
  ice which are enriched in deuterium, not that of the bulk of the
  ice. Indeed, observed HDO/H$_2$O values in hot cores, where the bulk
  ices sublimate, are up to an order of magnitude lower. A high
  D$_2$O/HDO ratio compared with HDO/H$_2$O is a signature of cold gas
  and ice chemistry in the dense pre-stellar phase.
  
  \item {\bf NH$_3$ does not follow H$_2$O:}
  In stark contrast with H$_2$O and HDO, the gaseous NH$_3$
  abundance profile inferred from {\it Herschel} data is flat with
  radius throughout pre-stellar cores and protostellar envelopes. This
  strengthens the conclusion that the bulk of the gaseous NH$_3$ does
  not result from photodesorption of ammonia ice, but from cold
  gas-phase processes.

\item {\bf Inner hot cores are dry:} inner hot core abundances derived
  with a step-function retrieval analysis show a large variation from
  $10^{-6}$ to a few$\times 10^{-4}$ for low- and high-mass sources,
  with only a few sources at the upper end of this range. This suggests
  that most hot cores are ``dry'', although not as dry as originally
  thought in analyses in which the small scale physical structure is
  ignored. {\it Herschel} data are not well suited to constrain hot
  core abundances due to high optical depths in THz continuum and
  lines. ALMA and NOEMA can make major advances on this topic.

\item {\bf Similarity of low- and high-mass sources:} all conclusions on
  abundances and chemistry hold across the luminosity range.
  
\item {\bf Puzzling oxygen budget strengthens evidence for UDO:} the
  combined analysis of water gas, water ice and O$_2$ limits in
  cold clouds indicates that
  a large fraction of the oxygen budget is unaccounted for.
  Within the simple water chemistry models, the only solution is to
  have a short pre-stellar stage of only 0.1 Myr to prevent all O
  being turned into H$_2$O. An alternative option is for dense cores
  to have a small fraction of large ($>$1 $\mu$m) grains which hide
  more than 50\% of the water ice from being observed through infrared
  absorption. This solution does not apply to hot cores and shocks,
  where the large icy grains should have sublimated and where a large
  fraction of oxygen is also missing. Another option is therefore that
  oxygen is in some refractory form called UDO, whose abundance
  increases from diffuse to dense clouds, and which consists of
  material that does not vaporize or atomize even in strong shocks up
  to 1000 K.

\item {\bf The water ice reservoir in disks is spatially confined and
    encompasses less of the disk than expected:} the weak water vapor
  emission from protoplanetary disks suggests either radial
  confinement of large icy grains in the inner disk or in several
  radial rings, or vertical settling of icy grains to the midplane
  below heights where UV photons can penetrate for
  photodesorption. This finding suggests that low gas-phase oxygen
  abundances in outer disks are common.
  
\end{itemize}

\subsection{Water from clouds to disks}

In conclusion, what have we learned about water in star-forming regions?

\begin{itemize}

\item {\bf Most water forms prior to star formation:} most water is
  formed as ice in dense molecular clouds before they collapse to form
  stars. However, not all oxygen may be turned into water ice,
  possibly implying a short pre-stellar phase {
    (\S~\ref{sec:prestellar}, \ref{sec:protocold})}.

\item {\bf Water formed through high-$T$ chemistry is mainly lost to
    space:} significant amounts of water are also produced or reformed
  by high temperature chemistry in warm outflows or shocks
  { (\S~\ref{sec:shocks})}. However, most of this water is lost to space
  in outflows and does not contribute to the water reservoir in
  planet-forming disks.

\item {\bf Water is mostly transported as ice during protostellar
    collapse and infall:} water enters the (forming) disk at large
  radii, with no observational evidence yet that it is affected by
  accretion shocks { (\S~\ref{sec:youngdisks})}.

\item {\bf Water locked up early in large grains in outer disks:} the
  water vapor content in the outer regions of protoplanetary disks
  indicates that the reservoir of available water ice is smaller than
  expected, being either radially or vertically confined or both.  The
  weakness of the HIFI lines implies that ice-coated grains grow
  quickly, settle to the midplane and drift inward when grown to mm/cm
  sizes. This process likely starts already in embedded phase. It also implies
  that disks with a low gas-phase oxygen abundance are common, in line
  with the findings of low volatile carbon abundances
  { (\S~\ref{sec:protoplanetary})}.

\item {\bf Inner disk water reservoir still to be probed:} warm water
  in the disk surface layers has been observed by {\it Herschel} and
  {\it Spitzer} but the inner disk midplane inside the water iceline
  is still invisible { (\S~\ref{sec:innerdisks})}.

\item {\bf New planetary systems are likely to be born with sufficient
    water to become habitable:} the {\it Herschel} data have shown
  that water gas and ice are commonly associated with star-forming
  regions and that this conclusion is independent of ``environment'' or
  location in our Galaxy. A key requirement is that the cloud out of
  which the star and disk are formed is cold enough ($\lesssim 25$~K)
  to enable water ice formation and has a long enough lifetime prior
  to collapse ($\gtrsim$0.1 Myr) to convert a significant fraction of atomic
  oxygen to water ice { (\S~\ref{sec:protocold}). Various loss and
    gain factors en route from cloud to disk and planet-forming
    sites then control the ultimate outcome of the water supply to a
    nascent planet (\S~\ref{sec:disktransport}).}

\end{itemize}

In conclusion, {\it Herschel} has shown that water has lived up to its
reputation of being a particularly interesting molecule for studying
the physics and chemistry of star-forming regions. The large abundance
changes of gas-phase water between cold and warm regions --now fully
quantified-- as well as the sensitivity of its line profiles to small
motions, make water a unique diagnostic probe among the suite of
interstellar molecules. {\it Herschel} has left a legacy for the
analysis and interpretation of future water observations in Galactic
and extragalactic sources. Also, even if {\it Herschel} has found that
not all available oxygen is locked up in water gas or ice, this does
not diminish its importance as a key ingredient for habitability on
other planetary systems: water is found to be abundantly present in
star- and planet-forming regions.

\begin{acknowledgements}

  The authors would like to thank all WISH team members over the years
  for their seminal contributions to this project, as well as the
  entire HIFI science consortium for two wonderful decades of working
  together to make this happen. They are particularly grateful to
  Malcolm Walmsley, who stimulated the WISH program from the very
  beginning and helped with numerous projects. He left a great
  legacy. Fruitful collaborations with the DIGIT and WILL teams, and
  with members of the HDO team, are also acknowledged. Detailed
  discussions with Kathrin Altwegg and Martin Rubin on the oxygen
  budget in comets and with Bruce Draine on interstellar clouds are
  appreciated. Constructive comments from the referee have helped to
  improve the paper.

  A big salute goes to the HIFI and PACS instrument teams and to ESA
  for designing, building and operating these two powerful instruments
  and the {\it Herschel} Space Observatory.  {\it Herschel} was an ESA
  space observatory with science instruments provided by the
  European-led Principal Investigator consortia and with important
  participation from NASA.
  
  HIFI was designed and built by a consortium of institutes and
  university de-partments from across Europe, Canada and the US under
  the leadership of SRON Netherlands Institute for Space Research,
  Groningen, The Netherlands with major contributions from Germany,
  France and the US. Consortium members are: Canada: CSA, U.\ Waterloo;
  France: CESR, LAB, LERMA, IRAM; Germany: KOSMA, MPIfR, MPS; Ireland,
  NUI Maynooth; Italy: ASI, IFSI-INAF, Arcetri-INAF; Netherlands:
  SRON, TUD; Poland: CAMK, CBK; Spain: Observatorio Astron\'omico
  Nacional (IGN), Centro de Astrobiolog{\i}a (CSIC-INTA);
  Sweden: Chalmers University of Technology - MC2, RSS \& GARD, Onsala
  Space Observatory, Swedish National Space Board, Stockholm University
  - Stockholm Observatory; Switzerland: ETH Z\"urich, FHNW; USA:
  Caltech, JPL, NHSC.

  PACS has been developed by a consortium of institutes led by MPE
  (Germany) and including UVIE (Austria); KUL, CSL, IMEC (Belgium);
  CEA, OAMP (France); MPIA (Germany); IFSI, OAP/OAT, OAA/CAISMI, LENS,
  SISSA (Italy); IAC (Spain). This development has been supported by
  the funding agencies BMVIT (Austria), ESA-PRODEX (Belgium), CEA/CNES
  (France), DLR (Germany), ASI (Italy), and CICYT/MCYT (Spain)
  
  Astrochemistry in Leiden is supported by the Netherlands Research
  School for Astronomy (NOVA). JRG thanks the Spanish MICIU for
  funding support under grants AYA2017-85111-P and
  PID2019-106110GB-I00.  Part of this research was carried out at the
  Jet Propulsion Laboratory, California Institute of Technology, under
  a contract with NASA. DF acknowledges financial support from the
  Italian Ministry of Education, Universities and Research, project
  SIR (RBSI14ZRHR) as well as project PRIN-INAF-MAIN-STREAM 2017.  AK
  acknowledges support from the Polish National Science Center grant
  2016/21/D/ST9/01098 and the First TEAM grant of the Foundation for
  Polish Science No.\ POIR.04.04.00-00-5D21/18-00.

\end{acknowledgements}




\newpage

\begin{appendix}

  \section{Overview of Herschel programs}
\label{app:programs}
Table~\ref{tab:programs} summarizes the various subprograms of the
WISH key program as well as related {\it Herschel} open time programs
whose data have been used in the analysis.

     \begin{table*}
      \caption[]{WISH key program and related OT programs}
{\footnotesize
         \label{tab:programs}
         \begin{tabular}{l l l r r l l }
            \hline 
            \hline
            \noalign{\smallskip}
 Program &  Subprogram & (co-)PIs & Hours & $\#$ sources 
    & Main char. \\
            \noalign{\smallskip}
            \hline
            \noalign{\smallskip}
 WISH &  & E.F. van Dishoeck & 499 && Mostly HIFI lines  \\
      &  &                   &     &                      
      & PACS lines + selected maps & \\
      & Pre-stellar cores  & P. Caselli &&2  &  \\
      & LM protostars & L. Kristensen/J. Mottram 
          &&29  &  \\
      & Outflows & R.\ Lisau/B. Nisini/M. Tafalla &&26&   \\
      & IM protostars & M. Fich/D. Johnstone &&6  &  \\
      & HM protostars & F. van der Tak/F. Herpin/F. Wyrowski &&24  &  \\
      & Disks & M. Hogerheijde/E. Bergin &&12  &  \\
      & Radiation diag. & A. Benz &&12  &  \\[7pt]
 WILL &  & E.F. van Dishoeck/J. Mottram & 134 & 49 &   
    LM selected HIFI+PACS lines & \\
 DIGIT &  & N.J. Evans & 250 & 30 &  LM full PACS scans  \\
 COPS-HIFI &  & L. Kristensen & 14 & 24  & LM HIFI CO 16-15 / OH \\
 LM-deep   &  & R. Visser & 20 & 5 &  Deep HIFI excited H$_2^{18}$O\\
 Outflows & & B. Nisini & 54 & 5 & LM PACS [O I], CO, H$_2$O maps\\
 Water maps & & R.\ Liseau & 19 & 6 & HIFI H$_2$O maps + PACS [O I] \\         
 IM Cygnus & & S. Bontemps & & &HIFI selected lines \\
HM ATLASGAL & & F. Wyrowski & & & HIFI selected lines \\
 Disks & & M. Hogerheijde & 135 &4 &  Very deep HIFI \\ 
 WatCH    & & S. Wampfler & 19 & 8 & HIFI HCO$^+$, OH, H$_3$O$^+$\\
            \noalign{\smallskip}
            \hline
         \end{tabular}
}
   \end{table*}

\section{Observational details and data reduction}
\label{app:reduction}

\subsubsection{HIFI}

HIFI consisted of a set of seven single-pixel dual-sideband heterodyne
receivers \citep{deGraauw10}.  All observations were taken in both
horizontal and vertical polarizations with simultaneously the Wide
Band Spectrometer (WBS) and High Resolution Spectrometer (HRS)
backends providing both wideband (WBS, 4 GHz bandwidth at 1.1MHz
resolution) and high-resolution (HRS, typically 230MHz bandwidth at
250 kHz resolution) frequency coverage. The sideband ratio is
approximately unity \citep{Roelfsema12}.

Observations were taken as single pointings in dual-beam-switch (DBS)
mode with a chop throw of 3$'$ using fast chopping. The only exception
are some of the H$_2$O $1_{10}-1_{01}$ observations of LM sources,
which were taken in position-switch mode to an emission-free position
\citep[see][for details]{Kristensen12}. The {\it Herschel}-HIFI beam
ranges from $12.7''$ to $38.7''$ over the frequency range of the
various water lines, set by the diffraction limit of the primary
mirror.

The data were reduced within the {\it Herschel Interactive Processing
  Environment} (HIPE; \citealt{Ott10,Shipman17}).  After initial
spectra production, first the instrumental standing waves were removed
where required. This is especially important at the highest
frequencies, see \citet{Kristensen17b} for details. This was followed
by baseline subtraction with a low-order ($\leq$2 in general)
polynomial in each sub-band. The fit to the baseline was then used to
calculate the continuum level, taking into account the dual-sideband
nature of the HIFI detectors. The initial continuum level is the
combination of emission in both the upper and lower sidebands, which
are assumed to be equal, so the final continuum level is half this
value.  Subsequently the WBS sub-bands were stitched together into a
continuous spectrum and all data were converted to the $T_{\rm MB}$
scale using efficiencies from \citet{Roelfsema12}.
Finally, all data were resampled to typically $\sim$0.3 km s$^{-1}$ on
the same velocity grid and either imported into CLASS or as FITS files
into custom-made python routines for further analysis.

Comparison of the two polarizations for each source revealed
insignificant differences in line shape or gain in most cases
(although occasional differences up to 30\% were seen, especially at
highest frequencies), so these were co-added to reduce the
noise. Comparison of peak and integrated intensities between the
original WISH observations and those obtained as part of open time
programs for the same source indicate that the calibration uncertainty
is $<10$\%.  The velocity calibration is better than 100 kHz, while
the pointing uncertainty is better than 2$''$ \citep{Roelfsema12}.
The beam positions of the H and V polarizations were, however, offset
by up to 6$''$ at the lowest frequencies, so although the overall
pointing accuracy was good, it did not mean that HIFI was always
perfectly pointed on-source. For specific peculiarities of individual
sources see discussion in
\citet{Kristensen12,Mottram14,vanderTak13,vanderTak19}.  The HIFI OTF
data reduction is described in \citet{Jacq16}.

\subsubsection{PACS}

The PACS integral field unit (IFU) spectroscopy mode was used
consisting of two photoconductor arrays with 16$\times$25 pixels,
allowing simultaneous observations in the red 1st order grating
(102--210 $\mu$m) and one of the 2nd or 3rd order blue gratings
(51--73 or 71--105 $\mu$m) \citep{Poglitsch10}. The IFU had
5$\times 5$ spatial pixels (called ``spaxels'') of $9.4''\times 9.4''$
each, which covered a 47$'' \times 47''$ field of view. It is
important to note that the {\it Herschel} beam changes with wavelength
whereas the spaxel size stays constant, so that corrections for the
wavelength-dependent loss of radiation need to be applied in case of a
well-centered point source.

The spectral resolving power $R$ increases with wavelength from about
1000 to 2000 (corresponding to velocity resolutions of 140 to 320 km
s$^{-1}$) in 1st order and from about 1500 to 3000 (100 to 210 km
s$^{-1}$) in 2nd order. At the shortest wavelengths, the velocity
resolution is as good as 75 km s$^{-1}$, so the [O I] 63 $\mu$m line
sometimes shows velocity-resolved line profiles
\citep{vanKempen10,Nisini15,Karska18}.

Most WISH observations obtained single footprint spectral maps in IFU
line-scan mode in which deep observations are obtained for a narrow
wavelength region around selected transitions (bandwidth
$\Delta \lambda=0.5-2$ $\mu$m).  All observations used a
chopping/nodding observing mode with off-positions within 6$'$ of the
target coordinates to subtract the background emission.

Data reduction was performed with HIPE including spectral
flat-fielding \citep[see][for full
details]{Herczeg12,Green13,Green16,Karska18}. The flux density was
normalized to the telescopic background and calibrated using
observations of Neptune.  The overall calibration uncertainty in flux
densities is estimated to be $\sim$20\% from cross comparisons of
sources observed in multiple programs or modes
\citep[e.g.,][]{Karska18}. 1D spectra were obtained by summing over a
number of spaxels chosen after inspection of the 2D spectral maps
\citep{Karska13}.  For sources with extended line emission, the
co-addition of spaxels with detected emission increases the $S / N$,
smooths the continuum, and enables correction for significant
differences in diffraction-limited beam sizes over the wide spectral
range covered by PACS.  For sources with point-like emission, only the
central spaxel was used, multiplied by wavelength-dependent
instrumental correction factors to account for the different point
source functions (PSF) (see PACS Observers Manual).

The PACS full range spectroscopy mode, in which the entire
far-infrared spectrum from 50 to 210 $\mu$m is obtained, was used for
a few WISH sources \citep{Herczeg12,Goicoechea12} and for all DIGIT
sources (which partially overlap with the WISH sample, 8 sources in
common) \citep{Green13}.  However, the spectral sampling within a
resolution element is about 3--4 times coarser in the full spectral
scan than in line spectroscopy mode, and integrations are generally
less deep.  Figure~\ref{fig:pacs_scans} shows part of the NGC 1333
IRAS4A spectral scan compared with that of the neighboring 4B source
\citep{Herczeg12}, with the latter clearly richer in
lines. Figure~\ref{fig:pacscomp} illustrates the importance of spectral
sampling for a few PACS lines observed in both line and range
spectroscopy modes.

\section{Additional water spectra}
\label{app:additionalspectra}

\begin{figure*}[t]
     \centering
    \includegraphics[width=16cm]{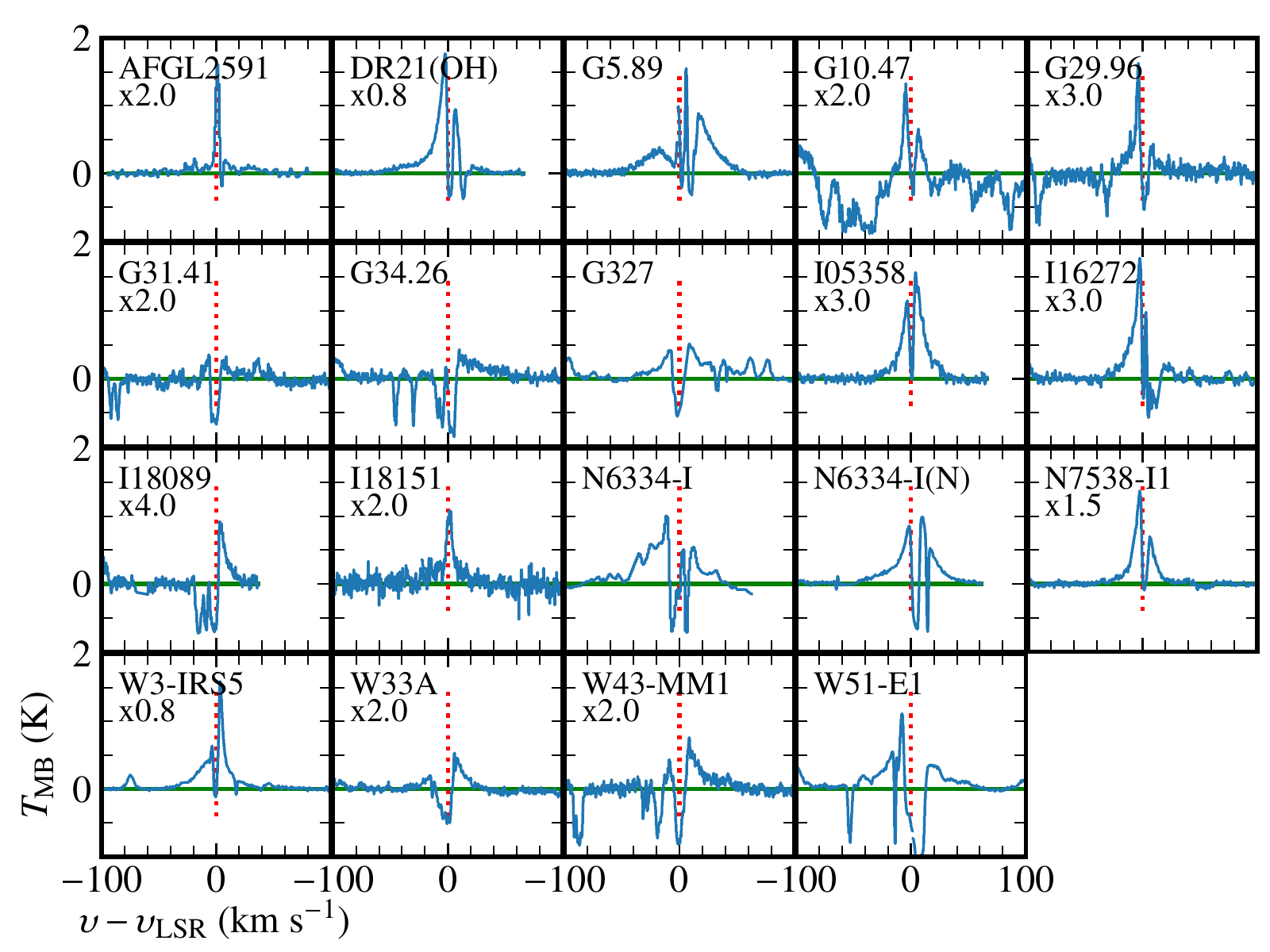}
    \caption{H$_2$O $1_{11}-0_{00}$ spectra toward all high-mass
      sources in the WISH program.}
         \label{fig:hm1113}
\end{figure*}

Figure~\ref{fig:hm1113} presents the p-H$_2$O $1_{11}-0_{00}$
spectra of all high-mass sources, illustrating the multitude of
foreground clouds along the lines of sight.

\begin{figure*}[tb]
     \centering
    \includegraphics[width=12cm]{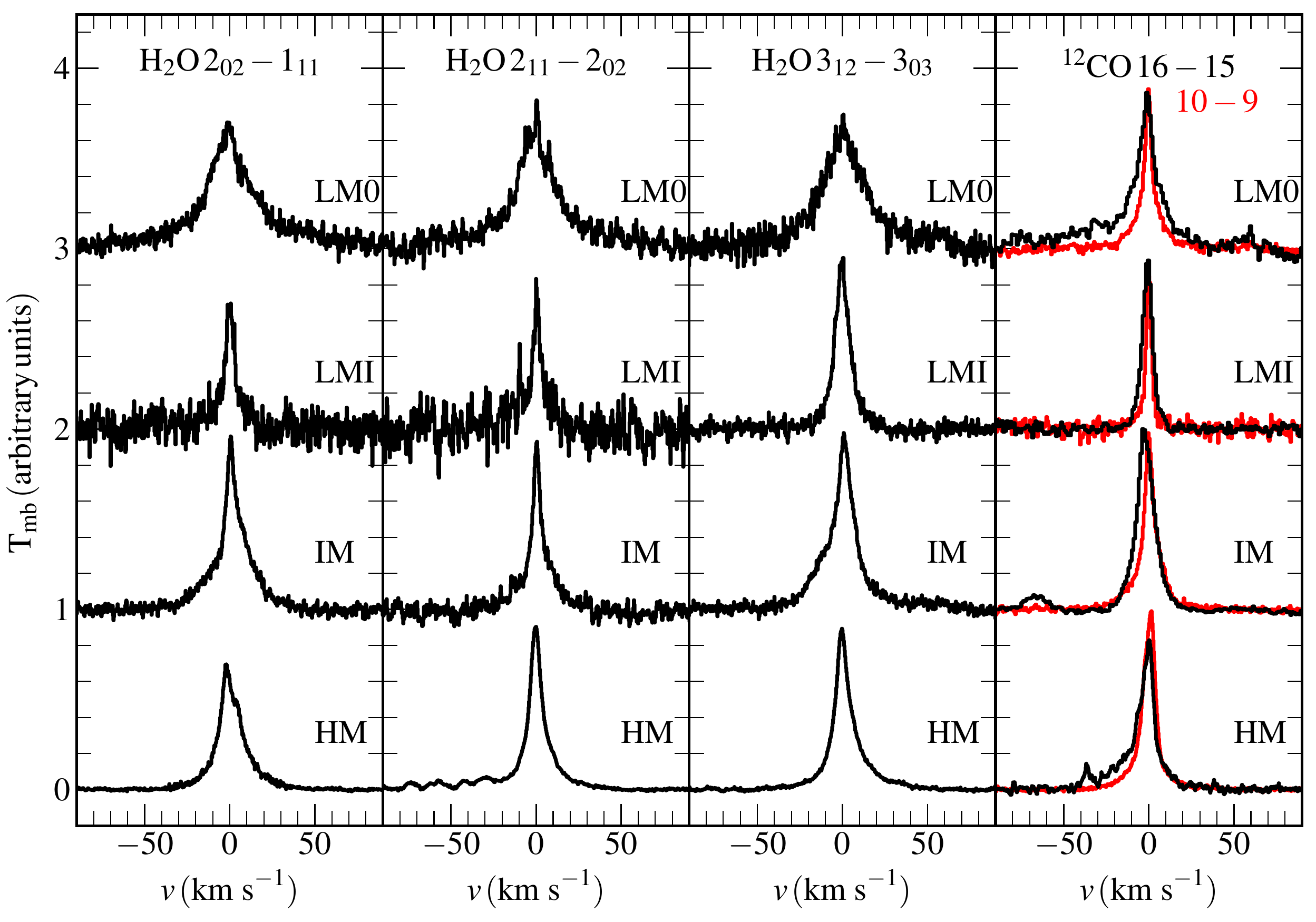}
    \caption{Averaged and normalized mid-$J$ water and CO spectra for
      low-mass Class 0 (LM0), low-mass Class I (LMI), intermediate
      mass (IM) and high mass (HM) sources. From left to right the
      H$_2$O $2_{02}-1_{11}$ line at 988 GHz, $2_{11}-2_{02}$ line at
      752 GHz, the $3_{12}-3_{03}$ line at 1097 GHz, and the CO
      $J$=16--15 together with the 10--9 profiles are presented. All
      spectra have been shifted to zero velocity. The low intensity
      features on the blue wing of the 752 GHz HM spectrum are due to
      methanol emission. Data from
      \citet{SanJose13,SanJose16,Kristensen17b}}
         \label{fig:averagespectra}
\end{figure*}

Figure~\ref{fig:averagespectra} presents the averaged and normalized
water spectra and high-$J$ CO spectra for each class of sources, from
low- to high mass. The water spectra arise from medium-$J$ levels that
do not connect with the ground state so that no absorption features
are present. The similarity in profiles within each class of sources
and between classes of sources is clearly seen, with only the low-mass
Class I profiles narrower than the others.

\begin{figure}[tb]
     \centering
    \includegraphics[width=7cm]{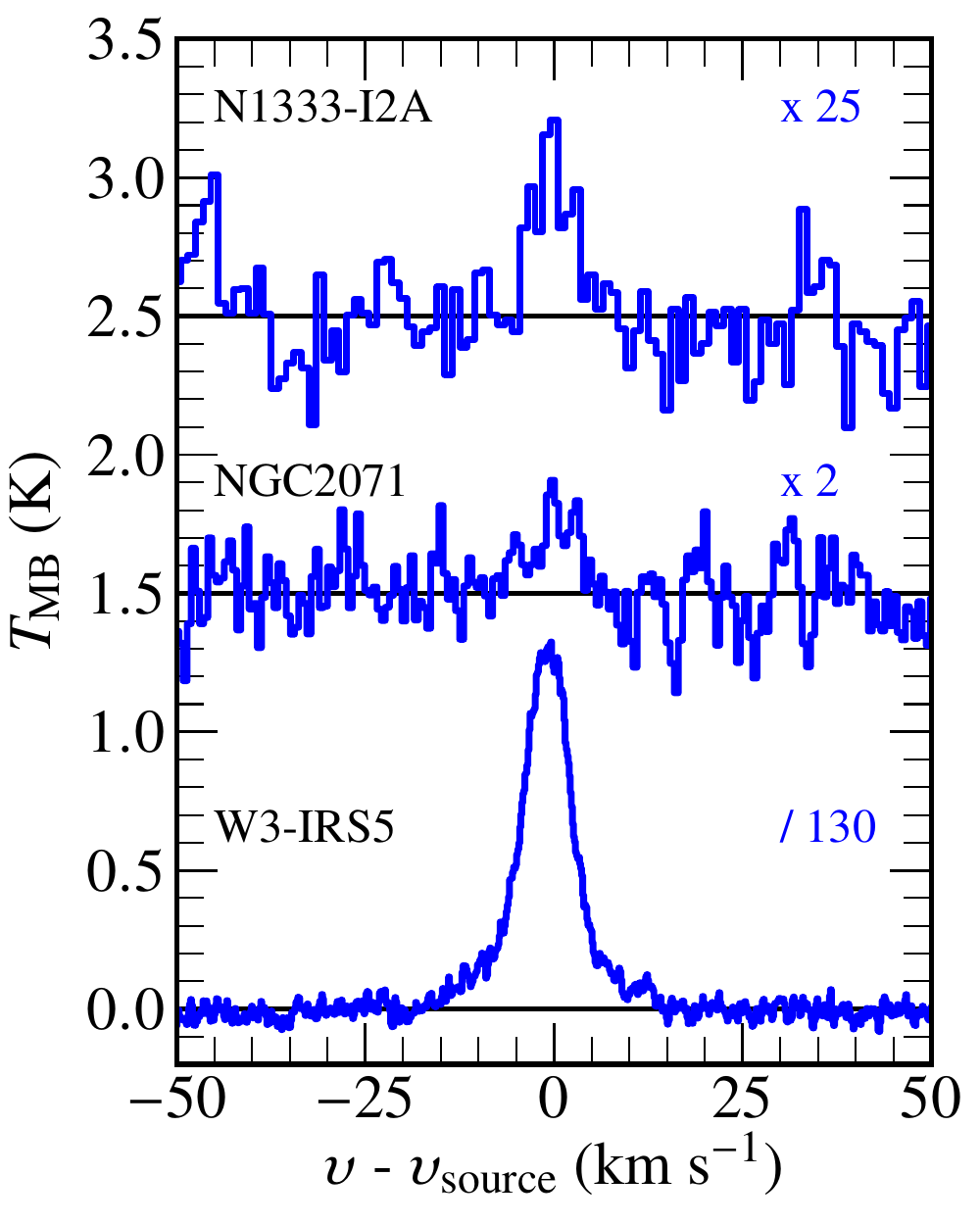}
    \caption{Deep H$_2^{18}$O $3_{12}-3_{03}$ spectra obtained toward
      a low-mass, intermediate-mass and high-mass protostar (from top
      to bottom) \citep[based on][]{Visser13}.}
         \label{fig:visserspectra}
\end{figure}

Figure~\ref{fig:visserspectra} presents deep integrations of the excited
$3_{12} - 3_{03}$ H$_2^{18}$O transition ($E_{\rm up}$ = 250 K) toward
low-, intermediate-, and high-mass sources. The line is barely
detected toward the low- and intermediate-mass sources, as opposed to
the high-mass source where it is prominent. However, in spite of the
difference in $S/N$, there is no indication that the line emission
traces different components toward the different sources: it probes
warm water in the inner envelope on small scales.

\section{Water chemistry: three routes}
\label{app:waterroutes}

\subsection{Water and related species}

The chemical reactions that form water in interstellar space have been
well described in various papers, details of which will not be
repeated here \citep[see recent overviews
in][]{Hollenbach09,vanDishoeck13,Lamberts13}. References to papers are
mostly those that have appeared since the 2013 reviews. Three main
routes can be distinguished, which are summarized in
Fig.~\ref{fig:waternetwork}: (i) gas-phase ion-molecule chemistry,
driven by cosmic ray ionization of H and H$_2$; (ii) high temperature
gas-phase neutral-neutral chemistry initiated by the O + H$_2$
reaction; and (iii) ice chemistry.

Route (i) dominates the formation of water in cold low-density diffuse
and translucent clouds with visual extinctions of a few
magnitudes. {\it Herschel}-HIFI has probed this chemistry in detail
through far-infrared absorption line observations of H$_2$O as well as
intermediates OH$^+$ and H$_2$O$^+$ as part of the PRISMAS key program
\citep{Flagey13,Gerin16}. Typical observed H$_2$O abundances are low,
only $(0.5-1.5)\times 10^{-8}$ with respect to total hydrogen, which
are well reproduced by models such as those by \citet{Hollenbach09}
within factors of 2. In the WISH sample, the intermediate species
OH$^+$, H$_2$O$^+$ and H$_3$O$^+$ have been detected together with
water in foreground clouds toward many protostars
\citep[e.g.,][]{Benz10,Wyrowski10,Benz16}. Analysis of these WISH +
PRISMAS data have resulted in a mean value of the cosmic ray
ionization rate of $\zeta_{\rm H}=1.8\times 10^{-16}$ s$^{-1}$ in
diffuse clouds \citep{Indriolo15}.

Route (ii) is much more relevant for star-forming regions, where
temperatures close to protostars and in winds can be up to a few
hundred K and in shocks as high as a few thousand K. The reaction O +
H$_2$ $\to$ OH + H is endothermic, whereas the subsequent reaction OH
+ H$_2$ $\to$ H$_2$O + H is exothermic but has a significant energy
barrier of $\sim$2100 K \citep{Baulch92}. The rates for these
reactions have been critically discussed by \citet{vanDishoeck13} and
start to dominate water formation at temperatures above $\sim$250
K. Water destruction can occur by the reverse reaction with atomic H,
but this reaction has a very high barrier of $\sim 10^4$ K and is
unlikely anyhow in dense molecular gas because of the low abundance of
atomic H. This leaves UV photons as the main route for water
destruction, with X-rays only effective in regions of high
extinction. Photodissociation of H$_2$O starts to be effective
shortward of 1800 \AA\ and continues down to the ionization threshold
at 983 \AA\ (12.61 eV), including Ly $\alpha$ at 1216 \AA. Its
lifetime in the general interstellar radiation field, as given by
\citet{Draine78}, is only 40 yr \citep{Heays17}. Close to
(proto)stars, the UV radiation field is enhanced by many orders of
magnitude, making the water vapor lifetime even shorter.

Route (iii) dominates the formation of water in cold dense clouds, and
is in fact the route that produces the bulk of water in space
\citep[][for review]{Bergin00,vanDishoeck13} since freeze-out of
gaseous water is insufficient \citep{Lee96b}. The timescale for an
atom or molecule to collide with a grain and stick to it is
$t_{\rm freeze}=3\times 10^9/n({\rm H_2})$ yr for normal 0.1 $\mu$m
size grains and sticking probabilities close to unity. Since the
freeze-out time scales as $1/n$ whereas the free-fall time of a cloud
(which is a lower limit to its lifetime) scales as $1/\sqrt{n}$, the
former becomes shorter than the latter for densities $>10^4$ cm$^{-3}$
so most species will end up on the grains
\citep[e.g.,][]{Bergin07}. The grain temperature plays a key role in
the ability to retain species on grains and for surface reactions to
occur. Light species such as H and H$_2$ are weakly bound so they have a
high mobility and short residence time on the grains. In fact, for
grain temperatures above $\sim$20 K, this residence time is so short
that hydrogenation ceases to be effective, except through the direct
Eley-Rideal mechanism. Heavier species such as O and OH are more strongly
bound and less mobile. These rates are usually parametrized through
binding energies $E_b$ and diffusion energies $E_d$, with the latter
often taken to be a fraction of 0.3--0.7 of $E_b$. At the lowest
temperatures, tunneling reactions of H and H$_2$ can be important as
well \citep{Meisner17,Lamberts17,Cuppen17}.

\citet{Tielens82} postulated that the formation of water from O atoms
on grain surfaces proceeds through three routes involving
hydrogenation of $s-$O, $s-$O$_2$ and $s-$O$_3$, respectively, where
$s-$X indicates a species on the surface. All three routes have
subsequently been studied and quantified in various laboratories
\citep[see][for summary]{Linnartz15} (Fig.~\ref{fig:waternetwork}).

Water ice can be returned to the gas by various desorption
processes. The most effective one is thermal desorption, which occurs
on timescales of years when the dust temperature rises above $\sim$100
K (precise value depends on pressure). This leads to gas-phase
abundances of H$_2$O as high as the original ice abundances, if there
is no subsequent destruction of gaseous H$_2$O. This process dominates
the production of gaseous water in the warm inner parts of high- and
low-mass protostellar envelopes (``hot cores'' or ``hot corinos'') and
inside the snow line in disks. Water ice can also be sputtered by
energetic particles in shocks, but not all water ice may be returned
to the gas and some of it may be dissociated (see \S~\ref{sec:shocks}
for details).

At dust temperatures below the sublimation limit, photodesorption is
an effective mechanism to get water back into the gas phase, even
though only a small fraction of the UV absorptions results in
desorption of intact H$_2$O molecules \citep{Andersson08}. The
efficiency is about $10^{-3}$ per incident photon within factors of a
few, as determined through laboratory experiments and theory
\citep[e.g.,][]{Arasa15,CruzDiaz18}.  Only the top few monolayers of
the ice contribute. Thus, icy clouds exposed to radiation from a
nearby star or the general interstellar radiation field will have a
layer of enhanced water vapor at the edge, the so-called
``photodesorption layer'' \citep{Hollenbach09}.

Deep inside clouds, cosmic rays produce a low level of UV flux, $\sim
10^4$ photons cm$^{-2}$ s$^{-1}$, through interaction with H$_2$
\citep{Prasad83,Shen04,Ivlev15}.  This process turns out to be crucial for
explaining the water observations of cold dense clouds. Other
nonthermal ice desorption processes include cosmic ray induced spot
heating (which may work for CO, but is generally not efficient for
strongly bound molecules such as H$_2$O) and desorption due to the energy
liberated by the formation reaction (called ``reactive'' or ``chemical''
desorption). The latter process is less well understood and
constrained than photodesorption and may not be as effective on water ices
as on other surfaces \citep{Dulieu13,Minissale16}.

Stimulated by the discovery of abundant O$_2$ ice in comet 67P but
lack of H$_2$O$_2$ and O$_3$ ice \citep{Bieler15}, a new critical look
has been taken at the ice formation network
\citep{Taquet16o2,Eistrup18o2}.  Important parameters whose values are
still uncertain and that are found to have an impact are (i) the
diffusion to binding energy ratios $E_d$/$E_b$; (ii) the binding
energy of atomic oxygen on ice, $E_b$(O), now found to be $\sim$1600
K, a factor of 2 higher than adopted in earlier work \citep{He15};
(iii) the activation barriers $E_a$ of the reactions O + O$_2$ and H +
O$_2$. \citet{Lamberts14} conclude that the O + H$_2$ reaction is not
important for water ice formation at low temperatures.

Taken together, it is gratifying to conclude that the advent of {\it
  Herschel} has stimulated a number of laboratory and theoretical
chemical physics studies of fundamental gas-phase and solid-state
processes involved in the water network.

\subsection{CO$_2$ chemistry}

Another molecule which could potentially be a significant oxygen
carrier is CO$_2$. Its chemistry is loosely connected with that of
H$_2$O, primarily through the OH radical.

Gas-phase formation of CO$_2$ takes place mostly through the reaction
CO + OH $\to$ CO$_2$ + H, which has a slight activation barrier of 176
K \citep{Smith04}. The CO$_2$ production rate is controlled by the
availability and fate of OH, since OH can also be consumed by the
H$_2$O formation route, OH + H$_2$ $\to$ H$_2$O + H. This route has a
higher activation energy, 1740 K, than that leading to CO$_2$, so
CO$_2$ production is favored at lower temperatures. However, since
H$_2$ is orders of magnitude more abundant than CO, the formation of
H$_2$O dominates at higher temperatures, typically above 150 K
\citep{Bosman18}.  Thus, gaseous CO$_2$ formation is only effective in
a quite narrow temperature range, 50-150 K, and then only when OH is
present as well. In dense regions, this requires UV photons or X-rays
to liberate O and OH from H$_2$O or CO.

The grain-surface formation of CO$_2$ is thought to proceed primarily
through the s-CO + s-OH $\to$ s-HOCO route, with most of the s-HOCO
subsequently converted into s-CO$_2$ as found in laboratory
experiments \citep[e.g.,][]{Ioppolo11} and theoretical calculations
\citep{Arasa13}. Again the reaction rate depends on the s-OH
abundance, usually created by cosmic-ray induced photodissociation of
s-H$_2$O. Alternative proposed CO$_2$ formation routes include the
energetic processing of ice mixtures \citep[e.g.,][]{Ioppolo09} and UV
irradiation of water-ice covered hydrogenated carbon grains
\citep{Mennella06}.

The s-CO$_2$ binding energy is about 2300 K \citep{Noble12}. This
implies that the s-CO$_2$ desorption temperature (its snow line) is
in between that of H$_2$O and CO.

\section{Excitation and radiative transfer of water vapor}
\label{app:excitation}

The populations of the H$_2$O and CO energy levels are determined by
the balance between the collisional and radiative excitation and
de-excitation rates.  A major effort was undertaken by the chemical
physics community in preparation for {\it Herschel} to calculate the
collisional rate coefficients of H$_2$O and CO with H$_2$ over a wide
range of temperatures. In WISH, the latest values from
\citet{Daniel11} for H$_2$O and \citet{Yang10} and \citet{Neufeld10}
for CO have been used, as tabulated in LAMDA \citep{Schoier05}. More
recently, rate coefficients for CO + H up to high temperatures have
become available which play a role in regions with high H/H$_2$
fractions \citep{Walker15}.

The water radiative transfer is particularly complex, because the
large Einstein $A$ values, subthermal excitation, (very) high optical
depths and large numbers of lines that need to be included. However,
many lower $J$ water lines are effectively thin, a regime in which the
line brightness still scales with column density
\citep[e.g.,][]{Snell00}. This assumption holds as long as the
collision rate is so slow that every excitation immediately leads to a
radiative de-excitation and the production of one photon which escapes
the cloud, possibly after many absorptions and re-emissions, before
another excitation occurs.
In formulae, this implies $C < A/\tau$ , where $C$ is the collision
rate and $\tau$ the optical depth \citep{Linke77,Keto14}.  To properly treat
water vapor excitation, many energy levels need to be taken into
account; while this slows down the computing speed, the advantage is
that some of the higher-lying lines, especially those in the PACS
domain, become optically thin so that they better the excitation
models. Another complication is that at high frequencies, the
continuum can become optically thick as well, thus limiting the depths
for looking into the innermost part of the source
\citep{Cernicharo06,vanKempen08,Visser13}. This far-infrared dust
continuum can also effectively couple with the water excitation
\citep{Gonzalez14}.

WISH team members developed and used a number of radiative transfer
codes to analyze the {\it Herschel} data, ranging from the simple
constant density and temperature non-LTE excitation program RADEX
using a local escape probability \citep{vanderTak07} to more
sophisticated 1D nonlocal Monte-Carlo codes RATRAN
\citep{Hogerheijde00} and MOLLIE \citep{Keto04} that can handle radial
temperature and density variations as expected in cold cores and
protostellar envelopes and that also follow the transport of continuum
photons. For nonspherical geometries, a 2D version of RATRAN and the
3D code LIME \citep{Brinch10} have been used, as well as a fast 3D
code using a local source approximation by \citet{Bruderer10mod}.
Also, team members had access to the Onsala Accelerated Lambda
Iteration (ALI) code for arbitrary source structures that can treat
line overlap \citep{Maercker08}. These codes have been tested against
each other in a 2004 workshop \citep{vanderTak05}
and subsequent WISH team meetings.

\section{Modeling approaches}
\label{app:modeling}

\subsubsection{Retrieval models}
\label{app:retrieval}

The simplest analysis method is to use a non-LTE excitation and
radiative transfer code such as RADEX to derive best-fitting densities
and temperatures from observed line ratios. Also column densities and
emitting areas can be derived from the observed line intensities,
although there is often some degeneracy between parameters
\citep[see][for examples]{vanderTak10,Herczeg12,Mottram14,Santangelo14b}. This
method is most appropriate for shock components seen in on-source line
profiles and at shock positions offset from the source.

To constrain the envelope abundance, a retrieval method is commonly
used in which a trial water abundance is chosen. Usually a step
function with a low outer and high inner abundance (when $T>100$~K) is
adopted (see also \S~\ref{sec:step}). The molecular excitation and
radiative transfer in the line is then computed at each position in
the envelope, using a temperature and density profile constrained by
other data. The resulting sky brightness distribution is convolved
with the {\it Herschel} beam profile. These inner and outer abundances
are then adjusted until best agreement with observational data is
reached. The velocity structure can be a constant (turbulent)
broadening as function of position in the envelope, and/or some other
function such as infall or expansion \citep[see][for
examples]{Johnstone10,Herpin12}.

For simplicity, spherically symmetric models are typically adopted but
{\it Herschel} data show that often UV-heated cavity walls need to be
added in a 2D geometry \citep[see][for
examples]{Bruderer09b,Visser12,Lee13}

\subsubsection{Forward models}
\label{app:forward}

The forward modeling approach uses a chemical model as illustrated in
Fig.~\ref{fig:waternetwork} coupled with a physical model of the
source to compute abundances as function of position and time.  These
are then fed into excitation and radiative transfer calculations to
obtain the line intensity maps, which are finally convolved with the
beam. These chemistry models can be run either in steady-state or in a
time-dependent mode at each position in the cloud.  Best fit
parameters in this case can be any parameter fed into the chemical
network such as overall (initial) abundances or rates of
reactions. Often, the chemical evolution time (``age'') or cosmic ray
ionization rate are adjusted to obtain the best fit \citep[see][for
  examples]{Doty06,Caselli12}. As noted above, often simplified
chemistry networks are used for computational speed and to allow key
processes to be identified.

Some of the above models keep the physical structure of the source
fixed even when the chemistry evolves in time, while others let the
physical structure evolve. Also, a parcel of gas often stays fixed at
a given location in the source. An alternative approach is to follow
the chemistry of a parcel of gas as it falls in from large radii to
smaller distances and eventually enters the (forming) protostellar
disk. In the context of WISH, such 2D models have been developed and
applied to water by \citet{Visser09,Visser11} and for deuterated water
by \citet{Furuya17}.

\section{Oxygen budget}
\label{app:obudget}

This section summarizes our current understanding of the oxygen budget
in various regions of the interstellar medium that are part of the
star- and planet formation cycle. The different oxygen reservoirs are
summarized in Table~\ref{tab:obudget} and visualized in percentages in
Figure~\ref{fig:obudget} in the main text. The following subsections
provide more background information on the numbers in this table and
figure. The numbers in the text plus their uncertainties are
  generally as stated in the original papers, with added text to
  motivate the uncertainties. The number of significant digits
  therefore varies and should not be taken to reflect the actual
  uncertainties. Very small numbers have been rounded off to ($<$)1
  ppm. 

\subsection{Overall and refractory oxygen budget}

The overall [O]/[H] abundance of oxygen in all forms in the ISM is
taken to be $5.75 \times 10^{-4}$ or 575 ppm, as measured for early
B-type stars \citep{Przybilla08}. This number, denoted in short as
[O], is somewhat higher than the solar oxygen abundance of
$4.9 \times 10^{-4}$ or 490 ppm \citep{Asplund09,Grevesse10}.  A
recent compilation by \citet{Lodders19} of solar and meteoritic
abundances gives a recommended value for the present solar system of
537 ppm (their Table 8), with values ranging from 512--660 ppm.  The
solar and solar system abundances presumably reflect the ISM
composition as it was 4.6 Gyr ago, whereas the higher
\citet{Przybilla08} value is representative of hotter and therefore
younger stars. Indeed, the difference in abundances is consistent with
galactic chemical evolution models \citep{Chiappini03}. In the following
analysis, [O]=575 ppm is used, but other studies
\citep[e.g.,][]{Pontoppidan14} use the lower [O] abundance in their
analysis of the oxygen budget in protoplanetary disks, decreasing the
need for unidentified forms of oxygen by nearly 100 ppm.

Oxygen can be partitioned into gas, ice and dust. The gas and ice are
considered as ``volatiles'', the dust as ``refractories'', with the
difference related to the sublimation temperature of the material. The
boundary between the two categories is traditionally taken to be
around 600 K, so that refractories include at least all the silicates,
metals (as chemically defined), and metal oxides.  The fraction of
oxygen locked up in refractory dust, [O$_{\rm refr}$], then comes from
counting the amount of [Mg], [Si] and [Fe] in the ISM, together
denoted as [M]. Here the analysis of \citet{Whittet10} is
followed. Depending on the type of mineral (e.g., olivines or
pyroxenes or metal oxides), the oxydation ratios differ slightly. The
maximum is [O$_{\rm refr}$]/[M]=1.5, whereas the minimum is [O$_{\rm
  refr}$]/[M]=1.2. This number could be lower only if a large fraction of
Fe would be in carbonaceous grains \citep{Weingartner99}.  Using the
Mg, Si and Fe abundances of \citet{Przybilla08}, this gives [O$_{\rm
  refr}$]=90--140 ppm.  The value of [O$_{\rm r}$]=140 ppm is assumed
here.  This number does not include any oxygen locked up in refractory
organics or any other types of refractory material.

\subsection{Diffuse clouds}

Diffuse interstellar clouds with extinctions $A_V<$1 mag have the
advantage that all hydrogen atoms (in H and H$_2$) as well as the
dominant form of oxygen, neutral atomic oxygen O, can be measured
directly with high accuracy through ultraviolet spectroscopy toward
background early-type stars. In this paper, the atomic oxygen
abundance of O/[H]=$3.2\times 10^{-4}$ or 320 ppm is used, as measured
by \citet{Meyer98} with HST-GHRS for 13 lines of sight in the local
ISM. The uncertainty of $\sim$10\% is due to the spread in
measurements as well as a small uncertainty in the oscillator strength
of the optically thin O I 1356 \AA\ line.  A subsequent survey of the
1356 \AA\ line in the Galactic disk with HST-STIS by
\citet{Cartledge04} finds atomic O abundances ranging from 390 ppm for
the lowest density clouds to 284 ppm in higher density regions,
consistent with the slight decrease with density seen by
\citet{Jenkins09} in diffuse clouds.

Oxygen-containing molecules such as H$_2$O and O$_2$ have abundances in
diffuse clouds that are at least 3 to 4 orders of magnitude lower than
that of atomic oxygen \citep[e.g.,][]{Spaans98,Flagey13}. Gaseous CO
\citep{Sheffer08} and ices \citep{Poteet15} are also negligible in
diffuse molecular clouds. Thus 320 ppm is taken as the amount of
volatile oxygen, [O$_{\rm vol}$] that can cycle between gas and ice in dense
clouds.

For our adopted [O] abundance of 575 ppm, the sum of the identified
refractory and volatile oxygen is 140 + 320 = 460 ppm, leaving 115 ppm
for Unidentified Depleted Oxygen, UDO. The estimated uncertainty in
the amount of UDO is significant, of order 50\% based on the above
discussion, but some UDO still seems required.

These average numbers for diffuse clouds are consistent with the
detailed analysis of a single well-studied line of sight, the diffuse
molecular cloud toward $\zeta$ Oph with $A_V\approx 1$ mag
\citep{Poteet15}. They derive [O$_{\rm refr}$] in silicates to be
126$\pm$45 ppm, with an additional $\leq$19 ppm in iron oxides such as
Fe$_3$O$_4$. Taking the measured atomic O toward $\zeta$ Oph at
307$\pm$30 ppm \citep{Jenkins09}, together with their upper limits on
H$_2$O ice ($\leq 9-23$ ppm depending on grain size) and organic
refractory material ($<7$ ppm in the form of polyoxymethylene), they
find that as much as 98--156 ppm of oxygen could be in UDO.

\subsection{Dense clouds}

Many oxygen-bearing species in gas, ice and dust can be measured
directly toward bright infrared sources in or behind dense
clouds. Because of their rising spectrum with wavelength, protostars
embedded in their natal envelope are particularly suitable for high
$S/N$ spectra. As for diffuse clouds, the advantage of absorption line
studies is that all species are measured for the same pencil beam line
of sight. Summaries of abundances and oxygen budget toward protostars
based on data from {\it ISO}, {\it Spitzer} and ground-based
instruments are presented in \citet{Gibb04,Oberg11} and
\citet{Boogert15}.

\subsubsection{Low mass protostars}

For low-mass protostars, the bulk of the envelope is cold so most of
the oxygen budget is in ice and dust. The numbers listed in
Table~\ref{tab:obudget} are taken from the \citet{Boogert15}
compilation. The median ice abundances in their Table 2 are listed
both with respect to water ice and with respect to hydrogen. The
latter values require a determination of the total hydrogen column
$N_{\rm H}$ along the line of sight. This column is obtained either
from the measured optical depth of the silicate feature
$\tau_{9.7\mu m}$ or from the near-IR extinction $A_K$, or both, with
relations between these observables and the total hydrogen column
density benchmarked for clouds in which both can be measured
\citep{Vuong03,Boogert13}. The amount of oxygen in silicates is taken
to be the same as for diffuse clouds, 140 ppm \citep{Whittet10}.

The bulk of the oxygen in ices is in H$_2$O, CO$_2$ and CO ices. Of
these three species, CO is the most volatile one and can show
significant variation from cloud to cloud. It is the only major ice
species that is not formed on the grains but frozen out from the gas,
with an ice abundance that depends on temperature and
density. However, when the CO ice abundance is low, its gas-phase
abundance is high, so this does not affect the total volatile oxygen
budget. Taken together, the amount of oxygen measured in ices sums up to only
76 ppm, or 13\% of [O].

The gas-phase CO column toward several low-mass protostars has been
measured through high spectral resolution infrared observations of CO
and its isotopologs by \citet{Smith15}. Their inferred CO
ice/(gas+ice) abundances are generally much less than 20\% (see their
Figure 11). With CO ice at 10 ppm, this implies a gas-phase CO
abundance of at least 40 ppm. Another independent measurement of
gaseous CO in cold clouds has been obtained by observing the infrared
absorption lines of both CO and H$_2$ \citep{Lacy17}.  Their inferred
CO/H$_2$ is $1.7\times 10^{-4}$ with a variation of $\sim$30\% between
lines of sight, which would correspond to CO locking up 83 ppm with
some spread. Based on these two studies, we take gaseous CO to account
for 100 ppm with a 50\% uncertainty.

No direct observational limits exist for gaseous atomic oxygen or
other oxygen-bearing molecules, but they should be negligible
according to dark cloud models \citep{McElroy13}.  Also, both
millimeter interferometry and {\it Herschel}-WISH+ studies of
H$_2^{18}$O emission lines show that the warm water abundance in the
inner regions of low-mass protostars is generally low, much less than
40 ppm \citep[][their Figure 11]{Visser13,Persson16}, and that O$_2$
is negligible \citep{Yildiz13} (see \S~\ref{sec:o2}).

Taken together, the silicate dust, ices and gas account for only 140 +
76 + 100 + 40 = 356 ppm of oxygen at most. This leaves at least 219
ppm or 38\% of the oxygen budget unaccounted for. If all in UDO, this
would imply a doubling of its abundance compared with diffuse clouds.

\subsubsection{High-mass protostars}

The abundances measured toward high-mass protostars listed in
Table~\ref{tab:obudget} are also taken from
\citet{Boogert15}. Overall, the pattern is similar as for low-mass
protostars, except that the ice abundances are even lower. In
particular, the CO$_2$ ice abundance, which counts double for oxygen,
is a factor of 2--3 lower in high-mass regions. Taken together, the
total amount of oxygen in ices is only 47 ppm.

The envelopes around high-mass protostars are more massive and warmer
due to their higher luminosities, so gas-phase column densities are
larger than for low-mass sources. The CO abundance in dense warm gas
has been measured directly by \citet{Lacy94} for once source through
infrared absorption lines of both CO and H$_2$ to be
CO/H$_2 = 2.5 \times 10^{-4}$, corresponding to 130 ppm. We take CO to
contain 100 ppm more generally. For high-mass protostars, CO$_2$ and
H$_2$O gas have been measured directly through infrared absorption
studies with $ISO$ for the same lines of sight as for the ices.
CO$_2$ gas has been detected with typical gas/ice ratios of 0.01--0.08
so it contributes negligibly \citep{Boonman03co2,vanDishoeck98fd}. In
contrast, the H$_2$O gas/ice ratio derived from infrared absorption is
more variable, with gas/ice ratios ranging from 0.01-2.2
\citep{vanDishoeck96,vanDishoeck98fd}. With H$_2$O ice accounting for
31 ppm, this gives H$_2$O gas at most 62 ppm, about 60\% of that of
CO. This number is consistent with the measured gaseous H$_2$O column
densities relative to those of hot H$_2$ in the inner region
\citep{Boonman03h2o}, with H$_2$ derived from CO infrared absorption
lines obtained with the CFHT-FTS by \citet{Mitchell90} assuming
CO/H$_2$=$2\times 10^{-4}$. Interestingly, the corresponding ratio
H$_2$O/CO$\approx$0.6 is similar to that found by
\citet{Indriolo15exes} for some of the same lines of sight using much
higher spectral resolution H$_2$O data obtained with SOFIA-EXES and
VLT-CRIRES.  Assuming that atomic O and O$_2$ are negligible, this
gives a total of 162 ppm accounted for in the gas.

Taken together, the dust, ice and gas account for 140 + 47 + 162 = 349
ppm, leaving 226 ppm for UDO (40\% of oxygen). This is very similar to
the case for low-mass protostars, even though very different
techniques and instruments are involved for measuring the gas
abundances.

\subsection{Shocks}

Shocks have the advantage that temperatures are so high that all 
volatile oxygen should be returned to the gas phase. The refractory
silicate part is again taken to be 140 ppm, and should be viewed as an
upper limit since some silicate dust could be sputtered by the
shocks, as evidenced by high gas-phase abundances of SiO
\citep{Guilloteau92}.

The most direct measurements of the gaseous H$_2$O and CO abundances
come from {\it Herschel}-PACS emission line observations of dense
shocks at positions offset from the protostars for which {\it Spitzer}
data on H$_2$ are available \citep{Neufeld14}. There is some
degeneracy in the fit, however, so not all column densities can be
determined independently. For the shock associated with the NGC 2071
outflow, H$_2$O/CO=0.8 is derived assuming
CO/H$_2$=$3.2\times 10^{-4}$ (that is, assuming most available carbon in
CO).  This would imply that CO accounts for 160 ppm and H$_2$O for 128
ppm.

As discussed in \S~\ref{sec:shocks} and Table~\ref{tab:shocks}, {\it
  Herschel} WISH observations of H$_2$O in shocks associated with
low-mass protostars generally give low abundances, down to 1 ppm. The
best determined case is that of HH 54 for which H$_2$ is obtained from
spectrally-resolved VLT-VISIR data, finding
H$_2$O/H$_2 <1.4\times 10^{-5}$, less than 7 ppm
\citep{Santangelo14a}. Similarly, \citet{Kristensen17b} find
H$_2$O/CO$\sim$0.02 for a range of protostars based on {\it
  Herschel}-HIFI H$_2$O and CO 16-15 data. Finally, fitting both CO
{\it Herschel}-PACS and H$_2$ {\it Spitzer} data simultaneously in a
number of shocks associated with low-mass protostars gives
CO/H$_2 \sim 2\times 10^{-5}$ with values up to $10^{-4}$,
corresponding to 10--50 ppm \citep{Dionatos13,Dionatos18}.

OH has been measured in some outflows associated with low- and
high-mass sources with {\it Herschel} but remains a small contributor
to the oxygen budget \citep{Wampfler11} (see \S~\ref{sec:oh}). The
same holds for O$_2$ \citep{Melnick11} (see \S~\ref{sec:o2}).

This leaves atomic oxygen as the only other plausible gaseous oxygen
reservoir. The few existing SOFIA-GREAT data of spectrally resolved [O
I] 63 $\mu$m lines imply low abundances, however, with atomic O
accounting for at most 15\% of the volatile oxygen
\citep{Kristensen17a}. For the NGC 1333 IRAS 4A shock, atomic O is at
most 30 ppm.

In summary, for shocks studied with {\it Herschel} and taking the
maximum CO abundance at 160 ppm, as assumed by \citet{Neufeld14} for
NGC 2071, gives a total oxygen budget in silicate dust and gas of 140
+ 160 + 128 + 30 = 458 ppm, leaving 117 in unidentified form. Taking
instead CO at 100 ppm would give 140 + 100 + 80 + 30 = 350 ppm,
leaving 225 ppm for UDO. For shocks associated with low-mass
protostars, the CO and H$_2$O abundances may be up to an order of
magnitude lower. Both numbers for UDO for NGC 2071 are remarkably
similar to the range found for diffuse clouds and protostars, even
though gas and dust in shocks experience much higher
temperatures. Apparently, the UDO material is refractory enough that
it is not sublimated or sputtered under typical shock conditions with
300--700 K gas probed with {\it Herschel} data.

There is some evidence that for the hottest gas, $>$1000 K, more
oxygen is returned to the gas and driven into H$_2$O.  Although the
best fit to the {\it Herschel} data of NGC 2071 gives H$_2$O/CO$<1$,
an acceptable fit can be found with a somewhat higher temperature
$T_w$ above which H$_2$O becomes abundant (1000 K rather than 300 K),
which results in H$_2$O/CO=1.2 \citep{Neufeld14}.  For the same NGC
2071 shock, \citet{Melnick08} find
H$_2$O/H$_2$=$(0.2-6)\times 10^{-4}$ using even higher excitation
H$_2$O lines coming from hotter gas observed with {\it Spitzer}. The
upper range of this study would account for most of the volatile
oxygen and leave no room for UDO. Similarly, \citet{Goicoechea15}
identify a hotter, $\sim$2500 K component of the Orion shock in the
very high $J$ CO and H$_2$O {\it Herschel} lines with
H$_2$O/CO$\sim 1.3$ and H$_2$O/H$_2 \gtrsim 2\times 10^{-4}$ and
CO/H$_2 \gtrsim 1.5\times 10^{-4}$. More such studies using
mid-infrared lines probed with JWST are warranted.

\subsection{Comets}

The {\it Rosetta} mission to comet 67P/Churyumov-Gerasimenko has
provided a unique opportunity to determine the composition of one icy
planetesimal in exquisite detail. Depending on its formation location
and history in the solar system, a significant fraction of its
composition could be inherited from the dense cloud phase
\citep{Drozdovskaya19}. The bulk abundances of the many molecules
measured by the {\it Rosetta} instruments in ice and dust are
summarized by \citet{Rubin19} \citep[see also][]{Altwegg19}. The ice
(volatile) abundances are derived primarily from data from the ROSINA
instrument \citep{Balsiger07} whereas the dust abundances come mostly
from the COSIMA instrument \citep{Kissel07}.  There is some ongoing
discussion of the overall bulk dust : ice ratio of the
comet as a whole, with values ranging from 1:1 \citep{Lauter19} to
$\sim$3:1 (by weight)  \citep{Rotundi15}.

The numbers listed in Table~\ref{tab:obudget} are based on those in
Table~5 of \citet{Rubin19} for the 1:1 dust:ice ratio, which is most
consistent with estimates for other comets \citep{Ahearn11}. For this
case, 66\% of [O] is in ice, 34\% is in dust. Overall, the abundances
of [O] and [C] in this comet are close to solar, as measured with
respect to [Si]. However, not all of the oxygen in dust is in
silicates: in fact, COSIMA finds that nearly 50\% of the dust by
weight consists of carbonaceous material \citep{Bardyn17,Fray16}.
\citet{Bardyn17} derive [O]/[C] $\approx$ 0.3 for the organic phase,
which they note is close to the average [O]/[C] ratio of $\sim$0.2
measured in Insoluble Organic Mattter (IOM) in carbonaceous chondrites
\citep{Alexander07}, but lower than [O]/[C]=0.5 found for carbon-rich
phases in Interplanetary Dust Particles (IDPs) \citep{Flynn01}. Given
that [O]/[C]$\approx$1 in the dust phase \citep{Bardyn17}, this leads
to a make up of [O$_{\rm refr}$] of about 25\% in silicate dust and 
9\% in refractory organics, for a total of 34\%.

The comet entries in Table~\ref{tab:obudget} assume that silicate dust
is unchanged at 140 ppm. The refractory organics then account for 50
ppm. For the ice phase, the abundances of CO, CO$_2$ and other
O-containing ices (mostly O$_2$) with respect to H$_2$O are taken from
Table~2 of \citet{Rubin19}, normalized to 66\% of total [O].

By definition, there is no room for UDO in the oxygen budget in comets
as summarized in \citet{Rubin19}. The question is whether ROSINA and
COSIMA measured all possible oxygen reservoirs: some intermediate
molecular mass or dust sizes or volatility ranges may have been missed
by the instruments. In particular the separation between volatiles and
refractories at 600 K may be too strict: there is likely a continuum
of semi-volatile or less refractory species which sublimate in the
300--600 K range and which could have escaped detection. Such an
oxygen reservoir could be in the form of large organic molecules of
intermediate size. It is important to note that large refractory
organics by themselves are not the full solution to the UDO problem,
since at 50 ppm they account for at most 25-50\% of the missing oxygen
in diffuse and dense clouds.  Another option could be semi-volatile
ammonium salts such as NH$_4^+$OCN$^-$ or NH$_4^+$HCOO$^-$ which have
recently been detected in comet 67P and which appear to be responsible
for the bulk of the NH$_3$ observed in comets
\citep{Altwegg20,Poch20}. If the maximum NH$_3$ abundance with respect
to H$_2$O of 4\% measured close to the Sun is used, and if the maximum
of two oxygen atoms per nitrogen atom is taken for salts, at most 5\%
of total volatile [O] can be accounted for in salts, about 19 ppm
(Altwegg \& Rubin, priv. comm.). The entry ``O-other ice'' for comets
takes this maximum value into account.

\end{appendix}
\end{document}